\documentclass[12pt,a4paper]{book}

\usepackage[spanish,activeacute]{babel}
\usepackage{graphicx,palatino,palatcm}
\usepackage{amsmath,amssymb,amsfonts} 
\usepackage{graphics}                 
\usepackage{caption2,topcapt}
\usepackage{rotating}
\usepackage{natbib}
\usepackage{wrapfig}
\usepackage{apjfonts}
\usepackage{amssymb}
\usepackage{multicol}
\usepackage{color}
\usepackage{fancyheadings}
\renewcommand{\chaptermark}[1]%
   {\markboth{\uppercase{\thechapter.\ #1}}{}}
\renewcommand{\sectionmark}[1]%
   {\markright{\uppercase{\thesection.\ #1}}}
\newcommand{\helv}{%
   \fontfamily{phv}\fontseries{b}\fontsize{9}{11}\selectfont}
\usepackage[Leonardo]{fncychap}

\ChNameVar{\fontsize{14}{16}\usefont{OT1}{phv}{m}{n}\selectfont} 
\ChNumVar{\fontsize{60}{62}\usefont{OT1}{ptm}{m}{n}\selectfont} 
\ChTitleVar{\Huge\bfseries\rm}
\ChRuleWidth{1pt}

\renewcommand{\sectionmark}[1]{\markright{\thesection.\ #1}}
\renewcommand{\chaptermark}[1]{\markboth{Cap\'itulo   \thechapter.  {#1}}{Cap\'itulo
		   \thechapter.  {#1}}}

\addto\captionsspanish{%
  \def\chaptername{Cap\'{\i}tulo}%
  \def\appendixname{Ap\'endice}%
}

\decimalpoint

\def\teff{$T_\mathrm{eff}$ }
\def\ebv{$E(B-V)$ }

\def\mpa{  $M_{\odot}$\;a$^{-1}$}

\def\cu{c\'umulo }
\def\cus{c\'umulos }
\def\cho{{\sc\footnotesize  CHORIZOS}}
\def\s99{{\sc\footnotesize  Starburst99}}
\def\hrd{ diagrama de Hertzsprung-Russell }
\def\mbol{$M_\mathrm{bol}$ }
\def\logoh{$12 + \log(\mathrm{O/H}) $ }  
\def\ecc{$E(4405-5495)$}
\def\rv{$R_{5495}$}

\def\sun{\odot}
\def\sp{secuencia principal }

\def\arcd{^{\circ}\,\!\!\!\!\!.}      
\def\arcs{''\,\!\!\!\!.\,} 
\def\asec{^{\prime\prime}} 
\def\amin{^{\prime}} 
\def\hii{H~{\scriptsize II }~{\normalsize }}

\def\hi{H{\scriptsize I }}

\def\ha{H$\alpha$}
\def\hb{H$\beta$}

\def\jnl@style{\it}
\def\aaref@jnl#1{{\jnl@style#1}}

\def\aaref@jnl#1{{\jnl@style#1}}

\def\aj{\aaref@jnl{AJ}}                   
\def\araa{\aaref@jnl{ARA\&A}}             
\def\apj{\aaref@jnl{ApJ}}                 
\def\apjl{\aaref@jnl{ApJ}}                
\def\apjs{\aaref@jnl{ApJS}}               
\def\ao{\aaref@jnl{Appl.~Opt.}}           
\def\apss{\aaref@jnl{Ap\&SS}}             
\def\aap{\aaref@jnl{A\&A}}                
\def\aapr{\aaref@jnl{A\&A~Rev.}}          
\def\aaps{\aaref@jnl{A\&AS}}              
\def\azh{\aaref@jnl{AZh}}                 
\def\baas{\aaref@jnl{BAAS}}               
\def\jrasc{\aaref@jnl{JRASC}}             
\def\memras{\aaref@jnl{MmRAS}}            
\def\mnras{\aaref@jnl{MNRAS}}             
\def\pra{\aaref@jnl{Phys.~Rev.~A}}        
\def\prb{\aaref@jnl{Phys.~Rev.~B}}        
\def\prc{\aaref@jnl{Phys.~Rev.~C}}        
\def\prd{\aaref@jnl{Phys.~Rev.~D}}        
\def\pre{\aaref@jnl{Phys.~Rev.~E}}        
\def\prl{\aaref@jnl{Phys.~Rev.~Lett.}}    
\def\pasp{\aaref@jnl{PASP}}               
\def\pasj{\aaref@jnl{PASJ}}               
\def\qjras{\aaref@jnl{QJRAS}}             
\def\skytel{\aaref@jnl{S\&T}}             
\def\solphys{\aaref@jnl{Sol.~Phys.}}      
\def\sovast{\aaref@jnl{Soviet~Ast.}}      
\def\ssr{\aaref@jnl{Space~Sci.~Rev.}}     
\def\zap{\aaref@jnl{ZAp}}                 
\def\nat{\aaref@jnl{Nature}}              

 \def\kms{$\rm km\;s^{-1}$}
\def\spose#1{\hbox to 0pt{#1\hss}}
\def\lta{\mathrel{\spose{\lower 3pt\hbox{$\sim$}}
    \raise 2.0pt\hbox{$<$}}}
\def\gta{\mathrel{\spose{\lower 3pt\hbox{$\sim$}}
    \raise 2.0pt\hbox{$>$}}}

\makeatletter
\newcommand\figcaption{\def\@captype{figure}\caption}
\newcommand\tabcaption{\def\@captype{table}\caption}
\makeatother
\makeindex

\begin{document}

\thispagestyle{empty}  
\begin{picture}(550,651.5)
   \put(-90,-90){\framebox(590,900){\ }}
   \put(-90,-90){
\includegraphics[width=210mm]{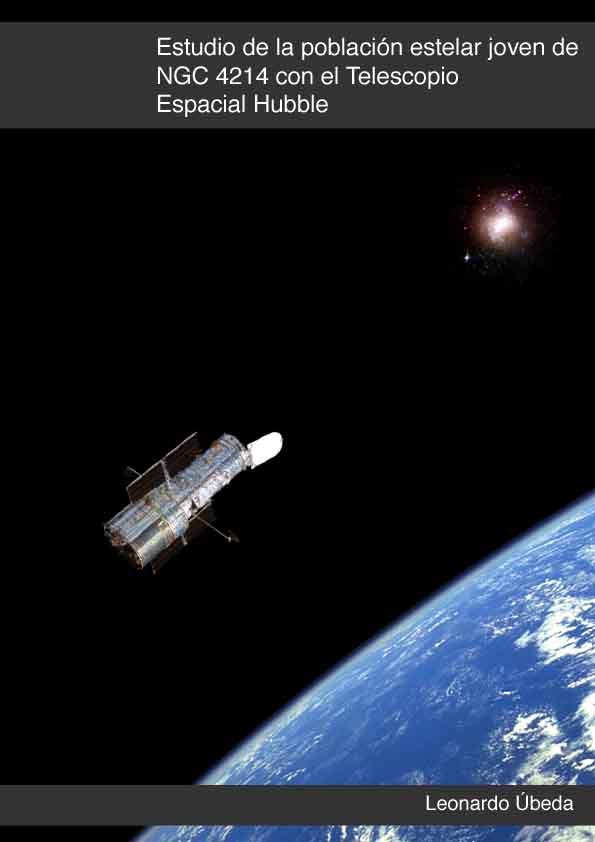}}
\end{picture}
\newpage

\thispagestyle{empty}
\begin{titlepage}
\begin{figure*}[ht!]
\centering
\includegraphics[width=30mm]{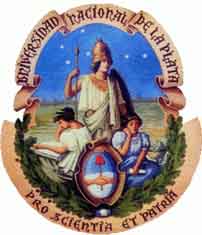}
\end{figure*}
\begin{center}
{\bf\Large Universidad Nacional de La Plata. Argentina}

{\bf\large Facultad de Ciencias Astron'omicas y Geof'isicas}
\end{center}

\vspace{2cm}

\begin{center}
{\bf\Huge Estudio de la poblaci\'on estelar  joven de  NGC~4214 con  el Telescopio Espacial Hubble}

{\bf\Large Lic. Leonardo 'Ubeda}
\end{center}

\vspace{2cm}

\begin{center}
{\bf\Large  Tesis presentada para optar por el grado de \\ Doctor 
en Astronom'ia}

{\bf\large Octubre 2006}
\end{center}

\vspace{4cm}

{\bf\large Director:  Dr. Jes\'us  Ma\'{i}z Apell\'aniz}

{\bf\large Co-Director:  Dr. Rodolfo H. Barb'a}
\end{titlepage}

\pagenumbering{roman} 
\setcounter{page}{1}
\chapter*{Resumen\markboth{Resumen}{Resumen}}
\addcontentsline{toc}{chapter}{Resumen}


\vspace{-30mm}
Se presenta un estudio original de la galaxia starburst enana NGC~4214 empleando 
 im'agenes obtenidas  
en filtros  ultravioletas y visibles.
Se usaron  im'agenes  de archivo  de  los instrumentos WFPC2 y STIS 
que est'an a bordo del
Telescopio Espacial Hubble. 

Se describe el proceso utilizado para obtener fotometr'ia de alta calidad y astrometr'ia
de la poblaci'on estelar resuelta  y de c'umulos de NGC~4214.
Se describe el m'etodo usado para transformar  las magnitudes y colores fotom'etricos
en par'ametros f'isicos mediante un novedoso c'odigo que compara datos observacionales
con modelos te'oricos.

Se analizaron diversos aspectos de las astrof'isica de NGC~4214: 
se describen brevemente las poblaciones
presentes en su estructura central con 'enfasis en la determinaci'on del cociente entre la cantidad 
de estrellas supergigantes azules y rojas. Se encuentra que los  modelos que mejor ajustan
a las observaciones son los asociados a la Nube Menor de Magallanes que posee una metalicidad 
semejante a NGC~4214. La distribuci'on morfol'ogica de las estrellas supergigantes es notable, present'andose
una alta concentraci'on de supergigantes azules en el complejo NGC~4214--I.

Se estudi'o la variaci'on de la extinci'on estelar y se hicieron comparaciones
con estudios previos de extinci'on nebular. 
Se encuentra un excelente acuerdo entre ambos. La extinci'on asociada a la
poblaci'on estelar difusa es, en general, 
baja ($E(B-V) < 0.1$ mag) y se observan valores mayores en las regiones de formaci'on estelar intensa.

Se hizo un estudio completo de la funci'on inicial de masa (IMF) de las estrellas de
 NGC~4214  en el rango $20-100 \,\,M_\odot$ con un m'etodo fotom'etrico.
  Se encontr'o que esta galaxia enana 
presenta una IMF m'as empinada que  $\gamma=-2.8$, lo cual est'a de acuerdo
con trabajos previos que determinaron la forma de la IMF por ajuste 
de espectros UV con modelos de s'intesis evolutiva.

Se analizaron varias estructuras extendidas y complejas 
distribuidas a lo largo del campo de observaci'on y se pudieron inferir varias de
sus importantes  propiedades como edad, masa
y extinci'on. Los resultados est'an garantizados por el m'etodo
fotom'etrico empleado abarcando un amplio rango de filtros adecuadamente elegidos.
 Se encuentran c'umulos candidatos a desaparecer por ``mortalidad
infantil'' y se discute la interacci'on entre las estrellas masivas y el medio interestelar, en particular,
la formaci'on estelar en etapas.
\newpage

\thispagestyle{empty}
\newenvironment{dedication}
  {\cleardoublepage \thispagestyle{empty} \vspace*{\stretch{1}} \begin{center} \em}
  {\end{center} \vspace*{\stretch{3}} \clearpage}
\begin{dedication}
{\large A mi mam'a y a mi pap'a, \\
 quienes
siempre me han  apoyado en mis elecciones.
}
\end{dedication}
\thispagestyle{empty} \cleardoublepage

\selectlanguage{spanish}
\title{NGC4214}

\author{Leonardo \'Ubeda}

\setcounter{tocdepth}{2}

\setcounter{chapter}{0}
\setcounter{secnumdepth}{4}
\setcounter{tocdepth}{3}






\markboth{\'Indice General}{\'Indice General}
\addcontentsline{toc}{chapter}{\'Indice General}
\tableofcontents
\newpage

\markboth{\'Indice de Figuras}{\'Indice de Figuras}
\addcontentsline{toc}{chapter}{\'Indice de Figuras}
\listoffigures
\newpage

\markboth{\'Indice de Tablas}{\'Indice de Tablas}
\addcontentsline{toc}{chapter}{\'Indice de Tablas}
\listoftables
\newpage

 
\pagenumbering{arabic} \setcounter{page}{1}

\chapter{Introducci'on}
\label{cha:introduction}
\thispagestyle{empty}
\hfill  {\em But remember: your hand at the level of
your eyes!}
 
\newpage

\section{Estrellas de gran masa}

\subsection{Definici'on e importancia}

Las estrellas se agrupan de acuerdo a su masa en tres tipos:
estrellas  masivas, estrellas  de masa intermedia y estrellas  de baja masa.
?`Cu'al ser'ia la menor masa necesaria  de una estrella para ser considerada masiva?
La divisi'on entre las dos primeras categor'ias se da seg'un
si la estrella  aislada evoluciona hacia una supernova o hacia
una enana blanca. Este l'imite depende de varios factores como 
la metalicidad y la rotaci'on entre otros y est'a en $\sim 8 \,M_{\odot}$.
Se  sabe que la caracter'istica
principal en la vida de una estrella masiva es su masa y su
metalicidad, las cuales a su vez determinan la p'erdida de masa. 
De acuerdo a \citet{Leitetal91}, en una estrella de $\sim 8 \, M_{\odot}$
la tasa de p'erdida de masa es lo suficientemente alta como para
afectar a la evoluci'on estelar y eso determinar'ia 
un l'imite inferior aproximado para llamar a una estrella ``masiva''.
Como consecuencia de esta  p'erdida, la masa de  las estrellas masivas
es una funci'on decreciente del tiempo. Esto implica que sus descendientes podr'ian 
tener masas menores que  $  10 \, M_{\odot}$, pero a pesar de eso
son consideradas masivas.  
De acuerdo a esta definici'on, las estrellas masivas comprenden
a las estrellas de tipo espectral O, a las de tipo espectral B tempranas, a
las supergigantes de todos los tipos, a  las estrellas Wolf--Rayet y  a algunas  supernovas.

Las observaciones espectrosc'opicas hechas desde Tierra,
mostraron que un n'umero peque'no de  estrellas peculiares
 presentan l'ineas con perfiles
P--Cygni, lo cual es una caracter'istica de p'erdida de masa en forma de
vientos muy densos. 
\cite{Mortetal68} y \cite{Carru68} descubrieron que todas las estrellas OB
tienen perfiles P--Cygni en las l'ineas de resonancia en UV haciendo mediciones
desde el espacio.  Entre otras, las l'ineas usadas para hacer este diagn'ostico son
las de C~{\scriptsize IV,}~{\normalsize} Si~{\scriptsize IV }~{\normalsize}
y N~{\scriptsize V. }~{\normalsize}
El mecanismo dominante de la p'erdida de masa en
las estrellas masivas son los vientos impulsados por radiaci'on.

Las estrellas de gran masa  son bastante raras si se compara
su abundancia con estrellas de tipo solar. Sin embargo, su importancia 
es considerable, ya que ellas proveen la mayor parte de la energ'ia mec'anica
al medio interestelar (ISM, del ingl'es: InterStellar Medium) en la forma
de vientos estelares y supernovas. Tambi'en generan la mayor parte
de la radiaci'on ultravioleta (UV) ionizante que modifica 
la apariencia del ISM, y son una  fuente   de enriquecimiento
del ISM en carbono, nitr'ogeno y ox'igeno \citep{Massey03}.

\subsection{Formaci'on y evoluci'on}
Las estrellas masivas  se
forman en un intervalo de tiempo relativamente corto ($< 1-2$ Ma)
dentro de Nubes Moleculares Gigantes.
Durante la mayor parte de su vida, las estrellas masivas emiten una fracci'on
significativa de su energ'ia
por debajo de $\lambda = 912$ \AA. 
Un fot'on estelar de longitud de onda inferior 
a   $912$ \AA\ es capaz de ionizar un 'atomo de H en su estado 
fundamental, produciendo un electr'on libre, el cual
reparte su energ'ia cin'etica con los electrones circundantes, calentando al
medio a una temperatura cercana a los 10\,000 K. La densidad
del gas circundante es del orden $n \sim 10-1000$ cm$^{-3}$.
La radiaci'on emitida  es atrapada en la nube que
dio origen a la estrella  y reemitida en longitudes de onda m'as largas.  
Hace tiempo se pensaba  \citep{Herb62}  que 
la formaci'on de estrellas masivas  detiene la formaci'on estelar. Hoy 
se entiende que la puede detener en sus alrededores inmediatos
pero tambi'en la puede estimular en nubes densas m'as alejadas
si 'estas existen \citep{ElmeLada77,WalbPark92}.

Las estrellas  m'as masivas comienzan su vida en la secuencia principal como 
estrellas de tipo espectral O y B tempranas
 y con temperaturas efectivas (\teff) entre $\sim$20\,000--50\,000 K.
Durante esta fase, convierten hidr'ogeno en helio a trav'es del ciclo CNO; su 
\teff  decrece levemente y su magnitud bolom'etrica (\mbol) 
permanece casi inalterada. 
La estructura interna de las estrellas masivas
est'a caracterizada por un n'ucleo convectivo que posee
la mayor parte de la masa y que determina la evoluci'on de la estrella.
Una vez que las reacciones termonucleares comienzan, la estrella empieza a emitir 
radiaci'on.  La nube progenitora  de gas y polvo intercepta esta energ'ia 
y la reemite en diferentes longitudes de onda.  En particular, los granos
de  polvo
reemiten  en longitudes de onda que dependen  de la temperatura de los mismos. 
Los procesos f'isicos que tienen lugar en el gas son descriptos por \cite{Oste89}.
En el caso de una distribuci'on infinita  y uniforme de gas, 
se genera una  regi'on ionizada
conocida como {\it esfera de Str\"omgren}. El radio de la misma
depende del n'umero de fotones ionizantes producidos por la estrella
y  del inverso del cuadrado de la densidad del gas. 

\begin{figure*}[h!]
 \centering
\includegraphics[width=\linewidth]{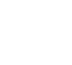}
\caption[Trayectorias evolutivas de estrellas masivas]{ {\sl\footnotesize Trayectorias evolutivas 
de estrellas masivas 
calculadas usando la metalicidad de las cercan'ias de Sol ($Z=0.02$).
Las  curvas rojas representan las trayectorias producidas por  \cite{Schaetal92}
para masas iniciales 120, 85, 60, 40, 25 y 20 $\, M_{\odot}$.
Los puntos azules marcan la posici'on de los puntos representativos de las Tablas~\ref{tbl024}
y \ref{tbl025}.
Se han indicado los tipos espectrales para cada instancia. N'otese que las estrellas con $M > 40 \,M_{\odot}$
no pasan por la fase de estrella supergigante roja.
    \label{fig087}}}
 \end{figure*}

Cuando finaliza la etapa de combusti'on de H, estas estrellas
son t'ipicamente supergigantes de tipo O o  B, ya que el n'ucleo se 
contrae y la envoltura se expande r'apidamente, mientras que las
 de mayor masa  saltean esta etapa. Esto da paso a la ignici'on del
n'ucleo de He. Durante esta nueva etapa, la estrella describe 
una trayectoria complicada  en el diagrama de Hertzsprung-Russell
 [$\log(T_{\mathrm{eff}}) $, $M_{\mathrm{bol}}$] 
 como se muestra en la Figura~\ref{fig087}.

En el caso de estrellas masivas, no se debe pensar en una relaci'on entre tipo espectral
y masa; es mejor pensar  que los  tipos espectrales son etapas por las que pasan  
las estrellas de diferentes
masas. Las Tablas~\ref{tbl024} y ~\ref{tbl025} muestran c'omo var'ia el tipo espectral  a medida que una
estrella masiva evoluciona.

 \begin{table}[htbp]
\centering
\begin{minipage}[u]{15cm}
\caption[Evoluci'on de estrellas de masas 120, 85 y 60 $M_{\odot}$ 
con metalicidad solar]{Evoluci'on de estrellas de masas 120, 85 y 60 $M_{\odot}$ 
con metalicidad solar ($Z=0.02$) calculadas
por  \cite{Schaetal92}. Notar la variaci'on del tipo espectral, la rapidez en la evoluci'on, as'i como la 
notable p'erdida de masa. Los valores est'an calculados cada 0.5 Ma. Esta Tabla est'a adaptada de \cite{Massey03}.  \label{tbl024}}
 \end{minipage}

\begin{tabular}{ lllllllll}    
\multicolumn{9}{c}{ \rule{161mm}{0.8mm}}      \\

    {\rule [0mm]{0mm}{5mm} }     Edad (Ma)                        &  &0.0& 0.5&1.0& 1.5 & 2.0 & 2.5 & 3.0  \\ 
\multicolumn{9}{c}{ \rule{161mm}{0.2mm}}      \\
    
     $120 \,M_{\odot}$    &   $\log(T_\mathrm{eff})$ &4.727& 4.706 &4.683& 4.648 & 4.635 & 4.420 &  \\
    &   $M_\mathrm{bol}$        &   -10.9& -10.9 &-11.0& -11.0 & -11.1 & -11.1 &  \\
      &   $M (M_{\odot})$        & 120.0 &117.2 &112.4& 103.5 & 90.5 &72.7 &  \\
     &   TE     &   O3 V& O3 V & O4 III& O5.5 III& O5 If & WNL &  \\ 
  \multicolumn{9}{c}{ \rule{161mm}{0.2mm}}      \\

  $\,\,\,85  \, M_{\odot}$        &   $\log(T_\mathrm{eff})$ &4.705& 4.691 &4.673& 4.645 & 4.577 & 4.449 &  \\
    &   $M_\mathrm{bol}$        &   -10.3& -10.3 &-10.4& -10.5 & -10.6 & -10.6 &  \\
          &   $M (M_{\odot})$        & 85.0 &84.2 &82.9& 80.7 & 75.9 &67.2 &  \\

     &  TE      &   O3 V& O4 V & O4 III& O5.5 III& O7 If & B0 I &  \\ 
  \multicolumn{9}{c}{ \rule{161mm}{0.2mm}}      \\

    $\,\,\,60 \, M_{\odot}$     &   $\log(T_\mathrm{eff})$ &4.683& 4.671 &4.660& 4.645 & 4.620 & 4.576 &4.464  \\
   &   $M_\mathrm{bol}$        &   -9.6& -9.6 &-9.7& -9.8 & -9.9 & -10.0 &-10.1  \\
       &   $M (M_{\odot})$        & 60.0 &59.7 &59.4& 58.9 & 58.0 &56.4 & 53.4 \\
    &   TE       &   O4 V& O5 V & O5 V& O5.5 III& O6.5 III & O7.5 If & B0 I \\ 
 \multicolumn{9}{c}{ \rule{161mm}{0.8mm}}      \\
\end{tabular}
\end{table}
 

 \begin{table}[htbp]
\centering
\begin{minipage}[u]{15cm}
\caption[Evoluci'on de estrellas de masas 40, 25 y 20 $M_{\odot}$ 
con metalicidad solar]{Evoluci'on de estrellas de masas 40, 25 y 20 $M_{\odot}$ 
con metalicidad solar ($Z=0.02$) calculadas
por  \cite{Schaetal92}. Notar la variaci'on del tipo espectral, la rapidez en la evoluci'on, as'i como la 
notable p'erdida de masa. Esta Tabla est'a adaptada de \cite{Massey03}. \label{tbl025}}
 \end{minipage}
\begin{tabular}{ llllllllll}    
\multicolumn{10}{l}{ \rule{161mm}{0.8mm}}      \\

    {\rule [0mm]{0mm}{5mm} }     Edad (Ma)                        &  &0.0& 1.0&2.0& 3.0 & 4.0 & 5.0 & 6.0&8.0  \\ 
\multicolumn{10}{l}{ \rule{161mm}{0.2mm}}      \\
    
  $40 \, M_{\odot}$    &   $\log(T_\mathrm{eff})$ &4.640& 4.626 &4.609& 4.570 & 4.443 && &  \\
    &   $M_\mathrm{bol}$        &   -8.7& -8.8 &-11.0& -9.0 & -9.1 & -9.3 & & \\
      &   $M (M_{\odot})$        & 40.0 &39.8 &39.4& 38.7 & 37.1 &  & & \\
     &   TE     &   O6 V& O6.5 V & O7 III& O8 III& B0.5 I & & & \\ 
  \multicolumn{10}{l}{ \rule{161mm}{0.2mm}}      \\

  $25\,  M_{\odot}$        &   $\log(T_\mathrm{eff})$ &4.579& 4.570 &4.563& 4.552 & 4.535 & 4.504 &4.438&  \\
    &   $M_\mathrm{bol}$        &   -7.5& -7.6 &-7.7& -7.8 & -7.9 & -8.1 & -8.3& \\
          &   $M (M_{\odot})$        & 25.0 &24.9 &24.8& 24.7 & 24.5 &24.3 &23.9 & \\

      &  TE      &   O8 V& O8 V & O9 V& O9 V& O9.5 III & O9.5 III &B0.5 I&  \\ 
  \multicolumn{10}{l}{ \rule{161mm}{0.2mm}}      \\

   $20 \, M_{\odot}$     &   $\log(T_\mathrm{eff})$ &4.640&  &4.626&  & 4.609 &  &4.570 &4.443 \\
  &   $M_\mathrm{bol}$        &   -6.9&  &-7.0&  & -7.2 & &-7.4 &-7.8 \\
       &   $M (M_{\odot})$        & 20.0 &  &19.9&   & 19.8 &  & 19.6& 19.3\\
    &   TE       &   O9.5 V&  & O9.5 V& & B0 V &  & B0 III& B1 I\\ 
 \multicolumn{10}{l}{ \rule{161mm}{0.8mm}}      \\
\end{tabular}
\end{table}
 

A pesar de que las estrellas masivas evolucionan a \mbol casi constante,
(al menos mientras est'an en la secuencia principal) se hacen 
considerablemente m'as brillantes en el visible, como lo
indica su magnitud absoluta en el visual $M_V$.  Esto es
una consecuencia de la dependencia de la distribuci'on de energ'ia
 con la $T_\mathrm{eff}$ y del aumento del radio que mantiene \mbol\
aproximadamente constante a pesar de la disminuci'on en la temperatura.

La evoluci'on de las estrellas masivas es muy dif'icil de modelar, 
debido principalmente a las complicaciones que provienen de la p'erdida de masa, su  caracter'istica  rotacion y el hecho de generarse en grupos \citep{Massey03}. 
Una estrella de tipo O puede eyectar  $\sim   10^{-8}- 10^{-5}$ \mpa  durante distintas
etapas de su vida.
A modo de comparaci'on, el Sol pierde $\sim   10^{-14} $ \mpa mientras
se encuentra  en la secuencia principal.

\begin{wrapfigure}{l}[0mm]{80mm}
  \centering
\includegraphics[width=\linewidth]{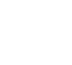}
\caption[Retorno de masa de estrellas masivas]{{\sl\footnotesize  Este gr'afico de \cite{Cast93}
  muestra el retorno de masa
de las estrellas masivas en funci'on de su masa inicial. N'otese que las estrellas con $M > 40   \, M_{\odot}$ no
pasan por la etapa de supergigante. Otro resultado importante es que la teor'ia 
predice que una estrella de masa inicial $ 100  \,  M_{\odot}$ debe  perder   un $\sim30$\% de su masa durante
su estad'ia en la secuencia principal.   \label{fig100}}}
\end{wrapfigure}

Casi toda la masa que forma a estrellas con  $M  > 8  \, M_{\odot}$  es retornada al medio interestelar
de una forma u otra.
Una estrella masiva puede perder gran parte  de su masa durante
su estad'ia en la secuencia principal como se muestra
en la    Figura~\ref{fig100}  y como explican  \cite{Chioetal78} y \cite{Cast93}.  
Luego de la secuencia principal, 
los procesos de p'erdida de masa son menos conocidos aun. 
Se cree que algunas  de las estrellas masivas
(con masa iniciales entre  $40 \, M_{\odot}$ y $80 \, M_{\odot}$)
 pasan por un estado
de variables azules luminosas (LBV, del ingl'es: Luminous Blue Variable).
S Dor en la Nube Mayor de Magallanes y $\eta$ Carinae en la
V'ia L'actea son  dos ejemplos.
Durante esta etapa la estrella sufre episodios de p'erdida de masa muy grandes \citep{LameLoor87}.
Similarmente, la p'erdida de masa durante la fase de estrella supergigante roja 
(RSG, del ingl'es: Red SuperGiant),  en estrellas de masa inicial menor que 
$30  \,  M_{\odot}$,
podr'ia ser m'as alta que la que se estima, afectando
seriamente la forma de las trayectorias evolutivas. Esta etapa 
se corresponde a estrellas masivas
de masa inicial en el rango $15 \lesssim  M   \lesssim 40   \, M_{\odot}$.
La 'ultima etapa en la evoluci'on de las estrellas masivas es la de estrella Wolf--Rayet 
(WR), durante la cual las p'erdidas de masa son las m'as altas.
Una estrella WR evolucionada puede perder $\sim   10^{-5}- 10^{-4}$~\mpa
\citep{LameCass99}.
Otros problemas en estudio de la f'isica de estos objetos
son la rotaci'on \citep{MaedMeyn00} y los procesos de mezclado y convecci'on.

Un  escenario aceptado para la evoluci'on de estos objetos
es el establecido por
\cite{Cont76}, aunque no ser'ia el 'unico.  Este autor propuso que la p'erdida de masa
que caracteriza a las estrellas masivas podr'ia explicar 
la existencia de estrellas WR, cuyos espectros se caracterizan por la presencia
de l'ineas de emisi'on muy intensas de helio y nitr'ogeno (tipo WN) o de helio,
carbono y ox'igeno (tipo WC), sin presentar vestigios de hidr'ogeno en sus espectros. 
Una estrella masiva de tipo O pierde una significativa parte de su masa a trav'es de 
los vientos  estelares, revelando primero los productos de la cadena CNO en su superficie,
y luego los productos de la combusti'on del helio.  Estas dos etapas se identifican
espectrosc'opicamente con los estados WN y WC (ver  Figura~\ref{fig100})

Todas las estrellas con masas iniciales mayores  que $30 \, M_{\odot} $ terminan
su vida con masas finales entre 5 y 10 $\, M_{\odot} $ \citep{Maed99,Maed99b}.
No es exagerado decir que la evaporaci'on de la envoltura de estas
estrellas es el factor dominante en su evoluci'on. Uno de los puntos m'as importantes
en la comprensi'on de la evoluci'on de estas estrellas es la relaci'on entre
la tasa de p'erdida de masa 
$(\dot{M})$ y la metalicidad $(Z)$; en general  $\dot{M}$ 
aumenta con $Z$ porque
el viento es impulsado por la interacci'on de la radiaci'on con iones met'alicos.

Una versi'on simplificada del escenario de la evoluci'on de estrellas masivas
se da en la Tabla~\ref{tbl030}. N'otese que los rangos de masas son una funci'on de la metalicidad,
por lo que los valores dados son s'olo  ilustrativos.

\begin{table}[htbp]
\centering
\caption[Escenario de Conti]{Evoluci'on de estrellas 
masivas de acuerdo al escenario de \cite{Cont76}. \label{tbl030}}
\vspace{2mm}

\begin{tabular}{@{} llllllllll@{}}   

$85 < M / M_{\odot}$      & O &$ \rightarrow$& LBV &$ \rightarrow$ & WN & $\rightarrow $& WC&$ \rightarrow$ & SN \\
$40 < M / M_{\odot}  <  85 $    & O &$ \rightarrow$& WN &$ \rightarrow$ & WC& $\rightarrow $& SN & &   \\
$25 < M / M_{\odot} <  40  $    & O &$ \rightarrow$& RSG &$ \rightarrow$ & WN& $\rightarrow $& WC&$ \rightarrow$ & SN  \\
$20 <   M / M_{\odot}  <  25  $    & O &$ \rightarrow$& RSG &$ \rightarrow$ & WN& $\rightarrow $& SN&  &    \\
$15 <   M / M_{\odot}  <  20  $    & OB &$ \rightarrow$& RSG &$ \rightarrow$ & BSG& $\rightarrow $& SN&  &    \\
\end{tabular}
\end{table}

\subsection{Detecci'on de estrellas masivas}

La observaci'on y estudio espectro/fotom'etrico de estrellas masivas
es de vital importancia para la Astrof'isica moderna.
Estos objetos son m'as f'acilmente detectables que otros grupos estelares por tratarse 
de los m'as luminosos (aunque no sean necesariamente  los m'as brillantes en el visible).

Cuando las estrellas masivas est'an muy alejadas o forman parte de una regi'on
que presenta mucha confusi'on, se las puede detectar mediante la presencia de l'ineas en el
espectro integrado.
Estas caracter'isticas espectrales pueden ser calculadas mediante modelos te'oricos
y pueden ser calibradas mediante observaciones de estrellas masivas locales.
\cite{WalbFitz90}  definen una grilla  de espectros de estrellas de referencia  de tipos espectrales
O y B.  Estos espectros cubren el rango 3\,800--5\,000 \AA\ a una resoluci'on
de 1.5 \AA. 
Lamentablemente, las l'ineas espectrales  de estrellas masivas
en el 'optico son muy d'ebiles en comparaci'on 
con  las l'ineas de otros  grupos estelares, y las pocas l'ineas  
que son algo intensas (como las de hidr'ogeno  y helio) est'an en general 
contaminadas por l'ineas nebulares de emisi'on del ISM. Esta situaci'on mejora si 
se dispone de espectros en el UV. Las estrellas masivas presentan l'ineas 
 resonantes caracter'isticas en el UV como Si~{\scriptsize IV}~{\normalsize}~$\lambda$1400 
y C~{\scriptsize IV }~{\normalsize}~$\lambda$1550
que se forman en los vientos estelares. 
En el IR cercano, las l'ineas m'as importantes de las estrellas RSG forman
el triplete de Ca~{\scriptsize II }~{\normalsize}~$\lambda\lambda$8500, 8540, 8660.
La l'inea de emisi'on de He~{\scriptsize II }~{\normalsize}~$\lambda$4686 es la caracter'istica
principal del espectro de las estrellas WR y su intensidad puede usarse
para inferir el n'umero de estrellas WR en galaxias cercanas \citep{Walb91}.
Hay que hacer notar que esta l'inea en emisi'on 
est'a formada por transiciones de varias especies qu'imicas,
siendo la principal el He~{\scriptsize II}~{\normalsize}. Las estrellas WR presentan
 muchas otras
l'ineas en emisi'on \citep{Miha78}.

La construcci'on de modelos de atm'osferas de estrellas masivas 
que se emplean para comparar con los espectros observados, presenta
varias complicaciones. Por empezar, las atm'osferas son extendidas, y por tal motivo
la geometr'ia plana paralela deja de ser v'alida. El equilibrio termodin'amico local (LTE, del ingl'es: Local 
Thermodynamic Equilibrium)
tampoco es v'alido en las atm'osferas calientes. Adem'as, estas atm'osferas est'an en 
continua expansi'on a altas velocidades.  Estos problemas han hecho que
el modelado de las atm'osferas haya sido tan  lento y todav'ia incompleto.

\subsection{Estrellas supergigantes}

De acuerdo a lo que se sabe de la evoluci'on de estrellas masivas, las estrellas con masas
$\lesssim 40 \, M_{\odot}$ se transforman en supergigates rojas luego de dejar la secuencia principal. 
Estas estrellas han evolucionado
a partir de estrellas de secuencia principal de tipo espectral O tard'io  y B. 
Estas estrellas expanden su envoltura convectiva  y se mueven hacia
la derecha en el  diagrama de Hertzsprung-Russell al final de la fase en la que 
transformaron H en He en su interior.

Las estrellas m'as masivas pierden tanta masa durante su vida como estrellas de 
secuencia principal    y como LBV que sus envolturas tienen insuficiente materia como para
hacerse convectivas.    Esto hace que estos objetos no puedan transformarse
en estrellas supergigantes rojas \citep{LameCass99}.

La identificaci'on tradicional de estrellas RSG en la Galaxia y las Nubes de Magallanes se hace 
principalmente mediante la identificaci'on
de  las l'ineas Fe~{\scriptsize II }~{\normalsize} y  Ca~{\scriptsize I, }~{\normalsize}
asi como bandas moleculares para las de tipo espectral M.  
En galaxias m'as alejadas, se emplea el triplete  de Ca~{\scriptsize II }~{\normalsize}~$\lambda$8500.
La identificaci'on mediante fotometr'ia es posible,
pero usando varias magnitudes en un 
amplio rango de  longitudes de onda. Un solo color no es suficiente debido a la posible contaminaci'on
por parte de  estrellas rojas gigantes y por
el enrojecimiento interestelar.


Hasta  el a~no  2005 exist'ia un problema significativo con los modelos evolutivos: ninguno de los
modelos se extend'ia lo suficiente hacia temperaturas bajas como para dar cuenta 
de la presencia de RSG de alta masa \citep{Mass02,Massey03,Masse03,MassOlse03}. 
Este problema fue atacado por \cite{Leveetal05}  quienes
presentan una nueva escala de temperatura efectiva para supergigantes
rojas de la  Galaxia obtenida al ajustar modelos de atm'osferas MARCS 
\citep{Gustetal75, Plezetal92}.
Esta nueva escala fue obtenida con 74 supergigantes rojas  gal'acticas 
$(Z=0.02)  $ 
 de distancia conocida. 
Usaron modelos te'oricos de \cite{MaedMeyn03} y hallaron un acuerdo 
mucho mejor entre teor'ia y observaci'on.

\subsection{Estrellas Wolf--Rayet}

En 1867, Wolf y Rayet realizaron un relevamiento de estrellas en la constelaci'on del Cisne usando 
un espectr'ometro visual  en el Observatorio de Paris. Hallaron tres estrellas cuyos espectros
est'an dominados por l'ineas de emisi'on muy intensas en lugar de l'ineas de absorci'on. 
Este espectro caracter'istico se debe a la presencia de  vientos estelares  muy intensos y 
 elementos pesados producto de la evoluci'on estelar.
 La p'erdida de masa
de estas estrellas es muy grande ($10^{-4}~M_{\odot}$~a$^{-1}$).
Estas estrellas son objetos evolucionados: las WR de tipo WN presentan muy poco o nada 
de hidr'ogeno, pero sus l'ineas de helio y nitr'ogeno son intensas; las WR del tipo WC
no presentan nitr'ogeno, pero se observa carbono y  ox'igeno a parte de helio
en sus atm'osferas. 

Estas estrellas son objetos  de suma importancia, pues permiten 
imponer condiciones sobre la teor'ia de la evoluci'on estelar.
Por tal motivo, es importante estudiar el contenido de WR en las galaxias cercanas: 
se estudian tanto 
la cantidad de WR como el n'umero relativo de WN y WC.
Estas estrellas se emplean para hacer mediciones de edad pues en los modelos de 
evoluci'on sint'etica el n'umero de estrellas WR es una funci'on dependiente del tiempo.
La detecci'on de la emisi'on  He~{\scriptsize II }~{\normalsize}~$\lambda$4686 en un starburst (ver Secci'on~\ref{sec:starburst}),
fija su edad en el rango 2--6 Ma, aunque los valores exactos dependen de la metalicidad
y otras complicaciones consideradas en los modelos.

\subsection{Funci'on inicial de masa}
Un resultado de  la teor'ia de evoluci'on estelar es que 
la estructura y la evoluci'on de una estrella de una dada composici'on qu'imica
est'a controlada principalmente (y entre otros par'ametros) por su masa.
La distribuci'on de masas en el momento del nacimiento de las estrellas
es una funci'on muy importante denominada {\em funci'on inicial
de masa}  (IMF, del ingl'es: Initial Mass Function). 
Esta  funci'on describe la proporci'on  de estrellas de diferentes tipos y masas.
La determinaci'on observacional de la IMF de una dada poblaci'on estelar
provee de restricciones fundamentales 
sobre las teor'ias de formaci'on estelar. La mayor'ia de las estrellas se forman en
grupos dentro de nubes moleculares \citep{LadaLada03}, y la
IMF
contiene informaci'on sobre de qu'e manera la masa original de la nube
se divide en fragmentos y se distribuye  en nubes protoestelares.

El estudio original de \cite{Salp55} consisti'o  en 
la determinaci'on de la IMF en el rango de masas $0.4$ -- 10 $\, M_{\odot}$, 
encontrando que 
la IMF pod'ia ser descripta mediante una ley de potencias
$  \left( \frac{dN}{dm}  = A \cdot m^{\gamma} \right)$  con un 'indice
 $\gamma = -2.35$.
 
\begin{wrapfigure}{l}[0mm]{80mm}
  \centering
\includegraphics[width=\linewidth]{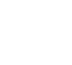}
\caption[ La IMF de estrellas de la  V'ia L'actea]{ 
{\sl\footnotesize  Gr'afico de la IMF construida con datos de estrellas de la V'ia L'actea por   
\cite{Scalo86}. \label{fig093}}}
\end{wrapfigure}


El estudio de la IMF  que incluye a  las estrellas masivas es complicado ya que la mayor parte de la energ'ia
es emitida en el UV lejano que no pueden detectarse desde Tierra, y porque
su vida en la secuencia principal es muy corta.
\cite{Garmetal82} usan un  n'umero limitado de estrellas O y estimaron que 
$\gamma = -2.6$. \cite{HumpMcel84}  usaron cat'alogos de 
estrellas O y B y obtuvieron $\gamma = -3.4$. Este problema no est'a agotado sin duda,
como se explica en  los trabajos de revisi'on realizados por  \cite{Scalo86} y 
\cite{Krou02}.
La IMF constru'ida a partir de datos de estrellas contenidos en varias referencias
est'a graficada en la Figura~\ref{fig093}  \citep{Scalo86}.
Nos interesa para este trabajo el valor de la pendiente de la curva 
cuando la masa de las estrellas es mayor que 10 $\, M_{\odot}$. En ese
intervalo,  la forma es todav'ia muy incierta, con  $\gamma $ entre  $-2.3$ y $-3.4$.

\clearpage
\section{C'umulos estelares}

\subsection{Origen}

Las Nubes Moleculares Gigantes (GMC, del ingl'es: Giant Molecular
Cloud)  representan una fase del medio interestelar (ISM) donde
se produce la formaci'on estelar. 
En nuestra Galaxia, se han identificado m'as de 4\,000 GMC con tama'nos
del orden $\sim 50$ pc, y masas entre $10^5$ y $10^6 \, M_{\odot}.$
Las dos galaxias sat'elite de la V'ia L'actea son las Nubes de Magallanes: la Mayor
(LMC, del ingl'es: Large Magellanic Cloud) y la Menor (SMC, del ingl'es: Small Magellanic Cloud). 
La distribuci'on de GMC en las Nubes de Magallanes fue determinada originalmente
 por \cite{Coheetal88} (LMC) y por \cite{Rubietal91} (SMC).
 Se han identificado unos 40 y 7 complejos moleculares respectivamente a una
 resoluci'on de 140 pc.

Estas nubes presentan estructuras internas  de diferentes formas y tama'nos. 
La formaci'on de estrellas  se debe principalmente
a un colapso gravitatorio en las regiones m'as densas de las nubes. 
La formaci'on de estrellas de baja masa se da en forma aislada
en gl'obulos y nubes oscuras, mientras que las estrellas
m'as masivas se forman {\em siempre} en c'umulos \citep{Rubi02}.
El nuevo c'umulo asi generado nace con propiedades relacionadas a las
de la nube de la cual se form'o.

Los c'umulos estelares j'ovenes  han sido siempre reconocidos como laboratorios
para realizar estudios de Astrof'isica por ser poblaciones estelares simples caracterizadas 
por una edad y metalicidad determinados.
Los c'umulos permiten  realizar estudios de din'amica estelar  y  proveen de
herramientas para determinar la IMF. Su distribuci'on espacial
se emplea para entender la estructura de las galaxias y se los usa para rastrear las
regiones de formaci'on estelar.

La evoluci'on de los c'umulos estelares es bastante m'as complicada que la
de una estrella individual, pero sin duda tambi'en es mucho m'as sencilla
que la de una galaxia entera. 
A pesar de las complicaciones, 
una caracter'istica fundamental de los c'umulos es que las estrellas
formadas en el mismo  fueron formadas  cuasi simult'aneamente a partir
de un mismo material y por lo tanto comparten la misma metalicidad.

Los \cus estelares se encuentran  en un amplio rango de ambientes como se 
deduce a partir de  muchos a'nos de observaciones, sin embargo, 
se sabe bastante poco acerca del origen de los mismos \citep{LadaLada03}.

\subsection{Clasificaci'on moderna}

Tradicionalmente, se clasifica a los c'umulos estelares en {\em globulares} y {\em abiertos}
\citep{Miha81,Shu82}, pero actualmente la clasificaci'on se ha extendido 
para dar cuenta de objetos m'as diversos. 
Las caracter'isticas y definiciones de los diferentes
tipos de \cus estelares que se considera en la actualidad es la siguiente \citep{Maiz01b,Greb02}:

\begin{description}
\item[ C'umulos globulares] Estos objetos tienen una estructura esf'erica, con una 
concentraci'on de estrellas hacia el centro. Sus masas cubren el rango
 $3 \times 10^4 \lesssim  M (M_{\odot})  \lesssim  3 \times 10^6$. Sus radios
varian entre $\sim 10 $  y 100 pc. En general son pobres en metales y muy viejos ($\sim 10^{10} $ a'nos). 
 Se han detectado c'umulos globulares en 12 de las 36 galaxias del Grupo Local, incluyendo las
 Nubes de Magallanes, NGC~6822 y WLM. 
Los c'umulos globulares de la V'ia L'actea  est'an catalogados en
``Globular Clusters in the Milky Way''  \citep{Harr96}.

 \item[ C'umulos abiertos] Estos objetos  j'ovenes ($\lesssim 1000$ Ma) 
 tienen     masas en el rango
$ 10^3 \lesssim  M (M_{\odot})  \lesssim   10^5$ y radios entre 1 y 20 pc.
Son objetos de metalicidad alta (en la Galaxia)  y no presentan simetr'ia alguna. Se los encuentra en
el disco de la Galaxia. Fueron identificados en $\sim 40$ \% de las galaxias del Grupo Local,
siendo principalmente habitantes de los brazos espirales de las galaxias espirales.
El cat'alogo   b'asico de c'umulos abiertos es el ``New catalog of 
optically visible open clusters and candidates''  \citep{Diasetal02}.

 \item[Superc'umulos estelares] Estos objetos (SSC, del ingl'es: Super Star Clusters)
  son c'umulos j'ovenes muy poblados, 
 masivos  y compactos. Usualmente contienen una masa superior a  
 $10^4 \, M_{\odot}$ en un radio de 1--2 pc. Se cree que estos objetos son
 los progenitores de los c'umulos globulares. Se los encuentra principalmente en galaxias
 starburst (ver Secci'on~\ref{sec:starburst}) y  en galaxias interactuantes.
 Se conocen pocos SSC en las galaxias del Grupo Local. El m'as masivo es R136
 ubicado en 30~Dor en la Nube Mayor de Magallanes. 
 M~82,  NGC~1569, NGC~1705 
 y las Antennae son otras galaxias que contienen SSC.
  Sus edades varian entre unos pocos
 a varios cientos de millones de a'nos.

 \item[Asociaciones] El resultado m'as com'un de formaci'on estelar a gran escala 
 son las asociaciones, las cuales contribuyen principalmente a las estrellas de campo
 de las galaxias. Las asociaciones son c'umulos extendidos, desligados
 gravitacionalmente y de formaci'on simult'anea de estrellas con un tama'no menor que 100 pc
 en radio en los cuales las fuerzas de marea son 
 muy eficaces. Se dispersan  completamente en $\sim 100$ Ma. Se las encuentra 
 en todo tipo de galaxias y se las puede identificar a distancias del orden de
 10 Mpc desde Tierra y 100 Mpc con observaciones hechas desde el espacio.
 
  \item[Asociaciones OB a gran escala] Existen c'umulos masivos  que no presentan un n'ucleo definido
  sino que poseen la apariencia de una asociaci'on de estrellas OB en t'erminos de tama'no 
  y forma pero son mucho m'as masivos que los conocidos en la Galaxia, y 
  para ellos   \cite{Maiz01b} propone el t'ermino {\em asociaci'on OB a gran escala } (SOBA, del ingl'es: Scaled
  OB Association).
  Los SOBA son objetos extendidos asim'etricos sin una definici'on precisa de su centro y parecen
  no estar ligados gravitacionalmente, o estar'ian muy d'ebilmente ligados.

\end{description}

\begin{figure*}[h!]
\centering
\includegraphics[width=\linewidth]{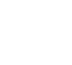}
\vspace{-5mm}

\begin{minipage}[c]{110mm}
\includegraphics[width=110mm]{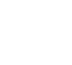}
\end{minipage}
\hfill
\begin{minipage}[u]{45mm}
\caption[Im'agenes de c'umulos gal'acticos y extragal'acticos]{{\sl\footnotesize 
[Arriba izquierda] Imagen 2MASS del c'umulo globular 47 Tuc en la Galaxia. 
[Arriba derecha] Imagen obtenida con la WFPC2 de las galaxias interactuantes
Antennae. Estas galaxias estan pobladas de SSC como fue estudiado por
\cite{Whitetal99b,Whitetal05}.
[Izquierda] Imagen en el filtro $V$ del c'umulo 
Gal'actico abierto M~45.
 \label{fig097}}}
\end{minipage}
\end{figure*}

Esta nueva clasificaci'on de c'umulos estelares incluye b'asicamente todos
los objetos que se han descubierto en diferentes galaxias. \cite{DaCo02}
describe con detalle las caracter'isticas de los c'umulos en las Nubes de Magallanes, 
algunos de los cuales no tienen su 
correspondiente representante en nuestra Galaxia.

\subsection{C'umulos masivos j'ovenes}
En la presente Tesis nos vamos a concentrar en los c'umulos masivos $(M > 10^4 \, M_{\odot})$,
y vamos a dar especial 'enfasis a los \cus  masivos j'ovenes  (MYC, del ingl'es: Massive Young
Cluster). Esta definici'on incluye a los c'umulos ligados gravitacionalmente (SSC) y los
no ligados gravitacionalmente (SOBA) arriba mencionados. 
Estos objetos han sido descubiertos cerca del centro Gal'actico \citep{Fige04}, 
en los n'ucleos 
de  galaxias tard'ias \citep{Bokeetal04}, en galaxias starburst cercanas \citep{deGr04}, 
en galaxias interactuantes \citep{Whitetal99b}, en
galaxias IR de alta luminosidad y tambi'en en galaxias espirales \citep{Larsetal04b}.

Desde el punto de vista observacional, un \cu masivo pasa por distintas etapas 
que resumimos en la Tabla~\ref{tbl026}. Esta Tabla muestra que durante los primeros 100 Ma
un c'umulo masivo presenta caracter'isticas en diferentes partes del espectro electromagn'etico,
desde emisi'on de radio durante su formaci'on hasta la emisi'on de rayos X asociados
con la emisi'on de vientos y supernovas.

 \begin{table}[htbp]
\centering
\begin{minipage}[u]{15cm}
\caption[Etapas de la evoluci'on de c'umulos masivos]{Etapas de la 
evoluci'on de c'umulos masivos. Esta Tabla est'a adaptada de \cite{GallaGreb02}.  \label{tbl026}}
 \end{minipage}

\begin{tabular}{ lrll}    
\multicolumn{4}{l}{ \rule{161mm}{0.8mm}}      \\

  Etapa       &Edad  &Propiedades & Observaciones   \\ 
\multicolumn{4}{l}{ \rule{161mm}{0.2mm}}      \\
    
Pre--formaci'on   &   $-1 - 0$ Ma & nube molecular densa & radio  \\
Nacimiento de estrellas   &   $0 - 1$ Ma & regi'on \hii / polvo & radio / IR lejano  \\
 Regi'on \hii gigante   &   $1 - 7$ Ma & regi'on \hii, estrellas $UV$ & $UV$-- 'optico -- IR  \\
 Supernovas   &   $3 - 30$ Ma & estrellas OB & $UV$-- 'optico -- IR-- rayos X  \\
 Supergigantes   &   $\sim 10$ Ma & m'aximo en 'optico & $UV$-- 'optico -- IR  \\
 AGB   &   $0.3 - 2$ 1000 Ma & estrellas AGB &  'optico -- IR -- IR lejano  \\
 RGB   &   $>2 - 3$ 1000 Ma & estrellas RGB &  'optico -- IR   \\
 \multicolumn{4}{l}{ \rule{161mm}{0.8mm}}      \\
\end{tabular}
\end{table}
 

De acuerdo a \cite{Elme04} deben existir al menos dos mecanismos para la formaci'on
de MYC: el primero implica la existencia de focos localizados y transitorios de alta presi'on en el ISM
que producen nubes masivas de alta densidad.
El segundo m'etodo act'ua en forma continua sobre grandes regiones
del ISM donde la presi'on del gas es alta por la presencia de estrellas preexistentes. 

\cite{Fall04}
estudia el espectro de masas, de edades y la distribuci'on espacial de \cus masivos,
llegando a interesantes conclusiones. Sostiene que la funci'on de masa de los \cus j'ovenes
(con edades $<100$ Ma) tiene la forma de una ley de potencias, $\frac{dN}{dM} \propto M^{-2}$
en el rango de masas $10^4 < M < 10^6 \,  M_{\odot}.$
La distribuci'on de edades de los c'umulos  es una funci'on decreciente 
del tiempo, $\frac{dN}{d\tau} \propto {\tau}^{-1}$, lo cual es un indicador 
de una r'apida ($<10 $ Ma) destrucci'on  de los c'umulos. Este fen'omeno,
llamado  ``moratilidad infantil''   fue introducido por  \cite*{Falletal05}.

Muchos de los \cus que se mantienen gravitacionalmente unidos
son destruidos en escalas de tiempo mayores por procesos que incluyen la p'erdida
de masa producto de la evoluci'on estelar y procesos din'amicos de la galaxia
que los contiene. Existen varios procesos que contribuyen a la destrucci'on de
los c'umulos, siendo los principales la evaporaci'on y la dispersi'on por
efectos de marea. El primer proceso domina la destrucci'on de los c'umulos 
muy compactos y densos, mientras que los procesos de marea son m'as importantes
en los c'umulos extendidos. Los SOBA tienden principalmente a dispersarse y
los c'umulos m'as compactos tienden a perder sus halos por efectos de marea
\citep{Maiz01b}.

La forma de estudiar MYC en la actualidad 
consta de dos m'etodos: El primero es
el de los modelos  de s'intesis evolutiva  (ver Secci'on~\ref{sec:models})
  los cuales contienen 
internamente la  informaci'on de modelos de evoluci'on estelar.
Este m'etodo se basa principalmente en mediciones precisas de magnitudes
en un amplio rango de longitudes de onda \citep{Alve04,Maiz04,deGretal05}. Los par'ametros 
de los c'umulos que se intenta
determinar  son, entre otros, la masa, la edad, el enrojecimiento, la metalicidad
 y la
ley de extinci'on en su linea de la visual.
Una segunda t'ecnica se basa en la comparaci'on entre espectros sint'eticos
y espectros observados \citep{Goleetal99,Trem04}.


Sin duda alguna, el empleo del Telescopio Espacial Hubble (HST, del ingl'es: Hubble Space Telescope)
es indispensable para
estudiar MYC. Su resoluci'on espacial y la amplia disponibilidad de filtros
que cubren desde el UV hasta el visible han sido fundamentales  
para entender las caracter'isticas de estos objetos en los 'ultimos 10 a'nos.
Pero se pueden emplear otros instrumentos tambi'en: observaciones de radio y en 
longitudes de onda milim'etricas proveen informaci'on invaluable del medio interestelar
donde los \cus m'as j'ovenes se forman.  El telescopio espacial Spitzer tiene una baja 
resoluci'on (s'olo $2\arcs5$); sin embargo, puede 
contribuir al estudio del polvo en regiones donde se forman los c'umulos, porque
es muy sensible a ondas de longitud IR mayores que otros sat'elites,
y porque no sufre de la absorci'on producida por la atm'osfera terrestre. 
Los telescopios de Tierra  de mayor tama'no ($6+$ m)
han mejorado enormemente la relaci'on entre la se'nal y el ruido 
permitiendo obtener espectroscop'ia de MYC. 
El telescopio espacial JWST estar'a formado
por un espejo de 6m segmentado. Realizar'a observaciones en el IR
con un resoluci'on superior a la de NICMOS en el HST. 
Sus observaciones permitir'an estudiar regiones de formaci'on de MYC
a alto corrimiento al rojo  
y regiones de formaci'on estelar.

\section{Starbursts}
\label{sec:starburst}
\subsection{Definici'on y caracter'isticas principales}
Un brote de formaci'on estelar (starburst) es un episodio corto e intenso de formaci'on
estelar que se da en una regi'on dada del Universo.
Los starbursts  generan  gran parte de las estrellas de 
mayor masa en el Universo Local  \citep{TerR97},  son bastante comunes y  los encontramos
en los discos, halos y n'ucleos de las galaxias.
Los starbursts son regiones en las que la energ'ia est'a dominada principalmente por 
procesos de formaci'on de estrellas y constituyen laboratorios ideales en los que se puede 
explorar la formaci'on y la evoluci'on de las galaxias.

Los starbursts de estrellas masivas  y sus regiones \hii asociadas son relativamente f'acilies de detectar.
Se los reconoce por tener colores azules, l'ineas de emisi'on nebulares  intensas y (si 
contiene polvo) una alta luminosidad en el IR \citep{Leit98}.

El t'ermino ``starburst'' evolucion'o de la frase ``flashing galaxies'' empleada por  \cite{Searetal73} 
para denotar un aumento en la luminosidad en regiones \hii extragal'acticas debido a
un incremento en la formaci'on de estrellas masivas. \cite{Balz83} introduce el t'ermino
``star--burst nuclei'' para describir un nuevo tipo morfol'ogico de galaxias. 

Con el objeto de definir ``galaxia starburst'', 
\cite{TerR97} sugiere por primera vez una clasificaci'on de galaxias que se basa
en la relaci'on entre la energ'ia emitida por el starburst  ($L_{SB}$) y la energ'ia emitida por
el resto de la galaxia que lo contiene ($L_{G}$).

Una galaxia con $L_{SB}  \gg L_{G} $ ser'ia una {\em galaxia starburst}, una
galaxia con $L_{SB}  \sim L_{G} $ ser'ia una {\em galaxia con  starbursts}
y una galaxia con $L_{SB}   \ll L_{G} $ ser'ia una {\em galaxia normal.}

Esto nos est'a diciendo que existen varios entornos donde se pueden presentar starbursts:
galaxias enanas en las que s'olo se observa el starburst, n'ucleos de galaxias, galaxias irregulares
ricas en gas, galaxias en colisi'on o regiones \hii gigantes localizadas en brazos espirales.
Existen otros t'erminos para describir a las galaxias starburst, como galaxias 
H~{\scriptsize II,}~{\normalsize} galaxias compactas
enanas azules, galaxias WR o regiones \hii extragal'acticas \citep{Brin97}.

En a'nos recientes, se ha incrementado el inter'es en las galaxias starburst debido a
su parecido con las galaxias de mediano y alto corrimiento al rojo. Tambi'en son importantes
pues se ha hecho cada vez m'as evidente que los  SSC  constituyen
el modo predominante de formaci'on estelar en esos sistemas.

Sus espectros integrados muestran en todos los casos que est'an dominados por
una poblaci'on estelar joven pues  se observan l'ineas de absorci'on en 
el UV. En los starbursts j'ovenes se observa la presencia de
      regiones \hii  mediante  l'ineas nebulares en emisi'on en el 'optico.
Los starbursts m'as viejos no presentan l'ineas de emisi'on. 
Esta dicotom'ia demuestra que  los starbursts j'ovenes  est'an energizados principalmente por 
estrellas muy calientes. 
Las l'ineas de absorci'on observadas en el UV se forman en el ISM y en los vientos de las estrellas masivas.
Las l'ineas nebulares se forman en el ISM que rodea a la regi'on de formaci'on estelar. 
\cite{TerR97} propone considerar las siguientes tres fases de starbursts:

 \begin{description}
\item[Fase nebular] Se caracteriza por la presencia de l'ineas de emisi'on muy intensas provenientes
de la ionizaci'on del gas por las estrellas masivas. El ancho equivalente de \ha\ es mayor que 100 \AA\
y el starburst  tiene una edad $< 10$ Ma.
\item[Fase temprana] Se caracteriza por la presencia de l'ineas de emisi'on d'ebiles 
salvo las de H$\alpha$ y de l'ineas de absorci'on de Balmer intensas.  
 El ancho equivalente de \ha\ es menor que 100 \AA\
y el starburst  tiene  una edad entre  10 y 100 Ma.
\item[Fase tard'ia] El espectro se caracteriza por l'ineas de absorci'on de Balmer.
Estos  starbursts  tiene  varios cientos de millones de a'nos y hasta 1000 Ma.

\end{description}

Una manera de estudiar a estas regiones es  empleando observaciones en 
un amplio rango de longitudes de onda que abarque, en lo posible,  desde el UV hasta el NIR,
y combinar esto con modelos de s'intesis evolutiva. Estas herramientas permiten derivar el contenido estelar
y restringir la historia de formaci'on estelar del starburst.

\subsection{?`Por qu'e son importantes los starbursts?}
La importancia del estudio de starbursts es may'uscula. Su an'alisis permite 
responder algunas preguntas fundamentales de la Astrof'isica y la Cosmolog'ia.

En un radio de 10 Mpc del Sol, las cuatro galaxias starburst m'as luminosas
(M~82, NGC~253, NGC~4945 y M~83)  representan el 25 \% de la formaci'on
de estrellas masivas en ese volumen. A pesar de que los mecanismos que dan origen
a la formaci'on estelar todav'ia son inciertos, se sabe con certeza
que los starbursts son las fuentes m'as importantes de formaci'on de estrellas masivas. 
Esto implica que es importante estudiar el espectro de masas de tales objetos astron'omicos
\citep{Leit98}.
Ser'ia interesante saber si las estrellas de baja masa se forman por los mismos mecanismos
que las de alta masa. Se intenta determinar si existe un  l'imite superior  de masa para una estrella
\citep{Figeetal98}.
Se intenta determinar de qu'e manera los starbursts modifican al ISM  y c'omo
evoluciona la composici'on qu'imica de la galaxia que los contiene \citep{Leitetal92}.

Se busca analizar diferencias (si es que existen) entre los starbursts originados 
en los brazos espirales de galaxias y aquellos generados en los n'ucleos. 
Se quiere determinar por qu'e los starbursts forman estrellas en dos tipos de 
c'umulos: las superasociaciones   y  los superc'umulos estelares  y si estos \'ultimos son los
progenitores de los c'umulos globulares \citep{deGretal05b}.

De particular importancia son los starbursts contenidos en  galaxias cercanas y que 
tienen una poblaci'on  estelar resuelta usando los instrumentos actuales.
Estos starbursts representan  una versi'on a escala (en tama'no) de 
starbursts a mayor corrimiento al rojo, que no podemos resolver ni siquiera
empleando los nuevos telescopios espaciales.

\subsection{Starbursts en la V'ia L'actea}

Las regiones con caracter'istica de 
starburst presentes en la Galaxia son los m'as cercanos y,
a pesar de no ser tan espectaculares como los presentes en otras galaxias, 
son objetos clave para entender c'omo funcionan los starbursts  en general.
En las siguientes secciones describimos brevemente a algunos de ellos y a
modo ilustrativo. 
\subsubsection{Westerlund 1}


El c'umulo abierto Westerlund 1 fue decubierto por \cite{West61} 
y est'a localizado en la constelaci'on Ara a una distancia entre 2 y 5 kpc. Se han realizado varios estudios
fotom'etricos y espectrosc'opicos de este   c'umulo. Sin embargo, la 
alta  absorci'on ($A_V \sim 12.9$, \cite{Piatetal98}) producto del polvo interestelar
hizo que no se le preste mucha atenci'on hasta recientemente. 
Este c'umulo est'a ubicado detr'as 
de una inmensa nube de gas y polvo que bloquea la mayor parte
de la luz visible. Se han identificado numerosas estrellas WR en su estructura \citep{ClarNegr02,NegrClar05}
\begin{figure*}[h!]
\centering
\includegraphics[width=\linewidth]{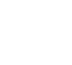}
\vspace{-5mm}

\begin{minipage}[c]{8.0cm}
\includegraphics[width=79mm]{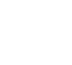}
\end{minipage}
\hfill
\begin{minipage}[u]{75mm}
\caption[Mosaicos de Westlund 1]{{\sl\footnotesize
[Arriba izquierda] Imagen 'optica de Westerlund 1 obtenida por ESO. Este mosaico 
 fue realizado con im'agenes obtenidas en los filtros $V, R$ e $I$. Las estrellas masivas
 y calientes del c'umulo aparecen de color naranja, y las m'as frias se muestran rojas.
 Este c'umulo est'a ubicado detr'as de una inmensa nube de gas y polvo
 interestelar. 
(ESO PR Photo 09a/05)
[Arriba derecha] Imagen de Westerlund 1 como se lo observa con el 2 Micron All-Sky Survey (2MASS). 
Cr'edito de la imagen: 2MASS, UMass, IPAC--Caltech, NASA, NSF.
[Izquierda] Imagen de Westerlund 1 obtenida con el telescopio espacial Chandra en rayos X. 
Cr'edito de la imagen: NASA, CXC, UCLA.
 \label{fig090}}}
\end{minipage}
\end{figure*}

\cite{Claretal05b} presentan el 'ultimo estudio espectro--fotom'etrico de Westerlund 1
el cual revela  una poblaci'on de estrellas masivas evolucionadas. 
Encuentran estrellas WR, supergigantes OB y objetos de vida breve como supergigantes B y LBV. 
Westerlund 1 es el c'umulo  m'as masivo dentro del Grupo Local, con una masa
mayor que la de los c'umulos que pueblan el Centro Gal'actico (Arches y Quintuplet).
Se cree que este c'umulo es m'as
masivo que cualquier c'umulo globular Gal'actico ($10^5 \, M_{\odot}$),
lo cual lo transforma en un candidato a proto-c'umulo globular.
\citeauthor{Claretal05b} establecen que las caracter'isticas de Westerlund 1 son comparables a las 
de un SSC con una edad estimada en $3.5-5$ Ma.
En particular, su densidad es comparable a la de Arches ($\log \rho = 5 \, M_{\odot} $ pc~$^{-3}$),
lo cual hace que las colisiones entre estrellas en el n'ucleo sean muy frecuentes: una cada $10^3-10^4$
a'nos.
  
  \subsubsection{W 49}

W~49  \citep{Mezgetal67} es una de las regiones m'as activas en cuanto a formaci'on estelar en la Galaxia. 
W~49 es una nube molecular gigante  con una masa estimada de  $10^6 \, M_{\odot}$ 
localizada a unos 11.4 kpc de distancia.  En su estructura, se encuentra
W~49A, una de las regiones que presentan  emisi'on en radio sumamente intensa
( $\sim 10^7  \, L_{\odot}$). Esta regi'on est'a energizada por el equivalente de
unas cien estrellas O7 \citep{ContBlum02}.

Trabajos recientes
\citep{AlveHome03,HomeAlve04,HomeAlve05}
 realizados con la c'amara  
infrarroja SOFI  en el telescopio NTT en La Silla 
han mostrado la 
presencia de c'umulos de estrellas masivas en esta regi'on.
Se han identificado mas de cien estrellas con masas mayores que $15 \,  M_{\odot}$. 
Estos autores proponen que la formaci'on estelar ha ocurrido en forma
casi simult'anea.

\begin{figure*}[h!]
 \centering
\includegraphics[width=0.99\linewidth]{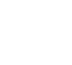}
\caption[Mosaicos de W~49]{{\sl\footnotesize [Izquierda]  La imagen presenta
un mosaico de W~49 obtenido con NTT en el IR. Cubre una regi'on de $5 \times 5$  arcmin~$^2$ en
las bandas $J$, $H$ y $K_S$. Los nombres indican fuentes de ondas de radio conocidas.  
(ESO PR Photo 21a/03)
[Derecha] Esta imagen muestra la regi'on central de W~49A,  en la que se combinan una imagen de
radio y dos en el IR. Las regiones rojas  representan zonas de hidr'ogeno ionizado
dentro de la nube molecular. 
(ESO PR Photo 21b/03)
   \label{fig089}}}
 \end{figure*}

\subsubsection{NGC~3603}

NGC~3603, localizado a 6 kpc \citep{Depretal99}
en el brazo espiral Carina,  es probablemente el starburst m'as cercano
y prominente \citep{Dris99,Stoletal04,Nurn04}. Se supone que contiene unas 2000 $\, M_{\odot}$ en
estrellas de tipos O y B y un masa total 
estimada en  $> 4000 \, M_{\odot}$. Una $L_\mathrm{bol} > 10^7 L_{\odot}$
  lo hace cien veces m'as luminoso que la Nebulosa de
Ori'on  aunque  bastante menos luminoso   que 30~Dor. 
Trabajos recientes \citep{Brandetal99,Stoletal04,Nurn04} estudian la formaci'on estelar
en esta regi'on de la Galaxia. Las observaciones muestran que en NGC~3603 se est'an
formando estrellas en un amplio rango de masas.
 Se est'a produciendo adem'as
formaci'on estelar secundaria en los pilares gaseosos. 
\cite{Branetal00} han descubierto tres discos protoplanetarios (proplyds) 
en esta regi'on starburst de la Galaxia, los cuales
son entre 5 y 10 veces m'as grandes que los  descubiertos en
Ori'on y m'as masivos. Estos objetos presentan un frente de ionizaci'on 
en la cara que enfrenta al starburst y una cola que apunta en direcci'on opuesta.
 
\begin{figure*}[h!]
 \centering
\includegraphics[width=0.99\linewidth]{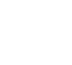}
\caption[Im'agenes de NGC~3603]{{\sl\footnotesize [Izquierda] Mosaico de NGC~3630 en el IR
compuesto por tres exposiciones ($J_S$, $H$ y $K_S$ ) obtenidas con el VLT en 1999. 
El campo mide $3.4 \times 3.4$  arcmin~$^2$.
El c'umulo central es una de las concentraciones de estrellas masivas m'as densas conocidas
en la Galaxia. Este mosaico muestra tres objetos de tipo ``proplyd'' recientemente descubiertos. 
(ESO PR Photo 38a/99)
[Derecha] Imagen WFPC2
de NGC~3603.
En el centro se observa el starburst dominado por estrellas j'ovenes, principalmente
WR y estrellas O tempranas. La cavidad en el ISM ha sido creada por la fuerte 
radiaci'on ionizante y los vientos estelares de estas estrellas. 
La interacci'on entre la radiaci'on y la nube  molecular se pone de manifiesto 
en los pilares gaseosos parecidos a los de la Nebulosa del 'Aguila.
Cr'edito de la imagen: W. Brandner, E. K. Grebel, Y. Chu y NASA.
   \label{fig088}}}
 \end{figure*}

\subsubsection{Cygnus OB2}

La asociaci'on Cygnus OB2  \citep{Reddetal66}
es una regi'on que contiene algunas de las estrellas m'as luminosas
conocidas en la Galaxia \citep{Torretal91}.
Se estima que Cygnus OB2 contiene unos $ 8\,000$ miembros
de los cuales $120\pm20$ son objetos de tipo espectral O, lo cual 
da una masa total de  $\sim 4-10 \times 10^4 \, M_{\odot}$.
Este c'umulo est'a ubicado a una distancia de 1.7 kpc detr'as de una
regi'on rica en gas y polvo que causa una absorci'on en el visual 
$A_V$ entre 4  y 10 magnitudes.
Su estructura contiene  una considerable cantidad de estrellas tempranas
 como por ejemplo las estrellas WR 145 y WR 146 \citep{Niemetal98},
 adem'as de supergigantes O muy tempranas  \citep{Walbetal00a,Walbetal02a}.
 El estudio de  \cite{Knod00},
 basado en observaciones de 2MASS, ha revelado que Cygnus OB2 
 es bastante m'as masivo que lo que se pensaba anteriormente y que la pendiente 
 de la IMF es $-2.6 \pm 0.1$.

\subsection{Starbursts extragal'acticos}

Existen muchos  starbursts  ubicados fuera de nuestra Galaxia, los cuales abarcan
un amplio rango de tama'nos y caracter'isticas.
Sin lugar a dudas, el mas importante es 30~Dor en la Nube Mayor de Magallanes. 

En esta regi'on \hii supergigante (N157; \citealt{Heni56}) y sus alrededores,  se est'a 
desarrollando la mayor parte de la  formaci'on estelar masiva de la Nube Mayor. 
La fuente de ionizaci'on principal de la nebulosa es un c'umulo 
compuesto por estrellas de tipo O muy tempranas 
\citep{Walb86}.
El objeto central R136 es el n'ucleo del c'umulo ionizante y est'a compuesto de muchas estrellas
brillantes \citep{Campetal92}.
Existen muchos estudios que abarcan todos los  aspectos de 30~Dor usando 
diferentes m'etodos y tipos de observaciones. Algunos  trabajos interesantes incluyen:
\citealt{Hunt95b}   analizaron im'agenes  WFPC2 para obtener la IMF de 30~Dor.
\citet{Rubietal98a}  proveen observaciones en el IR y las comparan con observaciones 
de la WFPC2.
Usando datos de NICMOS, 
\citet{Walbetal99c} hacen una descripci'on sumamente detallada 
 de las componentes de esta regi'on H~{\scriptsize II. }~{\normalsize} 
\citet{WalbBlad97}  presentan una  clasificaci'on espectral de 106 
estrellas OB presentes en esta regi'on.
\cite{MassHunt98}  hacen lo propio en el n'ucleo de R136 con datos de HST, revelando
la mayor concentraci'on resuelta de las estrellas m'as masivas. 
\cite{Boscetal01} hacen un estudio espectrosc'opico  detallado y presentan
 velocidades radiales confiables de 55 estrellas. 
\begin{figure*}[ht!]
\begin{minipage}[c]{15cm}
\centering
\includegraphics[width=15cm]{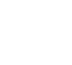}
\caption[Mosaicos de 30~Dor]{ {\sl\tiny Mosaicos de 30~Dor 
capturados en diferentes longitudes de onda. Las im'agenes
abarcan diferentes   regiones del starburst y sus  escalas son diferentes. La orientaci'on
de las mismas es semajante.
Se presentan estas im'agenes a modo ilustrativo solamente. 
[Arriba izquierda]
Imagen obtenida mediante el telescopio de rayos X Chandra. Las regiones 
rojas indican menor temperatura y las regiones en verde indican la presencia de 
objetos m'as calientes. La radiaci'on X m'as intensa est'a representada en azul. 
La fuente m'as brillante hacia el centro de la 
imagen es  Melnick~34. Cr'edito de la imagen: NASA, CXC, Penn State, L.Townsley. 
[Arriba centro]
Imagen obtenida con el  Ultraviolet Imaging Telescope (UIT), un telescopio de 38 cm.
La radiaci'on ultravioleta marca precisamente la posici'on de regiones de formaci'on estelar reciente.
Esta imagen muestra que la formaci'on est'a teniendo lugar en todo el starburst.
[Arriba derecha]
Imagen obtenida en luz visible del Digitized Sky Survey (DSS).
Cr'edito de la imagen: STScI.  
[Abajo izquierda] 
Imagen en el IR cercano obtenida del Two-Micron All-Sky Survey (2MASS). 
[Abajo centro]
Imagen de 30~Dor obtenida con el telescopio Spitzer. Este es un mosaico construido 
a partir de im'agenes en el IR cercano y medio. 
Estas observaciones penetran a las nubes de  polvo y permiten visualizar 
regiones escondidas de formaci'on estelar. 
Cr'edito de la imagen: NASA, JPL--Caltech, B. Brandl.
[Abajo derecha]
En esta imagen de baja resoluci'on obtenida con IRAS (InfraRed Astronomical Satellite), el rojo 
denota las zonas m'as brillantes en radio y el azul 
las regiones con emisi'on m'as d'ebil. 
La emisi'on a estas longitudes de onda (infrarrojo  lejano) resulta de la luz visible y UV 
absorbida por los granos de polvo
que luego es reemitida en longitudes de onda m'as largas. 
  \label{fig092}}}
\end{minipage}
\end{figure*}
Existe amplia evidencia  que favorece la presencia de regiones de formaci'on 
estelar en los alrededores del n'ucleo de 30~Dor \citep{Walbetal99c}.
Se supone que este evento ha sido disparado por el starburst original  \citep{WalbPark92}.
Se han encontrado muchos fen'omenos  de inter'es astrof'isico relacionados con episodios
de formaci'on estelar como estrellas masivas muy j'ovenes emergiendo
de la nube que los origin'o, pilares de polvo  y jets para nombrar algunos pocos 
\citep{Hylaetal92,Rubietal92,Rubietal98a,Walbetal02}

El complejo nebular N~11  \citep{Heni56}
es la segunda regi'on \hii m'as grande de la Nube Mayor de Magallanes 
luego de 30~Dor. Se ubica en el extremo opuesto de la 
barra que caracteriza a esta galaxia irregular.
Este complejo consiste de una cavidad principal
 rodeada de nueve  complejos nebulares de diversos tama'nos
 \citep{Davietal76,KennHodg86,Barbetal03a}. La asociaci'on
OB LH~9  est'a localizada en el centro de la cavidad y est'a
dominada por el objeto HD~32228 (Radcliffe 64), un 
c'umulo estelar masivo  evolucionado (estrellas O tard'ias y WC).
\cite{WalbPark92} remarcan la incre'ible analog'ia,  en  cuanto a la
morfolog'ia, de N~11 y 30~Dor. Estos autores proponen en ambos casos
un escenario de formaci'on estelar secuencial en al menos dos etapas
para explicar la distribuci'on espacial y temporal de estrellas
en ambos starbursts, pero con el proceso global m'as avanzado en N~11 que en 30~Dor.

El estudio de   objetos cercanos y bien resueltos 
nos permite entender la naturaleza de objetos m'as alejados y no tan
resueltos pero no por eso menos importantes.

Otros ejemplos de starbursts extragal'acticos estudiados en los 'ultimos a'nos 
incluyen: NGC~346 en la Nube Menor de Magallanes \citep{Notaetal04}, 
M~82 \citep{Meloetal05,deGretal05c}, NGC~604 \citep{Maizetal04a} en M~33,
 NGC~1705 \citep{Vazqetal04},  NGC~1569.
M'as ejemplos pueden hallarse en  \cite{Meln92}.


NGC~4214  es el prototipo
 de  galaxia starburst enana que analizamos con detalle en esta Tesis.
En el sentido de la clasificacion de \cite{TerR97}, en NGC~4214 se verifica que 
$L_{SB}  <  L_{G} $ y los efectos globales  del starburst son muy inferiores a los de, por ejemplo,
M~82.
 El objetivo de este trabajo es realizar un estudio detallado de
diferentes aspectos de la galaxia starburst NGC~4214.
Esta   es una galaxia irregular barrada de tipo Magallanes clasificada entonces como
IAB(s)m  \citep{Vauc91} y  ubicada a una distancia de 
$2.94 \pm 0.18 $ Mpc \citep{Maizetal02a}.
La misma presenta varios starbursts distribuidos principalmente a lo largo
de su barra. Es  una galaxia deficiente en metales  ($Z=0.006$) y rica en gas.

Los recientes e  importantes episodios de formaci'on estelar combinados
con la proximidad y la baja extinci'on hacen de esta galaxia un laboratorio ideal
para estudiar diversos aspectos de inter'es astrof'isico relacionados
con poblaciones estelares j'ovenes.  
En particular, el estudio de  este tipo de galaxias es importante
pues la mayor parte de la informaci'on que inferimos de las galaxias a alto corrimiento
al rojo (mayor distancia) 
se basa en nuestras observaciones de las galaxias en el Universo Local.
Siendo NGC~4214 una galaxia de baja metalicidad, tenemos la oportunidad de estudiar 
las condiciones f'isicas en un ambiente muy cercano al del Universo 
en sus comienzos. NGC~4214 es entonces un pelda'no esencial en la escalera de
distancias para descifrar el fen'omeno de starburst.

Para encarar nuestro estudio, empleamos im'agenes de alta resoluci'on
obtenidas con las c'amaras WFPC2 y STIS a bordo del Telescopio
Espacial Hubble  (HST).

Nuestro estudio abarca diferentes aspectos de NGC~4214:
(1) un an'alisis de las poblaciones estelares
(tanto estrellas j'ovenes como evolucionadas), (2) una estimaci'on de la cantidad 
de estrellas supergigantes, (3) un mapeo de
su extinci'on variable, (4) la determinaci'on de la pendiente de la IMF y (5) 
el estudio de una selecci'on de c'umulos distribuidos a lo largo 
de su estructura.
Para encarar este an'alisis, empleamos un c'odigo nuevo 
\citep{Maiz04}  que permite
inferir par'ametros f'isicos de estrellas y  c'umulos a partir
de la fotometr'ia observada de los mismos.

\section{Estructura de la Tesis}

La estructura de este trabajo de  Tesis es la siguiente: 
En el Cap'itulo 2 se describe brevemente al Telescopio Espacial Hubble haciendo 'enfasis 
sobre los detectores WFPC2 y STIS.
En el Cap'itulo 3 se hace una breve descripci'on de los trabajos sobre NGC~4214 que exist'ian
previamente a este trabajo de Tesis.
En el Cap'itulo 4 se hace una descripci'on exhaustiva de los datos empleados para este estudio
y de su reducci'on.
En el Cap'itulo 5 se introduce al c'odigo empleado para transformar observables en par'ametros te'oricos y
se da la justificaci'on del m'etodo.
En el Cap'itulo 6 analizamos  brevemente las poblaciones estelares de NGC~4214 y el cociente 
entre supergigantes azules y rojas.
En el Cap'itulo 7 hacemos un estudio in'edito de la extinci'on estelar de NGC~4214 y presentamos un
mapa de extinci'on.
En el Cap'itulo 8  establecemos un valor de la pendiente de la IMF luego de un an'alisis muy detallado
de la poblaci'on de estrellas de la secuencia principal.
En el Cap'itulo 9 obtenemos resultados muy interesantes de trece c'umulos  distribuidos a lo largo de 
la estructuras de NGC~4214. Comprobamos que mediante un estudio fotom'etrico se pueden inferir 
 importantes par'ametros  de los c'umulos.
  En el Cap'itulo 10 presentamos un resumen  y las conclusiones de este trabajo de Tesis.

\section{Publicaciones}
La totalidad de este trabajo de Tesis ha sido publicado en los siguientes art'iculos 
cient'ificos:

\begin{itemize}
\item{ {\bf The young stellar population of NGC 4214 as observed with the 
Hubble Space Telescope. I.  
 Data and methods.} Leonardo 'Ubeda, Jes'us Ma'iz-Apell'aniz and 
 John W. MacKenty.    {\it The Astronomical Journal,} {\bf 133}:917-931, 2007 }

\item{{\bf The young stellar population of NGC 4214 as observed with the 
Hubble Space Telescope. II. 
Results.} Leonardo 'Ubeda, Jes'us Ma'iz-Apell'aniz and 
John W. MacKenty.  {\it The Astronomical Journal,} {\bf 133}:932-951, 2007}

\item{ {\bf Numerical Biases on Initial Mass Function 
Determinations Created by Binning.}   
Jes'us Ma'iz-Apell'aniz and Leonardo 'Ubeda.   
{\it The Astrophysical Journal,} 
{\bf 629}:873-880, 2005}

\end{itemize}

\chapter{El Telescopio Espacial Hubble}
\label{cha:chapter1}
\thispagestyle{empty}
\begin{flushright}    {\em Nighttime sharpens, \\
  heightens each sensation  \\
  Darkness stirs and \\
   wakes imagination...  }
 \end{flushright}

\newpage

\section{El Telescopio Espacial Hubble}

El Telescopio Espacial Hubble (HST, del ingl'es: Hubble Space Telescope)  es un programa de cooperaci'on 
entre la European Space Agency (ESA) y la National Aeronautics and Space 
 Administration (NASA) para operar un observatorio espacial para el beneficio de
 la comunidad astron'omica internacional. 
El HST es un observatorio dise~nado y constru'ido en las d'ecadas del 70 y 80
y que est'a en funcionamiento desde 1990.
El HST es un telescopio reflector que est'a en una 'orbita 
baja (600 km) desde que la tripulaci'on del Space Shuttle Discovery
(STS-31) lo dej'o en el espacio el 25 de abril de 1990.

\begin{wrapfigure}[20]{r}[2mm]{100mm}
\includegraphics[width=70mm]{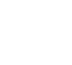}
\caption[Telescopio Espacial Hubble]{ {\sl\footnotesize   
Imagen del Telescopio Espacial Hubble con la Tierra como fondo.
 \label{fig101}}}
\end{wrapfigure}

La responsabilidad  de conducir y coordinar la ciencia del HST cae sobre el 
Space Telescope Science Institute (STScI)  ubicado en el
campus universitario de la  Universidad Johns Hopkins en Baltimore, Maryland.
La Association of Universities for Research
 in Astronomy, Inc. (AURA) opera STScI para la NASA.

Debido a la ubicaci'on priviligeada del HST por encima de 
la atm'osfera terrestre, los instrumentos pueden producir im'agenes  de objetos astron'omicos
de muy 
alta resoluci'on. Los mejores telescopios 'opticos--IR en Tierra
proveen una resoluci'on de  $0 \arcs 5$, que s'olo 
en las mejores condiciones de observaci'on, y mediante
t'ecnicas apropiadas ('optica activa, adaptiva, etc)  se puede mejorar.
La resoluci'on del HST es considerablemente  mejor ($ \approx 0 \arcs 1 $).

Cuando fue planeado originalmente, el programa Large Space Telescope
consideraba retornar al telescopio a Tierra y relanzarlo cada cinco a~nos,
luego de implementar mejoras y arreglos, con un servicio en 'orbita
cada 2.5 a~nos.
Problemas de contaminaci'on y estructurales asociados 
con el regreso a Tierra a bordo del Space Shuttle en 1985 determinaron que 
no se realizaran  esos retornos.
NASA determin'o que los servicios en 'orbita ser'ian suficientes
para mantener al HST operacional  durante 15 a\~nos, y se adopt'o
un ciclo de tres a'nos para los servicios. Tambi'en se tienen en cuenta posibles
contingencias. A continuaci'on se describen algunos de los detalles
de cada misi'on:

 \begin{figure*}[ht!]
 \centering
\includegraphics[width=\textwidth]{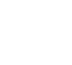}
 \caption[Misi'on de servicio 3A]{ {\sl\footnotesize  [Izquierda] Lanzamiento del Space Shuttle Discovery 
 para llevar a cabo la Misi'on de Servicio 3A (SM3A). [Derecha] Imagen del Hubble  orbitando la Tierra 
 una vez finalizada SM3A.
 Hubble orbita la Tierra 
a una altura de aproximadamente  600 km. 
Le lleva unos 95 minutos competar una 'orbita alrededor de la Tierra, con una
inclinaci'on de  $28\arcd5$.
La 'orbita es lo suficientemente elevada como para que el telescopio se encuentre
por encima de la atm'osfera terrestre y pueda asi conducir observaciones
sin el efecto nocivo de la misma.
  \label{fig103}}}
 \end{figure*}

\begin{description}
\item{\bf  Misi'on de servicio  1 (SM1)}  2 de diciembre de 1993: (STS-61) Lanzamiento
del Space Shuttle Endeavour; instalaci'on de COSTAR 
para corregir la aberraci'on esf'erica de la 'optica; WFPC2 reemplaz'o a la WFPC.

\item{\bf Misi'on de servicio 2 (SM2)}  11 de febrero de 1997: (STS-82) Lanzamiento 
del Space Shuttle Discovery; STIS reemplaz'o al instrumento FOS y NICMOS al  GHRS. 

\item{\bf Misi'on de servicio  3A (SM3A)}  19 de diciembre de  1999: (STS-103) Lanzamiento
del Space Shuttle  Discovery; reemplazo de RSU (Rate Sensing Units que contienen
gir'oscopos);  se instal'o una nueva computadora y se hizo mantenimiento general.

\item{\bf Misi'on de servicio  3B (SM3B)}  1 de marzo de  2002: (STS-109) Lanzamiento del
Space Shuttle  Columbia; se instalaron  la ACS y NCS, y se reemplazaron
paneles solares.

\item{\bf Misi'on de servicio 4 (SM4)} (aun sin fecha definitiva): Se instalar'an la  WFC3 y COS.

\end{description}

\subsection{Instrumentos a bordo del HST  (2006)}

\begin{description}
\item{\bf Wide Field/Planetary Camera 2:}
La original  Wide Field/Planetary Camera (WF/PC1) fue reemplazada por
la WF/PC2 en la misi'on Shuttle STS-61 en diciembre de 1993. WF/PC2 era
un instrumento de sobra desarrollado en 1985  
en el  Jet Propulsion Laboratory en  Pasadena, California.
WF/PC2  consiste de cuatro c'amaras. Los espejos en la WF/PC2 fueron modificados
de manera tal que puedan corregir la aberraci'on esf'erica 
del espejo principal del observatorio
 (el espejo primario del HST es dos micrones  demasiado chato en su borde,
 y la 'optica de la WFPC2 corrige este defecto en la misma cantidad.)
 La WFPC2 consiste de un trio de detectores en forma de letra ``L'' y un detector 
 de mayor resoluci'on ubicado en la esquina sobrante formando un cuadrado.

\item{\bf Space Telescope Imaging Spectrograph:}
El Space Telescope Imaging Spectrograph (STIS)  es un espectr'ografo
que puede estudiar objetos astron'omicos en un rango de longitudes de onda
desde el ultravioleta (UV)  (115 nan'ometros), pasando por el visible y hasta el  infrarojo cercano
(1000 nan'ometros).
STIS usa tres detectores: un Multi-Anode 
Microchannel Array (MAMA) para el rango  115 a 170 nm, 
un  MAMA para el rango 
115 a 310 nm, 
 y un CCD para el rango  2\,000 a 10\,300 \AA. 
Los tres detectores tienen un formato de 1024 $\times$ 1024 p'ixeles. 
El campo de observaci'on de cada MAMA es  $25 \asec \times 25 \asec$ 
y el del CCD es   $50 \asec \times 50 \asec$.

La caracter'istica principal de STIS
es la capacidad de realizar espectroscop'ia bidimensional. Por ejemplo, 
es posible grabar el espectro de varias fuentes en una galaxia simultaneamente, en
lugar de observar una posici'on cada vez.
STIS puede, adem'as,  grabar un rango mas ancho de longitudes de onda
en el espectro de una estrella, lo cual lo hace un instrumento cient'ifico mas eficiente 
que los espectr'ografos que previamente volaron a bordo del HST.

STIS fue instalado a bordo del HST el 14 de febrero de 1997,
reemplazando el espectr'ografo GHRS.
El 3 de agosto de 2003, STIS dej'o de operar y actualmente est'a desactivado.

\item{\bf Near Infrared Camera and Multi-Object Spectrometer:}
El Near Infrared Camera and Multi-Object Spectrometer (NICMOS) es un instrumento
que tiene la capacidad de obtener im'agenes y espectros en el IR.
NICMOS detecta luz con longitudes de onda entre 0.8 y 2.5 micr'ometros.
Los detectores de HgCdTe que forman NICMOS son muy sensibles y deben
operar a muy bajas temperaturas, lo cual  se logra conteniendo a los mismos
dentro de recipientes aislados t'ermicamente con nitr'ogeno helado.
NICMOS es el primer instrumento criog'enico a bordo del HST.

\item{\bf Advanced Camera for Surveys:}
La Advanced Camera for Surveys (ACS) es una c'amara que fue dise~nada
para proveer al HST de la capacidad de realizar surveys desde el visible 
hasta el  IR cercano, asi como la posibilidad de capturar im'agenes de alta resoluci'on
desde el UV cercano hasta el IR cercano.

\end{description}
\begin{figure*}[ht!]
\centering
\includegraphics[width=120mm]{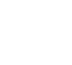}
\begin{minipage}{120mm}
\caption[Instrumentos]{ {\sl\footnotesize   
Esquema del interior del Telescopio Espacial Hubble. Se muestra la disposici'on 
actual (2006) 
de los instrumentos. \label{fig104}}}
 \end{minipage}
 \end{figure*}

\subsection{Operaci'on de la misi'on y observaciones}
Aunque el HST est'a en funcionamiento todo el tiempo, s'olo parte
del mismo se dedica a observaciones.
Cada 'orbita dura alrededor de 95 minutos, de los cuales algunos minutos 
se usan para tareas de mantenimiento y otros a las observaciones.
El mantenimiento incluye cambios de orientaci'on del telescopio para
acceder al nuevo objetivo o evitar al Sol o la Luna, cambios en las
antenas de comunicaci'on y modos de transmisi'on de datos, calibraci'on 
y actividades similares.

Cuando STScI completa el plan maestro de observaciones cient'ificas, 
se lo envia
al Goddard's Space Telescope Operations Control Center (STOCC), 
donde se mezcla con las tareas de mantenimiento y se crea un plan detallado.
Cada evento es traducido en una serie de comandos que deben enviarse
a las computadoras de a bordo.
Cuando es posible, se emplean dos instrumentos cient'ificos simult'aneamente
para observar regiones del cielo adyacentes. Durante las observaciones, los 
Fine Guidance 
Sensors (FGS)
siguen a las estrellas de referencia para mantener al telescopio apuntando 
correctamente al objeto seleccionado.

\begin{wrapfigure}[19]{r}[2mm]{80mm}
\includegraphics[width=80mm]{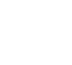}
\caption[Space Telescope Imaging Spectrograph (STIS)]{ {\sl\footnotesize   
Esta fotograf'ia muestra al Space Telescope Imaging Spectrograph (STIS)
dentro de un laboratorio en el Ball Aerospace in Boulder, Colorado, donde 
fue construido.
STIS se ubica en el HST, detr'as del espejo principal. Este instrumento 
fue dise\~nado para trabajar en tres regiones espectrales, cada una con su propio
detector.
 \label{fig021}}}
\end{wrapfigure}

Si un astr'onomo desea estar presente durante la observaci'on, existe una consola
en STScI y otra en STOCC, donde hay monitores que despliegan im'agenes y otros datos
a medida que las observaciones se van desarrollando. Algunos comandos pueden enviarse
en tiempo real desde estas estaciones, principalmente los que implican cambios de  filtros.

Los datos cient'ificos y los comandos son transmitidos a trav'es del  sistema
Tracking Data Relay Satellite y una estaci'on de tierra en White Sands, New Mexico.
Un m'aximo de 24 horas de comandos pueden almacenarse en las
computadores de a bordo.

Los observadores pueden examinar las im'agenes crudas en muy pocos minutos 
luego de la observaci'on para un an'alisis r'apido. Dentro de las 24 horas, 
Goddard Space Flight Center (GSFC) formatea los datos 
para ser entregados al STScI. 
STScI es el responsable del procesamiento de los datos (calibraci'on, edici'on,
distribuci'on y mantenimiento de los datos para la comunidad cient'ifica).


\subsection{Instrumentos que volaron anteriormente}

\begin{description}

\item{\bf Faint Object Spectrograph:}
El  Faint Object Spectrograph (FOS) pod'ia estudiar objetos en un amplio rango
espectral: desde el UV  (1\,150 \AA\ ) hasta el IR cercano (8\,000 \AA\ ).
El FOS usaba dos sensores: el azul era sensible desde 1\,150 hasta 5\,500 \AA\ y 
el rojo era sensible desde 1\,800 hasta 8\,000 \AA. La luz pod'ia entrar
al instrumento a trav'es de cualquiera de las 11 aperturas que ten'ian un 
tama\~no variable  entre 
 $0 \arcs 1$ y  $1 \arcs 0$ en di'ametro. 
 Un peque~no dispositivo permit'ia bloquear la luz del centro del objeto
 observado y dejaba pasar a la luz de los bordes. Esto era muy 'util al estudiar,
 por ejemplo, 
la envoltura de gas que rodea a estrellas gigantes.
El FOS soportaba dos modos de operaci'on: en baja y en alta
resoluci'on. A baja resoluci'on, pod'ia alcanzar magnitud 26 en una hora
de exposici'on con un poder de resoluci'on de 250. En alta resoluci'on,
el FOS pod'ia alcanzar s'olo magnitud 22 en una hora,
pero el poder de resoluci'on  aumentaba en este caso a 1\,300.

 \item{\bf Goddard High Resolution Spectrograph:}
El  High Resolution Spectrograph  era un espectr'ografo
que permit'ia el estudio de la composici'on, temperatura, 
velocidad y otras propiedades f'isicas y qu'imicas de 
objetos astron'omicos. Su diferencia con el FOS consist'ia en que 
este instrumento se especializaba en espectroscopia UV. 
Los detectores empleados eran los mismos que en el FOS,
pero los de HRS fueron deliberadamente enceguecidos en luz visible.
Un detector era sensible en el rango  1\,050--1\,700\AA, mientras que el otro lo era entre
1\,150 y 3\,200 \AA.
El HRS dispon'ia de tres modos de resoluci'on: bajo, medio y alto.

\item{\bf Corrective Optics Space Telescope Axial Replacement:}
COSTAR no era un instrumento cient'ifico;  este  paquete reemplaz'o 
el High Speed Photometer durante la primera misi'on de servicio
al HST. COSTAR fue dise~nado para corregir los problemas de aberraci'on
del espejo principal de la Faint Object Camera (FOC), la FOS y el GHRS. 
El resto de los instrumentos, instalados
desde el lanzamiento del HST, fueron dise\~nados con sus 
propios m'etodos para corregir la 'optica.

\item{\bf Faint Object Camera:}
La Faint Object Camera (FOC) fue construida por la 
Agencia Espacial Europea.
'Esta contaba con dos sistemas de detectores. Cada uno usaba un amplificador
para producir una imagen sobre una pantalla de f'osforo que era 100\,000
veces m'as brillante que la luz original recibida. Esta im'agen era luego escaneada
por una c'amara. El sistema era tan sensible que los objetos mas brillantes
que magnitud 21 debian ser debilitados para no saturar a los
detectores.
La FOC trabajaba en tres modos, con diferentes resoluciones: 
 $0 \arcs 043$ / p'ixel,   $0\arcs 022$ / p'ixel  y $0 \arcs 0072$ / p'ixel.

\end{description}

\section{WFPC2: descripci'on  del instrumento e im'agenes}
\subsection{WFPC2}

El  instrumento Wide Field and Planetary Camera 2 (WFPC2)  es un fot'ometro
bidimensional localizado en el centro del plano focal del HST y que cubre 
el rango espectral entre aproximadamente 1\,150 \AA\ y 10\,500 \AA. 
Simult'anemante obtiene las  im'agenes de una regi'on del cielo 
con forma de L de $150 \asec \times  150\asec$ con una
resoluci'on de 
$0 \arcs 1$ y de otra regi'on cuadrada mas peque~na  $( 34\asec \times  34\asec )$ 
con resoluci'on $0 \arcs 0455$ por p'ixel. 
Fue construido en el  Jet Propulsion Laboratory e instalado 
en diciembre de  1993 \citep{Bireetal02}.

El objetivo  cient'ifico de la WFPC2 es proveer im'agenes 
fotom'etrica y geom'etricamente precisas de objetos astron'omicos, 
con una resoluci'on alta
y en un amplio rango de longitudes de onda.

La  WFPC2  permite observar objetos mucho mas d'ebiles
que los observados desde Tierra, y resolver la estructura de los mismos
con mayor confiabilidad. Otro importante objetivo cumplido por la
WFPC2 fue corregir 
la aberraci'on esf'erica del HST.

El campo de observaci'on de la  WFPC2 est'a dividido en 
cuatro c'amaras por medio de un espejo piramidal cerca del plano
focal del HST.
Cada una de las cuatro c'amaras contiene un detector CCD Loral
de 800 $\times$ 800 p'ixeles. Tres de las c'amaras operan en una escala
de $0 \arcs 1$ por p'ixel (f/12.9) y componen la Wide Field Camera (WFC). 
La cuarta c'amara opera con la resoluci'on $0 \arcs 0455$ por p'ixel
(f/28.3) y se la denomina Planetary Camera (PC).

\subsection{Filtros espectrales}

La  WFPC2 contiene 48 filtros montados sobre 12
ruedas en el Selectable Optical 
Filter Assembly. Estos incluyen un conjunto de filtros 
que se asemejan a los filtros  $UBVRI$  del  sistema Johnson--Cousins,
un conjunto de filtros de banda ancha $U, B, V$ y  $R$
y filtros  Str\"{o}mgren  de banda intermedia  $u, v, b$ e  $y$.
Los filtros de banda angosta incluyen, entre otros, 
[O~{\scriptsize III}~{\normalsize] } (4\,363\AA  ~y 5\,007\AA), \ha ~(6\,563\AA), 
[S~{\scriptsize II}~{\normalsize] } ~(6\,716\AA ~y 6\,731\AA).

La Tabla~\ref{tbl008}  describe algunos de los filtros 
a bordo del HST que hemos empleado en nuestro estudio.

\begin{table*}[ht!] 
\centering
\begin{minipage}[u]{10cm}
\topcaption[Filtros usados en este trabajo]{Detalles de los filtros en HST  empleados en este trabajo. Se incluyen
la longitud de onda efectiva  $\lambda $ y el ancho    $\Delta \lambda  $.  }
\label{tbl008}
\end{minipage}

\begin{tabular}{llcc} 
\multicolumn{4}{c}{ \rule{100mm}{0.8mm}}      \\
Filtro &  Descripci'on &  $\lambda $(\AA) &   $\Delta \lambda  $(\AA)  \\ 
\multicolumn{4}{c}{ \rule{100mm}{0.2mm}}      \\
F170W & Continuo en UV  lejano & 1\,666 & 434.6   \\  
F336W & WFPC2 $U$  & 3\,317 & 370.5   \\  
F502N &  [O~{\scriptsize III}~{\normalsize ]}   & 5\,012 & 26.9   \\  
F555W & WFPC2 $V$ & 5\,202 & 1\,222.6  \\ 
F656N & \ha & 6\,564 & 21.5   \\  
F702W & Wide $R$  & 6\,940 & 1\,480.6   \\  
F814W & WFPC2 $I$  & 8\,203 & 1\,758.0   \\  
F25CN182 &  Continuo en 1\,800 \AA &  1\,820 &350    \\  
F25CN270 & Continuo en 2\,700 \AA &  2\,700 &350    \\ 
\multicolumn{4}{c}{ \rule{100mm}{0.8mm}}      \\
\end{tabular} 
\end{table*}

\subsection{Eficiencia cu'antica y tiempos de exposici'on}

La eficiencia cu'antica (QE, del ingl'es: Quantum Efficiency) es el t'ermino empleado para 
indicar la capacidad de un detector para transformar a los fotones entrantes
en una cantidad medible. Generalmente, se la define como el cociente entre 
la cantidad de fotones  entrantes y aquellos realmente detectados por el
instrumento. Un detector ideal debe tener una QE de 100\% \citep{Howe00}.
La QE del sistema WFPC2+HST  tienen un m'aximo de  40\% en el rojo
y se mantiene por encima del 5\% sobre todo el espectro visible. 
La respuesta en el UV es a'un m'as d'ebil, con una QE de
aproximadamente 0.1\%.

 \begin{wrapfigure}[18]{l}[1mm]{80mm}
\includegraphics[width=80mm]{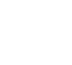}
\caption[Wide Field and Planetary Camera 2 (WFPC2)]{ {\sl\footnotesize  Esta fotograf'ia muestra a un astronauta
en el proceso de reemplazar a la  Wide Field and Planetary 
Camera 1 (WFPC1) por la 
 WFPC2. La WFPC2 tiene la forma aproximada de un piano de cola.
La c'amara obtiene im'agenes  a trav'es de una selecci'on de 48 filtros
que abarcan un rango espectral  que va desde el UV lejano hasta longitudes de
onda en el rojo.
\label{fig020}}}
\end{wrapfigure}

Una de las mayores ventajas de un CCD es la linealidad de su respuesta
sobre un amplio rango de valores de entrada. El hecho de que sea lineal
significa que existe una simple relaci'on lineal entre el valor de entrada (la
carga colectada en cada p'ixel) y el valor de salida (el valor guardado  en la
imagen de salida).

El  n'umero m'aximo de cuentas en un solo p'ixel de un detector CCD
queda determinado por varios factores:
la capacidad del p'ixel, la saturaci'on y la zona de nolinearidad.

La cantidad de carga que un p'ixel puede contener durante una operaci'on
se denomina su {\it capacidad}. El n'umero de cuentas m'as grande que un CCD
puede producir queda determinado por el n'umero de bits en el transformador que
traduce la se\~nal anal'ogica en digital, y var'ia entre 0 y  $2^{\rm\scriptsize \mbox{(n'umero de  bits)} }-1$.
El transformador de la WFPC2 tiene un nivel de saturaci'on de aproximadamente
53\,000$e^{-}$ (con una ganancia de  14) o 27\,000$e^{-}$ (con una ganancia de 7) por p'ixel.

La exposici'on de objetos brillantes est'a limitada por efectos
de saturaci'on, los cuales aparecen por encima de  ~53\,000 fotones
detectados por p'ixel.

\subsection{La tecnolog'ia del detector CCD }
Los CCD de la  WFPC2 son detectores de silicio iluminados por su frente hechos
por  Loral Aerospace. Tienen una cubierta  especial que los hace sensibles a la
radiaci'on UV. 
Algunas de sus caracter'isticas   \citep{Bireetal02}  se pueden resumir as'i:

\begin{itemize}
	\item {\bf Ruido de lectura:} el ruido de lectura de los CCD es  aproximadamente cinco electrones.
	\item {\bf Ruido de corriente oscura:} los CCD est'an operando a $-88 ^{\circ}$C 
	y la corriente oscura media es de aproximadamentre 0.0045 $e^{-}$ p'ixel$^{-1}$ s$^{-1}$.
	\item {\bf Flat field:} los CCD de la  WFPC2 tienen una respuesta
	uniforme p'ixel--a--p'ixel, con una no--uniformidad  $< 2$ \% p'ixel--a--p'ixel.
	\item {\bf Eficiencia en la transferencia de carga (CTE):}  
	objetos brillantes sufren un p'erdida de se~nal
	de $\approx$ 4\% cuando 
	la imagen de la estrella se lee a trav'es de todas las filas del CCD.
	
	Las im'agenes m'as d'ebiles muestran un efecto mayor que parece ser una funci'on
	del tiempo. Para la mayor'ia de las aplicaciones, CTE es despreciable
	y calibrable. Este efecto no s'olo depende de la se\~nal, sino adem'as
	del fondo y  de la localizaci'on de la fuente entre otros par'ametros.
	 Ver Secci'on 2.2.9 para m'as detalles sobre el problema CTE.
	\item{\bf Ganancia:} los CCD tienen dos modos posibles de 
	ganancia: la primera, un canal de  7  
	$e^{-}$ DN$^{-1}$ que satura a aproximadamente 27\,000 $e^{-}$ y la segunda 
	de  14 $e^{-}$ DN$^{-1}$ que satura a 53\,000 $e^{-}$. 
	La capacidad de estos instrumentos es de  $\approx$90\,000 $e^{-}$ y son lineales
	hasta 4\,096 DN en ambos canales.
\end{itemize}

\subsection{Orientaci'on y lectura de los CCD}
 \begin{wrapfigure}[22]{l}[1mm]{80mm}
\includegraphics[width=80mm]{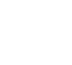}
\caption[Campo de observaci'on de la WFPC2]{ {\sl\footnotesize  
 Esta figura muestra el campo de observaci'on de la WFPC2, asi como
la orientaci'on (mediante ejes {\it x} e {\it y}) de cada chip.
\label{fig039}}}
\end{wrapfigure}

La orientaci'on de los ejes de los cuatro CCD est'an definidos
por una rotaci'on de $90^{\circ}$ respecto de los CCD adyacentes
como se muestra en la Figura~\ref{fig039}.
Si se obtienen una imagen con los 4 CCD y luego cada subimagen 
es desplegada con lineas en la direcci'on  ``{\it x}'' y columnas
a lo largo de la direcci'on  ``{\it y}'', cada imagen apareceria
rotada en $90^{\circ}$ respecto de la anterior.
Luego de pasar por los filtros, la luz del objeto observado es dividida 
en cuatro rayos mediante un espejo con forma de pir'amide localizado
en el plano focal del HST. Para asegurase que la imagen pueda reconstruirse
sin espacios entre los cuatro CCD, existe una peque~na regi'on
de intersecci'on entre cada CCD y sus vecinos.
Debido a que el rayo en el espejo con forma de pir'amide est'a aberrado,
los bordes de la regi'on de intersecci'on no quedan bien definidos,
sino que hay una transici'on gradual desde iluminaci'on cero a iluminaci'on m'axima
en cada CCD. Los objetos astron'omicos en esta regi'on de los chips 
aparecen en dos o mas CCD.

\subsection{Im'agenes de la WFPC2 }

Las im'agenes del HST son distribuidas por el Archivo del STScI
en formato FITS (del ingl'es: Flexible Image 
Transport System). Las im'agenes WFPC2 que empleamos para
nuestro an'alisis son las de sufijo {\sc c0f} (datos cient'ificos calibrados)
y las {\sc c1f} (datos de calidad).
Estas im'agenes fueron convertidas a formato GEIS (del ingl'es: Generic Edited Information Set) 
con lo cual obtuvimos  dos componentes: una   cabecera y un
archivo binario con datos.
La cabecera consiste en un archivo de texto ASCII con los registros  que especifican
las propiedades de la imagen y los par'ametros usados 
durante la observaci'on y el procesamineto de las mismas.
El archivo binario de datos, contiene uno o m'as grupos de datos.
Una exposici'on simple del instrumento  WFPC2  consiste de cuatro 
im'agenes, una por cada chip del CCD  \citep{Bireetal02}.
Los archivos GEIS usan un formato de grupos para
mantener a todas las exposiciones de una misma imagen juntas. La cabecera
de una imagen contiene la informaci'on que se aplica a la observaci'on como un todo.

\subsection{Procesamiento de datos en STScI}

Una vez que una observaci'on de la WFPC2 fue completada,
los datos pasan a trav'es del datoducto (pipeline) del STScI, donde
son procesados y calibrados.
Cuando una imagen es solicitada del archivo, se la reprocesa completamente
mediante un sistema denominado {\it On--the--Fly reprocessing} (OTFR) 
antes de ser entregada al astr'onomo. Durante el reprocesado,
se emplean los archivos de referencia m'as actualizados asi como
el mejor software disponible en ese momento \citep{wfpc2data}.

El paquete de programas usados
tanto en el pipeline como en el OTFR  es el mismo que est'a provisto
dentro de  {\sc\small  IRAF/STSDAS}  ({\sc\small  calwp2} en el 
paquete  {\sc\small  hst\_calib} ).
Lo que sigue es un resumen de los pasos de calibraci'on:

\begin{enumerate}
        \item {\bf M'ascara est'atica:}  una m'ascara est'atica  de referencia
        contiene un mapa de los p'ixeles malos y de las
        columnas con p'ixeles malos en cada chip de la WFPC2.

	\item {\bf Conversi'on anal'ogico--digital (A/D):} 
	el conversor A/D toma la carga observada en cada p'ixel del
	CCD y la convierte en un n'umero digital. Existen dos modos 
	o ganancias. El primero convierte una carga de aproximadamente
	siete electrones en una 'unica cuenta  (DN, del ingl'es:  Data Number),
	y el segundo convierte aproximadamente 14 electrones. Los
	conversores  producen peque~nos errores en la conversi'on entre electrones
	y DN.

	\item {\bf Determinaci'on del bias:} las cargas que est'an en cada p'ixel,
	se asientan sobre un pedestal electr'onico o   ``bias'', dise~nado  para
	mantener los nivedes de lectura y conversi'on por encima de cero todo
	el tiempo. El valor medio de este bias debe determinarse emp'iricamente
	usando el registro de p'ixeles que no son expuestos a la observaci'on.

	\item {\bf Remoci'on del bias:}  el valor del pedestal puede variar
	con la posici'on sobre el chip. El archivo de referencia de bias se genera
	sumando varios archivos con tiempo de exposici'on cero. 
	La correcci'on por bias consiste en substraer este archivo de referencia
	de la observaci'on.

	\item  {\bf Remoci'on de la imagen oscura:} una correcci'on  
	es necesaria para dar cuenta del efecto producido 
	por una corriente oscura inducida t'ermicamente.
	Existe un archivo de referencia que contiene la informaci'on
	de esta correcci'on y que se sustrae de la observaci'on.
	Todas las observaciones de mas de 10 segundos deben corregirse.

	\item {\bf Multiplicaci'on por flatfield:} el n'umero de electrones generados
	en un dado p'ixel por una estrella de una dada magnitud depende
	de la eficiencia cu'antica del mismo. Los flatfields de la WFPC2
	se generan a partir de una combinaci'on de datos de 'orbita
	(los llamados Earthflats que son im'agenes de la Tierra) e 
	im'agenes obtenidas en Tierra antes del lanzamiento. 
	Los flats de la WFPC2 son muy uniformes y no varian considerablemente
	con la  longitud de onda.

	\item {\bf Correcci'on por empleo del obturador:} como el obturador tiene
	una velocidad finita, cada vez que se abre/cierra, se produce una
	iluminaci'on despareja del chip. Este efecto es s'olo significativo en exposiciones
	de unos pocos segundos.

	\item {\bf Creaci'on de palabras clave:} las palabras clave (keywords), que proveen la 
	conversi'on entre cuentas calibradas (DN) y magnitud astron'omica,
	se calculan con la tarea {\sc\small  IRAF} {\sc\small  calwp2}    en el paquete 
	{\sc\small  synphot}. Los p'ixeles de la imagen con datos cient'ificos 
	no se cambian; los datos permanecen en DN;  {\sc\small  calwp2} simplemente
	calcula
	los par'ametros fotom'etricos y cambia las palabras clave en la cabecera.

	\item {\bf Creaci'on del archivo de datos cient'ificos:} 
	este archivo contiene la combinaci'on de los archivos creados previamente
	y es el resultado final del pipeline.
	
\end{enumerate}

\section{STIS: descripci'on  del instrumento e im'agenes}

\subsection{STIS  }
El  Space Telescope Imaging Spectrograph (STIS)  fue construido 
por la  Ball Aerospace 
Corporation  para el  Laboratorio de Astronom'ia y  F'isica Solar
del  Goddard Space Flight Center (GSFC).
Fue instalado durante la seguna misi'on de servicio el 14 de febrero de 1997 en lugar
del espectr'ografo GHRS. El 3 de agosto del 2004, STIS dej'o de funcionar
y se encuentra actualmente en modo  ``seguro''.

STIS es un instrumento vers'atil que permite obtener  tanto im'agenes como 
espectros con tres detectores bidimensionales que operan desde el UV hasta el IR cercano.
Este instrumento tiene la capacidad de obtener espectroscop'ia de rendija larga
en un amplio rango de longitudes de onda, asi como redes Echelle 
para obtener espectros en el UV.
Este instrumento puede usarse para obtener, entre otras cosas:
(1) espectroscop'ia de  red larga o sin red en el rango  1\,150--10\,300 \AA;
(2) espectroscop'ia   con redes Echelle con resoluci'on media y alta; (3)  
im'agenes CCD usando:  un detector en el UV lejano de   CsI ( Multi-Anode 
Microchannel Array, MAMA)  ($1\,150-1\,700$ \AA\ ); otro detector MAMA de  Cs$_2$Te 
en el UV cercano  ($1\,650-3\,100 $\AA ); y un CCD que act'ua en el 'optico
($2\,000-11\,000$ \AA).

 \subsection{Detectores de STIS}
 
STIS usa tres detectores de tama~no grande: $1024 \times 1024$  p'ixeles:
  
\begin{description}

\item[CCD]  detector  CCD Scientific Image Technologies (SITe),   de $0\arcs05$ p'ixeles$^{-1}$, 
que cubren un campo cuadrado de   $51 \asec \times  51 \asec $  y que opera en el
rango  $2\,000 - 10\,300$ \AA

\item[NUV--MAMA]  detector: CCD MAMA de  Cs$_2$Te   de  $0\arcs024$ p'ixeles$^{-1}$, 
que cubren un campo cuadrado de   $25 \asec \times  25 \asec $  y que opera en el
rango  $1\,650 - 3\,100$ \AA 
\item[FUV--MAMA]  detector: un MAMA  CsI,  de  $0\arcs024$ p'ixeles$^{-1}$, 
que cubren un campo cuadrado de   $25 \asec \times  25 \asec $  y que opera en el
rango  $1\,150 - 1\,700$ \AA 

\end{description}

Los detectores de STIS no operan en paralelo; s'olo uno puede ser usado a 
la vez.
La descripci'on t'ecnica completa de STIS puede hallarse en 
\cite{Woodetal98}.

\subsection{Estructura de los datos de STIS}
Los datos crudos de STIS son calibrados mediante la datoducto de STScI. 
En este proceso, se realiza la reducci'on de las im'agenes y de los espectros
para obtener como resultado archivos que pueden usarse directamente 
en  ciencia.

Todos los productos de STIS (im'agenes y espectros) son archivos  FITS que pueden
ser manipulados sin conversi'on alguna por  {\sc\small  IRAF/STSDAS}.
La estructura de un archivo STIS consiste de: una cabecera principal
que almacena la informaci'on de propiedades globales de todas las exposiciones  de
un archivo, y una serie de im'agenes cada una con su cabecera correspondiente
con datos propios de la misma.
Cada lectura de STIS produce tres im'agenes FITS de extensiones {\sc SCI, ERR} y {\sc DQ}.
La primera guarda los datos cient'ificos, la segunda contiene errores estad'isticos
y la tercera contiene los datos que dan la calidad de los datos para hacer ciencia.

 \subsection{Procesado de los datos en STScI }

 
Los datos obtenidos con el instrumento STIS son recogidos por
la  Space 
Telescope Data Distribution Facility
y enviados  a la datoducto del STScI, la cual extrae datos de las cabeceras
y crea  ficheros crudos, no calibrados pero cient'ificamente interpretables.
Cuando un usuario solicita datos de STIS del archivo, 'estos son procesados
por el sistema denominado {\it On--The--Fly Reprocessing }
(OTFR) que es un software de calibraci'on id'entico al 
provisto por {\sc\small  IRAF/STSDAS}  ({\sc\small  calstis}   en el paquete  {\sc\small  calib}).

El proceso incluye, como en el caso de la WFPC2, tales procesos como 
sustracci'on del bias, sustracci'on del dark y  aplicaci'on del flatfield, entre otros.

Los datos originales del MAMA tienen un formato crudo de $2048 \times 2048$ p'ixeles (los 
denominados p'ixeles de alta resoluci'on), mientras que los datos calibrados 
sufren un proceso de binning y son llevados al tama'no $1024 \times 1024$ p'ixeles.

\begin{figure*}[ht!]
\centering
\includegraphics[width=150mm]{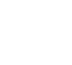}
\begin{minipage}{120mm}
\caption[Actividad extra vehicular]{ {\sl\footnotesize   
Astronautas desarrollando actividades extravehiculares durante la Misi'on de Servicio 
SM3A.  \label{fig105}}}
 \end{minipage}
\end{figure*}

%
\chapter{NGC 4214: una galaxia starburst enana prototipo}
\label{cha:ngc4214}
\thispagestyle{empty}
\hfill {\em  What's all this nonsense?}
\newpage 

\section{Ubicaci'on y clasificaci'on}
\begin{figure*}[ht!]
\centering
\includegraphics[width=\textwidth]{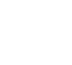}
\caption[Ubicaci'on de NGC~4214]{  {\sl\footnotesize  Mapa  de la constelaci'on 
Canes Venatici
en el hemisferio norte \citep{Tirietal01}. Arriba a la derecha, se muestra una im'agen
blanco y negro del 
Digitized Sky Survey, con NGC~4244 y NGC~4214.
Abajo a la derecha,  se muestra un mosaico de  NGC~4214 
obtenido con la combinaci'on de im'agenes en filtros
 {\it B, V} y {\it R} obtenidas con el Telescopio  Isaac 
Newton.  \label{fig011}}}
\end{figure*}

NGC~4214 es una galaxia de tipo Magallanes en gran parte  resuelta ubicada en 
el complejo 	Osa Mayor, en una regi'on denominada Nube  Canes Venatici (CVn) I, 
la cual contiene, adem'as, a las galaxias NGC 4190, NGC 4395, NGC 4449 e IC 4182, 
entre otras galaxias bien resueltas de acuerdo a   \cite{Sand94}. 
Est'a ubicada en 
 $\alpha_{2000} = 12^{h} 15^{m} 38\arcs 9 $ , 
$\delta_{2000} = +36 ^{\circ} 19' 39\arcs0 $ de acuerdo al cat'alogo de 451 galaxias 
cercanas de   \citet{Karaetal04}  y a una distancia de $2.94 \pm 0.18 $ Mpc
 \citep{Maizetal02a}. 
La Figura~\ref{fig011}  presenta un mapa obtenido de \cite{Tirietal01}  que muestra la posici'on de 
la misma en coordenadas ecuatoriales.
 
El panel 21 del cat'alogo de 95 galaxias de \citet{SandBedk85} 
contiene una fotograf'ia de  NGC~4214, y ah'i  fue clasificada como
SBmIII. \citet{Vauc91} la clasifican como IAB(s)m.
En el {\it Hubble Atlas of Galaxies}  \citep{Sand61}  se 
 clasifica a NGC~4214 como  Irr I, mientras que en  el  {\it Carnegie Atlas of Galaxies}
se la lista como  SBmIII  en el Panel  330 \citep{Sand94}.

Su falta de estructura global definida hace que en algunas referencias se
la clasifique como irregular o amorfa. 
En luz visible aparece una barra central y unos brazos espirales incipientes parecidos a 
los que presenta la Nube Mayor de Magallanes. 

\section{Observaciones de NGC 4214  }  

NGC~4214  ha sido estudiada en varias longitudes de onda.
Cada una de ellas provee informaci'on sobre su 
estructura y composici'on:

J. Gallagher observ'o a  NGC 4214 en septiembre de 1979 y junio de 1980 
usando filtros  
{\it B, V,  R}, \ha\  e {\it I} con el sistema de video-c'amaras del KPNO 
usando el  telescopio de 2.1 m  con el detector  RCA 4849 ISIT  \citep{Hunt82}. 
Las im'agenes resultantes tienen  256 $\times$ 256 p'ixeles, con una escala de
 $0 \arcs 55$ p'ixel$^{-1}$ 
en un campo de aproximadamente $2\amin$. En marzo de  1980,  usaron el 
espectr'ografo Echelle del telescopio de 4 m de KPNO 
para obtener espectros de alta resoluci'on de 
las l'ineas de emisi'on  \citep{Hunt82}.

 \cite{Huchetal83a} recolectaron fotometr'ia $UBV$
  e infraroja de  NGC 4214   usando diferentes aperturas  
 de acuerdo a   \citet{Vauc61},  \citet{Neff70} y  
 \citet{Aaro77}. En marzo de 1981, se obtuvieron im'agenes  UV de esta galaxia
 usando la c'amara  SWP del  International Ultraviolet 
 Explorer (IUE) en el modo de baja dispersi'on entre 1\,150 \AA\ y 
 1\,950 \AA\ .

\cite{Hartetal86} obtuvieron sus datos con el detector  Reticon combinado con
el espectr'ografo Echelle  del reflector  Tillinghast de 1.5 m 
en Mount Hopkins. Se obtuvieron espectros  en regiones   \ha  ~y 
[O~{\scriptsize III}~{\normalsize ]} 
 con exposiciones en el rango de 5  a 390 min. 
Obtuvieron espectros de  22  regiones \hii\  en  \ha\ y de diez regiones 
 en  [O~{\scriptsize III}~{\normalsize ]}.

 \cite{Throetal88} realizaron observaciones en el IR y en CO. Las observaciones 
 en el IR lejano fueron realizadas en $160 \mu$m
en 1986 usando el telescopio de 0.9 m abordo del Kuiper Airborne Observatory.
Las observaciones en el IR cercano en $K$ ($2.2 \mu$m) se realizaron con el telescopio
de 2.3 m del Wyoming Infrared Observatory (WIRO) durante mayo de 1986.  En octubre de 1986, 
NGC~4214 fue observada con el telescopio de 12 m NRAO en Kitt Peak en Arizona.
El sistema fue sintonizado para detectar la transici'on $J=1\rightarrow0$ de $^{12}$C\,\,$^{16}$O
a 115 GHz.
 
\cite{SargFili91} obtuvieron espectros 'opticos en febrero de 1987  
con el telescopio reflector  Hale  de 5 m en el  Palomar Observatory.  Emplearon el 
espectr'ografo doble montado sobre el foco Cassegrain con dos 
CCD Texas Instruments de  800$\times 800$ p'ixeles. Las c'amaras roja y azul
grabaron luz en los rangos  de  longitud de onda 
6\,200--6\,860 \AA\ y
4\,210--5\,100 \AA\, respectivamente.
Tambi'en obtuvieron im'agenes directas de NGC~4214 en enero de 1991
con el reflector de 1 m en el  Lick 
Observatory. Usaron un CCD Texas Instruments de 500$\times$500 p'ixeles.

En abril de  1987 y en diciembre de  1988, \cite{Becketal95}  realizaron 
mediciones en CO usando  la antena
IRAM de 30 m  en  Pico Veleta en Espa\~na. 
Tambi'en obtuvieron im'agenes CCD de   NGC~4214 en el Hoher List 
Observatorium de la Universidad de  Bonn en  mayo de  1993 con el telescopio de  
1 m.  El detector usado fue una c'amara CCD  
equipada con un chip  Loral  de 2048$\times $2048 p'ixeles.

\begin{figure*}
\begin{minipage}[h]{7cm}
\includegraphics[height=7cm]{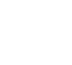}
\caption[ Imagen UV  (2200 \AA ) de NGC 4214--I]{ 
 {\sl\footnotesize   Imagen UV  (2200 \AA ) de NGC 4214--I  observada con el instrumento  Faint  
Object
  Camera (FOC)  a bordo del $HST$. 
La im'agen tiene   $\approx 22'' \times 22''$;  la escala de placa es  
$0 \arcs 022497$ p'ixel$^{-1}$   \citep{Leitetal96}. 
\label{fig014}}}
\end{minipage}%
\hspace{1cm}%
\begin{minipage}[h]{7cm}
\centering
\includegraphics[height =7cm]{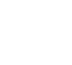}
\caption[Imagen CCD de NGC~4214]{ {\sl\footnotesize  Imagen CCD de 
NGC~4214. Representa la diferencia entre 
una imagen obtenida a trav'es de un filtro angosto centrado en  $\lambda = $ 6606 \AA\ 
y la imagen del continuo centrada en  $\lambda = $ 6477 \AA\ \citep{SargFili91}.
\label{fig013}}}
\end{minipage}
\end{figure*}

\cite{Leitetal96} obtuvieron una imagen de  NGC 4214 con el instrumento  Faint  Object
Camera (FOC) a bordo del $HST$ en marzo de 1993 (Figura~\ref{fig014}).  
La imagen original consiste de
512 $\times$ 1024 p'ixeles y un campo de  $\approx 22 \asec \times 22 \asec.$ 
 La observaci'on fue realizada con los  filtros F220W   
  y F2ND.
Obtuvieron espectros del n'ucleo brillante con el instrumento
Faint Object Spectrograph  (FOS) en noviembre de  1993 usando las grillas  G130H y 
G190H. El rango  de longitud de ondas cubierto por las dos 
mediciones es 
1\,200--2\,300 \AA .

\begin{figure*}[ht!]
\begin{minipage}[c]{16cm}
\centering
\includegraphics[width=16cm]{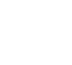}
\caption[Im\'agenes de NGC~4214: UIT y Observatorio Palomar]{{\sl\footnotesize
[Izquierda] Contornos en el filtro $I$ dibujados sobre una imagen en el FUV de 
NGC~4214. La im'agen de fondo fue obtenida con el UIT usando el  filtro 
B1. 
[Derecha] Imagen en el filtro $I$ de  NGC~4214 obtenida con el telescopio de 5 m 
del  Palomar Observatory.  \label{fig023}}}
\end{minipage}
\end{figure*}

 \cite{KobuSkil96}  obtuvieron espectros en un rango de longitudes de onda
de  3\,650--7\,100 \AA\ con el  CCD de 3000 $\times$ 1000  p'ixeles  en el telescopio de 2.1 m en Kitt Peak
durante enero y febrero de 1995.
 \cite{Faneetal97} observaron  
 a  NGC~4214 usando el Ultraviolet Imaging Telescope  (UIT), 
durante la misi'on  Astro--2 
en marzo de 1995. El  UIT obtuvo exposiciones de  
 1060 segundos en el filtro ancho  B1.
Las im'agenes del UIT fueron grabadas en film y luego digitalizadas
con un microdensit'ometro, 
 generando asi una imagen con una resoluci'on de $ 1\arcs 14$ p'ixel$^{-1}$. 
Tambi'en obtuvieron im'agenes directas de NGC~4214
en los filtros  Johnson $B$ y  Cousins $I$ 
en el telescopio de 5 m del  Palomar Observatory en febrero de  
1994 que se muestra en la Figura~\ref{fig023}.

 \cite{McIn98}  observ'o a NGC~4214 usando las configuraciones B, C y 
 D del VLA en 1993 y 1994 para
estudiar detalladamente la cinem'atica del gas neutro \hi.

 \cite{Maizetal98}  obtuvieron espectros 'opticos de  NGC~4214 en abril
 de 1992 con el espectr'ografo del telescopio William Herschel. 
Los espectros fueron obtenidos en doce posiciones diferentes.
Se obtuvieron dos espectros simult'anemante de cada posici'on:
uno en el rango  6\,390--6\,840 \AA\  y el otro en el 
rango entre  4\,665 
y 5\,065 \AA\ , ambos con una dispersi'on de aproximadamente 0.4 \AA\ / p'ixel. 
Los tiempos de exposici'on variaron entre 900 y 1200 segundos. 
Con estas observaciones produjeron un conjunto de datos tridimensionales   
 (2D espacial + 1D longitud de onda) que pudo ser arreglado en forma de mapas.

\begin{figure*}[ht!]
\begin{minipage}[c]{16cm}
\centering
\includegraphics[width=16cm]{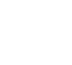}
\caption[Im'agenes de NGC~4214 de \citep{MacKetal00} ]{ {\sl\footnotesize  
[Izquierda] Imagen \ha\  de NGC~4214 en escala logar'itmica
con el continuo sustra'ido.  
[Derecha] Mosaico a color de  NGC~4214. El canal rojo corresponde a 
l'ineas de emisi'on en \ha\, el canal verde a  la l'inea de emisi'on
 [O~{\scriptsize III}~{\footnotesize]}, y el canal azul al continuo en \ha.
 La intensidad en cada caso es logar'itmica.  
Las marcas se muestran cada cien p'ixeles  \citep{MacKetal00}. \label{fig015}}}
\end{minipage}
\end{figure*}

\cite{MacKetal00} obtuvieron im'agenes de muy alta resoluci'on en
varios filtros   con el instrumento  WFPC2 abordo del HST 
(propuesta   6569) en julio de 1997.  
Usaron cuatro filtros de banda ancha (F336W, WFPC2 $U$; F555W, 
WFPC2 $V$; F702W, WFPC2 $R$ ancho; y F814W, WFPC2 $I$) 
y dos filtros nebulares (F656N, \ha\ y F502N,  [O~{\scriptsize III}~{\normalsize]} ) para  
estudiar a la galaxia en 15 exposiciones.
La posici'on de las c'amaras fue seleccionada para minimizar los 
defectos producidos por  CTE.
Las im'agenes de banda ancha consisten en dos exposiciones
largas y una corta; esta 'ultima se emplea principalmente para
corregir por posible saturaci'on de alguna fuente intensa.
Ver Figura~\ref{fig015}.
Estos autores disponen adem'as de observaciones en radio realizadas
con la NRAO Very Large Array (VLA)  en abril de 1988 en la configuraci'on C
en 6 cm y en marzo de 1989 en la configuraci'on B en 20 cm.

\cite{Waltetal01}  observaron NGC~4214 en la   
 transici'on $J=1\rightarrow0$ de CO
  usando un interfer'ometro en  el Owens Valley Radio Observatory (OVRO)
Ellos presentan el primer an'alisis interferom'etrico del gas molecular
en la galaxia, donde descubrieron  tres regiones de emisi'on molecular que son muy diferentes
entre s'i y que se encuentran en estados de evoluci'on diferentes.

\begin{figure*}[htdp]
\centering
\includegraphics[width=16cm]{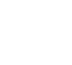}
\caption[Mosaico de NGC~4214 construido
con  im'agenes del   telescopio Isaac 
Newton]{ {\sl\footnotesize  Mosaico de NGC~4214 ($ 9' \times 9' $) obtenido con una combinaci'on RGB 
de im'agenes p'ublicas en los filtros $B, V$ y $R$ provenientes del telescopio Isaac 
Newton. Se muestran los campos de la  WFPC2 (azul; propuesta 6569) y NICMOS 
(rojo; propuesta 7859) \citep{Droz02}. 
\label{fig016}}}
\end{figure*}

\cite{Droz02} hicieron uso de las observaciones WFPC2 de archivo
en filtros  $V, R, I, $ \ha, y  [O~{\scriptsize III}~{\normalsize ]}  de la propuesta  6569 del HST. 
Tambi'en emplearon 
im'agenes NICMOS de las propuesta 7859 en filtros
 F110W  ($J$) y F160W ($H$).
Las observaciones NICMOS fueron obtenidas en julio de 1998. Este
instrumento  tiene tres c'amaras:   NIC1, NIC2 y  NIC3 ordenadas linealmente
como puede verse en la Figura~\ref{fig016}.  La c'amara
NIC2 fue centrada en NGC~4214-II, la m'as peque\~na
y joven de las dos regiones m'as importantes de formaci'on estelar.

\cite{Hartetal04} observ'o a  NGC~4214 usando el instrumento
 Advanced CCD Imaging 
Spectrometer (ACIS) a bordo del telescopio  Chandra en septiembre  de 2001.
Toda la galaxia fue incluida en un 'unico chip. En noviembre  de 2001, 
observaron  a NGC~4214 usando el  telescopio   $XMM-Newton$.

\cite{Calzetal04}  emplearon im'agenes de archivo  de  NGC~4214 obtenidas con la  WFPC2
en las  propuestas  6569 y  9144. Para la 'ultima propuesta, se usaron filtros
 F487N y F673N.
En octubre de 2004 se realizaron observaciones de NGC~4214 empleando la ACS a bordo del
HST de acuerdo a la propuesta de observaci'on 10332 (PI: Holland Ford, The Johns Hopkins University).
Se obtuvieron im'agenes en los filtros F330W, F555W y F814W.

\cite{Kiucetal04}  realizaron observaciones en el la regi'on 
submilim'etrica del espectro electromagn'etico (450 y 850 $\mu$m).
Su objetivo era estudiar la distribuci'on  de energia y las propiedades de la emisi'on t'ermica en un
objeto de baja metalicidad como NGC~4214.

\citet{Vukoetal05} buscaron remanentes de supernovas
en NGC~4214 usando otras observaciones realizadas con el VLA. 
Una observaci'on en 20 cm fue realizada en mayo de 1986; las observaciones
en 6 cm y  3.5 cm
fueron realizadas en abril de 1997 en la configuraci'on B del instrumento.

\section{Metalicidad de NGC~4214}
 \citet{KobuSkil96} realizaron un an'alisis  espectrosc'opico  de  NGC~4214
\begin{wrapfigure}[35]{l}[0mm]{80mm}
\includegraphics[width=80mm]{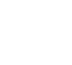}
\caption[Estimaci'on de la metalicidad de NGC~4214]{  {\sl\footnotesize  
Valores de metalicidad ($Z$)  graficados en funci'on de \logoh seg'un  \cite{Massey03}.
Se obtiene el valor $Z  =  0.006 $ para NGC~4214 mediante interpolaci'on lineal.
 \label{fig012}}}
\end{wrapfigure}
 para determinar
 de qu'e manera los vientos
y explosiones de supernovas contribuyen 
 al enriquecimiento localizado del medio interestelar en galaxias de baja
 metalicidad. Estudiando la abundancia de elementos
 encontraron que NGC~4214 tiene una deficiencia de metales, y una abundancia de
\logoh $= 8.15-8.28$, y consideran que una galaxia es pobre en metales si 
\logoh $ < 8.4 $. Suponiendo que  \logoh $= 8.22$ para
NGC~4214, e interpolando linealmente en los resultados provistos por  
\cite{Massey03} para las Nubes de Magallanes y el entorno solar,
obtenemos $Z = 0.34 Z_\odot  =  0.006 $.

\bigskip	
\bigskip	
\section{Distancia}

El estudio de la distancia a NGC~4214 fue tratado en varias ocasiones
y se han empleado varios m'etodos para su determinaci'on.

 \cite{SandTamm74a}  calibraron un m'etodo para determinar 
 distancias a galaxias tard'ias usando el tama\~no de las regiones \hii y las
 estrellas  resueltas m'as brillantes.
 En  \cite{SandTamm74b}   usan este m'etodo para determinar
 la distancia a 39 galaxias, incluyendo a NGC~4214.
 Midieron el tama\~no angular de las tres regiones \hii mas grandes sobre im'agenes
 obtenidas con el telescopio Hale de 5 m.
 Con este m'etodo obtienen una  distancia  de 6.3 Mpc para NGC~4214.

\cite{Leitetal96}  calcularon la distancia a NGC~4214
usando la velocidad Galactoc'entrica observada de   313 km~s$^{-1}$,
la velocidad del c'umulo de Virgo referida al centroide del Grupo Local
 (976 km~s$^{-1}$ ),  la velocidad   del Grupo Local hacia 
 Virgo  (220 km~s$^{-1}$ )  y la distancia al c'umulo de Virgo (17.6 Mpc).  
La distancia obtenida por este m'etodo es 4.1 Mpc. 
 
 \cite{Rowa85} establecieron que la distancia al grupo  de galaxias
 Canes Venatici I  es 5.1 Mpc, mientras que  \cite{Saha94}
determinaron una  distancia de  4.7 Mpc  a este  grupo de galaxias
a partir de la presencia de Cefeidas en un miembro del grupo:  IC 4182.

 \cite{Maizetal02a}  emplearon el m'etodo del extremo  superior
 de la rama de las gigantes rojas
(TRGB, del ingl'es: Tip of the Red Giant Branch)  para derivar la distancia a  NGC~4214. 
Este m'etodo se ha transformado en un m'etodo muy confiable para la 
determinaci'on de distancias  a galaxias con poblaciones estelares 
resueltas \citep{Leeetal93}; esto es posible debido a la 
d'ebil dependencia de la magnitud absoluta de la TRGB con la metalicidad
\citep{DaCoArma90,Belletal01}.
Otra ventaja de este m'etodo (en contraposici'on al uso de estrellas variables
como  Cefeidas or RR Lyrae) es que s'olo se necesita un 'unica observaci'on 
para estimar la distancia.  Su confiabilidad ha sido probada
recientemente obteniendo distancias consistentes a las Nubes de Magallanes e IC~1613
usando diferentes m'etodos: Cefeidas, RR Lyrae, red clump
y TRGB \citep{Dolpetal01a}.
El uso de datos de la WFPC2 permite hacer mediciones de distancia 
con el m'etodo TRGB,  hasta 
 $\approx 5$ Mpc usando una sola 'orbita  \citep{Karaetal01,Tosietal01,Dolpetal01b} 
 y  el uso de la nueva c'amara ACS  ha extendido este rango de distancias
 en un factor de dos.
 El valor obtenido por este m'etodo para NGC~4214 es $2.94 \pm 0.18 $ Mpc.

\cite{Droz02} presentaron una estimaci'on de la distancia a NGC~4214
usando este mismo m'etodo.
El valor obtenido por ellos es $2.7 \pm 0.3 $ Mpc.

\begin{table}[h]
\caption[Distancia a  NGC~4214]{Determinaci'on de la distancia a  NGC~4214}
\begin{center}
\begin{tabular}{lll} 
\multicolumn{3}{c}{ \rule{130mm}{0.8mm}}      \\
 Paper  & M'etodo & Distancia    {\rule [-3mm]{0mm}{8mm}  }\\ 
 \multicolumn{3}{c}{ \rule[2mm]{130mm}{0.2mm}}      \\
      \cite{SandTamm74a}         & tama\~no de regiones  \hii  & 6.3 Mpc  \\ 
    \cite{Leitetal96}                   & din'amica estelar al grupo Virgo   & 4.1 Mpc  \\ 
    \cite{Rowa85}                   & diferentes m'etodos   & 5.1 Mpc  \\ 
    \cite{Saha94}                   & Cefeidas en IC 4182   & 4.7 Mpc  \\  
      \cite{Maizetal02a}             & TRGB  & 2.94  Mpc  \\ 
    \cite{Droz02}           & TRGB  & 2.7 Mpc  \\  
\multicolumn{3}{c}{ \rule{130mm}{0.8mm}}      \\    
\end{tabular}
\end{center}
\label{tbl007}
\end{table}%

La distancia a NGC~4214 adoptada para este trabajo es 
 $2.94 \pm 0.18 $ Mpc.

\section{Formaci'on estelar y la IMF de NGC~4214}

Las propiedades de la formaci'on estelar, como la forma
de las funci'on de masa inicial (IMF) y  la velocidad de
formaci'on estelar (SFR, del ingl'es: Star Formation Rate)  
juegan un papel muy importante
en la evoluci'on de las galaxias.
Desafortunadamente, los procesos de formaci'on estelar
son muy complejos y    todav'ia no se entiende muy bien
c'omo se relacionan 
con las propiedades globales de las galaxias.

La velocidad de formaci'on estelar 
es uno de los par'ametros m'as importantes
en el estudio de la evoluci'on de las galaxias.

Cuando las estrellas j'ovenes no pueden ser resueltas, las propiedades
de formaci'on estelar en la galaxia se obtienen, principalmente, a partir de mediciones
de luz en el ultravioleta y/o  infrarrojo lejano.
Estos m'etodos se basan principalmente en modelos te'oricos de poblaciones sint'eticas.
La SFR de NGC~4214 ha sido estudiada con cierto detalle por varios autores:

\cite{Huchetal83a} usaron modelos evolutivos simples
para ajustar  espectros observados. 
Hallaron que diferentes modelos con varias caracter'isticas pueden 
responder a los espectros observados.
Descartaron modelos con una SFR decreciente y con SFR constante
pues estos no producen el tipo de estrellas observadas.
Los modelos que  proveen un ajuste razonable generaron  un 5\% de su masa 
en un brote de formaci'on estelar reciente
($ 20  < t < 100 $  Ma), y est'a superpuesto a una poblaci'on m'as vieja
que ha experimentado una SFR aproximadamente constante en el pasado.
suponen adem'as que la galaxia ha experimentado un brote de
formaci'on estelar hace algunos $10^7$ a'nos.

 \cite{Galletal84} exploraron la historia de formaci'on estelar (SFH, del ingl'es:
 Star Formation History) de NGC~4214
 usando tres  par'ametros: (1) la masa de la galaxia en forma de estrellas,
 que mide la SFR integrada sobre la vida completa de la galaxia; (2) la luminosidad en el azul,
 que est'a dominada principalmente por  estrellas 
 generadas en los  'ultimos millones de a\~nos; y (3) 
 el flujo de fotones Lyman derivado a partir de la luminosidad \ha\, que da 
 la SFR actual.
 Estos autores encontraron que NGC~4214 es una galaxia irregular en un estado de 
 brote o post--brote de formaci'on estelar.
  
  \cite{Hartetal86} proponen a partir de un estudio del campo de velocidades
  de las galaxias irregulares NGC~4214 y NGC~4449 
que ambas deben haber experimentado una reciente colisi'on
con algun sistema peque'no externo o con alguna nube de gas.
Esta colisi'on  explicar'ia la compresi'on del gas en NGC~4214 y el resultante
brote de formaci'on estelar.
 
 \cite{Throetal88}  sugirieron que la formaci'on estelar 
 en NGC~4214 ha migrado de posici'on en posici'on  y que en al actualidad
 est'a concentrada en su centro.
Emplearon la luminosidad en el IR lejano para estimar la 
SFR presente, asumiendo una IMF del tipo Salpeter entre    0.1 y 100 $M_\odot  $, 
y  sostienen que  la velocidad de  
$0.5-1 M_\odot  \rm{a}^{-1},$
se ha mantenido constante en los 'ultimos millones de a\~nos
dentro de un factor de 2 o 3. 
Teniendo en cuenta las incertezas, estos autores interpretan que
la SFR de la galaxia ha sido pr'acticamente constante
desde su formaci'on.
Sin embargo, varias partes de la misma, como su centro, pueden haber
experimentado per'iodos de aumento en la creaci'on estelar.
Sus resultados no concuerdan con el modelo de brotes 
de \cite{Huchetal83a}.

Las observaciones de   \cite{SargFili91}  han revelado la presencia de 
l'ineas en  emisi'on  en  4660 \AA\  en los espectros 'opticos de dos
regiones centrales de la galaxia NGC~4214. Esta l'inea corresponde
a He~{\scriptsize II }~{\normalsize }~$\lambda$4686  a la distancia de NGC~4214.
Esta caracter'istica espectral es propia de las estrellas WR.
Su an'alisis sugiri'o que estas regiones deben contener n'umeros semejantes de
estrellas O y WR. Interpretaron este fen'omeno como el resultado de un brote muy
intenso de formaci'on estelar, con una duraci'on de menos de 
$10^6$ a\~nos,  y que ha ocurrido hace menos de 5 Ma.

\cite{Leitetal96} modelaron el espectro UV de  NGC~4214--I, 
usando la t'ecnica de modelos sint'eticos descripto por 
  \cite{Leitetal95b}. 
Sus ajustes  sugieren que el modelo de brotes es m'as apropiado 
que el modelo de formaci'on estelar continua, y 
tambi'en infieren que esta galaxia debe tener una IMF
que sigue una ley de potencias cuyo exponente est'a entre 2.35 y 3.00.
Sus modelos sugieren una edad de $4-5$ Ma para   NGC~4214--I

 Para interpretar los espectros en el FUV de NGC~4214, 
\cite{Faneetal97}  construyeron modelos  simples  para
dos casos extremos: un brote de 3 Ma con una IMF de pendiente
$\gamma=-2.35$, un rango de masas  $0.1 < M_\odot < 100 $,
$Z = 0.4  Z_{\odot}$,  y otro modelo con los mismos par'ametros
pero con formaci'on estelar continua.
Determinaron una SFR de   $0.05  M_\odot  \rm{a}^{-1}$, 
consistente con  el derivado por
 \cite{Galletal84} dentro de los errores.

\cite{Droz02}  presentan un resumen de los valores de la SFR
referidos a la nueva distancia determinada por ellos usando el 
m'etodo de la TRGB.
Usando los resultados de  \cite{MacKetal00} en \ha,
derivan una SFR de $ 0.08-0.99
M_\odot  \rm{a}^{-1}.$
Los resultados de \cite{Faneetal97} dan  $0.08 \, M_\odot  \rm{a}^{-1},$
y est'an de acuerdo con los valores observados de \ha\ .

Resultados de estudios en radio (flujo en 1,4 GHz proveniente
del centro de la galaxia obtenido por  \cite{Alls79} ) 
dan una SFR de $0.06 M_\odot  \rm{a}^{-1},$ 
mientras que  observaciones con {\it IRAS}  en  60 
$\mu$m,  da una SFR de 0.17 $M_\odot  \rm{a}^{-1}.$

Existen muy pocos art'iculos en los que se haga referencia a la IMF de NGC~4214.
Todos estos  estiman a  la pendiente de la IMF ajustando el espectro integrado de 
parte de la galaxia
mediante modelos evolutivos sint'eticos.
\cite{Leitetal96}  modelaron el espectro UV y concluyeron que 
la IMF sigue una ley de potencias con un exponente comprendido entre
 $-2.35$  y  $-3.00.$
\cite{Chanetal05} 
compararon espectros de NGC~4214 con modelos provistos por \s99 \citep{Leitetal99}, e
 infieren el valor   $\gamma =  -3.5 $ para  NGC~4214.
Usando un enfoque similar, pero con espectros IUE
\cite{MasHKunt99} pudieron restringir la pendiente de la IMF de NGC~4214 a
$ \gamma = -3.0$. Claramente, sus espectros incluyen  estrellas en c'umulos as'i
como estrellas de campo.

\section{Estructura de objetos j'ovenes en NGC~4214}    

De acuerdo a \citet{MacKetal00}   la morfolog'ia asociada  con las im'agenes 
de banda angosta de NGC~4214 muestran tres componentes diferenciadas: 	

\begin{enumerate}
\item  Dos complejos H~{\scriptsize II, }~{\normalsize} conocidos en la literatura como NGC 4214--I (o
el complejo NW) y NGC 4214--II (o  el complejo SE).

\item Una cierta cantidad de n'ucleos aislados d'ebiles secundarios  dispersos por todo el campo.

\item Una estructura extendida y amorfa de gas interestelar  (DIG, del ingl'es: 
 Diffuse Interstellar Gas)  que rodea a los dos
complejos principales y a algunos de los secundarios. 
\end{enumerate}

Las estructuras individuales visibles en  \ha\ han sido identificadas
en el pasado por varios autores y  \citet{MacKetal00}  desarrollaron 
una nomenclatura para las  13 unidades  principales
de NGC~4214 que representan la poblaci'on joven
detectable en el visible.  En este trabajo adoptamos esta nomenclatura.

\begin{figure*}[ht!]
\begin{minipage}[c]{15cm}
\centering
\includegraphics[width=15cm]{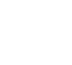}
\caption[Nomenclatura de complejos
principales]{ {\sl\footnotesize   Imagen \ha\  de NGC~4214 en escala logar'itmica
con el continuo sustra'ido.  Se muestra la nomenclatura de algunos complejos
as'i como su correspondiente apertura circular \citep{MacKetal00}. 
\label{fig017}}}
\end{minipage}
\end{figure*}

\begin{figure*}[ht!]
\begin{sideways}
\begin{minipage}[c]{15cm}
\centering
\includegraphics[width=15cm]{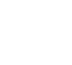}
\caption[Nomenclatura de los complejos
en la regi'on central de NGC~4214]{{\sl\footnotesize   Area 
central de la imagen \ha\ con el continuo sustra'ido.
La escala de intensidad es logar'itmica entre 20 y 1000 $10^{-16}$ erg s$^{-1}$ 
cm$^{-2}$ arcsec$^{-2}$. Se muestran las aperturas con excepci'on de las regiones
I y II. Se muestran ademas los nombres de todas las regiones
con excepci'on de las peque\~nas  I--A1n y I--A3n.
Las coordenadas se miden en p'ixeles comenzando desde el extremo inferior izquierdo
del campo de la WFPC2.
\label{fig018}}}
\end{minipage}
\end{sideways}
\end{figure*}

Las primeras dos unidades, NGC 4214--I y NGC 4214--II,
corresponden a los complejos \hii\ principales.
NGC 4214--III y NGC 4214--IV son dos fuentes compactas ubicadas hacia el NO.
Luego, desde NGC 4214--V a NGC 4214--XIII 
son estructuras cada vez m'as d'ebiles. Cuando es apropiado, se definieron 
subestructuras dentro de cada unidad asign'andoles letras y nombr'andolas
 I--A, I--B, etc.
Si  se encuentran unidades aun m'as peque\~nas, se les asigna un n'umero (I--A1,
 I--A2, etc). Finalmente, se agrega una letra  {\em n}  al nombre si la apertura fue definida
usando la imagen de banda angosta o una letra {\em s}
si la apertura de la regi'on fue definida con una im'agen de banda ancha.
La posici'on de cada unidad y sus componentes esta incluida en las Figuras~\ref{fig017} y 
\ref{fig018}. En esta Secci'on hacemos una breve descripci'on de las unidades en NGC~4214.

\subsection{NGC~4214--I}     
	
De acuerdo a  \citet{MacKetal00}  	NGC 4214--I es el complejo \hii\ m'as grande
en la galaxia y tambi'en el que tiene una estructura m'as compleja.
Incluye varios c'umulos estelares y una estructura visible en  \ha\ dominada por la presencia
de dos cavidades o m'inimos de intensidad.

I--As  es un   superc'umulo estelar (SSC)  joven  ($3.0-3.5$ Ma) que fue estudiado
previamente usando  observaciones  del HST obtenidas con el FOC por  \citet{Leitetal96}. 
Estos autores encontraron que el SSC  es peque\~no  (di\'ametro $\lesssim$~5~pc)
y contiene varios cientos de estrellas O as'i como estrellas WR.
Adem'as se encuentra que su espectro contiene l'ineas en emisi'on
como es de esperarse  \citep{SargFili91,Maizetal98}. 
I-As est'a localizado en una cavidad \ha\ con forma de coraz'on.
Se detecta muy poca emisi'on \ha\ dentro de la cavidad con la excepci'on de la 
proveniente del n'ucleo de la SSC.
La emisi'on  alrededor de I-As tiene una estructura sin simetr'ia 
respecto al SSC. 
En el borde norte de la cavidad, I--A2n  contiene una delgada pared recta, 
que es probablemente el frente de ionizaci'on producido por I--As.
La estructura del borde oriental, I--A4n, es bastante similar.
Ambos se juntan formando un 'angulo recto lo cual define la forma de coraz'on.
Los bordes del sur y oeste (I--A3n, I--A1n y  I--A5n)  
presentan un estructura mucho m'as
compleja.
En esta regi'on el borde no est'a tan bien definido y contiene
algunos n'ucleos compactos muy intensos de \ha.

I--Bs es una asociaci'on OB a gran escala (SOBA, 
\cite{Hunt99}) de unos $3.0-3.5$ Ma.
El n'umero de estrellas O en I-Bs es un 
$\sim$ 30\% de las que contiene I--As (\cite{Maiz99}) y  contiene algunas
estrellas WR \citep{SargFili91}. 
Esta estructura est'a contenida dentro de una cavidad \ha\ no tan bien
definida como la de I--As. Sin embargo, el gas dentro de la cavidad muestra
que es una estructura  en expansi'on de acuerdo a estudios
espectrosc'opicos \citep{Maizetal99a}.
Las estructuras \ha\ que rodean a la cavidad parecen  estar formadas por
una serie de estructuras lineales junto a un n'ucleo intenso (I--B1n)
en el borde SO. En la imagen \ha\ la cavidad aparece rota 
entre  I--B1n y I--B2n. En las cercan'ias de esta ruptura  \cite{Maizetal99a} 
detectaron una burbuja en expansi'on que podr'ia ser la causante
de la ruptura misma.

Entre I--A y I--B se encuentran dos c'umulos m'as peque\~nos: I--Es y I--F.
I--Es tiene una estructura bastante compacta, por lo que se piensa que es un candidato a 
SSC (aunque seguramente m'as peque\~no que I--As). La emisi'on \ha\ que la rodea no es muy intensa
y parece ser parte del fondo de NGC~4214--I. Adem'as, los colores del continuo en los datos de 
\cite{Maizetal98}  son significativamente m'as rojos que aquellos de I--As o I--Bs,
lo cual indica una edad m'as avanzada  ($\gtrsim$ 10 Ma). 
Por otro lado, I--F debe ser un c'umulo joven ya que existe emisi'on \ha\ asociada
con 'este y sus colores en el continuo son similares a los de  I--As o I--Bs.

No se observa ning'un c'umulo grande en la apertura I--Cn y la emisi'on en
\ha\ se debe producir por un grupo de objetos localizados 
en el SO de I--A.
I--D es un SOBA m'as peque\~no que I--Bs  (\cite{Maiz99}). 
Se aprecia muy poca emisi'on \ha\ en comparaci'on con I--A y I--B,
habiendo s'olo dos  (I--D1n and 
I--D2n)  o posiblemente tres estructuras de \ha\ asociadas con 'el.
I--Gn es un n'ucleo \ha\ compacto altamente enrojecido ( \cite{Maizetal98}).

\subsection{NGC~4214--II}

NGC~4214--II es el segundo complejo m'as grande de la galaxia
en tama\~no  y contiene las regiones con los picos de intensidad \ha\ m'as altos.
Su morfolog'ia es muy diferente a la de NGC~4214--I. En primer lugar, 
no contiene un SSC dominante sino varios SOBA m'as peque\~nos responsables
por la ionizaci'on del gas. Adem'as no contiene cavidades de ning'un tipo
(a la escala observable desde Tierra) 
y los c'umulos est'an localizados muy cerca (o en la misma posici'on) que los 
n'ucleos que emiten \ha\ (\cite{Maizetal98}).

II--A y II--B son los  dos n'ucleos m'as brillantes en NGC 4214--II.
Ambos est'an localizados sobre SOBA que, a la resoluci'on de la WFPC2,
aparecen centrados en la misma posici'on.
Los centros de las estructuras sufren de baja extinci'on pero 
la emisi'on \ha\ en los alrededores est'a altamente extinguida (\cite{Maizetal98}). 
Se observa una estructura de polvo que corre de S a N al oeste del n'ucleo
II--A hasta el II--B. Otra diferente separa a 'este del c'umulo II--C.

La SOBA en II--B  parece ser m'as compacta que la hallada en II--A.
Desde Tierra, II-C aparece como una estructura simple, levemente extendida
y desde el espacio, se la  puede resolver en un cascar'on incompleto
dividido en dos fragmentos, II--C1n hacia el SO y  II-C2n hacia el NE. 
%
II--Dn y II--En  son dos n'ucleos \ha\ m'as d'ebiles con muchas menos estrellas.

\subsection{Otras estructuras}

NGC~4214--IIIs  es una fuente compacta con emisi'on \ha\ asociada muy d'ebil.
Se piensa que es probablemente un SSC o el n'ucleo de la galaxia, como fue sugerido por
el estudio de \citet{Faneetal97}  en  el filtro $I$ y el FUV. En la imagen  $I$
de esos autores, NGC~4214--IIIs  aparece en el centro de un disco muy d'ebil
de la galaxia. Las observaciones en \hi de \citet{McIn98} ubican al 
centro de rotaci'on de la galaxia muy cerca de NGC~4214--IIIs.
NGC~4214--IVs tiene una estructura similar 
a  NGC~4214--IIIs  y tambi'en aparece en la imagen $I$ de   \citet{Faneetal97}. 
Sin embargo, es menos brillante en las im'agenes del continuo.
El resto de las estructuras visibles en la imgen de \ha\
pueden ser clasificadas como n'ucleos difusos (VIIn, VIIIn, IXn, XIIn, XIIIn),
una serie de n'ucleos (Xn, XIn) o complejos \ha\ + cascarones  (Vn, VIn).

\chapter{Observaciones y reducci'on de los datos    }
\label{cha:chapter2}
\thispagestyle{empty}

\begin{flushright} 
{\em   Masquerade! \\
Take your fill, \\
let the spectacle \\
astound you!}
 \end{flushright}
\newpage
 
\section{Propuestas }
Para este trabajo  hicimos uso extensivo de   im'agenes de NGC~4214 de alta resoluci'on 
obtenidas con  la WFPC2 en varios filtros  asi como de   im'agenes  STIS en el UV cercano
obtenidas con el NUV--MAMA. Ambos instrumentos se encuentran a bordo del 
Telescopio Espacial Hubble. Las im'agenes fueron obtenidas en varias propuestas:

\begin{description}
	
	\item{\bf Propuesta 6569}:   (PI: John  MacKenty)  En este caso se 
	 explot'o la resoluci'on de la WFPC2 para obtener 
	im'agenes en filtros $U, V, R, I$ de banda ancha y  en filtros nebulares    \ha\ y 
	[O~{\scriptsize III}~{\normalsize}]  de tres 
	galaxias enanas starburst locales para entender los mecanismos de formaci'on estelar
	en estos sitemas. Las im'agenes de NGC~4214 fueron obtenidas en julio de 1997.

	\item{\bf Propuesta 6716}:  (PI: Theodore Stecher) Para esta propuesta
	se decidi'o obtener im'agenes en diferentes filtros 
	de las poblaciones estelares de cuatro galaxias cercanas  que se sab'ia est'an formando
	estrellas masivas de tipos O y B.
	Su objetivo principal era determinar la distribuci'on espacial de regiones  de formaci'on estelar
	y confirmar la presencia de  superc'umulos estelares  y medir sus propiedades b'asicas.
	Las im'agenes de NGC~4214 fueron obtenidas en junio y diciembre de 1997

	\item{\bf Propuesta 9096}: (PI: Jes\'us  
Ma\'{i}z-Apell\'aniz) Para esta propuesta se   obtuvieron im'agenes de NGC~4214 con el   STIS NUV-MAMA
	usando filtros F25CN182 
	y F25CN270.

\end{description}
\section{M'etodo de reducci'on de datos}
\label{sechst}
\subsection{ Paquete  {\sc HSTphot}}

La reducci'on de los archivos {\sc c0f} (datos cient'ificos) fue realizada
empleando el conjunto de programas  {\sc\small  HSTphot} \citep{Dolp00b,Dolp00a,Dolp02}. 
La WFPC2 es un instrumento capaz de proveer im'agenes de alta
resoluci'on sobre las que se puede realizar fotometr'ia de alta calidad.

Sin embargo, la  funci'on de imagen puntual (PSF, del ingl'es: point--spread function) 
de este instrumento sufre de un problema: es inframuestreada,
lo cual significa que la FWHM tiene un tama~no comparable
al tama~no de un p'ixel de la PC y a aproximadamente
medio p'ixel de la WFC. Esto implica que  se dificulta el obtener fotometr'ia
sobre estos CCD \citep{Dolp00a}.
{\sc\small  HSTphot} usa una biblioteca de PSF muy completa
espcialmente dise~nada que permite centrar estrellas y determinar
su fotometr'ia con alta precisi'on.
Se denomina {\it Point Spread Functions} o PSF a los perfiles
de fuentes astron'omicas puntuales en im'agenes de matrices bidimensionales
como CCD. 
Un m'etodo  \citep{Howe00} para hacer mediciones sobre esas im'agenes
consiste en ajustar perfiles, los cuales pueden 
ser modelados mediante varias funciones matem'aticas como gaussianas,
lorentzianas y otras.
En lugar de usar funciones matem'aticas, tambien se pueden emplear
matrices con valores  de modelos  calculados para un dado filtro
y para cada posici'on dentro del CCD  \citep{Hasa95}.
El paquete {\sc\small  HSTphot} usa una grilla de PSF que fueron calculadas
con  Tiny TIM  \citep{Kris01} que  es un software que se emplea para 
generar PSF para los instrumentos a bordo del HST. 
{\sc\small  HSTphot} determina el brillo de una estrella y su posici'on
calculando una combinaci'on de estos par'ametros que mejor 
se ajustan a los datos observados. Existe una biblioteca de PSF de la WFPC2
que puede emplearse como una semilla inicial  de iteraciones.
Para solucionar el problema del  inframuestreo, las PSF
sint'eticas fueron calculadas para una gran variedad de 
centrados en subp'ixeles. Exite un total de 1\,600 PSF para la PC
y 6\,400 para la WFC.

Los archivos necesarios para al  reducci'on son los archivos con los
datos cient'ificos ({\sc c0f}) y los que contienen la calidad de la imagen ({\sc c1f}).
Para obtener fotometr'ia calibrada de una imagen, se deben completar los 
siguientes pasos:
(1) Preparaci'on de la imagen, (2) Determinaci'on de la PSF, (3) Detecci'on de 
las estrellas,
(4) B'usqueda de la soluci'on fotom'etrica, (5) Correcciones de apertura.


El primer paso en la preparaci'on de la imagen es el enmascarado 
de los p'ixeles malos y de las columnas  malas, lo cual se lleva a cabo
con la rutina  {\sc\small  mask}. 
Esta rutina lee la imagen de los datos y la que contiene la calidad
de los datos 
y enmascara todos los p'ixeles que son considerados  como malos. 
La columna 800 y la linea 800 tambien son descartadas completamente.
Todos los p'ixeles con 3\,500 o m'as cuentas se marcan como 
saturados (4\,095 DN) para evitar cualquier ambig\"uedad
en la reducci'on.
Los p'ixeles enmascarados son marcados con el valor $-100$ DN, y 
son ignorados en el resto de la fotometr'ia.


Las im'agenes de la WFPC2 usualmente contienen
muchos   rayos c'osmicos, que son causados 
por la interacci'on entre  part'iculas c'osmicas y protones del cintur'on 
de radiaci'on de la Tierra con el detector CCD.
Esta interacci'on ocurre en un promedio de 1.8 eventos por segundo por 
CCD.
La mayor'ia de estos eventos depositan una cantidad significativa
de carga sobre los CCD, incluso en exposiciones cortas. 
Para eliminar  rayos c'osmicos de las im'agenes, se requiere de 
varias im'agenes alineadas, comparando el flujo del mismo p'ixel
en diferentes im'agenes, suponiendo 
que cualquier diferencia por encima del ruido son desviaciones debidas
a rayos c'osmicos  \citep{wfpc2data}.
La imagen enmascarada   obtenida est'a lista para 
la limpieza por rayos cosmicos. Para esto se emplea 
 {\sc\small  crmask}   dentro de  {\sc\small  HSTphot}.
Este procedimiento tiene la capacidad de limpiar im'agenes que no est'an
perfectamente alineadas, as'i como de im'agenes de distintos filtros.
La identificaci'on de rayos c'osmicos se hace 
estudiando cada p'ixel, uno por uno. Se determinan los valores
m'inimo y m'aximo de cada p'ixel. Si un p'ixel de una imagen tiene un valor
entre el m'inimo y el m'aximo,  se lo considera; si no, se lo
enmascara.

Luego de la remoci'on de los rayos c'osmicos, 
 se combinan las im'agenes del mismo filtro usando
 la rutina  {\sc\small  coadd}

El siguiente paso consiste en la remoci'on de p'ixeles calientes que no
fueron descartados por el datoducto  en el STScI. 
El valor de estos p'ixeles calientes, es proporcional al tiempo de 
exposici'on y a la cantidad de carga en el mismo, de manera
tal que  {\sc\small  crmask} no podr'ia enmascararlos. 
La rutina  empleada es {\sc\small  hotp'ixels}.  


La determinaci'on del fondo es un paso importante y obligatorio de
{\sc\small  HSTphot}, y se realiza con la tarea   {\sc\small  getsky}.
El valor del cielo se calcula en cada p'ixel, usando el promedio de los valores 
en un anillo cuadrado centrado en tal p'ixel. 
Para la PC, el cuadrado interno tiene 33 p'ixeles en cada lado  
y el externo tiene 45, lo cual da un total de 1\,064 p'ixeles para
la determinaci'on del cielo.   
En el caso de la WFC, los cuadrados tienen la mitad de ese tama~no. 
El programa calcula un promedio pesado con todos los p'ixeles 
que no fueron enmascarados y los que no est'an saturados dentro de esos
anillos en forma recursiva. Cuando el proceso converge,
el valor medio del cielo en la 'ultima iteraci'on es el valor del cielo asociado
al p'ixel. 

Luego de sustraer la imagen del cielo de la imagen con los datos, se buscan 
picos residuales en la imagen diferencial, leyendo esta imagen en varios pasos.
Cuando se detecta un pico, se hace una estimaci'on de la posici'on del centro 
de la estrella. 
Usando esa posici'on como semilla, {\sc\small  hstphot}  calcula la fotometr'ia 
para mejorar la determinaci'on del centro y del brillo de la estrella. 
La fotometr'ia  se obtiene mediante un proceso iterativo. Luego de la
convergencia, si la se'nal ruido (S/N) de la estrella es mayor o igual
que el definido por el usuario, se guardan estos valores; sino, se la descarta.
El procedimiento  determina un valor ($\chi^2$) que indica la calidad del ajuste.

%
El proceso que calcula la fotometr'ia usa un m'etodo iterativo. 
Luego de la convergencia o si el m'aximo n'umero de iteraciones fue alcanzado,
se realiza una 'ultima iteraci'on para mejorar la precisi'on de los c'alculos.
En todas las iteraciones anteriores, las posiciones de las estrellas
se obtienen a la resoluci'on 
de la biblioteca  de PSFs: cada 0.2 p'ixeles en la PC y 0.1 p'ixeles en las WFC.
Para mejorar la precisi'on astrom'etrica y fotom'etrica, la 'ultima iteraci'on 
determina la mejor posici'on con una precisi'on de 0.01 p'ixeles, 
mediante interpolaci'on lineal.

\subsection{Correcciones  adicionales a la fotometr'ia}
A los efectos de obtener la mejor fotometr'ia, es necesario
hacer algunas correcciones a los datos de la WFPC2.

{\bf Contaminaci'on:  una correcci'on dependiente del tiempo } 

La correcci'on m'as importante dependiente del tiempo es 
la de  {\it contaminaci'on} de las ventanas de los CCD, que afectan principalmente
a las observaciones en el ultravioleta.
Los contaminantes se adhieren a las ventanas fr'ias de la WFPC2. 'Estos tiene
un efecto despreciable sobre la sensibilidad  en el visible y en el 
infrarojo cercano  de las c'amaras, pero el efecto sobre el UV puede ser apreciable.
Estas part'iculas se remueven durante calentamientos peri'odicos
(descontaminaciones)  de las c'amaras, y el efecto sobre la fotometr'ia 
es lineal y estable y puede ser removido usando valores del 
programa de calibraci'on  de la WFPC2 \citep{McMaWhit02}. Aproximadamente
una vez al mes, los contaminantes son evaporados de las c'amaras, lo cual 
restaura la transmisi'on de las mismas a sus valores 'optimos. 
Una vez al mes, se calientan los CCD aproximadamente a $20^{\circ}$C 
por unas seis horas. Este procedimiento remueve completamente los 
contaminantes. Los CCD de la WFPC2  fueron
dise~nados para operar entre  $-70^{\circ}$C y $-90^{\circ}$C. Luego
del primer calentamiento el 23 de abril de 1994, se fij'o la temperatura
de los CCD a $-88^{\circ}$C  \citep{Holtetal95a}. 
Los valores de la  correcci'on por contaminaci'on 
se publican anualmente en un  Instrument Science Report.

{\bf Eficiencia en la transferencia de carga (CTE): una correcci'on que depende de la posici'on}

 \citet{Holtetal95a} realizaron un test para analizar la precisi'on de los flat--fields
 de la WFPC2: observaron un campo de estrellas en 
  $\omega$ Centauri con  diferentes orientaciones, de manera tal
  que una dada estrella se observa en unas 30 posiciones en el mismo chip.
Cuando redujeron los datos, encontraron que la misma estrella aparec'ia m'as d'ebil
cuando se encontraba en una zona particular del chip. 
El efecto es m'as pronunciado sobre objetos
ubicados cerca del borde superior del chip donde los datos tienen que 
pasar por m'as p'ixeles durante la lectura del chip y se pierde por
tal motivo m'as carga.
Este efecto est'a presente en los cuatro chips de la WFPC2
 y  es causado por problemas de transferencia de carga dentro de 
cada CCD. Se piensa que este efecto surge  por la presencia de trampas
en  el material  con que se fabrica el CCD el cual retiene a los electrones 
en lugar de transferirlos durante la lectura.

Reduciendo la temperatura a la cual trabajan las c'amaras, se pudo
reducir el efecto, pero todav'ia sigue presente.
Para corregir este problema, se deben emplear las coordenadas
$x$ e $y$ del objeto observado, el n'umero de cuentas en el fondo,
el brillo de objeto, la fecha de observaci'on y una f'ormula dada por
 \cite{Whitetal99a}.

{\bf Distorsi'on  geom'etrica: correcci'on dependiente de la posici'on}

Las c'amaras de la WFPC2 sufren de una distorsi'on geom'etrica
como resultado de su dise~no  \citep{Holtetal95a}.
Por este motivo, el tama~no efectivo de los p'ixeles var'ia sobre cada chip.
El caso de  la WFPC2 es complicado pues la imagen final est'a
dividida en cuatro campos diferentes, de manera tal que se necesitan
cuatro transformaciones individuales, una por cada chip.

La correcci'on es importante no solo para la astrometr'ia sino para
el registro preciso de im'agenes obtenidas con diferentes orientaciones.
Este efecto puede ser muy marcado, con diferencias de posiciones
 de hasta varios p'ixeles en los 
costados de los chips.
Esta distorsi'on produce una variaci'on en el area de cada p'ixel 
sobre el campo de observaci'on, lo cual debe tenerse en cuenta 
cuando se efectua fotometr'ia. La distorsi'on depende de la longitud de onda en
el UV y es independiente en el visible.

La correcci'on por distorsi'on geom'etrica de la WFPC2 queda definida
por cuatro transformaciones entre las coordenadas de cada detector
y un sistema est'andar de coordenadas  \citep{Koek02}. 
La soluci'on de  \cite{Trau95} se deriva a partir de 
un modelo de rayos que se desplazan a trav'es de la 'optica de la WFPC2,
e incluye correcciones dependientes de la longitud de onda. 
Las soluciones de \citet{Gilm95} y  \citet{Holtetal95b}
se basan en observaciones de c'umulos globulares y se han hecho para una 'unica
longitud de onda  ($\lambda = 555$ nm). 
La soluci'on de \citet{CaseWigg01} se basa en observaciones recientes de 
$\omega$ Centauri.

{\bf Dependencia de la escala de placa con el filtro usado: correcci'on dependiente de la posici'on}

La escala de los p'ixeles es 
$0 \arcs 04554 $  /p'ixel en la PC y  $0 \arcs 09961, 
 0 \arcs 09958  $ y $0 \arcs 09964 $ / p'ixel en  WF2, 3 y  4 respectivamente. 
Las escalas de placa se refieren al centro de cada chip en el filtro F555W.
La escala verdadera es, en realidad, menor en el resto de cada chip
debido a la distorsi'on geom'etrica, y existe una dependencia con
la longitud de onda que es corregible  \citep{Bireetal02}.

{\bf Efecto cada 34 l'ineas: correcci'on dependiente de la posici'on}

Los CCD de la WFPC2 tienen adem'as una particularidad cada 34 l'ineas:
la sensibilidad es menor en esas lineas en un 3\%. 
Este defecto es causado por un error en el manufacturado al producir los CCD.
El efecto es que cada 34 l'ineas, las mismas son un 3\% m'as angostas.

{\bf Correcci'on de apertura: una correcci'on dependiente de la posici'on}

Cuando se hace fotometr'ia, es importante referir a las mediciones
a un cierto tama~no de apertura. 
En trabajos hechos desde  Tierra, las mediciones fotom'etricas se extrapolan
a un radio  ``infinito'' usando observaciones de estrellas brillantes \citep{Holtetal95b}.
Cuando se usa la WFPC2, la medici'on de una correcci'on hasta una apertura
de radio  ``infinito'' se complica por el peque~no tama~no del p'ixel y por 
la naturaleza extensa de las alas de la  PSF.
Para minimizar estos problemas, Holtzman decidi'o elegir una apertura de tama~no intermedio
de radio $0 \arcs 5$ para todas las mediciones de calibraci'on, lo cual se corresponde
a 11 p'ixeles de la PC y a 5 p'ixeles de la WFC.

La correcci'on por apertura es la diferencia entre la fotometr'ia de apertura
con un radio de $0 \arcs 5$  y las magnitudes obtenidas por ajuste de PSF
con  {\sc\small  hstphot}. Esta correcci'on nos permite determinar la
magnitud de un objeto, como si hubi'esimos usado una apertura de radio  ``infinito''.

 {\sc\small  hstphot} determina las correcciones de apertura midiendo 
 magnitudes con un radio de $0 \arcs 5$  de varias de las estrellas m'as brillantes
de cada chip en una imagen de la que el resto de los objetos fueron sustraidos. 
Luego, un valor promedio de la correcci'on por apertura para cada combinaci'on
de chip y filtro se
determina al promediar las diferencias entre el ajuste de la PSF y las magnitudes de apertura.

\subsection{Rutina   {\sc  hstphot}}
La rutina que realiza la fotometr'ia dentro del paquete {\sc\small  HSTphot} se denomina
  {\sc\small  hstphot}.
Este c'odigo toma las im'agenes de entrada y produce una lista de magnitudes en el sistema 
fotom'etrico de vuelo (WFPC2 flight system) definido por \cite{Holtetal95b}. 
Las magnitudes son calibradas  usando las recetas de calibraci'on de    \cite{Holtetal95b} 
con correcciones por CTE y determinaci'on de 
los puntos cero provistas por  \cite{Dolp00b}. 
 {\sc\small  hstphot} incluye adem'as alineaci'on y correcci'on de apertura.

El archivo de salida de   {\sc\small  hstphot} contiene la fotometr'ia de todos los objetos
hallados en la imagen. Toda la informaci'on de una estrella est'a contenida
en una sola linea, y los  datos provistos son los siguientes:

{\bf chip} (0 (Planetary Camera), 1 (Wide Field Camera 2), 2 (Wide Field Camera 3)
 y 3 (Wide Field Camera 4)), {\bf posici'on} (valores  {\it x} e  {\it y} del centro del objeto en el chip), 
 {$\chi^2$:} ( ver  \cite{Dolp00a}  para la definici'on ), {\bf se~nal ruido}
 ( ver  \cite{Dolp00a}  para la definici'on ), {\bf sharpness} ( este valor es cero si
 la estrella tiene un ajuste perfecto, es negativo si es muy ancha y es positivo si
 es muy angosta ), {\bf tipo de objeto} (1 (estrella), 2 
 (posible binaria no resuelta), 3 (estrella con problemas: centrada sobre un
 p'ixel saturado o una columna mala),
  4 (rayo c'osmico o p'ixel caliente) y  5 (objeto extendido)), {\bf cuentas} 
  (el n'umero de cuentas empleado para calcular la magnitud), {\bf fondo} 
  (el valor del cielo calculado para el objeto), {\bf magnitud} 
  y  {\bf error en la magnitud}.

\section{Descripci'on de im'agenes de la WFPC2 y su reducci'on}

En la Tabla~\ref{tbl001} listamos los datos de la WFPC2 empleados
para este estudio.
En el caso de las im'agenes  de la propuesta 6569, la orientaci'on del telescopio se eligi'o
de manera tal que se   minimizara el efecto de transferencia de carga (CTE)
 \citep{MacKetal00}. 
En el caso de las im'agenes de la propuesta 6716, la orientaci'on fue seleccionada para capturar
las regiones de mayor inter'es de NGC~4214 con la PC, o sea, la c'amara que provee
la mayor resoluci'on.

\begin{figure*}[ht!]
 \centering
 \includegraphics[width=\textwidth]{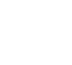}
 \caption[Campos de observaci'on de
 la c'amara WFPC2]{  {\sl\footnotesize  Mosaico  de NGC~4214 de $9' \times 9'$ construida  combinando im'agenes
 RGB p'ublicas en los filtros  {\it B, V} y  {\it R}  obtenidas con el  telescopio Isaac 
Newton.   Se han dibujado los campos de observaci'on HST/WFPC2  de la propuesta 6716 en verde.
El campo de la izquierda corresponde a la segunda visita (09 Dec 1997) y el de la derecha 
corresponde a la primera visita   (29 Jun 1997). 
El campo de observaci'on de la propuesta 6569 est'a dibujado en rojo.
\label{fig003}}}
\end{figure*}

La Figura~\ref{fig003} muestra una imagen de NGC~4214 creada con un combinaci'on de
im'agenes  p'ublicas  en filtros {\it B, V} y {\it R} obtenidas con el Telescopio Isaac Newton. 
Superpusimos los campos de HST/WFPC2 (verde) correspondientes a 
la propuesta 6716. El campo de la izquierda corresponde a la segunda visita (09 Dec 1997)
y el campo de la derecha corresponde a la primera visita (29 Jun 1997).
El campo HST/WFPC2  correspondiente a la propuesta 6569 est'a marcado en rojo.

\begin{figure*}[ht!]
\begin{sideways}
\begin{minipage}[c]{15cm}
\centering
\includegraphics[width=15cm]{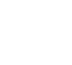}
\caption[Mosaicos WFPC2 de NGC~4214 en filtros F336W,
 F814W y F656N]{ {\sl\footnotesize    Mosaicos de NGC~4214 obtenidos con im'agenes de la WFPC2
en tres filtros:
F336W [izquierda], F814W [centro] y F656N (\ha) \,\,\, [derecha]. 
La imagen \ha\ muestra claramente las regiones de formaci'on estelar
a lo largo de la barra.
 \label{fig009}}}
\end{minipage}
\end{sideways}
\end{figure*}

La  Figura~\ref{fig009} muestra tres mosaicos de NGC~4214 obtenidos
con la WFPC2. Estas im'agenes representan a la galaxias en los filtros  F336W {\it U},
F814W {\it I} y F656N (\ha).

\begin{table*}[ht!] 
\begin{minipage}[h]{16cm}
 {\small
\centering
\caption[Observaciones de  NGC4214 con WFPC2 y 
STIS]{Observaciones de  NGC4214 con WFPC2 y STIS. Se listan los filtros usados 
y los tiempos de exposici'on.
\label{tbl001}}
\begin{center} 
\vspace{-6mm}
 \begin{tabular}{cllll} 
\multicolumn{5}{c}{ \rule{161mm}{0.8mm}}      \\
Propuesta & Filtro &  Comentario  & Imagen\footnote{nomenclatura de STScI }  &  {\rule [-3mm]{0mm}{8mm} Exposici'on (seg) }\\ 
\multicolumn{5}{c}{ \rule[2mm]{161mm}{0.2mm}}      \\
{\rule [0mm]{0mm}{4mm} }			     & F170W  & {\it UV}             & u4190101r + 102r + 201m + 202m & 400 + 400 + 400 + 400 \\
   						     & F336W   & WFPC2  {\it U}       & u4190103r + 104r + 203m + 204m & 260 + 260 + 260 + 260 \\
    	 					     & F555W   & WFPC2 {\it V}        & u4190105r + 205m    & 200 + 200  \\
{\rule [-2mm]{0mm}{0mm} }  \raisebox{4ex}[0pt]{6716} & F814W  & WFPC2  {\it I}       & u4190106r + 206m   & 200 + 200  \\ 
\multicolumn{5}{c}{ \rule{161mm}{0.2mm}}      \\
{\rule [0mm]{0mm}{4mm} }  	     & F336W  & WFPC2  {\it U}       & u3n80101m + 2m + 3m & 260 + 900 + 900  \\
      & F502N  & [O~{\scriptsize III}~{\small}] $\lambda5007$ & u3n8010dm + em    & 700 + 800  \\
     & F555W  & WFPC2 {\it V}        & u3n80104m + 5m + 6m  & 100 + 600 + 600  \\
\raisebox{2ex}[0pt]{6569}       & F656N   & \ha           & u3n8010fm + gm   & 800 + 800  \\
           & F702W  & WFPC2 {\it R}   & u3n80107m + 8m & 500 + 500  \\
{\rule [-2mm]{0mm}{6mm} }     & F814W   & WFPC2 {\it I}        & u3n8010am + bm + cm & 100 + 600 + 600  \\
\multicolumn{5}{c}{ \rule{161mm}{0.2mm}}      \\
{\rule [0mm]{0mm}{4mm} }  	  & F25CN182  &   \dotfill    & o6bz02isq + 02iwq + 01afq   & 288 + 288 + 288  \\
 & F25CN182                               & \dotfill      & o6bz04r8q + 04raq + 03afq   & 288 + 288 + 288  \\
    \raisebox{2ex}[0pt]{9096}	  & F25CN270   &  \dotfill     & o6bz02j7q + 02jbq + 03awq  & 288 + 288 + 288  \\
{\rule [-2mm]{0mm}{0mm} } 	  & F25CN270    &  \dotfill     & o6bz04ruq + 04ryq              & 288 + 288  \\ 
\multicolumn{5}{c}{ \rule{161mm}{0.8mm}}      \\

\end{tabular}  
\end{center}
}
\end{minipage}
\end{table*}


La reducci'on b'asica de los datos se llev'o a cabo
en el datoducto   de STScI. Los datos del archivo est'an parcialmente reducidos,
habiendo sido ya corregidos del efecto umbral (bias), y de la exposici'on homog'enea
(flatfield). 
La reducci'on de las im'agenes pre--procesadas   {\sc c0f}   fue realizada usando
los  c'odigos  dentro del paquete  {\sc\footnotesize HSTphot}. Este paquete
fue dise\~nado espec'ificamente para analizar los datos de la WFPC2, como
se explica en la Secci'on~\ref{sechst}.
 La principal innovaci'on de este software
es que tiene en cuenta la naturaleza de la PSF caracter'istica de la WFPC2 \citep{Dolp00a}. 
Las Secciones \ref{sec:datared1} hasta   \ref{sec:datared2}   contienen los
pasos empleados que seguimos para obtener la fotometr'ia calibrada de nuestras observaciones.

\subsection[Primeros pasos en la reducci'on]{M'ascara de p'ixeles malos, 
correcci'on por rayos c'osmicos y combinaci'on de im'agenes}
\label{sec:datared1}

El primer paso en la reducci'on de los datos
consiste en construir una m'ascara para descartar los p'ixeles y columnas malas, lo
cual se realiza con la rutina  {\sc\footnotesize mask}. Esta rutina lee cada im'agen con los datos cient'ificos
y la correspondiente imagen con los datos de calidad. Luego procede a construir una m'ascara de todos 
los p'ixeles que son considerados malos. La columna 800 y la fila 800 son enmascaradas completamente.
Todos los p'ixeles con 3500 o mas cuentas se consideran como saturados (4095 DN) para evitar cualquier ambig\"uedad.
A los p'ixeles malos se les asigna el valor -100 DN y se los descarta para el resto de la fotometr'ia. 
La imagen resultante est'a asi lista para ser limpiada de rayos c'osmicos  y para su combinaci'on;
para eso, usamos la rutina    {\sc\footnotesize crmask}.

Las im'agenes  u4190105r/205r (filtro F555W)  y u4190106m/206m (filtro F814W)  carecen de una im'agen equivalente para
hacer la combinaci'on 
y por tal raz'on tuvimos que comparar dos im'agenes de diferentes filtros para deshacernos de los rayos c'osmicos.
Luego de ejecutar  {\sc\footnotesize crmask}, desplegamos la im'agen original y su correspondiente corregida para 
decidir visualmente si todos los rayos c'osmicos fueron enmascarados en  estas
im'agenes.
Descubrimos que en algunos casos los p'ixeles centrales de estrellas brillantes fueron 
enmascarados innecesariamente, mientras que en otros casos, p'ixeles que obviamente representan
rayos c'osmicos no fueron descartados.
Decidimos que la mejor opci'on ser'ia dejar algunos rayos c'osmicos en la im'agen ya que la rutina {\sc\footnotesize hstphot} que aplicamos 
m'as adelante tiene la capacidad de distinguir p'ixeles con apariencia de rayo c'osmico y marcarlos como tales.
Luego de descartar los rayos c'osmicos de nuestras im'agenes, combinamos las im'agenes obtenidas con el  mismo filtro
usando la rutina {\sc\footnotesize coadd}.
La Tabla~\ref{tbl002} lista los nombres de las im'agenes que fueron combinadas.


\begin{table*}[ht!] 
\centering
\begin{minipage}[u]{12cm}
\caption[Datos empleados obtenidos con la WFPC2]{Datos 
de la WFPC2 combinados para remover rayos c'osmicos empleando la rutina  
{\sc\footnotesize coadd}
\label{tbl002}}
\end{minipage}

\begin{center}
\vspace{-4mm}
\begin{tabular}{cllc} 
\multicolumn{4}{c}{ \rule{121mm}{0.8mm}}      \\
 Propuesta & Filtro &   Datos &  {\rule [-3mm]{0mm}{8mm} Fecha }\\
 \multicolumn{4}{c}{ \rule[2mm]{121mm}{0.2mm}}      \\
{\rule [0mm]{0mm}{4mm} }                             & F170W...... &  u4190101r + 102r         & $1997-12-09$   \\
                                                     & F170W...... &  u4190201m + 202m              & $1997-06-29$\\ 
                                                     & F336W...... &  u4190103r + 104r          & $1997-12-09$   \\
                                                     & F336W...... &  u4190203m + 204m              & $1997-06-29$  \\
                                                     & F555W...... &  u4190105r     & $1997-12-09  $  \\
                                                     & F555W...... &  u4190205m       & $1997-06-29 $ \\
                                                     & F814W...... &  u4190106r    &$ 1997-12-09 $  \\ 
{\rule [-2mm]{0mm}{0mm} }  \raisebox{10ex}[0pt]{6716}& F814W...... &  u4190206m       & $1997-06-29 $ \\ 
 \multicolumn{4}{c}{ \rule[2mm]{121mm}{0.2mm}}      \\
{\rule [0mm]{0mm}{4mm} }                             & F336W...... &  u3n80101m  + 2m + 3m             &$ 1997-07-22 $ \\
                     				     & F502N...... &  u3n8010dm  + em       & $1997-07-22$ \\
   						     & F555W...... &  u3n80104m  + 5m + 6m             &$1997-07-22$  \\
\raisebox{2ex}[0pt]{6569}			     & F656N...... &  u3n8010fm  + gm       & $1997-07-22$ \\
    						     & F702W...... &  u3n80107m  + 8m        &$1997-07-22  $\\
{\rule [-2mm]{0mm}{0mm} } 			     & F814W...... &  u3n8010am  + bm + cm             &$1997-07-22 $ \\ 
 \multicolumn{4}{c}{ \rule[2mm]{121mm}{0.8mm}}      \\
\end{tabular}  \end{center} 

\end{table*}

\subsection{Determinaci'on del cielo y  remoci'on   de los p'ixeles calientes}
La determinaci'on del cielo fue realizada con la rutina {\sc\footnotesize getsky}  que realiza una simple c'alculo
antes de la fotometr'ia.
La rutina  {\sc\footnotesize getsky} determina un valor del cielo a priori  en cada p'ixel  y genera un archivo  {\sc sky}
de cada im'agen combinada.
El 'ultimo paso de limpieza  consiste en la remoci'on de  los hotpixels que no fueron enmascarados
usando la im'agen de calidad de los datos ni durante el proceso del pipeline en STScI.
Ejecutamos la utilidad  {\sc\footnotesize hotpixels} cuatro veces, hasta que ning'un p'ixel fue descartado en nuestras im'agenes.

\subsection{Ejecuci'on de  {\sc hstphot}}

Para obtener la magnitud calibrada de todas las im'agenes, empleamos la rutina 
{\sc\footnotesize hstphot} en el paquete {\sc\footnotesize HSTphot}.
Esta rutina incluye la alineaci'on de las im'agenes y 
las correcciones por CTE y apertura.
Siguiendo la Guia del Usuario de {\sc\footnotesize HSTphot} usamos los valores 3.5 y 5.0
como los umbrales independiente (m'inimo S/N para la detecci'on de las estrellas) y total 
(m'inimo S/N para conservar a las estrellas).
{\sc\footnotesize ~hstphot}  fue ejectutado cuatro veces sobre todas
las im'agenes usando  varios valores de la bandera de 
 opciones.

\begin{description}
	\item{\bf 2    }  Esta opci'on activa la determinaci'on local del cielo,
	lo cual significa que el cielo local es calculado usando valores inmediatamente fuera 
	de la apertura usada para la fotometr'ia. Esta opci'on es 'util en campos con cielo 
	altamente cambiante como el observado en las im'agenes de NGC~4214.

	\item{\bf 10   } Esta opci'on activa la determinaci'on local del cielo,
	y emplea una  correcci'on de apertura por defecto. Esta opci'on es 'util cuando no se 
	pueden elegir buenas estrellas de apertura.
	
	\item{\bf 512   } Esta opci'on hace un nuevo ajuste del cielo durante la fotometr'ia.
	\item{\bf 520   }  Esta opci'on hace un nuevo ajuste del cielo durante la fotometr'ia emplea una 
	correcci'on por apertura default para cada filtro.
	\end{description}

\subsection{Objetos saturados}
Estudiamos la presencia de objetos saturados
en todas las im'agenes y hallamos que  el c'umulo I--As
aparece saturado en la im'agen obtenida con el filtro F555W de la
propuesta 6716 y en las im'agenes obtenidas con los filtros 
F336W, F555W, y F702W de la propuesta 6569.
El c'umulo  IIIs  est'a saturado  en todas las im'agenes 
obtenidas con el filtro F555W.
En la Secci'on 4.4 se explica c'omo se obtuvo la fotometr'ia de estos c'umulos.

\subsection{Registro de las listas fotom'etricas}

Debido a que las im'agenes de NGC~4214 no proveen buenas estrellas para 
la determinaci'on de la correcci'on de apertura,
decidimos quedarnos con las salidas de  {\sc\footnotesize ~hstphot}  correspondientes a las opciones
10 y 520.  Esto significa que en ambos casos la correcci'on de  apertura  de
nuestra fotometr'ia fue la por defecto  de cada filtro provista por {\sc\footnotesize HSTphot}.
Empleamos un procedimiento IDL para combinar las listas de salida de 
{\sc\footnotesize hstphot}, donde la magnitud final  $m$  de un objeto se obtiene como el promedio
 $(m = (m_1 + m_2)/2) $ de las magnitudes originales 
 $m_1$ y $m_2$.
El error en la magnitud $\sigma_{m}$ 
se obtuvo calculando el m'aximo valor entre los errores indivivules
 ($\sigma_1$ y $\sigma_2$)  y el valor $ |m_1 - m_2| /2 $.
 
  \subsection{Correcci'on por  escala de placa y contaminaci'on}
\label{sec:datared2}
Como resultado de la distorsi'on geom'etrica que sufre la WFPC2,
la escala de placa es una funci'on  del tipo de filtro empleado en
la observaci'on.

\citet{Dolp00a} emplea la correcci'on por distorsi'on determinada
emp'iricamente por  \citet{Holtetal95a}, la cual  fue calculada para el filtro F555W.
Para referir esta correcci'on a los otros filtros que usamos, debimos hacer una peque\~na
correcci'on:

La posici'on  $(x,y)$  de cada p'ixel es desplazada una cantidad
que depende de su posici'on relativa respecto al centro del chip
 [posici'on $(400,400)$] y de un valor $\mathrm{fdps}$ que depende del filtro usado.
Las f'ormulas usadas para esta correcci'on son: 

\[
\vspace{0.5cm}
\begin{array}{rcl}
dx &= &  x - 400 \\
dy & = &  y - 400 \\
r2 & = & ((dx+6.2)^2+(dy+6.2)^2)  \hspace{3cm} \quad\mbox{para la WFC}\quad \\
r2 & = & ((dx+15.8)^2+(dy+15.8)^2)  \hspace{3cm}  \quad\mbox{para la PC}\quad \\
dps & = & \mathrm {fdps} \cdot (1+(7 \cdot 10^{-7}) \cdot r2) \\
xx & =& x - dx \cdot dps \\
yy & =& y - dy \cdot dps 
\end{array}
\]


\begin{wraptable}[13]{r}[-1mm]{45mm}
 \begin{tabular}[h]{cc} 
 \multicolumn{2}{c}{ \rule{45mm}{0.8mm}}  \\
Filtro & correcci'on fdps \\  
 \multicolumn{2}{c}{ \rule[2mm]{45mm}{0.2mm}}  \\

F170W & 45.1 \\  
F336W & 4.2 \\  
F555W & 0.0 \\  
F702W & -5.2 \\  
F814W & -5.0 \\   
     \multicolumn{2}{c}{ \rule{45mm}{0.8mm}}  \\
\end{tabular}
 \begin{minipage}[u]{45mm}
\caption{Valores de la correcci'on por escala de placa seg'un el filtro}
\label{tbl004}
\end{minipage}
\end{wraptable}
 

Aqu'i $x$ e $y$ representan las coordenadas originales del p'ixel, $dx$ y $dy$ 
representan  su desplazamiento con respecto  al centro del chip, y  $\mathrm{fdps}$   es un 
coeficiente que depende del filtro y cuyos valores est'an dados en la  Tabla~\ref{tbl004} 
a la derecha. Est'an  expresados en unidades de  
$10^{-4}$ p'ixeles por p'ixel
y tienen incertidumbres t'ipicas de  $2 \times 10^{-5}$. 
 $xx$ e $yy$ son las coordenadas  $x$ e $y$ originales corregidas.

Se sabe que ciertos contaminantes se adhieren a las ventanas fr'ias del CCD de la WFPC2.
Esta contaminaci'on tiene poco efecto sobre las observaciones 
realizadas en el visible e infrarrojo, pero su efecto en im'agenes 
en el UV pueden ser apreciables.
Estas impurezas  son eliminadas peri'odicamente al realizar un leve calentamiento de
las c'amaras, y el efecto sobre la fotometr'ia es lineal y estable y
 puede ser eliminado usando valores
medidos regularmente  por el programa de calibraci'on de la WFPC2  \citep{McMaWhit02}. 
Nosostros realizamos correcciones por contaminaci'on de todas las im'agenes 
obtenidas en los  filtros F170W y F336W.

\subsection{Astrometr'ia}
Usamos la tarea   {\sc\footnotesize metric} de  {\sc\footnotesize IRAF} para completar la correcci'on por 
distorsi'on geom'etrica y para traducir las coordenadas de cada 
objeto a ascenci'on recta ($\alpha$) y declinaci'on ($\delta$). 
Esta transformaci'on se realiza con los par'ametros 
 CRVAL,  CRPIX, 
y  con la matriz de  coeficientes CD que est'an en la cabecera de 
la imagen de referencia. La 'epoca de 
 $\alpha$ y $\delta$ es la misma que la de estos par'ametros que, en nuestro
 caso, es J2000.

La distorsi'on geom'etrica de la WFPC2 est'a muy bien estudiada \citep{CaseWigg01}, 
lo cual permite obtener una precisa astrometr'ia.
Sin embargo, la astrometr'ia absoluta tiene dos problemas: Primero, el Cat'alogo 
de Estrellas Gu'ia (GSC2), que es empleado por el HST,
tiene errores t'ipicos de $\sim 0\arcs$3 \citep{Russetal90}. 
Segundo, si s'olo se  puede determinar un punto sobre los  cuatro chips de la c'amara
como referencia, el error promedio en la orientaci'on produce un error inducido de
$\sim 0\arcs03$  en la posici'on de los otros tres chips.

Nosotros buscamos objetos en el campo de NGC~4214 usando  el
Aladin Interactive Sky Atlas. Desplegamos la im'agen de NGC~4214 del relevamiento POSII
junto al cat'alogo  USN0--A2.0. S'olo encontramos dos objetos en nuestro campo:
1200-06870167 
y 1200-06870199.

Usamos el sevicio VizieR \citep{Ochsetal00} para obtener las coordenadas de estos objetos 
en la 'epoca J2000.
El objeto 1200-06870167 ($\alpha_{2000} = 12^{h} 15^{m} 38\arcs 571 $ , 
$\delta_{2000} = +36^{\circ} 18'   47\arcs09 $ )  puede identificarse 
como  el centro de la regi'on VIn de acuerdo a la nomenclatura de 
\cite{MacKetal00}, mientras que el objeto 
1200-06870199 ($\alpha_{2000} = 12^{h}15^{m}39\arcs213$, 
$\delta_{2000} = +36^{\circ}19'  38\arcs67$)  est'a claramente asociado a la regi'on
I-A in la misma nomenclatura.
Esta 'ultima es una regi'on muy compleja, llena de objetos en nuestras im'agenes de alta resoluci'on
y, por tal motivo, las coordenadas  provistas por 
USNO-A2.0  corresponden a una mezcla de fuentes puntuales, haciendo que 
este objeto sea in'util para nuestro prop'osito.

Empleamos entonces al  objeto 1200-06870167  como nuestro punto de
referencia astrom'etrico para determinar las 
coordenadas ecuatoriales.
Comparamos las posiciones de varias estrellas en varias regiones de im'agenes
obtenidas con diferentes filtros y, como era de esperarse, observamos diferencias de algunas cuantas 
cent'esimas de segundos de arco lo cual consideramos como la precisi'on  de nuestra astrometr'ia.

\subsection{Listas fotom'etricas}
La galaxia NGC~4214 fue visitada dos veces para la propuesta 6716:
primero en junio de 1997 (im'agenes u41902*) y luego en diciembre de 1997 (im'agenes u41901*). 
La orientaci'on de las c'amaras en el cielo es diferente, como puede claramente apreciarse en 
la   Figura~\ref{fig003}, donde marcamos los campos en verde.
Para cada uno de los filtros de esta propuesta obtuvimos dos listas fotom'etricas,
una por cada visita. Para combinarlas en una sola, desarrollamos un c'odigo IDL
que compara las coordenadas ecuatoriales de los objetos 
y busca concidencias dentro de  $0 \arcs 25$  como tolerancia.
Cuando un objeto se encuentra en ambas listas, se le asigna la magnitud

\begin{equation}  
m= \frac{\frac{m_1}{\sigma_1^2} +
\frac{m_2}{\sigma_2^2} }{\frac{1}{\sigma_1^2} 
+ \frac{1}{\sigma_2^2}   },
\end{equation} 

donde   $m_1$ y $m_2$ son las magnitudes originales, 
y $\sigma_1$ y $\sigma_2$ son sus errores. 
El valor asignado a la incerteza est'a dado por:

\begin{equation}  
\sigma_{m} = \frac{1 }{ \sqrt{ \frac{1}{\sigma_1^2} 
+ \frac{1}{\sigma_2^2}}}   
\end{equation}

Tambi'en obtuvimos las listas de objetos  detectados en las im'agenes
de la propuesta 6569.


El 'ultimo paso en la reducci'on de los datos, 
consisti'o en relacionar las diferentes listas fotom'etricas en una 'unica 
lista final.
Desarrollamos un c'odigo IDL para identificar la posici'on de una misma estrella
en diferentes im'agenes.
Creamos as'i dos listas principales usando los filtros F336W y F814W como
referencia.
Nos referiremos a estas listas como  {\sc\footnotesize LISTA336} y   {\sc\footnotesize LISTA814}.  
Para combinar la lista producida a partir de las im'agenes de la propuesta 6569
con la lista proveniente de la propuesta 6716, desarrollamos un procedimiento
con las siguientes caracter'isticas:

\begin{description}
\item[regiones  con confusi'on]  En las regiones m'as pobladas de nuestras im'agenes  (NGC~4214-I) 
consideramos s'olo objetos de la propuesta 6716 ya que estas im'agenes tienen
una mayor resoluci'on.
\item[regiones  sin confusi'on]  Fuera de las regiones anteriores, se pueden presentar varios casos:

\begin{enumerate}

\item  un objeto es hallado en la propuesta 6569 pero no en la propuesta 6716
\item  un objeto es hallado en ambas propuestas
\item  un objeto hallado en la propuesta 6569 se corresponde a 
dos o mas objetos en la propuesta 6716

\end{enumerate}

\end{description}

Nuestra lista final incluye todos los objetos en el caso (1); si el objeto cae en 
el caso (2), consideramos los datos de la propuesta 6569 debido a su mayor se\~nal ruido.
Finalmente, si el objeto   corresponde al caso (3)
nos quedamos con los objetos de la lista proveniente de la propuesta
6716 debido a su mayor resoluci'on. 
Con este procedimiento, construimos dos listas fotom'etricas:
Primero tomamos  al filtro F336W como referencia
y buscamos todas las coincidencias posibles con
el resto de los filtros.
Con los objetos que no fueron asociados construimos otra lista
tomando al filtro F814W como referencia.
 La Tabla~\ref{tbl031} muestra los primeros 25 objetos de la  {\sc\footnotesize LISTA336}
 ordenados por magnitud ascendente en el filtro F336W. Se dan las coordenadas 
 ecuatoriales de los mismos y todas las coordenadas halladas en los filtros
 F336W, F170W,  F555W, F702W, F814W, CN182 y CN270.


\begin{sidewaystable}
\centering
\begin{tabular}{  lcccccccc  }
\multicolumn{9}{c}{ \rule{220mm}{0.8mm}}      \\
 
$\alpha$ (J2000) & $\delta$ (J2000)&F336W&F170W &  F555W &F702W&F814W&CN182&CN270  \\ 
  $12^h 15^m +$  &  $36^{\circ} +$  & &  &    & & & &   \\ 
\multicolumn{9}{c}{ \rule{220mm}{0.2mm}}      \\

                         37.802                                       &                             19:44.200                                       &                       $17.55 \pm  0.01$                                       &                                \dotfill                                       &                       $19.06 \pm  0.01$                                       &                       $18.90 \pm  0.01$                                       &                       $18.94 \pm  0.04$                                       &                                \dotfill                                       &                                \dotfill                                      \\
                             39.584                                       &                             19:35.376                                       &                       $17.66 \pm  0.04$                                       &                       $17.67 \pm  0.10$                                       &                       $19.08 \pm  0.07$                                       &                                \dotfill                                       &                       $19.03 \pm  0.04$                                       &                       $17.43 \pm  0.12$                                       &                       $17.22 \pm  0.11$                                      \\
                             40.561                                       &                             19:33.604                                       &                       $17.86 \pm  0.02$                                       &                       $18.00 \pm  0.10$                                       &                       $19.24 \pm  0.01$                                       &                                \dotfill                                       &                       $19.16 \pm  0.02$                                       &                       $18.04 \pm  0.14$                                       &                       $17.73 \pm  0.12$                                      \\
                             39.586                                       &                             19:29.245                                       &                       $17.87 \pm  0.01$                                       &                       $17.25 \pm  0.06$                                       &                       $19.57 \pm  0.05$                                       &                       $19.57 \pm  0.05$                                       &                                \dotfill                                       &                       $17.04 \pm  0.11$                                       &                       $17.18 \pm  0.11$                                      \\
                             40.574                                       &                             19:33.422                                       &                       $17.88 \pm  0.06$                                       &                       $16.78 \pm  0.08$                                       &                       $19.62 \pm  0.07$                                       &                                \dotfill                                       &                       $19.74 \pm  0.06$                                       &                       $17.28 \pm  0.12$                                       &                       $17.65 \pm  0.11$                                      \\
                             40.096                                       &                             19:36.976                                       &                       $17.88 \pm  0.01$                                       &                       $18.78 \pm  0.18$                                       &                       $18.95 \pm  0.00$                                       &                       $18.87 \pm  0.00$                                       &                       $18.65 \pm  0.01$                                       &                       $18.80 \pm  0.17$                                       &                       $18.23 \pm  0.12$                                      \\
                             39.515                                       &                             19:34.940                                       &                       $17.89 \pm  0.09$                                       &                       $18.05 \pm  0.10$                                       &                       $19.12 \pm  0.06$                                       &                                \dotfill                                       &                       $19.11 \pm  0.06$                                       &                       $17.89 \pm  0.13$                                       &                       $17.42 \pm  0.11$                                      \\
                             41.019                                       &                             19:28.766                                       &                       $17.98 \pm  0.02$                                       &                                \dotfill                                       &                       $19.64 \pm  0.02$                                       &                       $19.75 \pm  0.04$                                       &                       $19.77 \pm  0.03$                                       &                                \dotfill                                       &                                \dotfill                                      \\
                             40.711                                       &                             19:15.166                                       &                       $17.99 \pm  0.02$                                       &                       $16.64 \pm  0.04$                                       &                       $19.84 \pm  0.04$                                       &                       $20.02 \pm  0.06$                                       &                                \dotfill                                       &                       $17.17 \pm  0.18$                                       &                       $17.74 \pm  0.16$                                      \\
                             40.503                                       &                             19:32.866                                       &                       $18.01 \pm  0.04$                                       &                       $16.79 \pm  0.07$                                       &                       $19.61 \pm  0.03$                                       &                                \dotfill                                       &                       $19.73 \pm  0.05$                                       &                       $17.24 \pm  0.11$                                       &                       $17.63 \pm  0.12$                                      \\
                             41.149                                       &                             19:24.004                                       &                       $18.02 \pm  0.01$                                       &                       $18.55 \pm  0.10$                                       &                       $19.06 \pm  0.00$                                       &                       $18.91 \pm  0.00$                                       &                       $18.83 \pm  0.00$                                       &                       $19.00 \pm  0.12$                                       &                       $18.70 \pm  0.10$                                      \\
                             34.115                                       &                             20:31.083                                       &                       $18.10 \pm  0.01$                                       &                       $17.42 \pm  0.05$                                       &                       $19.88 \pm  0.01$                                       &                                \dotfill                                       &                       $19.78 \pm  0.02$                                       &                                \dotfill                                       &                                \dotfill                                      \\
                             39.215                                       &                             19:53.069                                       &                       $18.16 \pm  0.05$                                       &                       $18.42 \pm  0.17$                                       &                       $19.43 \pm  0.01$                                       &                       $19.25 \pm  0.02$                                       &                                \dotfill                                       &                                \dotfill                                       &                                \dotfill                                      \\
                             40.786                                       &                             19:11.591                                       &                       $18.17 \pm  0.05$                                       &                       $17.16 \pm  0.05$                                       &                       $19.66 \pm  0.11$                                       &                       $19.53 \pm  0.14$                                       &                       $19.57 \pm  0.12$                                       &                       $17.42 \pm  0.18$                                       &                       $17.93 \pm  0.16$                                      \\
                             40.459                                       &                             19:31.576                                       &                       $18.19 \pm  0.12$                                       &                       $16.96 \pm  0.13$                                       &                       $19.92 \pm  0.08$                                       &                                \dotfill                                       &                       $20.05 \pm  0.08$                                       &                                \dotfill                                       &                       $17.26 \pm  0.08$                                      \\
                             41.012                                       &                             19:28.795                                       &                       $18.31 \pm  0.10$                                       &                       $17.34 \pm  0.13$                                       &                       $19.90 \pm  0.07$                                       &                                \dotfill                                       &                       $19.78 \pm  0.36$                                       &                                \dotfill                                       &                                \dotfill                                      \\
                             39.686                                       &                             19:08.985                                       &                       $18.31 \pm  0.01$                                       &                       $18.91 \pm  0.12$                                       &                       $19.20 \pm  0.00$                                       &                                \dotfill                                       &                       $19.00 \pm  0.00$                                       &                       $18.84 \pm  0.12$                                       &                       $18.57 \pm  0.10$                                      \\
                             40.691                                       &                             19:14.928                                       &                       $18.37 \pm  0.06$                                       &                                \dotfill                                       &                       $20.30 \pm  0.10$                                       &                       $20.36 \pm  0.13$                                       &                                \dotfill                                       &                       $17.30 \pm  0.18$                                       &                       $17.94 \pm  0.16$                                      \\
                             39.447                                       &                             19:36.325                                       &                       $18.38 \pm  0.05$                                       &                                \dotfill                                       &                       $19.22 \pm  0.02$                                       &                                \dotfill                                       &                       $18.90 \pm  0.05$                                       &                       $18.76 \pm  0.18$                                       &                       $18.44 \pm  0.13$                                      \\
                             40.426                                       &                             19:31.109                                       &                       $18.47 \pm  0.02$                                       &                       $18.26 \pm  0.10$                                       &                       $19.93 \pm  0.02$                                       &                                \dotfill                                       &                       $19.96 \pm  0.05$                                       &                       $18.28 \pm  0.15$                                       &                       $18.23 \pm  0.12$                                      \\
                             39.497                                       &                             19:36.110                                       &                       $18.47 \pm  0.06$                                       &                       $17.62 \pm  0.09$                                       &                       $20.17 \pm  0.06$                                       &                                \dotfill                                       &                       $20.20 \pm  0.08$                                       &                       $17.86 \pm  0.13$                                       &                       $18.15 \pm  0.12$                                      \\
                             38.354                                       &                             19:26.346                                       &                       $18.47 \pm  0.01$                                       &                       $18.37 \pm  0.06$                                       &                       $20.02 \pm  0.04$                                       &                       $20.02 \pm  0.04$                                       &                                \dotfill                                       &                                \dotfill                                       &                                \dotfill                                      \\
                             39.162                                       &                             19:53.032                                       &                       $18.51 \pm  0.06$                                       &                                \dotfill                                       &                       $19.98 \pm  0.02$                                       &                       $20.06 \pm  0.07$                                       &                       $20.00 \pm  0.01$                                       &                                \dotfill                                       &                                \dotfill                                      \\
                             39.638                                       &                             19:37.031                                       &                       $18.51 \pm  0.03$                                       &                       $17.71 \pm  0.08$                                       &                       $20.24 \pm  0.02$                                       &                                \dotfill                                       &                       $20.58 \pm  0.04$                                       &                       $17.71 \pm  0.12$                                       &                       $18.02 \pm  0.12$                                      \\
                             36.297                                       &                             19:50.079                                       &                       $18.52 \pm  0.01$                                       &                       $17.84 \pm  0.05$                                       &                       $20.19 \pm  0.13$                                       &                       $20.31 \pm  0.00$                                       &                                \dotfill                                       &                                \dotfill                                       &                                \dotfill                                      \\

\multicolumn{9}{c}{ \rule{220mm}{0.8mm}}      \\
 \end{tabular}
\caption[Primeros 25 objetos de la  {\sc\footnotesize LISTA336}]{Primeros 
25 objetos de la  {\sc\footnotesize LISTA336}
 ordenados por magnitud ascendente en el filtro F336W. Se dan las coordenadas 
 ecuatoriales de los mismos en J2000 y todas las coordenadas halladas en los filtros
 F336W, F170W,  F555W, F702W, F814W, CN182 y CN270.
\label{tbl031}}
\end{sidewaystable}

\begin{figure*}[ht!]
\centering
\includegraphics[width=\textwidth]{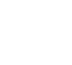}
\caption[ Campos de observaci'on de HST/STIS]{ {\sl\footnotesize  Campos 
de observaci'on de HST/STIS superpuestos sobre un
mosaico de NGC~4214.
Cada campo corresponde a una visita diferente.
El mosaico fue construido usando las im'agenes originales en los filtros
 F336W (canal azul), 
F814W+\ha~(canal rojo) y 
F555W+[O~{\scriptsize III}~{\normalsize]}~(canal verde).  \label{fig010}}}
\end{figure*}

 
La regi'on de la galaxia empleada para nuestro an'alisis se muestra
en la Figura~\ref{fig010}.  Esta im'agen fue construida usando 
los filtros F336W (canal azul), F814W+\ha~(canal rojo), y 
F555W+  [O~{\scriptsize III}~{\normalsize]}~(canal verde).

\subsection{Experimentos de completitud}
\label{sec:compl}
La mayor'ia de los estudios propuestos para este trabajo
requieren una evaluaci'on cuantitativa de la completitud de los datos.
Por tal motivo, realizamos varios experimentos en los que agregamos estrellas artificiales
de magnitud y posici'on conocidas sobre las im'agenes 
y luego tratamos de detectarlas empleando los mismos procedimientos usados para
las estrellas reales.

Esto fue hecho con rutinas dentro del paquete {\sc\footnotesize HSTphot}. Las estrellas artificiales
agregadas tienen magnitudes y colores fotom'etricos
en el rango $14 \le $ F336W $  \le    26$ 
y $-2.0 \le$ F336W -- F555W $   \le    4.0$. 
Agregamos aproximadamente 50\,000 estrellas artificiales en cada
im'agen.
El cociente entre el n'umero de objetos recuperados y el n'umero de
objetos insertados artificialmente en cada bin de magnitudes da el factor
de completitud buscado como funci'on de la magnitud.

\begin{figure*}[ht!]
\begin{minipage}[c]{15cm}
\centering
\includegraphics[width=\textwidth]{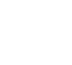}
\caption[Valores de completitud de la fotometr'ia de la  WFPC2]{  {\sl\footnotesize  
 Valores de completitud de la fotometr'ia de la  WFPC2  basados
en experimentos  artificiales ejecutados con el paquete de programas {\sc\footnotesize HSTphot}. 
[Izquierda] valores de completitud de la fotometr'ia  de las
im'agenes de la propuesta  6716 de los filtros  
F336W, F555W y F814W.
[Derecha] Lo mismo que en los paneles de la izquierda pero para la 
propuesta 6569.
\label{fig004}}}
\end{minipage}
\end{figure*}

La Figura~\ref{fig004}  representa los resultados de estos experimentos
realizados sobre los filtros F336W, F555W y
F814W en  las propuestas  6569 y 6717. Se observa que 
en el extremo m'as brillante, la completitud 
es de aproximadamente 95\% o mayor en todos los filtros.
Tambien se observa que en los objetos m'as d'ebiles,
la completitud cae m'as r'apidamente en las im'agenes de la propuesta
6716 que en las im'agenes de la propuesta 6569, como es de esperar
debido al tiempo de exposici'on m'as corto de las primeras.

 
\begin{figure*}[ht!]
\centering
\includegraphics[width=15cm]{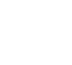}
\begin{minipage}{15cm}
\caption[ Mapa de contornos con los  valores de
completitud ]{ {\sl\footnotesize  Mapa de contornos  que muestra los valores de
completitud para cada punto del diagrama de Hertzsprung-Russell.  
Se han marcado tambi'en (para referencia) porciones de 
algunas trayectorias evolutivas obtenidas de \cite{LeSc01}
   \label{fig005}}}
\end{minipage}
\end{figure*}

Empleando \cho\ (ver  Cap'itulo 5) obtuvimos 
dos relaciones muy 'utiles:  la relaci'on entre 
F336W -- F555W y \teff; y la relaci'on entre
F336W -- F555W y \mbol\ calculado para un valor t'ipico de 
la extinci'on de NGC~4214.
Mapeamos el \hrd\ como una matriz muy fina, y calculamos
el valor de la completitud para cada punto. Esto est'a representado
en la   Figura~\ref{fig005}.  Este procedimiento fue empleado solamente para estimar
valores de completitud, no para determinar propiedades f'isicas de 
estrellas o c'umulos. Un estudio independiente de la completitud de los datos de STIS no fue necesario
ya que nuestras listas relacionan a todos los objetos detectados con  STIS
 con objetos de la   {\sc\footnotesize LISTA336}.
 Por tal motivo, los experimentos de completitud realizados para la   {\sc\footnotesize LISTA336}
 incluyen a los datos de la WFPC2 y los de STIS. 
 
\newpage
 
\subsection{Consistencia de la fotometr'ia estelar}
\begin{figure}
\centerline{\includegraphics*[width=0.47\linewidth]{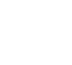}
                  \includegraphics*[width=0.47\linewidth]{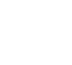}}
\centerline{\includegraphics*[width=0.47\linewidth]{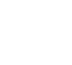}
                  \includegraphics*[width=0.47\linewidth]{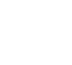}}

\caption[Consistencia de la fotometr'ia ]{{\sl\footnotesize   Histogramas de las magnitudes medidas
normalizadas por las incertidumbres experimentales
 ( $  \Delta m /   \sqrt{ \sigma_{1}^{2} + \sigma_{2}^{2} } $ ). 
aqui comparamos las magnitudes de las estrellas observadas en
dos visitas de la propuesta 6716.
Ver el texto para mas detalles. La curva continua representa la distribuci'on 
normal esperada.
\label{fig006}}}
\end{figure}

Los campos observados durante las dos visitas
de la  propuesta 6716 contienen una regi'on en com'un que nos permiti'o
estudiar la consistencia de nuestra fotometr'ia y chequear por posibles
incertidumbres  introducidas por la CTE y las correcciones por contaminaci'on.
Obtuvimos dos listas fotom'etricas (una por cada visita)
de cada filtro. Luego realizamos un estudio estad'istico 
sobre los objetos que fueron identificados como la misma
estrella en las dos listas.
Para cada objeto tenemos dos valores de magnitud con sus correspondientes errores
$m_1 \pm \sigma_1$ y $m_2 \pm \sigma_2$.  
Calculamos la distribuci'on $(m_1 -  m_2) /   
\sqrt{  \sigma_{1}^{2} + \sigma_2^{2}} $, donde  
$m_i$ y $ \sigma_i$  
son las magnitud y su incertidumbre en la lista $i$. 
 Esta distribuci'on se asemeja
a una distribuci'on normal en todos los filtros estudiados, 
como puede observarse en los histogramas de la Figura~\ref{fig006}. El n'umero
de puntos (N)  considerados para los histogramas, el promedio de
  $  \Delta m /   \sqrt{ \sigma_1^2 + \sigma_2^2 } $
y su desviaci'on estandar est'an dados en la  
Tabla~\ref{tbl005}. Notar que, dentro de los errores, 
obtenemos una distribuci'on Gaussiana con promedio aproximadamente nulo y 
desviaci'on estandar de aproximadamente uno como es de esperarse.

\begin{figure*}[ht!]
\begin{minipage}[c]{8cm}
\includegraphics[width=8cm]{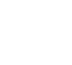}
\vspace{2cm}
\end{minipage}
\begin{minipage}[b]{6cm}
\caption[Consistencia de la
fotometr'ia]{ Experimento para analizar la consistencia de la
fotometr'ia.
Comparamos las magnitudes en el filtro F336W en las dos listas
y graficamos la diferencia  $  \Delta  \mathrm{F336W} = \mathrm{F336W}_1 - 
\mathrm{F336W}_2  $ como funci'on de 
$\mathrm{F336W}_1$.  \label{fig007}}\end{minipage}
\end{figure*}

Tambi'en comparamos la magnitud obtenida para  un mismo
objeto en las dos listas para asegurarnos la calidad de la fotometr'ia.
Graficamos una de las magnitudes versus la diferencia en magnitudes en la 
Figura~\ref{fig007} para el filtro F336W de la propuesta 6716.
Como puede verse, la coincidencia es muy buena, con una leve dispersi'on
aumentando hacia las magnitudes m'as d'ebiles donde los errores
son mayores como es de esperarse.

\begin{figure}
\centerline{\includegraphics*[width=0.47\linewidth]{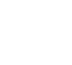}
                  \includegraphics*[width=0.47\linewidth]{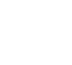}}
\centerline{\includegraphics*[width=0.47\linewidth]{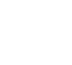}
                  \includegraphics*[width=0.47\linewidth]{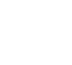}}
\centerline{\includegraphics*[width=0.47\linewidth]{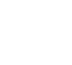}
                  \includegraphics*[width=0.47\linewidth]{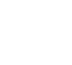}}

\caption[Residuos en las magnitudes]{ {\sl\footnotesize Residuos en las magnitudes 
obtenidas con los filtros 
 F336W, F555W y F814W. Los resultados de la columna de la izquierda corresponden a
 la lista constru'ida con objetos de la propuesta 6569.
 Los gr'aficos de la columna de la derecha son los de la propuesta 6716. \label{fig043}}}
\end{figure}

Analizamos los errores en la fotometr'ia  como una funci'on de la magnitud
para todos los filtros usados. Estas incertidumbres fueron determinadas para cada estrella
en base a 
errores estad'isticos, errores en la determinaci'on del cielo y en las correcci'on por
apertura. Hallamos que algunos de los puntos exceden a los errores estad'isticos,
especialmente en los filtros F555W y F814W. Confirmamos  que estos objetos
se encuentran dentro de regiones \hii donde las l'ineas de emisi'on  afectan considerablemente
al cielo local. La Figura~\ref{fig043}  presenta seis gr'aficos de los errores en la magnitud en
funci'on de las magnitudes F336W, F555W y F814W para las propuestas 6716 y 6569.

\vspace{2cm}
\begin{table}[htbp]
\centering
\begin{tabular}[h]{lcccc} 
\multicolumn{5}{l}{ \rule{85mm}{0.8mm}}  \\
  Filtro &  F170W &  F336W   {\rule [0mm]{0mm}{5mm} }   &  F555W & F814W   {\rule [-2mm]{0mm}{0mm}}  \\
 \multicolumn{5}{l}{ \rule[2mm]{85mm}{0.2mm}}  \\
 N       & 189  & 1\,114  & 3\,219  & 4\,116  {\rule [0mm]{0mm}{5mm} }   \\  
promedio &$ -0.120$&$-0.013$  & 0.119 & 0.049  {\rule [0mm]{0mm}{5mm} }  \\  
$\sigma$  & 0.839 & 0.915 & 0.841 & 0.847  {\rule [0mm]{0mm}{5mm} }  \\  
   
     \multicolumn{5}{l}{ \rule{85mm}{0.8mm}}  \\

 \end{tabular}
 
 \begin{minipage}[u]{85mm}
\caption[Resultados del estudio estad'istico]{Resultados del estudio estad'istico.
N representa el n'umero de estrellas en com'un entre las dos
listas; promedio y $\sigma$  son el valor medio
y la desviaci'on estandar de la cantidad
$  \Delta m /   \sqrt{ \sigma_1^2 + \sigma_2^2 } $}
\label{tbl005}
\end{minipage}
\end{table}

\vspace{2cm}

\newpage


\section{Descripci'on de im'agenes de STIS y su reducci'on}
La Tabla~\ref{tbl001} contiene el resumen de las im'agenes STIS
utilizadas en este trabajo.
Todas las observaciones emplean el detector 
Cs$_2$Te Multianode Microchannel Array (MAMA) en el UV cercano.
Este detector tiene un campo de $25\asec \times  25\asec $  y un tama\~no de p'ixel
de $\approx 0 \arcs 02468.$ 
Las im'agenes fueron obtenidas en  la propuesta 9096 (P.I.: Jes\'us  
Ma\'{i}z Apell\'aniz) y muestran a las regiones 
NGC4214--I and NGC4214--II  con una mayor resoluci'on que las im'agenes
de la WFPC2. La misma fuente fue observada en varias im'agenes con un 
desplazamiento  levemente diferente (dithering) como puede
apreciarse en la Figura~\ref{fig010}.
Los filtros empleados son F25CN182 y F25CN270.
El filtro F25CN182 provee im'agenes de banda intermedia en  una longitud de 
onda central de 1\,820 \AA\ y un  FWHM de  350 \AA. 
El filtro F25CN270  provee im'agenes de banda intermedia en  una longitud de 
onda central de 2\,700 \AA\ y un  FWHM de  350 \AA.

Los datos de STIS fueron reducidos por el datoducto de STScI mediante el
proceso  {\it On--the--Fly}. El programa empleado para la
calibraci'on es el {\sc\footnotesize calstis}.
Esta calibraci'on realiza correcciones por p'ixeles malos, 
la conversi'on de an'alogo a digital, la sustracci'on del bias y
de la corriente oscura y la correcci'on por flats.

El proceso que nosotros realizamos fue el c'alculo del
valor de {\sc\small  PHOTFLAM} correspondiente a nuestras observaciones
empleando los 'ultimos archivos de calibraci'on accesibles.

Las im'agenes calibradas      (archivos {\sc flt})    est'an expresados
en cuentas. La conversi'on a flujo ($erg \times cm^{-2} \times  sec^{-1} \times   $  {\AA}$^{-1}$)
est'a dada por:

 \begin{equation}
f_{\lambda} = \frac{\mathrm{{\sc\small  COUNTS}}  \times  \mathrm{{\sc\small  PHOTFLAM}}}{\mathrm{{\sc\small  EXPTIME}}} 
\end{equation}

donde {\sc\small  PHOTFLAM} es la sensibilidad del modo
de observaci'on, y {\sc\small   EXPTIME } es el tiempo de exposici'on en segundos.
Ambos par'ametros est'an dados en la cabecera de cada imagen.
Estos flujos pueden convertirse a magnitudes en el sistema {\sc\footnotesize  STMAG} mediante
la relaci'on

\begin{equation}
\mathrm{{\sc\small   STMAG}} = -2.5 \times  \log_{10}( f_{\lambda}) - 21.1
\end{equation}

El paquete {\sc\footnotesize  STSDAS} de fotometr'ia sint'etica llamado
{\sc\footnotesize synphot} puede simular las observaciones 
HST de fuentes astron'omicas. Contiene curvas de transmisi'on de todos los componentes
del HST como espejos, filtros, aperturas y detectores.
Con 'estos se puede generar cualquier combinaci'on posible.

Nosotros empleamos    {\sc\footnotesize synphot}  para recalcular
el valor de {\sc\small   PHOTFLAM}  para cada imagen en las fechas correspondientes.
%
Editamos las cabeceras de las im'agenes y agregamos los nuevos
valores de  {\sc\small   PHOTFLAM.}
Para corregir nuestras im'agenes por distorsi'on geom'etrica, usamos
la correcci'on m'as reciente provista por
\citet{MaizUbed04}.


Comparando las Figuras~\ref{fig003} y~\ref{fig010}  observamos que los campos de observaci'on
de STIS est'an situados dentro de los de la WFPC2.
Usando el procedimiento    {\sc\footnotesize  FIND } de IDL
que es el equivalente a    {\sc\footnotesize  DAOFIND } en el paquete  
  {\sc\footnotesize  DAOPHOT} de    {\sc\footnotesize IRAF}, hallamos
las coordenadas de los objetos en las 11 im'agenes disponibles de STIS
que tenemos.
Luego, realizamos fotometr'ia de apertura y asociamos estos objetos 
con los  objetos identificados con la WFPC2 en la    {\sc\footnotesize LISTA336}. 
Algunos de los objetos en la   {\sc\footnotesize LISTA336} tienen un 'unico 
objeto asociado en la lista de STIS, y algunos pocos tiene dos o mas
objetos asociados en STIS. Este 'ultimo caso se debe a la mayor resoluci'on
del instrumento STIS. Por tal raz'on, decidimos combinar la magnitud de estos objetos
antes de incluirlos en nuestra lista final.
La magnitud y su error fueron calculados mediante:

\begin{equation}
 m    =   -2.5 \times  \log_{10} \left(    10^{   -0.4  \times m_1}       +  
  10^{  -0.4 \times  m_2}            \right)
\end{equation}
\begin{equation}
\sigma_{m}  =     10^{ 0.4\times  m  }    \times  
\sqrt{ 10^{-0.8\times  m_1  }     
\times  \sigma_{m_{1}}^{2}   +   10^{-0.8  \times  m_2   }   
\times  \sigma_{m_{2}} ^{2}      }    
\end{equation}

Este problema surge pues estamos trabajando con detectores  que tienen diferente resoluci'on.
A una distancia comparable a la de NGC~4214 (2.94 Mpc), los  problemas inherentes a 
sistemas m'ultiples y resoluci'on son inevitables.
Por ejemplo, 
 \citet{Maizetal05} estudian los  problemas derivados de 
 sistemas m'ultiples  en la determinaci'on de la funci'on inicial  de 
 masa.
Presentan los resultados de Trumpler~14, un c'umulo masivo muy joven
en la Nebulosa de Carina que contiene al menos tres estrellas muy tempranas de tipo 
espectral O.
Si este c'umulo estuviese localizado a una distancia similar a la de 
NGC~4214, aparecer'ia como una fuente puntual, aunque en realidad
sabemos con certeza de que se trata de un sistema m'ultiple.
Nosotros empleamos este resultado para justificar el procedimiento empleado 
en este trabajo.


Algunas de las im'agenes STIS comparten el mismo campo de observaci'on.
Cuando una misma fuente fue identificada en dos listas,
le asignamos la magnitud

\begin{equation}  m= \frac{\frac{m_1}{{\sigma_{m_{1}}^{2}}} +
\frac{m_2}{{\sigma_{m_{2}}^{2}}} }{\frac{1}{\sigma_{m_{1}}^{2}} + 
\frac{1}{\sigma_{m_{2}}^{2}}   }  , 
\end{equation} 

donde   $m_1$ y $m_2$  son las dos magnitudes originales,
y $\sigma_{m_{1}}$ y  $\sigma_{m_{2}}$ son sus respectivos errores. 
El valor asignado a la incertidumbre est'a dado por:

\begin{equation} \sigma_{m} = \frac{1 }{ \sqrt{ \frac{1}{{\sigma^{2}_{m_{1}}}} 
+ \frac{1}{{\sigma^{2}_{m_{2}}}}}}   
\end{equation} 

siempre que    $(m_1 -  m_2)/\sqrt{{\sigma^2_{m_{1}}} + {\sigma^2_{m_{2}}}}  \le 2.0 $. 
Por otro lado, si   
$(m_1 -  m_2)/\sqrt{{\sigma^2_{m_{1}}} + {\sigma^2_{m_{2}}}}  >  2.0 $
la magnitud se calcul'o mediante 
\begin{equation}  m = \frac{m_1 + m_2}{ 2}    \end{equation} 
y la incertidumbre con
\begin{equation}  \sigma_{m} = \frac{|m_1  -  m_2|}{ 2}    \end{equation} 


\newpage

\section{Fotometr'ia de objetos extendidos}

\begin{figure*}[ht!]
\centering
\includegraphics[width=\textwidth]{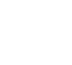}
\caption[Aperturas circulares usadas
para la fotometr'ia de los c'umulos]{ {\sl\footnotesize   
Mosaico de NGC~4214 constru'ido con im'agenes en filtros  F336W (azul), 
F814W~(rojo) y 
F555W~(verde). Se muestran los c'umulos estudiados y sus aperturas circulares.
La orientaci'on es: Norte  hacia arriba y Este  hacia la izquierda.
Las dimensiones del campo son 875 pc $\times$ 972 pc 'o $ 61 \arcs 4 \times 68 \arcs 3$.
\label{fig027}}}
\end{figure*}

\begin{figure*}[ht!]
\centering
\includegraphics[width=\textwidth]{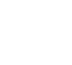}
\caption[Detalle de los c'umulos
 estelares I--As, I--Es, IIIs y IVs.]{ {\sl\footnotesize   Im'agenes HST de los c'umulos
 estelares I--As, I--Es, IIIs y IVs. La escala de  
  las im'agenes  fue ajustada para que todas tengan el mismo tama\~no lineal
  de 50 pc  $\times$ 50 pc. El Norte  hacia arriba y
  el Este hacia la izquierda. Los c'irculos muestran las aperturas empleadas en
  nuestra fotometr'ia. Mostramos al c'umulo I--As sobre una im'agen de STIS en
  el filtro F25CN270.  El c'umulo I--Es est'a representado sobre una imagen WFPC2/PC en
  el filtro F555W. Los c'umulos IIIs y IVs se muestran sobre im'agenes de la WFPC2/WF
  en el filtro F814W. Se observa claramente el  tama\~no diferente de los
  p'ixeles de las tres c'amaras que empleamos.   \label{fig028}}}
\end{figure*}

El paquete {\sc\footnotesize  HSTphot}   \citep{Dolp00a} 
fue dise\~nado espec'ificamente para obtener  la fotometr'ia
de fuentes puntuales en im'agenes obtenidas con las c'amaras HST/WFPC2.
Nuestras im'agenes WFPC2 y STIS contienen algunas fuentes 
(I--As, I--Es, IIIs  y  IVs) que se sabe son
\cus estelares no resueltos, y que aparecen
como fuentes extendidas en lugar de como objetos puntuales.
La posici'on de estos objetos   se muestra en la 
Figura~\ref{fig027}  y sus correspondientes ampliaciones
en la 
Figura~\ref{fig028}.

 
Otros \cus est'an resueltos  y tambi'en fueron analizados
empleando  fotometr'ia de apertura.
A estos \cus los organizamos en dos grupos: 
los complejos ( que incluyen \cus m'as peque\~nos)
 I--A, I--B, y  II; y los \cus resueltos 
 I--Ds, II--A, II--B, II--C, II--D, y II--E.
Sus posiciones en la galaxia se muestran en la 
 Figura~\ref{fig027}.  

Hicimos fotometr'ia de apertura de estos c'umulos, seleccionando
cuidadosamente los radios de apertura y los radios del cielo.
En todos los casos incluimos el 100\% de la luz
proveniente de cada c'umulo, con excepci'on del \cu I--As
que presenta un halo muy extendido.
Usamos c'odigos IDL para desarrollar esta tarea sobre las im'agenes 
de la WFPC2 y STIS.
Los valores de la correcci'on de apertura  (en magnitudes)  para las
im'agenes STIS  las  obtuvimos de  \citet{Prof03}.
En algunos casos particulares (cuando
el \cu era suficientemente brillante
y no encontramos problemas  de confusi'on con otras fuentes), 
pudimos determinar  las magnitudes $JHK_s$ 
de los \cus  a partir de las im'agenes 2MASS empleando los mismos
radios de apertura que usamos para las im'agenes de la WFPC2 y STIS.
%
La fotometr'ia de los  \cus se encuentra detallada en las Tablas~\ref{tbl006} and \ref{tbl014}.


\begin{sidewaystable}
\centering
\begin{tabular}{  lccccccc  }
\multicolumn{8}{l}{ \rule{210mm}{0.8mm}}      \\
Filtro &  C\'umulo I--As &  C\'umulo I--Es &  C\'umulo IIIs &  C\'umulo IVs 
&  C\'umulo I--A &   C\'umulo I--B &  C\'umulo II   {\rule [-3mm]{0mm}{8mm} }  \\ 
\multicolumn{8}{l}{ \rule{210mm}{0.2mm}}      \\
F170W   \dotfill  & 13.47 $\pm$ 0.07  & 18.82 $\pm$ 0.52  & 16.77 $\pm$ 0.15  & \dotfill    &11.81 $\pm$ 0.02  & 13.54 $\pm$ 0.03  & 12.45 $\pm$ 0.04  
 {\rule{0mm}{5mm} }  \\
F25CN182  &  13.50 $\pm$ 0.02  &  19.27 $\pm$ 0.10  & \dotfill  & \dotfill  &11.99 $\pm$ 0.02  &13.60 $\pm$ 0.02&  12.91 $\pm$ 0.02   {\rule{0mm}{0mm} } \\
F25CN270  &  14.00 $\pm$ 0.03   & 19.56 $\pm$ 0.11  &  \dotfill  & \dotfill  & 12.51 $\pm$ 0.01  & 14.17 $\pm$ 0.02 & 13.20 $\pm$ 0.02  {\rule{0mm}{0mm} }\\
F336W  \dotfill &  14.46 $\pm$ 0.04  &  19.30 $\pm$ 0.03  & 16.82 $\pm$ 0.05  & 17.17 $\pm$ 0.03  & 12.96 $\pm$ 0.01  & 14.62 $\pm$ 0.01 &13.59 $\pm$ 0.01 {\rule{0mm}{0mm} }  \\
F555W  \dotfill & \dotfill  & 18.98 $\pm$ 0.05  & 16.59 $\pm$ 0.03  &17.30 $\pm$ 0.01  & \dotfill  &15.83 $\pm$ 0.23   & 14.37 $\pm$ 0.04   {\rule{0mm}{0mm} } \\
F702W  \dotfill & \dotfill  & 18.27 $\pm$ 0.04&  \dotfill  & 16.97 $\pm$ 0.01 & \dotfill  & \dotfill    &   14.04 $\pm$ 0.08  {\rule{0mm}{0mm} }\\
F814W  \dotfill & 15.98 $\pm$ 0.02  & 17.90 $\pm$ 0.03  & 15.67 $\pm$ 0.03  & 16.75 $\pm$ 0.02 & 13.93 $\pm$ 0.01 & 15.55 $\pm$ 0.01&  14.24 $\pm$ 0.01    {\rule{0mm}{0mm} }\\
$J$ \dotfill & \dotfill  & \dotfill  &  14.74 $\pm$ 0.10 &15.93 $\pm$ 0.13   & 14.43 $\pm$ 0.50  &  15.69 $\pm$ 0.25 &  14.27 $\pm$ 0.63 {\rule{0mm}{0mm} }  \\
$H$  \dotfill  & \dotfill  & \dotfill  & 14.30 $\pm$ 0.11 &  15.55 $\pm$ 0.19 & 14.31 $\pm$  0.88 &15.05 $\pm$ 0.27  &13.65 $\pm$ 0.70  {\rule{0mm}{0mm} }  \\
$K_s$  \dotfill   &  \dotfill  & \dotfill  &13.87 $\pm$ 0.10   & 15.39 $\pm$ 0.20  &  13.54 $\pm$  0.66  & 15.27 $\pm$ 0.42  &    13.71 $\pm$ 1.02    {\rule [-3mm]{0mm}{0mm} }   \\ 
\multicolumn{8}{l}{ \rule{210mm}{0.8mm}}      \\
 \end{tabular}
\caption{Fotometr'ia de los c'umulos no resueltos y de los complejos grandes. \label{tbl006}}
\end{sidewaystable}

\begin{sidewaystable}
\centering
\begin{tabular}{  lcccccc}
\multicolumn{7}{l}{ \rule{185mm}{0.8mm}}      \\
 Filtro &C\'umulo I--Ds&  C\'umulo II--A&  C\'umulo II--B&  C\'umulo II--C 
& C\'umulo II--D &  C\'umulo II--E     {\rule [-3mm]{0mm}{8mm} } \\    \multicolumn{7}{l}{ \rule{185mm}{0.2mm}}      \\

F170W   \dotfill  & 15.51 $\pm$ 0.03  &15.54 $\pm$ 0.18  & 15.26  $\pm$ 0.05  & 14.23 $\pm$ 0.04 &16.39 $\pm$  0.16&15.97 $\pm$ 0.13  {\rule [0mm]{0mm}{5mm} }  \\
  F25CN182  &  \dotfill  &15.63 $\pm$ 0.06  & 15.50 $\pm$ 0.02  & 14.67 $\pm$  0.02 &16.66 $\pm$ 0.06 &16.12 $\pm$  0.04{\rule{0mm}{0mm} } \\
 F25CN270  &  \dotfill   &15.66  $\pm$ 0.06  &15.70 $\pm$  0.02    &15.07 $\pm$ 0.02&16.96 $\pm$  0.07&  16.33 $\pm$  0.04{\rule{0mm}{0mm} }\\  
 F336W  \dotfill &  16.39 $\pm$ 0.01  &15.87  $\pm$ 0.01  &16.04 $\pm$ 0.01  & 15.52 $\pm$ 0.01& 17.30 $\pm$  0.02 &16.67 $\pm$  0.01{\rule{0mm}{0mm} } \\ 
 F555W  \dotfill &17.62 $\pm$ 0.02&16.46 $\pm$ 0.08  & 16.87 $\pm$ 0.03  & 16.76 $\pm$ 0.06 &18.12 $\pm$  0.12&17.75 $\pm$  0.06{\rule{0mm}{0mm} }\\
 F702W  \dotfill & 17.54 $\pm$ 0.04  & 16.09  $\pm$ 0.07& 16.56 $\pm$  0.06  & 16.50 $\pm$  0.12&18.00 $\pm$  0.22& 17.48 $\pm$ 0.05{\rule{0mm}{0mm} }\\

  F814W  \dotfill & 17.43 $\pm$ 0.03  & 16.46  $\pm$ 0.08  & 17.02 $\pm$ 0.01  & 16.72 $\pm$ 0.01&18.21 $\pm$  0.01&17.93 $\pm$  0.01  {\rule [-3mm]{0mm}{0mm}} \\ 
  \multicolumn{7}{l}{ \rule{185mm}{0.8mm}}      \\
 \end{tabular}
\caption{Fotometr'ia de los c'umulos resueltos \label{tbl014}}
\end{sidewaystable}


\subsection{C'umulos no resueltos}

Las im'agenes de la propuesta 6716 muestran al \cu  I--As
en la PC de la WFPC2, mientras que los de la propuesta 6569 contienen a esta regi'on 
en la WF3.
Este c'umulo es el m'as brillante en NGC~4214 en el 'optico, y algunos de sus p'ixeles
est'an saturados en algunas im'agenes
(filtro F555W en la propuesta 6716; filtros  F336W, F555W, y F702W en  la propuesta
 6569.)
Este objeto tambien aparece en algunas 
de nuestra im'agenes de STIS. Desafortunadamente, las im'agenes del programa 6716
de diciembre de 1997 tienen una columna de p'ixeles malos
que pasan a trav'es de I--As; para deshacernos de este problema, interpolamos
el n'umero de cuentas en cada p'ixel a lo largo de la columna, usando los p'ixeles de las columnas
adyacentes del chip.
Realizamos fotometr'ia de apertura con un radio de 9 p'ixeles de la PC
lo cual es equivalente a  $5.7 $ pc  a una distancia de 2.94 Mpc,
o  $0\arcs 41. $ 
Las coordenadas ecuatoriales del c'umulo est'an dadas en la Tabla~\ref{tbl015}.
Luego de obtener las magnitudes en todas las 
im'agenes disponibles, calculamos un valor medio 
pesado para cada filtro usando a los errores como pesos.

Entre los \cus  I--A y I--B se observa un peque\~no c'umulo:  I--Es.
Lo econtramos en las im'agenes de los filtros F170W, F336W, F555W
y  F814W en la  PC de la propuesta 6716.
Tambien se encuentra en la WF3 de las im'agenes 
de la propuesta  6569 y en algunas de las im'agenes de STIS.
En algunas de estas im'agenes  encontramos el mismo problema 
de la columna de p'ixeles malos y lo solucionamos  siguiendo la
misma t'ecnica que usamos con el \cu  I--As.

El \cu IIIs se encuentra en la WF4 en las im'agenes de la propuesta 6716.
Algunos de sus p'ixeles est'an saturados  en el filtro F555W. Lo mismo
pasa en el filtro F702W de la propuesta 6569.
Este \cu aparece muy cerca de la zona del vi\~neteado entre la PC
y la WF4 en las im'agenes de
la propuesta 6716; esto nos llev'o a pensar que 
la magnitud obtenida se ver'ia afectada y las descartamos, excepto la magnitud
obtenida con el filtro F170W.
Usamos un radio de apertura de 12 p'ixeles  de la WF
que es equivalente a  $16.8 $ pc. 
Este objeto est'a presente en el 2MASS All--Sky Catalog of Point Sources  \citep{Cutretal06} 
y est'a  relativamante aislado en esas im'agenes lo cual nos permiti'o
obtener as'i las magnitudes en los filtros  $J$, $H$ y $K_s$.
Adem'as, una correcci'on por apertura fue requerida
para que las magnitudes sean comparables con las de la WFPC2, teniendo 
en cuenta la diferente resoluci'on espacial.
Esta correcci'on fue calculada comparando las magnitudes   $J$, $H$ y $K_s$
del c'umulo IIIs
provistas por el Cat'alogo 2MASS  con las magnitudes calculadas del mismo objeto 
usando las im'agenes 2MASS.

\begin{wraptable}[20]{l}[2mm]{75mm}
\begin{tabular}{lcccc }
\multicolumn{5}{c}{ \rule{70mm}{0.8mm}}  \\
\raisebox{-0.9ex}[0pt]{C'umulo}   &   $\alpha$ (J2000) &   $\delta$ (J2000)   {\rule [0mm]{0mm}{5mm} }   & 
Radio   \\
 & $12^h 15^m +$  &  $36^{\circ} +$  &  {["]}   {\rule [-3mm]{0mm}{0mm}}  \\ \multicolumn{5}{c}{ \rule[2mm]{70mm}{0.2mm}}  \\
      I--As      &39.54&      19  36.49& 0.41   {\rule [0mm]{0mm}{5mm} }   \\ 
      I--Es     &40.26&       19  39.94& 0.36   {\rule [0mm]{0mm}{0mm} }   \\ 
      IIIs     &38.27&        19  46.29& 1.20 {\rule [0mm]{0mm}{0mm} }\\ 
       IVs     &37.34&       19  58.20& 0.80  {\rule [0mm]{0mm}{0mm} }\\  
       
   I--A     &  39.53      &  19   37.00  &    6.83    {\rule [0mm]{0mm}{0mm} } \\
       I--B     &  40.40  &  19   33.18  & 2.96 {\rule [0mm]{0mm}{0mm} } \\
        II    &   40.79    &  19  09.26  & 9.10 {\rule [0mm]{0mm}{0mm} }\\ 
        
       I--Ds     &39.22  &   19  52.96  & 0.91 {\rule [0mm]{0mm}{0mm} }\\ 
      II--A     &41.20    &   19   05.12& 2.28 {\rule [0mm]{0mm}{0mm} }\\ 
      II--B      &40.77&      19  11.06& 1.68 {\rule [0mm]{0mm}{0mm} }\\ 
      II--C      &40.72&      19  15.11& 2.59 {\rule [0mm]{0mm}{0mm} } \\ 
      II--D      &40.80&      19   08.25& 1.46 {\rule [0mm]{0mm}{0mm} }\\ 
      II--E     &40.87&       19   04.71& 1.59    {\rule [-3mm]{0mm}{0mm}} \\  \multicolumn{5}{c}{ \rule{70mm}{0.8mm}}  \\

 \end{tabular}
 
 \begin{minipage}[u]{7cm}
\caption{Astrometr'ia de los c'umulos y radio de apertura empleado en la fotometr'ia. \label{tbl015}}
\end{minipage}

\end{wraptable}


El \cu IVs cae sobre la zona del vi\~neteado  entre las c'amaras WF3 y WF4 de la
propuesta 6716, lo cual hizo que sea imposible
calcular  su magnitud en esas im'agenes.
Empleamos un radio de apertura de $0\arcs 8 $ para la
fotometr'ia.
Este \cu est'a listado en el Cat'alogo 2MASS y usamos  sus magnitudes en los filtros
  $J$, $H$ y $K_s$ para nuestro estudio, luego de corregirlas
  por efectos de apertura como hicimos con
  el \cu IIIs.
La Tabla~\ref{tbl006} contiene la fotometr'ia de estos dos c'umulos.

\subsection{Complejos compuestos}

Realizamos fotometr'ia de apertura de los complejos
I--A, I--B, y  II  usando los 
radios de apertura  que  \citet{MacKetal00} 
usa para describir a esos  c'umulos. 
  
Centramos al \cu  I--A en una posici'on levemente desplazada
respecto al \cu  I--As. Cuando nos encontramos con columnas malas
las corregimos siguiendo el mismo m'etodo empleado  con el \cu  I--As.
Algunas de nuestras im'agenes contienen algunos 
p'ixeles saturados dentro del radio de apertura de 
150 p'ixeles  de la PC y sus magnitudes fueron 
descartadas.

Otro \cu que es parte del complejo  NGC~4214--I es el  I--B, que 
est'a localizado al SE del \cu I--A. Empleamos un 
radio de apertura de 65 p'ixeles  de la PC  ($ 2\arcs 96 $) en este
caso.
No pudimos usar las im'agenes de la propuesta 6569 para 
este \cu ya que el mismo se encuentra muy cerca de la  zona
de vi\~neteado de los chips. Algunas de las im'agenes de la propuesta 6716
contienen  columnas con p'ixeles malos.

La otra estructura nebular grande en la galaxia es  NGC~4214--II. 
Esta interesante regi'on de  NGC~4214  se encuentra en
la c'amara WF4 en las im'agenes de la propuesta 6716, en 
el chip WF2 de la propuesta 6569 y en algunas de nuestras im'agenes
de STIS.
El radio de apertura empleado para la fotometr'ia de apertura
fue  de 200 p'ixeles  de la PC (o 91 p'ixeles  de la WF).
Nos encontramos con algunas columnas de p'ixeles malos que se extienden
a lo largo de este c'umulo, especialmente a trav'es de las regiones II--A, y  II--C.

Usando las im'agenes provistas por el cat'alogo 2MASS, obtuvimos la fotometr'ia de
apertura en los filtros $J$, $H$  y $K_s$ de los \cus  I--A, I--B y  II.
Empleamos los centros y radios de apertura listados en Tabla~\ref{tbl015}.
Para calibrar nuestras mediciones, usamos las magnitudes provistas por
el Cat'alogo 2MASS de los \cus  IIIs y IVs, que fueron calculados
con un radio de $ 4\arcs 0 $.
La fotometr'ia de estas im'agenes da errores relativos grandes
pues el cielo en estas im'agenes es  extremadamente ruidoso. 
La Tabla~\ref{tbl006} contiene la fotometr'ia de estos   \cus.

\subsection{C'umulos resueltos }

El \cu  I--Ds est'a compuesto por un peque\~no grupo de estrellas
y est'a  localizado hacia el norte del \cu I--A. Contiene poco gas en sus 
alrededores \citep{MacKetal00}. 
Usamos una apertura de 20 p'ixeles de la  PC para nuestra fotometr'ia sobre
las im'agenes de la WFPC2. Este \cu cae fuera del campo de
observaci'on de  STIS.


Decidimos analizar con detalle las diferentes estructuras que componen
al complejo  II  (II--A, II--B, II--C, II--D y  II--E.)
Estos \cus est'an 'intimamente relacionados con varias regiones \ha\ 
como se puede apreciar al comparar las Figuras~\ref{fig010} y  ~\ref{fig027}.
La fotometr'ia de apertura  fue realizadas siguiendo la misma rutina
que usamos para el complejo  II, teniendo en cuenta las
columnas  de p'ixeles malos. Los radios de apertura para estos
\cus son 50, 37, 57, 32 y  35 p'ixeles de la PC o
32.49, 23.94, 36.91, 20.81, 22.66 pc  respectivamente.
La Tabla~\ref{tbl006} contiene la fotometr'ia de estos  c'umulos.

\chapter{Transformaci'on de observables en par'ametros f'isicos}
\label{cha:chorizos}
\thispagestyle{empty}

\begin{flushright} 
 {\em  Let your mind \\
start a journey through a \\
strange new world! \\
Leave all thoughts \\
of the world \\
you knew before!}
 \end{flushright}
 
\newpage

\section{Fotometr'ia en varios filtros frente a  espectroscop'ia}

El empleo de espectroscop'ia consume mucho tiempo de observaci'on.
Una alternativa muy 'util e interesante 
es obtener fotometr'ia en varios filtros
que cubran principalmente longitudes de onda entre el infrarrojo cercano
y el ultravioleta.
La idea es, a partir de mediciones fotom'etricas, tratar de  determinar param'etros (intr'insecos y extr'insecos)
 relacionados 
a estrellas y c'umulos estelares.
En el caso de estrellas, los par'ametros que se desean determinar son, entre otros,
temperatura efectiva (intr'inseco), magnitud bolom'etrica (intr'inseco), enrojecimiento (extr'inseco) y
ley de extinci'on  (extr'inseco).
Para los c'umulos estelares  se desean determinar su edad (intr'inseco), masa (intr'inseco),
 enrojecimiento (extr'inseco), y 
ley de extinci'on  (extr'inseco).
Este trabajo se hace comparando magnitudes obtenidas usando diferentes
filtros con modelos te'oricos de estrellas y c'umulos. 
Existen  limitaciones  sobre este procedimiento que provienen principalmente
de degeneraciones en los par'ametros a determinar. Sin embargo, es posible hacer una determinaci'on
confiable de los par'ametros bajo ciertas circunstancias  \citep{deGretal03a,Maiz04,Alve04,deGretal05}.

La precisi'on con la que este trabajo puede ser realizado depende del n'umero de
filtros (de banda ancha) disponibles y, cr'iticamente, de las longitudes de onda
empleadas, del rango de longitudes de onda cubierto por las observaciones y de la precisi'on de
la fotometr'ia. 

Se ha demostrado que eligiendo un grupo peque\~no pero adecuado de filtros 
se obtienen  resultados m'as confiables que si se hacen muchas mediciones en filtros 
inapropiados o redundantes  \citep{Andeetal04}.

\section{Modelos te'oricos}
\label{sec:models}
La manera ideal de estudiar un starburst ser'ia tener suficiente
resoluci'on espacial y espectral para analizar
cada estrella individualmente. Ni siquiera empleando
im'agenes provistas por el HST podemos tener tal resoluci'on
a la distancia (2.94 Mpc) de NGC~4214.

Por este motivo, es posible estudiar las propiedades integradas de 
los starbursts. Una t'ecnica muy poderosa para el estudio de starbursts
se basa en la construcci'on de modelos de s'intesis, que fueron introducidos por 
\citet{Tins68,Tins72}.
La idea detr'as de estos modelos consiste en suponer una tasa de formaci'on estelar
y una IMF y seguir la evoluci'on de las estrellas mediante trazas evolutivas.
Esto permite obtener predicciones para los observables integrados  en funci'on de la edad. 
 
Estos c'odigos son diferentes en varios aspectos. En particular,  en lo que respecta a la f'isica
(trayectorias evolutivas  de diferentes grupos, leyes de extinci'on, etc).
Tambien son diferentes en cuanto a  los m'etodos empleados para
su implementaci'on (rutinas de interpolaci'on, lenguages, etc.) 

Sin lugar a dudas, uno de los componentes  fundamentales de los modelos de s'intesis
evolutiva son las trazas estelares. En la actualidad existen trayectorias evolutivas
desarrolladas por distintos grupos, destac'andose los de Ginebra y Padua 
\citep{Chioetal92,GarV96}.

Los modelos de evoluci'on estelar proveen
la variaci'on de cantidades f'isicas y qu'imicas desde el centro  
hasta la fot'osfera de una estrella de masa y composici'on iniciales dadas.
Tambi'en proveen la evoluci'on de estas cantidades con el tiempo.
Los par'ametros fundamentales son la 
temperatura efectiva \teff ~y la magnitud bolom\'etrica \mbol\  o
su equivalente la luminosidad.
La informaci'on que nosotros podemos obtener  a partir de las observaciones
consiste en mediciones de los fotones 
que las estrellas emiten. Estas mediciones se hacen en forma de espectros
e im'agenes.

Cuando uno transforma las predicciones de los modelos estelares en 
magnitudes observables, necesita determinar el flujo emitido 
por la superficie estelar a partir de la luminosidad y la temperatura efectiva.
Esto se logra mediante los modelos de atm'osferas.

Los modelos de atm'osferas estelares quedan determinados por la composici'on qu'imica,
 la gravedad y la temperatura efectiva y proveen del flujo monocrom'atico
 (el espectro estelar o SED, del ingl'es: Spectral Energy Distribution)  en la 
 superficie estelar.

Varios grupos han introducido c'odigos te'oricos de modelos de atm'osferas estelares
y modelos de s'intesis evolutiva. 
La siguiente es una lista que no pretende ser completa:

\vspace{10mm}
{\bf Modelos de atm'osferas estelares}
\begin{description}
\item[Kurucz]  http://kurucz.harvard.edu
\item[Lejeune:]   
Lejeune, T., {Cuisinier}, F. y {Buser}, R. 1997 Standard Stellar Library for Evolutionary Synthesis. I. 
Calibration of Theoretical Spectra. {\it A\&AS} {\bf 125}, 229--246. 

\item[{\sc\footnotesize {\bf TLUSTY:}}]  
{Lanz}, T. y  {Hubeny}, I. 2003 A Grid of Non-LTE Line-blanketed Model Atmospheres of 
O--Type Stars. {\it ApJS} {\bf 146}, 417--441.

\end{description}

{\bf Modelos de s'intesis evolutiva}
\begin{description}
\item[Bruzual: ] {Bruzual}, G. and {Charlot}, S.  2003 Stellar Population Synthesis at the Resolution of 2003. 
{\it MNRAS} {\bf 344}, 1000--1028. 

\item[{\sc\footnotesize  {\bf GALEV:} } ]{Fritze-v.~A.}, U. y {Gerhard}, O.~E. 1994 Star Formation in Merging Galaxies. {\it A\&A} {\bf 285}, 
751--774.

\item[\sc\footnotesize  {\bf Starburst99:}] 
{Leitherer}, C., {Schaerer}, D., {Goldader}, J.~D., 
	{Delgado}, R.~M.~G., {Robert}, C., {Kune}, D.~F., {de Mello}, D.~F., 
	{Devost}, D.  y  {Heckman}, T.~M.  1999 Starburst99: Synthesis 
Models for Galaxies with Active Star Formation. {\it ApJS} {\bf 123}, 3--40.

\end{description}

\vspace{10mm}

Desafortunadamente, no existen grillas de modelos de atm'osferas de estrellas
masivas suficientemente realistas.  Sin duda los modelos m'as completos  de la actualidad 
son los de Kurucz. Estos modelos  consideran un tratamiento muy riguroso de la opacidad
por 'atomos y mol'eculas, pero asume una atm'osfera est'atica plana bajo 
equilibrio termodin'amico local y se sabe que las atm'osferas de las estrellas de gran masa est'an
en expansi'on y estas estrellas tienen intensos vientos durante toda su vida \citep{LameCass99}.

\begin{table}[htbp]
\centering
\begin{minipage}[u]{10cm}
\caption[Par'ametros de atm'osferas estelares ]{Par'ametros de atm'osferas estelares 
de algunos modelos te'oricos disponibles en la literatura.  \label{tbl017}
\vspace{5mm}}
\end{minipage}

\begin{tabular}{  lccc}   
\multicolumn{4}{c}{ \rule{100mm}{0.8mm}}      \\

 Modelo          &\teff     &$\log({\mathrm{g/ 1cgs}}) $      & $\log(Z/Z_{\sun})$        \\  
  \multicolumn{4}{c}{ \rule{100mm}{0.2mm}}      \\

Kurucz        & $3\,500 -50\,000$& $0.0 -5.0$&$ -1.5 -0.0 $    \\
Lejeune        &$ 3\,500 -50\,000$& $0.0 -5.0$& $-1.5 -0.0$ \\
{\sc\footnotesize  TLUSTY}      & $27\,500 -55\,000$& $3.00 -4.75$& $-0.7 -0.0$    \\ 
\multicolumn{4}{c}{ \rule{100mm}{0.8mm}}      \\

\end{tabular}

\end{table}


\begin{table}[htbp]
\centering
\begin{minipage}[u]{8cm}
\caption[Rangos de valores de edad y metalicidad de c'umulos]{Rangos de 
valores de edad y metalicidad de c'umulos 
provistos por  algunos modelos te'oricos disponibles en la literatura.  \label{tbl018}
\vspace{5mm}}
\end{minipage}

\begin{tabular}{ lcc}  
\multicolumn{3}{c}{ \rule{80mm}{0.8mm}}      \\

 Modelo       &$\log({\mathrm{edad/a}}) $      & $Z (Z_{\odot}=0.02)$      \\   
 \multicolumn{3}{c}{ \rule{80mm}{0.2mm}}      \\

\s99        & $6.0-9.0$&$ 0.004-0.02$     \\
{\sc\footnotesize GALEV   }    & $6.6-10.2$&$ 0.0004-0.05$ \\
Bruzual    & $5.1-10.3$& $0.0004-0.1  $\\
\multicolumn{3}{c}{ \rule{80mm}{0.8mm}}      \\

\end{tabular}

\end{table}


\section{Extinci'on}
Una correcci'on que debe tenerse en cuenta 
en el momento de transformar observables en par'ametros te'oricos es aquella debido a la 
presencia del medio interestelar  (ISM). El espacio entre las estrellas no est'a
vacio, y el material que lo forma interact'ua con la radiaci'on estelar, afectando
las magnitudes observadas. El gas tiende a absorber y reemitir la radiaci'on
en una longitud de onda y direcci'on diferente, mientras que el polvo 
tiende a dispersar  a la radiaci'on.  En el Cap'itulo~\ref{cha:ext} hacemos una descripci'on m'as detallada 
de la extinci'on.

La relaci'on entre la magnitud aparente observada ($m_{\lambda}$) y la magnitud aparente real 
($m_{\lambda,0}$) es

 \begin{equation}
\label{eqn-10}  
m_{\lambda} = m_{\lambda,0} + A_{\lambda}
\end{equation}

donde $A_{\lambda}$ es la absorci'on en magnitudes  debida al ISM para  la
longitud de onda $\lambda$. 

La magnitud bolom'etrica ser'a entonces

 \begin{equation}
\label{eqnmbol}  
M_\mathrm{bol} = m_{\lambda} - A_{\lambda} - 5 \log(d) + 5 + BC_{\lambda} 
\end{equation}

El color fotom'etrico $m_A-m_B$ se define como la diferencia 
entre dos magnitudes aparentes en dos filtros diferentes $A  $ y $B$.
El efecto de la extinci'on sobre un color $m_A-m_B$ es:

\begin{equation}
\label{eqn-10}  
m_A-m_B = m_{A,0} - m_{B,0} + A_A -A_B = (m_A-m_B)_0 + E(A-B)
\end{equation}

donde definimos: $E(A-B)  \equiv A_A -A_B$  como  el exceso de color o enrojecimiento.
 \begin{wrapfigure}{l}[0mm]{80mm}
\includegraphics[width=0.5\textwidth]{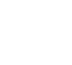}
\caption[Relaci'on entre  $E(B-V)$ y \ecc ]{\sl\footnotesize   
En este gr'afico se muestra la variaci'on de $E(B-V)-E(4405-5495)$ con  \ecc\  calculados 
con 
modelos de atm'osferas estelares de Kurucz
de distintas temperaturas.}
\label{fig107}
\end{wrapfigure}
 
Se define como ley de extinci'on a  los valores de  $A(\lambda)/A(V)$ 
en funci'on de la longitud de onda $\lambda$, la cual
se  determina emp'iricamente. El cociente 
 $R_V \equiv A(V)/E(B-V) $ suele emplearse como valor para
 parametrizar a las leyes observacionales  de extinci'on \citep{Cardetal89}.
Nosotros empleamos la definici'on monocrom'atica de los par'ametros  \ebv  
y, $R_V$ que denominamos \ecc\ y $R_{5495}$.
 Los valores  4405 y 5495 son las longitudes de onda centrales
 (en \AA\ ) de los filtros Johnson  $B$ y $V$  respectivamente. 
Empleamos cantidades monocrom\'aticas pues  $E(B-V)$ y  $R_V$
dependen, no s\'olo de la cantidad y tipo de polvo, sino  tambien de la atm\'osfera estelar.
La diferencia entre $E(B-V)$ y \ecc\ se pone en evidencia en la  Figura~\ref{fig107}
La relaci'on entre $R_{5495}$ y \ecc\  est'a dada por 

 \begin{equation}
\label{eqn-10}  
R_{5495} = \frac{A_{5495}}{E(4405-5495)}
\end{equation}

\section{C'odigo de ajuste de modelos: {\sc CHORIZOS }}

Nuestras im'agenes nos permiten obtener magnitudes fotom'etricas en varios filtros
que representan a los observables principales.
Para traducir estos observables en par\'ametros estelares fundamentales
(principalmente temperatura efectiva \teff ~y magnitud bolom\'etrica $M_\mathrm{bol}$)
usamos \cho ~\citep{Maiz04},  un algoritmo que ajusta una familia arbitraria
de   espectros sint\'eticos   a datos espectrosc\'opicos y/o
fotometr\'ia en varios colores.

\cho\ consta de tres m'odulos programados en IDL:   {\sc\footnotesize  GENSYNPHOT}
que realiza la fotometr'ia sint'etica sobre los modelos te'oricos; {\sc\footnotesize  CHORIZOS},
donde se realizan los c'alculos principales del ajuste y {\sc\footnotesize  STATPLOTS},
  que calcula la verosimilitud  final de los modelos y determina varias cantidades
  derivadas de la fotometr'ia. Este 'ultimo procedimiento genera adem'as 
 varias salidas gr'aficas.

Los par\'ametros fundamentales se determinan en forma simult\'anea 
por medio de una t\'ecnica  de cuadrados m\'inimos  $\chi^2 $
a partir de los colores fotom\'etricos calculados con  las magnitudes
originales.
Los errores asociados con las medidas de los par\'ametros se obtienen a partir
del ajuste  $\chi^2 $.

 Dado un conjunto de $M+1$ magnitudes $(m_1, m_2, ..., m_{M+1})$
 y $N$ par\'ametros, uno puede definir dos problemas diferentes:
 (1)  $M=N$:  en este caso uno puede establecer tantas ecuaciones
 como inc\'ognitas. (2) $M>N$: en este caso, el problema tiene m'as ecuaciones que
 inc\'ognitas y no se puede obtener una soluci\'on exacta. Sin embargo,
 se pueden hallar soluciones aproximadas que sean compatibles con las incertezas medidas.
 En cualquiera de los dos casos, $M-N$  da el n\'umero de grados de libertad del ajuste.

La versi\'on empleada de \cho\ maneja  $N=4$  par\'ametros:
dos par\'ametros provenientes de la familia de propiedades intr\'insecas de la
familia de SEDs (temperatura y gravedad de los modelos estelares, y edad y metalicidad 
para modelos de c\'umulos), adem\'as de dos par\'ametros  extr\'insecos relacionados
con la extinci\'on     (\ecc y la ley de extinci\'on).

\subsection{Par'ametros de entrada de  {\sc CHORIZOS}}

El archivo de entrada de \cho\ es un archivo de texto que incluye las  magnitudes de cada
objeto junto con  sus respectivos errores, expresados en un formato particular. Se deben indicar tambi'en
los filtros empleados asociados a las magnitudes. Esto se aplica tanto a estrellas como a 
c'umulos.
   
Los modelos de atm'osferas estelares 
 precalculados e incluidos en \cho ~son: modelos de Kurucz   \citep{Kuru04}, 
de Lejeune  \citep{Lejeetal97} y   TLUSTY \citep{LaHu03}.
Los modelos de c'umulos incluidos por defecto son los 
modelos de  \s99 \citep{Leitetal99}. \cho\ incluye un total de   88 filtros predefinidos entre 
los que se encuentran los filtros con los que se obtuvieron las im'agenes de NGC~4214
que empleamos para este trabajo.

Las leyes de extinci'on consideradas son: la familia de leyes que tienen a 
$R_{5495}$ como par'ametro principal \citep{Cardetal89} y que corresponden a
la Galaxia; las leyes promedio   de la Nube Mayor de Magallanes y la ley LMC2
provistas por  \citet{Missetal99}; y por 'ultimo la ley asociada a la
Nube Menor de Magallanes de  \citet{GordClay98}.
El usuario puede decidir qu'e modelos emplear en cada ejecuci'on as'i como fijar
los rangos en los que var'ia cada par'ametro.

\subsection{Archivos de salida de  {\sc CHORIZOS}}

Cuando aplicamos \cho\ a objetos estelares, obtenemos una salida de texto que provee, para 
cada estrella, su  temperatura efectiva, gravedad superficial, el exceso de color y la ley de extinci'on que mejor se
ajusta a los datos.
 
 Cuando aplicamos \cho\ a c'umulos, obtenemos para cada uno su edad, metalicidad,
el exceso de color y la ley de extinci'on que mejor se
ajusta a los datos.

En ambos casos, los valores se obtienen con sus
respectivos errores provenientes del ajuste de cuadrados m'inimos.
 \cho\ tambi'en provee un valor ($\chi^2 $) 
que indica qu'e tan bueno es el ajuste.

\subsection{Archivos de salida de  {\sc  STATPLOTS}  }

{\sc\footnotesize  STATPLOTS}   genera archivos de texto con varios valores deducidos
a partir de la fotometr'ia de entrada.
Entre otros, {\sc\footnotesize  STATPLOTS}  calcula la absorci'on ($A_r$) en el filtro de 
referencia elegido $r$ y la correcci'on bolom'etrica
$(BC_r)$, a partir de los cuales se puede obtener la magnitud bolom'etrica absoluta
(suponiendo conocida la distancia al objeto) usando  la Ecuaci'on~\ref{eqnmbol}.

La correcci'on bolom'etrica de \cho\ es la provista
por  \cite{Bessetal98}. 
Estos autores han calculado las correcciones bolom'etricas en el sistema 
Johnson--Cousins--Glass a partir de espectros sint'eticos de varios modelos de atm'osferas.
Los valores te'oricos obtenidos permiten una transformaci'on 
confiable entre diagramas color magnitud observables y
diagramas te'oricos de  Hertzsprung-Russell.

Cuando se ajustan modelos de c'umulos, 
 {\sc\footnotesize  STATPLOTS} provee  la magnitud aparente en el filtro de referencia $r$
corregida por extinci'on y a edad cero  $(m_r-C_r-A_r)$ que permite inferir la masa del c'umulo en cuesti'on.
$C_r$ es un factor que tiene en cuenta la variaci'on del brillo del c'umulo con el tiempo.

\cite{Maiz04}  sostiene que cuando uno mide
magnitudes y luego calcula dos colores que incluyen un filtro en com'un,
la distribuci'on de probabilidad 
de los dos colores est'a correlacionada.
Lo mismo sucede con la magnitud bolom'etrica \mbol\ y la temperatura efectiva \teff\ de una
estrella.
 {\sc\footnotesize  STATPLOTS} da la correlaci'on entre \teff y \mbol para cada estrella
 analizada.
Cuando dos variables con distribuci'on normal  $x_1 $ y $x_2 $ est'an correlacionadas,
los contornos de igual densidad de probabilidad son elipses caracterizadas por 
dos ejes y un 'angulo de orientaci'on \citep{Bran99}.
Estas elipses se denominan {\em elipses de covarianza} de la distribuci'on 
normal en dos variables.
Las tres cantidades mencionadas son funci'on de las desviaciones est'andar
individuales ($\sigma_1 $ y $\sigma_2 $) de cada variable y del coeficiente de correlaci'on
$\rho$.  La elipse est'a centrada sobre el punto 
cuyas coordenadas son los valores medios $\overline{x_1} $ y $\overline{x_2} $.

La aplicaci'on directa de esto es la determinaci'on de las elipses que marcan la incerteza
en la posici'on de un objeto en el \hrd. La Figura~\ref{fig001} (arriba izquierda) muestra  como ejemplo
dos objetos con sus respectivas elipses de incertidumbre graficados sobre un 
\hrd con trayectorias evolutivas de  \cite{LeSc01}. Las elipses dibujadas corresponden a una
probabilidad del 68\%.

\subsection{Salidas gr'aficas de   {\sc STATPLOTS}  }

{\sc\footnotesize  STATPLOTS}  produce dos tipos de salidas gr'aficas:
un espectro sint'etico y diagramas de contornos. 

\begin{figure*}
\centerline{\includegraphics*[width=0.47\linewidth]{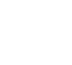}
                  \includegraphics*[width=0.47\linewidth]{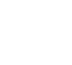}}
\centerline{\includegraphics*[width=0.47\linewidth]{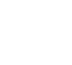}
                  \includegraphics*[width=0.47\linewidth]{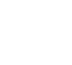}}

\caption[Salidas gr'aficas producto de   {\sc\footnotesize  STATPLOTS}]{ {\sl\footnotesize  Salidas 
gr'aficas producto de   {\sc\footnotesize  STATPLOTS}.
[Arriba izquierda]  {\sc\footnotesize  STATPLOTS} genera el valor de la correlaci'on entre
la \teff\ y la \mbol\ lo cual permite estimar la elipses de incertidumbre de un punto sobre
el diagrama de Hertzsprung-Russell.
Se han graficado algunas trayectorias evolutivas  de \cite{LeSc01}  como guia. Las elipses 
marcadas corresponden a un 68\% de probabilidad. 
[Arriba derecha] Espectro sint'etico que mejor se ajusta a las magnitudes
observadas de un c'umulo. Se muestran los filtros usados para el ajuste. La ordenada
es la magnitud AB (m$_\mathrm{AB} = 48.60-2.5 \log(f_{\nu}) $) y la abscisa
es la inversa de la longitud de onda. La explicaci'on del significado de las barras
marcadas en los puntos se da en el texto. 
[Abajo izquierda]  Ejemplo de diagrama de contornos asociado al ajuste 
de magnitudes a un espectro estelar.
[Abajo derecha] Ejemplo de diagrama de contornos asociado al ajuste 
de magnitudes al espectro de un c'umulo.
En ambos gr'aficos de abajo, los contornos est'an normalizados al valor m'aximo
y el rango de variaci'on mostrado es $0.05-0.95$.
\label{fig001}}}
\end{figure*}

Cuando se ejecuta  {\sc\footnotesize  STATPLOTS}  (tanto para estrellas como para c'umulos)
se crea un gr'afico que contiene el   espectro  que mejor se ajusta a los datos de entrada,
la fotometr'ia  sint'etica (indicada con el s'imbolo estrella) y la fotometr'ia observada (indicada
con s'imbolos con barras de error; la barra horizontal muestra la extenci'on del filtro en 
longitud de onda y la barra vertical muestra la incerteza en la magnitud.)
La escala horizontal del espectro es la inversa de la longitud de onda (expresada en $\mu^{-1}$ )
y la escala vertical est'a expresada en magnitudes AB. 
La Figura~\ref{fig001} (arriba  derecha) muestra  un ejemplo de espectro 
sint'etico con las magnitudes medidas marcadas.

 Para cada par de par'ametros de salida,   {\sc\footnotesize  STATPLOTS} crea un diagrama de contornos 
 que permite la visualizaci'on de los resultados del ajuste. Un s'imbolo blanco 
 muestra  los valores de los par'ametros usados para la gr'afica del espectro. 
En la Figura~\ref{fig001}  (abajo) se muestran dos ejemplos del tipo de diagramas de 
contornos que produce  {\sc\footnotesize  STATPLOTS}.
El gr'afico de abajo a la izquierda es un ejemplo t'ipico de ajuste de magnitudes 
 a  espectros estelares. Los distintos contornos dan una medida de la probabilidad en \teff\
 y \ecc\
El gr'afico de abajo a la derecha es un ejemplo de un ajuste de magnitudes
a espectros integrados de c'umulos.  En este caso, se puede inferir
la edad media del c'umulo asi como su enrojecimiento medio y sus
respectivas incertezas.

\subsection{Aplicaci'on a la poblaci'on de estrellas}

Aplicamos \cho\   para obtener temperatura,
enrojecimiento y, cuando fuera posible, la ley de extinci'on
de las estrellas de nuestras listas.
Empleamos como  modelos de estrellas a los de baja gravedad de Kurucz
con $\log(Z/Z_{\odot})= -0.5$. Un estudio de comparaci'on aplicando 
los modelos de Lejeune no present'o diferencias significativas.
Las estrellas de la {\sc\footnotesize LISTA336} fueron analizadas as'i:

\begin{description}
\item[seis y cinco filtros ]  Ejecutamos \cho\ sobre
el conjunto de objetos de los que ten'iamos magnitudes en cinco y seis  filtros
(cuatro y  cinco colores independientes respectivamente). 
Dejamos  $T_\mathrm{eff}$, \ecc\  y la ley de extinci'on como par'ametros libres.

\item[cuatro filtros ]  Para los objetos con cuatro magnitudes (tres colores
independientes y dos par'ametros para 
determinar), restringimos la ley de extinci'on a 
tres valores:  primero elegimos la ley de la Nube Menor de Magallanes; luego
experimentamos con el valor cl'asico  $R_{5495}= 3.1$,
y por 'ultimo fijamos a  $R_{5495}= 5.0$. 
Comparamos para cada estrella el valor del ajuste $(\chi^2) $ 
y en cada caso nos quedamos con el ajuste que diera el menor valor de $\chi^2.$

\item[tres filtros ] El resto de los objetos 
con s'olo tres magnitudes medidas fueron procesados 
por \cho\ dejando \teff\ y \ecc\
como par'ametros libres y usando al ley de extinci'on 
de \citet{Cardetal89},  con $R_{5495}= 3.1$. 
En este caso nos quedamos con ning'un grado de libertad 
para el ajuste, pero esto es lo mejor que se puede hacer 
con s'olo tres magnitudes.

\end{description}

Se sabe a partir de estudios en nuestra Galaxia y 
en otras cercanas que la ley de extinci'on puede variar en escalas
espaciales muy peque\~nas  (del orden de  1 pc)
(ver   \cite{Ariaetal05}  donde se presenta un estudio reciente). 

Decidimos fijar el valor  $R_{5495}= 3.1$ para los objetos con tres y cuatro filtros,
debido a que este valor es muy cercano al valor promedio  de $R_{5495}$ 
calculado para las estrellas con cinco y seis filtros, donde dejamos a $R_{5495}$ 
como un par'ametro libre del ajuste y debido a que $R_{5495}= 3.1$ es
el valor est'andar medido en la Galaxia.

Luego de ejecutar \cho\ para los objetos de la {\sc\footnotesize LISTA336},
hicimos una estad'istica sobre los valores de \ecc\  entre los objetos
de la poblaci'on estelar difusa y lejos de las regiones de formaci'on estelar intensa
y encontramos que 
NGC~4214 presenta valores  muy bajos de la extinci'on  (\ecc  $\lesssim 0.1$ mag).  
Los objetos en la {\sc\footnotesize LISTA814} tienen magnitudes en tres filtros:  
  F555W, F702W y F814W, con los que pudimos construir s'olo dos colores
F555W--F702W y F555W--F814W.  
Con dos colores, el n'umero m'aximo de par'ametros libres permitidos es dos
(cero grado de libertad).  
Decidimos fijar $R_{5495}= 3.1$,  y restringir \ecc\ en el rango 
 $-0.05 \leqslant  E(4405-5495) \leqslant 0.25  $
usando el resultado mecionado m'as arriba.
\teff fue dejada como par'ametro  libre.

Para determinar qu'e tan bueno es un ajuste, confiamos en el valor de  $\chi^2$.
Todos los objetos con   $\chi^2 \leq  4.0 $  fueron considerados como
buenos ajustes para nuestro trabajo.
Hicimos un an'alisis detallado de aquellos objetos con seis y cinco magnitudes
con $\chi^2 >  4.0$.  
Estudiamos cada uno de los ajustes de espectros a las magnitudes observadas de cada estrella.
En algunos casos hallamos que una o dos de las magnitudes no se corresponden
con el resto. Esto puede ser debido a varias razones, como sistemas
m'ultiples no resueltos o contaminaci'on nebular en el filtro F555W.
Estos problemas fueron analizados uno por uno. En todos estos casos
descartamos una o dos magnitudes de la lista y volvimos a ejecutar \cho\ con el nuevo conjunto de
magnitudes. 

Esta es una de las ventajas de usar un c'odigo como \cho.
El c'odigo busca autom'aticamente el espectro que mejor se 
ajusta a los colores fotom'etricos de entrada, y es extremadamente 'util 
cuando uno tiene listas con cientos de estrellas.
\cho\  provee adem'as una manera r'apida de identificar sistemas no resueltos,
simplemente desplegando el espectro de salida superpuesto a los datos de entrada.
En el caso de objetos m'ultiples  compuestos por  dos objetos indistinguibles
(como una componente roja y una azul), \cho\ muestra un 
espectro caracter'istico.  Este m'etodo nos permiti'o identificar
algunas estrellas binarias con componentes contrastadas en nuestra muestra.
Tambien pudimos identificar casos en los que alguna magnitud fue 
asignada en forma err'onea.

\subsection{Aplicaci'on a la poblaci'on de c'umulos estelares}
Las fuentes extendidas en nuestra muestra (tanto  resueltas como
no resueltas) fueron analizadas de otra manera.
\cho\ incluye los modelos precalculados de c'umulos de  \s99, donde
los par'ametros intr'insecos son la edad del \cu y su 
metalicidad \, y los par'ametros externos son los mismos que en el
caso estelar:  \ecc\ y $R_{5495}$.
Estos modelos se ecuentran tabulados usando una 
funci'on inicial  de masa  \citet{Salp55} en el rango 
$1 - 100 \, M_\sun$.

Este rango de masas es apropiado para la  determinaci'on 
de edades de los c'umulos (debido a que los colores integrados de los \cus 
m'as j'ovenes que $10^8$  a~nos est'an dominados principalmente
por masas mayores que 1 $M_\sun$), 
de sus  magnitudes  y de su extinci'on entre otros
par'ametros, pero no es el adecuado para la estimaci'on de la masa ya que
la mayor parte de la masa de un c'umulo estar'ia contenida en su poblaci'on
de estrellas de menor masa.
Esto significa que usando los resultados directos 
de \s99\ obtendr'iamos una subestimaci'on de la masa real de un c'umulo.
En el Apendice~B se explica el procedimiento empleado para solucionar 
este problema.

En todos los casos, usamos modelos \s99 con metalicidad  $Z/Z_{\sun}= 0.4$,
que es la apropiada para NGC~4214, y modelos de edades que varian entre 
$\log({\rm edad}) = 6.0$ (1 Ma)
y $\log({\rm edad}) = 9$ (1 Ga). Los valores de enrojecimiento \ecc\ usados
est'an en el rango $0.0 - 5.0.$ 

Al igual que en el caso estelar, las 
leyes de extinci'on consideradas son: la familia de leyes que tienen a 
$R_{5495}$ como par'ametro principal \citet{Cardetal89} y que corresponden a
la Galaxia; las leyes promedio   de la Nube Mayor de Magallanes y la ley LMC2
provistas por  \citet{Missetal99}; y por 'ultimo la ley asociada a la
Nube Menor de Magallanes de  \citet{GordClay98}.

\subsection{Justificaci'on del m'etodo   }

Existen varios trabajos que analizan hasta qu'e punto se pueden determinar
propiedades f'isicas de estrellas y c'umulos estelares a partir de un conjunto de 
mediciones fotom'etricas en distintos filtros \citep{deGretal03a,Maiz04,Alve04,deGretal05,deGrAnde06}.

En  \cite{deGretal03a} los autores analizaron la precisi'on con la que es posible determinar 
edad, masa y extinci'on de c'umulos estelares extragal'acticos.
Estudiaron  $\sim 300$ c'umulos en NGC~3310  con observaciones de archivo del HST
que abarcan desde el UV hasta el IR cercano.
Dispusieron de observaciones en un total de siete filtros. 
Realizaron varios experimentos interesantes para ajustar la edad, metalicidad y extinci'on 
de los \cus usando primero cuatro filtros y luego fueron agregando otros en el siguiente orden:
(a) combinaci'on de filtros azules: F300W, F336W, F439W y F606W
(b) filtros 'opticos: F336W, F439W, F606W y F814W
(c) combinaci'on de filtros rojos: F606W, F814W, F110W, F160W
(d) combinaci'on de cinco filtros 'opticos+IR: F439W, F606W, F814W, F110W, F160W
(e) seis filtros: F336W, F439W, F606W, F814W, F110W, F160W
(f) siete filtros:  F300W, F336W, F439W, F606W, F814W, F110W, F160W

\begin{figure*}
\includegraphics[width=\textwidth]{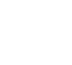}
\caption[ Distribuci'on de edades de c'umulos en NGC~3310 \citep{deGretal03a}]{{\sl\footnotesize   
Distribuci'on de edades de c'umulos en NGC~3310 en base
a seis combinaciones de filtros. Los histogramas sombreados muestran
las soluciones en com'un (60 c'umulos) y los histogramas
abiertos  muestran las soluciones adicionales. Para guiar al ojo
se ha graficado la distribuci'on Gaussiana correspondiente 
a la distribuci'on 'optima obtenida con siete filtros (f)  \citep{deGretal03a}.}}
\label{fig049}
\end{figure*}

La Figura~\ref{fig049} muestra la dependencia de la distribuci'on de edades
de acuerdo a la combinaci'on de filtros empleados. Los histogramas  sombreados 
muestran  las soluciones aceptables en com'un, mientras que los histogramas abiertos muestran
las soluciones adicionales en cada caso. 
Los autores infieren varias conclusiones a partir de este estudio:

\begin{enumerate}
\item  La distribuci'on de edades de los c'umulos es una funci'on sensible de los
filtros empleados en las observaciones; el pico de la distribuci'on queda bien reproducido en 
todos los casos, excepto cuando se usa una combinaci'on de filtros 
rojos.

\item  El empleo de filtros rojos da una soluci'on mayor para la edad
 (ver Figura~\ref{fig049} c y d) lo cual se debe a la  d'ebil dependencia de las magnitudes en 
 el IR cercano con el tiempo.

 \item  Cuando se emplea una combinaci'on de filtros en el azul (ver Figura~\ref{fig049} a y b) 
 la distribuci'on de edades tiende hacia valores menores que los reales; esto es as'i 
 por efectos de selecci'on observacionales: los \cus m'as j'ovenes producen flujos
 mayores en el UV y filtros $U$ y por lo tanto 
 son m'as f'acilmente detectables en esas longitudes de onda.

\end{enumerate}

 Los autores concluyen que para determinar edades 
  de \cus en forma confiable, es necesario
 emplear una combinaci'on
  de al menos cuatro filtros en el 'optico que incluyan tanto un filtro azul como 
  uno rojo.
  
En nuestro estudio sobre c'umulos en NGC~4214, empleamos en todos los casos 
un m'inimo de cinco filtros abarcando un amplio rango de longitudes de onda.

\begin{figure*}
\includegraphics[width=\textwidth]{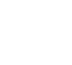}
\caption[Determinaci'on de la edad de un c'umulo]{{\sl\footnotesize   Determinaci'on de la edad de un c'umulo empleando dos combinaciones
de filtros. (a) En este caso se emplearon los filtros F435W ($U$), F555W ($V$) y F814W($I$).
(b) Para esta ejecuci'on se agrega un filtro UV: F330W.
Claramente, es necesaria una combinaci'on de al menos cuatro filtros en el 'optico con un filtro 
azul y otro rojo para obtener un valor confiable de la edad.    }}
\label{fig050}
\end{figure*}

Un ejemplo que muestra en forma dram'atica  de qu'e manera var'ia
la determinaci'on de la edad de un c'umulo ficticio empleando diferentes filtros
se presenta en la Figura~\ref{fig050}.
El gr'afico (a) representa el diagrama de contornos  [ $\log({\mathrm{edad/a}}) $,  \ecc] 
de un c'umulo ficticio con magnitudes calculadas en los filtros F435W, F555W y F814W
de la ACS. Se observa una gran dispersi'on en edades, con un valor promedio de
$\log({\mathrm{edad/a}}) = 8.09 \pm 0.52$. 
El gr'afico (b) se obtuvo agregando la magnitud del c'umulo en el filtro F330W,
lo cual extiende el rango de longitudes de onda desde el  UV hasta $I$. La edad del c'umulo 
queda entonces muy bien definida, con un valor medio de
$\log({\mathrm{edad/a}}) = 7.99 \pm 0.06$.

\citet{Maiz04} analiza la aplicaci'on de este m'etodo fotom'etrico 
principalmente  sobre  estrellas individuales. Por ejemplo, se comparan valores deducidos de la 
\teff de estrellas est'andar de la secuencia principal, usando ajustes con modelos
de atm'osferas de Kurucz y Lejeune. Encuentran que para estrellas de tipo
A--K ambas ejecuciones producen buenos resultados y son compatibles entre s'i. 
En el caso de estrellas B, ambos modelos producen
resultados id'enticos, pero hay una diferencia de $\sim 3000$ K
en los tipos m'as tempranos, lo cual podr'ia ser un efecto similar al observado 
para estrellas O por varios autores usando datos espectrosc'opicos.
Cuando se analizan estrellas de tipo M, las atm'osferas de Kurucz dan resultados inaceptables
para la temperatura. Esto significa que estos modelos te'oricos no proveen 
una buena representaci'on de los colores de estrellas M enanas.

Otro an'alisis interesante  que realizaron estos autores implica la determinaci'on de
la ley de extinci'on en la l'inea de la visual de varias estrellas que \citet{Cardetal89} 
emplearon para su determinaci'on de las famosas leyes.
Los resultados obtenidos por el m'etodo fotom'etrico 
est'an de acuerdo con los de \citet{Cardetal89}. \citeauthor{Maiz04} 
concluye que el m'etodo multifotom'etrico puede ser empleado
para determinar enrojecimiento y leyes de extinci'on con alta precisi'on,
incluso cuando no se dispone de  los tipos espectrales  de las estrellas a analizar.

\begin{center}\begin{figure*}[ht!]
 \begin{minipage}[c]{12cm}
\centering
\centerline{\includegraphics[width=\textwidth]{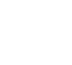}    }
\caption[ Diagrama color--color con modelos de Kurucz]{{\sl\footnotesize  Diagrama color--color
 [$U-B$, $B-V$ ] de modelos de Kurucz de secuencia
principal con $Z=0.0$.  La l'inea negra representa los colores no enrojecidos
como funci'on de la temperatura. Las l'ineas de colores  indican los colores enrojecidos 
usando las leyes de  \citet{Cardetal89}. Se han marcado cuatro objetos y en
el texto se analizan
sus curvas de probabilidad.   }}
\label{fig047}
\end{minipage}
 \end{figure*}
 \end{center}

\begin{figure*}
\includegraphics[width=\textwidth]{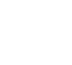}
\caption[Diagramas 
 de contornos de las cuatro estrellas estudiadas 
 en la Figura~\ref{fig047}]{ {\sl\footnotesize  Diagramas  [ \teff,  \ecc] 
 de contornos de las cuatro estrellas estudiadas en la Figura~\ref{fig047}. 
 Se observa un ejemplo  (1) con una soluci'on 'unica y dos ejemplos (2,3) con
 soluciones m'ultiples. El ejemplo (4) representa una caso especial donde
 los colores observados no se corresponden a colores de modelos te'oricos. Ver el texto para
 una explicaci'on.}}
\label{fig048}
\end{figure*}

Con el objeto de estudiar la presencia de m'ultiples soluciones, \citeauthor{Maiz04} 
experiment'o con  colores Johnson $U-B$ y $B-V$ para determinar \teff\ y  enrojecimiento. 
La Figura~\ref{fig047} muestra el diagrama color--color de objetos de secuencia principal
obtenidos usando  modelos de Kurucz. La l'inea  negra representa los valores no enrojecidos como funci'on
de la temperatura. Las otras  l'ineas
de colores representan los colores como funci'on   del enrojecimiento 
obtenidos con la ley de \citet{Cardetal89}.
Se han marcado cuatro objetos. El primer objeto representa un caso ideal
ya que existe una sola posible soluci'on (dentro de los errores en la
determinaci'on de las magnitudes) como puede observarse en el diagrama de contornos  
de la  Figura~\ref{fig048} (1).
El ejemplo (2) cae en una regi'on del diagrama tal que aparecen dos posibles
soluciones. Notar el cambio en la direcci'on de la linea no enrojecida de la 
Figura~\ref{fig047}. Estas dos soluciones aparecen claramente en la
Figura~\ref{fig048} (2).
Similarmente, el objeto (3) produce tres soluciones factibles, lo cual se
traduce en tres picos marcados en el correspondiente diagrama de contornos.
Ver Figura~\ref{fig048} (3).
El objeto (4) muestra un ejemplo en el que 
los colores medidos son incompatibles con los modelos,
pero los errores en los colores hacen que  se obtenga una
posible soluci'on alrededor del valor m'aximo de la temperatura.
Ver \citet{Maiz04} para m'as ejemplos ilustrativos.
 
En \citet{deGretal05}, los autores analizan  las incertezas en edad y masa de 
c'umulos estelares  as'i como en la determinaci'on de la extinci'on y la metalicidad.
Esto lo hacen empleando varios modelos de poblaciones estelares simples
 ({\sc\footnotesize  GALEV}, \s99, y
varios modelos de Bruzual \& Charlot) sobre im'agenes de NGC~3310 y NGC~4038/9.
Encuentran que los picos en la distribuci'on de edad y masa de c'umulos estelares j'ovenes
($\lesssim 10^9$ a\~nos) pueden determinarse con errores de $\sigma_T \leqslant 0.35$
y  $\sigma_M \leqslant 0.14$ respectivamente.
Sugieren adem'as que para obtener  edades y masas consistentes y con 
incertezas razonables,  se requiere del uso de  un rango 
de longitudes de onda lo m'as extenso posible. 

\chapter{Poblaciones estelares y cociente de supergigantes}
\thispagestyle{empty}

\begin{flushright} 
   {\em  Mystified!\\
baffled S\^uret\'e say, \\
we are mystified! \\
we suspect foul play!}
 \end{flushright}
\newpage

\section{Descripci'on de la poblaci'on estelar}

Con los valores de \teff\ y \mbol\  de las estrellas obtenidos
mediante el c'odigo  \cho\
pudimos construir un \hrd te'orico para NGC~4214.
Estos diagramas nos permiten comparar a nuestras
observaciones
con modelos evolutivos e interpretarlos 
en t'erminos f'isicos como masa, edad, composici'on, etc.
Usamos  el conjunto de trayectorias 
evolutivas provistas por \cite{LeSc01} que contiene los modelos
de Ginebra sin rotaci'on y que cubren masas en el rango 
$0.8$ a $120 \, M_{\sun}$.

NGC~4214 es una galaxia deficiente en metales,
con una abundancia de 
$12 + \log(\mathrm{O/H}) = 8.15 - 8.28$ \citep{KobuSkil96} 
o su equivalente de $Z=0.34 Z_\odot  =  0.006 $ de acuerdo a 
los resultados de \cite{Massey03} para la Galaxia y las Nubes de Magallanes.

Para analizar  nuestros datos observacionales  empleamos 
los modelos disponibles en la literatura: usamos modelos con  metalicidades
 $Z=0.004$  y $Z=0.008$,  que son los valores
   m'as cercanos a la metalicidad estimada y adaptada de 
 NGC~4214 $(Z\approx 0.006).$ 

\begin{figure}
\includegraphics[width=0.9\textwidth]{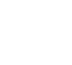}
\caption[Diagrama HR de  las estrellas en la  {\sc\footnotesize  LISTA814} ]{ {\sl\footnotesize 
Diagrama HR de todas las estrellas en la  {\sc\footnotesize  LISTA814} 
construido a partir de los valores de magnitud bolom'etrica \mbol\ y  
temperatura efectiva $T_{\mathrm{eff}}$  deducidos mediante el
usa de  \cho\  Los contornos corresponden a
10, 15, 20, 40, 50, 100, 300, 600, 1200, 
y 2000    estrellas por elemento. Cada elemento mide  0.0875 mag en el eje vertical
($M_{\mathrm{bol}}$) y 0.0075 en el eje horizontal ($\log(T_{\mathrm{eff}}) $).
Las estrellas individuales se dibujaron mediante cruces
en aquellas regiones del diagrama en las que la densidad era menor 
a 10 estrellas.
Las l'ineas rojas representan las trayectorias evolutivas de estrellas de masas
 5, 
7, 10, 12, 15, 20, 25, 40 y  120 $ M_{\sun}$. Las l'ineas cont'inuas negras
representan isocronas calculadas para las  edades  0, 1, 2, 3, 5, 10 y 15 Ma. 
Este diagrama tambi'en muestra  dos pol'igonos entre las trayectorias
de  15 y  25  $ M_{\sun}$. El pol'igono de la derecha 
marca la regi'on del diagrama donde calculamos la cantidad de
supergigantes rojas.
  }}
\label{fig035}
\end{figure}

\begin{figure}
\includegraphics[width=0.9\textwidth]{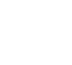}
\caption[Diagrama HR de  las estrellas en la  {\sc\footnotesize  LISTA336} ]{ {\sl\footnotesize 
Diagrama HR de todas las estrellas en la  {\sc\footnotesize  LISTA336} 
construido a partir de los valores de magnitud bolom'etrica \mbol\ y  
temperatura efectiva $T_{\mathrm{eff}}$  deducidos mediante el
usa de  \cho\ Los contornos corresponden a   5, 10, 15, 20, 40, 50, 75, 100, 150 
y 200  estrellas por elemento. Cada elemento mide  0.0875 mag en el eje vertical
($M_{\mathrm{bol}}$) y 0.0075 en el eje horizontal ($\log(T_{\mathrm{eff}}) $).
Las estrellas individuales se dibujaron mediante cruces
en aquellas regiones del diagrama en las que la densidad era menor 
a 5 estrellas.
Las l'ineas rojas representan las trayectorias evolutivas de estrellas de masas
 5, 
7, 10, 12, 15, 20, 25, 40 y  120 $ M_{\sun}$. Las l'ineas cont'inuas negras
representan isocronas calculadas para las  edades  0, 1, 2, 3, 5, 10 y 15 Ma. 
Este diagrama tambi'en muestra  dos pol'igonos entre las trayectorias
de  15 y  25  $ M_{\sun}$. El pol'igono de la izquierda 
marca la regi'on del diagrama donde calculamos la cantidad de
supergigantes azules.}}
\label{fig036}
\end{figure}

Los modelos te'oricos  nos dan la  luminosidad, edad y temperatura
efectiva de puntos representativos de las trayectorias evolutivas.
Calculamos siete isocronas interpolando en las magnitudes bolom'etricas
como funci'on del logaritmo de la edad a lo largo 
de cada trayectoria en pasos de 
$1-5$
millones de a~nos (ZAMS, 1Ma, 2Ma, 3Ma, 5Ma, 10Ma y 15Ma)  comenzando por la 
secuencia principal de edad cero (ZAMS, del ingl'es: Zero Age
Main Sequence).
Presentamos los diagramas de Hertzsprung$-$Russell en las 
Figuras~~\ref{fig035} y  \ref{fig036}.

Separamos a las estrellas en dos diagramas:
La Figura~\ref{fig035}  es un gr'afico de objetos de la {\sc\footnotesize  LISTA814} y la
 Figura~\ref{fig036} contiene los objetos de la  
  {\sc\footnotesize  LISTA336}. En lugar de representar objetos individuales,
usamos  diagramas de contornos para mostrar la densidad  variable de estrellas
en el plano  [$\log(T_{\mathrm{eff}}) $, $M_{\mathrm{bol}}$]. 
Los niveles de los contornos corresponden a las cantidades 
10, 15, 20, 40, 50, 100, 300, 600, 1200
y  2000 objetos por elemento en la Figura~\ref{fig035}  y 
 5, 10, 15, 20, 40, 50, 75, 100, 
150 y 200 objetos por elemento en la  Figura~\ref{fig036}.  
Cada elemento mide 0.0875 mag en el eje  $M_{\mathrm{bol}}$ y 
0.0075 en el eje de $\log(T_{\mathrm{eff}}) $.
Mostramos estrellas individuales como peque\~nas cruces
en las regiones de los diagramas donde la densidad es menor que 10 y 5 estrellas
respectivamente.

Al ser  una galaxia irregular enana, NGC~4214  es un laboratorio muy interesante
para analizar procesos de formaci'on estelar.
Esta galaxia est'a a'un procesando gas en estrellas. Es un sistema relativamente
rico en gas, con una poblaci'on joven de estrellas brillantes y azules
y regiones H~{\scriptsize II. }~{\normalsize  }
Las regiones de formaci'on estelar est'an concentradas 
a lo largo de una barra rodeada de un disco de gas. 

El medio interestelar  de NGC~4214 muestra regiones altamente ionizadas por
radiaci'on UV 
proveniente de las estrellas m'as j'ovenes. Estas regiones  se pueden
localizar usando la  emisi'on en \ha\  como indicador de 
regiones de formaci'on estelar reciente.

La Figuras~\ref{fig017} y \ref{fig018}  muestran las regiones de formaci'on estelar 
con emisi'on en H$\alpha$.
En las siguientes secciones
usamos las Figuras~\ref{fig035} y \ref{fig036}  para  analizar 
la poblaci'on estelar de NGC~4214.

\subsection{Poblaci'on estelar a partir de la LISTA814 }

La Figura~\ref{fig035}  muestra los objetos de la  {\sc\footnotesize  LISTA814}. 
La caracter'istica m'as importante en este diagrama es la concentraci'on 
de objetos en la parte roja del mismo en el rango
$ 3.50 \leqslant   \log(T_{\mathrm{eff}}) \leqslant 3.75$.  
Esta regi'on contiene gigantes rojas, estrellas gigantes  de
la rama asint'otica (AGB, del ingl'es: Asymptotic Giant Branch)
 y posiblemente estrellas del lazo azul  de edad
intermedia.
Esta parte del diagrama contiene una poblaci'on de estrellas de varias edades
que han evolucionado fuera de la secuencia principal.

Un grupo de supergigantes rojas se puede ver en el
extremo superior 
de la concentraci'on de estrellas rojas.
'Estas son estrellas  con masas iniciales 
$\sim  20-25 M_{\odot}$  que han dejado la secuencia principal.
Un resultado importante es que algunas de estas estrellas est'an ubicadas
a la derecha del extremo de las trayectorias evolutivas.
Este resultado ha sido observado por  \cite{MassOlse03} 
cuando estudiaron supergigantes rojas en las Nubes de Magallanes.
Las trayectorias evolutivas no se extienden lo suficiente a la derecha
del diagrama para producir a las estrellas supergigantes que se observan
en la realidad.
Este mismo problema ya fue discutido  por 
 \cite{Massey03}  al estudiar supergigantes rojas en la Galaxia.
 La concentraci'on de estrellas en la parte roja del diagrama presentado en
la Figura~\ref{fig035}
sugiere que la galaxia posee una poblaci'on de edad intermedia--vieja subyacente.
Este hecho es reforzado morfol'ogicamente por el prominente halo
rojizo observado en la extensi'on del disco de la galaxia (Figura~\ref{fig003}).

\subsection{Poblaci'on estelar a partir de la LISTA336 }

La Figura~\ref{fig036}  contiene  a  las estrellas masivas de la secuencia principal.
Estos objetos se ubican por encima de la trayectoria
evolutiva de  $5 M_{\odot}$, y se extiende hasta  $M_{\mathrm{bol}}  \approx -14. $ 
Esta regi'on est'a muy poblada en el rango  
$\approx 12 M_{\odot} - 60 M_{\odot}$, lo cual indica 
que existe formaci'on estelar reciente. 

En esa figura, se observa claramante un grupo de objetos
sobre la secuencia principal,  lo cual indica que son objetos j'ovenes. 
Sin embargo, a  masas elevadas, la posici'on de las estrellas est'a desplazada
hacia la derecha (menores temperaturas) de esta secuencia.
Esto puede deberse a una combinaci'on de efectos: la dificultad de
observar estrellas de la ZAMS a alta masa (debido  a que se encuentran
en regiones caracterizadas por una alta extinci'on  como se explic'o en el Cap'itulo~\ref{cha:introduction}) 
y la presencia de sistemas m'ultiples que forman 
grupos compactos no resueltos 
(que se observan como puntos por encima de la
trayectoria evolutiva de   $120 M_{\odot}$).

La otra caracter'istica en este diagrama es el grupo de estrellas 
localizado  entre  las trayectorias evolutivas
de $5 M_{\odot}$  y  $20 M_{\odot}$, y con 
$\log(T_{\mathrm{eff}}) \lesssim 4.25$. Estos  objetos son estrellas
masivas evolucionadas con edades mayores  que 15 Ma.

Estamos de  acuerdo con  \cite{Droz02}  en que las estrellas parecen poblar 
el \hrd de NGC~4214 en todas las fases de evoluci'on estelar, lo cual
nos lleva a concluir que la formaci'on estelar de esta galaxia ha sido mas o menos continua
en tiempos recientes, aunque puede haber habido algunos brotes presentes.

La mayor'ia de las estrellas m'as brillantes de la galaxia se concentran
en dos regiones principales de formaci'on estelar: 
NGC~4214--I y NGC~4214--II.
Estas regiones se caracterizan por tener una morfolog'ia diferente
que \cite{MacKetal00} interpretan como un efecto de evoluci'on.
Las componentes  estelares de NGC~4214--II  comparten la misma posici'on
que el gas, y las estrellas masivas individuales 
pueden observarse en diferentes estadios evolutivos.
Esta regi'on de  NGC~4214 es probablemente la m'as joven  (edad $\approx$ 2 Ma)
de la galaxia. 
Por otro lado,  NGC~4214--I  (edad $\approx 3-5$  Ma) presenta ciertos 
signos de evoluci'on, con cavidades producidas por vientos provenientes de
estrellas masivas y explosiones de supernovas. 
Las estrellas m'as j'ovenes tienden a concentrarse cerca de los c'umulos 
principales (I--A y I--B) en esta parte de la galaxia.
Tambi'en se observan concentraciones de estrellas brillantes fuera de la
regi'on central, marcando una estructura de barra que es una caracter'istica 
de NGC~4214.

\section{Cociente de estrellas supergigantes azules y rojas }

El cociente B/R de estrellas supergigantes azules (B) y rojas (R)
 es un observable muy importante 
de la poblaci'on m'as brillante de las galaxias cercanas. Esta cantidad  depende fuertemente
de los par'ametros de los modelos de estructura de evoluci'on estelar
y se  puede emplear para restringir la f'isica de los modelos con mucha precisi'on.

Este cociente es una caracter'istica de la poblaci'on de estrellas m'as brillantes 
en cada galaxia. \cite{Bergh68}  sugiri'o que B/R var'ia entre las galaxias cercanas
como un resultado   del efecto  que produce la metalicidad sobre la evoluci'on
de las estrellas masivas. 
Este trabajo fue continuado por otros realizados sobre la Via L'actea,
las Nubes de Magallanes (SMC y LMC) y otras galaxias cercanas.

\cite{HumpDavi79}  analizaron el cociente  B/R en la Nube Mayor de Magallanes 
y en la Galaxia.
Estudiaron la variaci'on de B/R con la distancia al centro de cada galaxia
y con la luminosidad, y mostraron que el cociente disminuye con la luminosidad
de manera semejante en cada galaxia y que existe una variaci'on de B/R con la distancia
que podr'ia ser debida a un gradiente en la composici'on qu'imica.
Este trabajo fue complementado por  \cite{Hump83}.  En este trabajo, encontraron
que el cociente B/R en la SMC no presenta el mismo decrecimiento  gradual
con la luminosidad observado en la Galaxia y en la LMC.
  \cite{HumpMcel84} reexaminaron este problema en las mismas galaxias
y encuentran el mismo decrecimiento, pero 
no tan pronunciado como fue sugerido por  \cite{HumpDavi79}.
\cite{Free85}  analizaron el cociente en M~33 y hallaron que  presenta
una variaci'on muy d'ebil con el radio, contrario a resultados previos.

El resultado principal de trabajos observacionales anteriores es que, para un dado
rango de luminosidad, B/R aumenta con  la metalicidad $Z$, en un factor
de 10 entre la SMC y las regiones internas de la Galaxia.
Ning'un  modelo te'orico actual de estrellas masivas
puede reproducir correctamente los cambios de B/R con la metalicidad, desde la solar ($Z_{\odot}=0.02$)
hasta la de la SMC ($Z=0.004$).

Se sabe que el cociente B/R es una cantidad sensible a la p'erdida de masa, la rotaci'on, la convecci'on
y los procesos de mezclado en el interior estelar; por tal motivo,
la determinaci'on de B/R constituye una prueba  muy sensible e importante sobre los modelos de evoluci'on estelar
\citep{MaedCont94,LaMa94,Eggeetal02}. 
En particular,  \cite{Eggeetal02} proponen una relaci'on promedio entre el cociente B/R
y la metalicidad dada por:

\begin{equation}  
 \frac{B/R}{(B/R)_\sun} \cong  0.05 \times e^{ \frac{3\,Z}{Z_{\sun}} }
\end{equation} 

donde  $Z_{\sun} = 0.02 $ y  $(B/R)_\sun$ es el valor de  B/R a $Z=Z_{\sun}$.

Ellos concluyen que el aumento de B/R con la metalicidad no parece ser un
artefacto debido a, por ejemplo, la incompletitud de
los datos, sino que es una caracter'istica que  los modelos de evoluci'on estelar
deben poder reproducir.

El cociente B/R es muy dependiente de c'omo se cuentan las estrellas, y por lo tanto,
las comparaciones  con las predicciones de los modelos evolutivos tiene que ser realizados
con mucho cuidado.
Es importante notar que la definici'on de B/R no es siempre la misma. Nosotros
seguimos el m'etodo sugerido por \cite{MaedMeyn01}.
Los modelos  de estrellas que se cuentan como B son aquellos ubicados desde
el l'imite derecho de la \sp  hasta el tipo espectral  B9.5 I, que corresponde
a $\log(T_{\mathrm{eff}})  =3.99 $ de acuerdo a la calibraci'on de \cite{Flow96}.
Las supergigantes rojas, en cambio, son aquellos modelos debajo de 
 $\log(T_{\mathrm{eff}})  = 3.70 $.  Las Figuras~\ref{fig035}   y ~\ref{fig036} 
 muestran las trayectorias evolutivas de
 modelos estelares sin rotaci'on para las  masas individuales 
  5, 7, 10, 12, 15, 20, 25, 40 y  120   $M_{\sun}$,  metalicidad de la LMC $(Z=0.008) $
obtenidos de \cite{Schaetal93}.
Las dos l'ineas verticales [en $\log(T_{\mathrm{eff}})  =3.99 $ y
$\log(T_{\mathrm{eff}})  =3.70 $]
marcan los l'imites entre la regi'on de supergigantes rojas y la regi'on de
supergigantes azules.
La ubicaci'on de estas estrellas est'a marcada  en esas figuras por dos pol'igonos entre
las trayectorias de 15 y 25 $M_{\sun}$.

Varios estudios identifican a las supergigantes al considerar a los objetos m'as brillantes
que $M_{\mathrm{bol}} = -7.5 $ mag lo cual corresponde a masas m'as grandes que 
15 $M_{\sun}$ para supergigantes rojas. 
Si se emplea un l'imite inferior de luminosidad, existe una posibilidad de que algunas estrellas de la
rama asint'otica de las gigantes (AGB) contaminen a la muestra \citep{Brunetal86}.
Este problema se evita al seleccionar objetos localizados encima de la trayectoria evolutiva
de 15 $M_{\sun}$ en todos los casos.
Realizamos un estudio para analizar la posible contaminaci'on generando 
una distribuci'on aleatoria de 1000
puntos distribuidos alrededor de los valores promedio de \teff~   y \mbol~ 
para un grupo arbitrario de estrellas
ubicadas debajo de la trayectoria evolutiva de 15 $M_{\sun}$ y calculamos 
el n'umero de puntos en cada  rango de masas. De ac'a obtuvimos un factor de contaminaci'on 
que consideramos como la correcci'on. Este efecto fue despreciable 
(menor que el 1\%) y decidimos no aplicarlo.

Para comparar nuestros resultados observacionales con la teor'ia, 
calculamos los cocientes de B/R usando
tres grillas de modelos te'oricos de Ginebra:

\begin{itemize}
\item Las trayectorias evolutivas de modelos no rotantes ($v_{inicial} = 0$ km~s$^{-1}$) 
con masas iniciales en el rango 9 a 60 $M_{\sun}$ y metalicidad de la SMC 
$(Z=0.004) $  \citep{MaedMeyn01}

\item  Las trayectorias evolutivas de modelos rotantes ($v_{inicial} = 300$ km~s$^{-1}$) 
con masas iniciales en el rango 9 a 60 $M_{\sun}$ y metalicidad de la SMC 
$(Z=0.004) $  \citep{MaedMeyn01}

\item  Las trayectorias evolutivas de modelos no rotantes ($v_{inicial} = 0$ km~s$^{-1}$) 
con masas iniciales en el rango 10 a 120 $M_{\sun}$ y metalicidad de la LMC 
$(Z=0.008) $   \citep{Schaetal93}.

\end{itemize}

No hallamos mayor diferencia para nuestro trabajo entre los modelos
con alta p'erdida de masa y p'erdida de masa normal, por lo que decidimos adoptar
 las trayectorias con p'erdida de masa normal.
 Desafortunadamente, los modelos evolutivos de trayectorias calculados con rotaci'on
 para una metalicidad de la LMC no est'an disponibles en la literatura.

El cociente de densidad de las supergigantes azules con respecto a 
las supergigantes rojas para una dada masa inicial es igual al cociente
de las  edades a lo largo de los trayectorias evolutivas en el intervalo
correspondiente de  $\log(T_{\mathrm{eff}})  $.
Los resultados te'oricos de B/R se encuentran dados en la Tabla~\ref{tbl009}.


\begin{table}[htbp]
\centering
\begin{minipage}[u]{15cm}
\caption[Valores te'oricos  del cociente B/R]{{\sl  Valores te'oricos  
del cociente B/R calculados a partir de 
modelos sin rotaci'on ($v_{ini} =  $ 0 \kms) y con rotaci'on ($v_{ini} =  $ 300 \kms). 
 B significa supergigantes de tipo espectral B
y R  significa supergigantes de tipos K y M.
La unidad de tiempo es  $10^6$ a\~nos. } \label{tbl009}
\vspace{5mm}}
\end{minipage}

\begin{tabular}{ cccccc} 
\multicolumn{6}{l}{ \rule{160mm}{0.8mm}}      \\
 
      $M_{\sun}$     & $v_{ini} $   & $Z$ &  fin de combusti'on del H --   & $\log(T_{\mathrm{eff}})  =  3.70  $ --   & B/R te'orico     \\   
          &  (km~s$^{-1})$ &    &  $\log(T_{\mathrm{eff}})  =  3.99 $ &  fin de combusti'on del C &     \\   \multicolumn{6}{l}{ \rule{160mm}{0.2mm}}      \\

      12    & 0 & 0.004 & 5.592 & 0.113 & 24.743   \\
             & 300 & 0.004 & 3.296& 0.022 & 74.910   \\
             & 0 & 0.008 & 3.154 & 0.754 & 4.183   \\ \multicolumn{6}{l}{ \rule{160mm}{0.2mm}}      \\
    
     15    & 0 & 0.004 & 1.304 & 0.210 & 6.210   \\
             & 300 & 0.004 & 0.248 & 0.815 & 0.304   \\
             & 0 & 0.008 &0.526  & 1.288 & 0.408   \\  \multicolumn{6}{l}{ \rule{160mm}{0.2mm}}      \\
    
     20    & 0 & 0.004 & 1.070 & 0.026 & 41.154   \\
             & 300 & 0.004 & 0.261 & 0.548 & 0.476   \\
             & 0 & 0.008 & 0.504 & 0.310 & 1.626   \\  \multicolumn{6}{l}{ \rule{160mm}{0.2mm}}      \\
    
     25    & 0 & 0.004 & 0.859 & 0.016 & 53.688   \\
             & 300 & 0.004 & 0.141 & 0.397 & 0.354   \\
             & 0 & 0.008 & 0.485 & 0.132 & 3.674   \\  \multicolumn{6}{l}{ \rule{160mm}{0.8mm}}      \\
    
   \end{tabular}

\end{table}


Para contar el n'umero de supergigantes rojas y azules de nuestras
listas, debimos aplicar las correcciones basadas en los estudios  de incompletitud del 
Cap'itulo~4.
Calculamos los valores de incompletitud de dos intervalos de 
$\log(T_{\mathrm{eff}}) $  a lo largo  de cada trayectoria evolutiva:
desde el l'imite derecho de la \sp hasta 
 $\log(T_{\mathrm{eff}}) = 3.99 $ y sobre el intervalo 
 $\log(T_{\mathrm{eff}})  \le  3.70 $.
Para cada punto que define las trayectorias evolutivas,
interpolamos linealmente en las tablas de completitud para obtener este valor. 
Tambi'en computamos un peso definido como la diferencia entre la edad de un punto 
y la edad del mismo punto en la trayectoria evolutiva
inmediatamente m'as masiva. El valor de la completitud adoptado es el promedio pesado
en cada intervalo.

Para calcular el n'umero de  estrellas azules en nuestra muestra, simplemente contamos
el n'umero de objetos dentro de cada rango de masas que proveen las
trayectorias evolutivas. Si analizamos  la distribuci'on de 
supergigantes rojas en el \hrd~ se observa que ninguno de los modelos produce RSG tan fr'ias 
y luminosas como las que se observan en la realidad.
Esto ya fue notado en estudios recientes por 
\cite{Mass02,Massey03,Masse03,MassOlse03}; 
estos autores encuentran el mismo problema al estudiar el contenido de
estrellas supergigantes rojas en las Nubes de Magallanes.

\begin{figure*}[ht!]
 \centering
 \begin{minipage}[u]{10cm}
 \includegraphics[width=95mm]{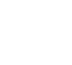}
 \end{minipage}
  \begin{minipage}[u]{6cm}
 \caption[Posici'on de supergigantes seg'un \citep{MassOlse03}]{
{\sl\footnotesize En este \hrd   se compara la posici'on
de las supergigantes rojas de la Nube Mayor de Magallanes
con un conjunto de trayectorias evolutivas 
calculadas para   $Z = 0.008$.
y que incluyen  p'erdidas de masa normales  \citep{Schaetal93}.
Se observa que los modelos no producen
supergigantes rojas los suficientemente
fr'ias y luminosas como las  que se observan las Nubes.
Este gr'afico est'a sacado de \cite{MassOlse03}.} \label{fig025}}
 \end{minipage}

\end{figure*}

El  trabajo de   \citeauthor{MassOlse03}, 
contiene estrellas supergigantes rojas con fotometr'ia  y 
con  tipo espectral conocido en algunos objetos.
Graficaron estas estrellas sobre un plano 
[ $\log(T_{\mathrm{eff}}) $, $M_{\mathrm{bol}}$ ]  y superpusieron
varios grupos de trayectorias evolutivas. 
Utilizan una nueva calibraci'on entre tipo espectral y 
$\log(T_{\mathrm{eff}}) $ para ubicar a las estrellas con tipo espectral conocido.
Para el resto de las estrellas usaron el color intr'insico   $(V-R)_0$.
Su Figura  5(a) (ver  Figura~\ref{fig025}) de las estrellas supergigantes rojas
de la LMC es muy similar a nuestra Figura~\ref{fig035}.
Es importante hacer notar que nuestro m'etodo es estrictamente 
fotom'etrico, debido a que \cho\ ajusta los espectros de    \cite{Kuru04} a un 
conjunto de colores fotom'etricos.

Para contar a las supergigantes rojas  en nuestra muestra, debimos extender artificialmente
las trayectorias evolutivas a valores constantes de $M_{\mathrm{bol}}$, 
y usamos la regi'on marcada por un rect'angulo peque\~no entre las
trayectorias de 15 y 25  $M_{\sun}$ en la Figura~\ref{fig035}
a la derecha de $\log(T_{\mathrm{eff}})  = 3.70 $.

El paso siguiente al conteo fue la correcci'on por incompletitud. Un resumen de 
este proceso est'a presentado en la   Tabla~\ref{tbl010}, donde
las columnas marcadas ``teor'ia'' muestran los n'umeros
derivados a partir del cociente del tiempo que una estrella pasa en cada
regi'on, y donde estamos suponiendo una velocidad de formaci'on estelar
constante. Las columnas ``obs'' muestran los valores que observamos.

\begin{figure*}[ht!]
 \centering
 \includegraphics[width=15cm]{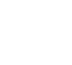}
 \caption[Distribuci'on espacial de las supergigantes]{
{\sl\footnotesize Distribuci'on de supergigantes rojas y azules
desplegadas sobre un mosaico de NGC~4214 construido
con im'agenes obtenidas con el filtro F814W. 
Los c'irculos llenos representan a las supergigantes confirmadas, rojas 
y azules, que usamos para calcular el cociente B/R.
Con c'irculos abiertos representamos al grupo de estrellas que 
siguen el criterio  
$\log(T_{\mathrm{eff}})  \leqslant 3.70 $ y  $M_{\mathrm{bol}}  \leqslant  -6.0$.
Estos objetos podr'ian ser supergigantes rojas, estrellas
de la rama asint'otica gigante, e incluso gigantes rojas brillantes 
se han dibujado las aperturas de algunos 
c'umulos como referencia.  } \label{fig066}}
\end{figure*}

La Figura~\ref{fig066} muestra la distribuci'on de supergigantes rojas y azules
sobre un mosaico construido con im'agenes obtenidas en el filtro F814W.
Los c'irculos llenos representan a las supergigantes confirmadas, rojas 
y azules, que usamos para calcular el cociente B/R.
Con c'irculos abiertos representamos al grupo de estrellas que 
siguen el criterio  
$\log(T_{\mathrm{eff}})  \leqslant 3.70 $ y  $M_{\mathrm{bol}}  \leqslant  -6.0$.
Estos objetos podr'ian ser supergigantes rojas, estrellas
de la rama asint'otica gigante, e incluso gigantes rojas brillantes que se
han pasado a regiones de mayor masa del \hrd debido a 
incertezas en las observaciones.

La morfolog'ia de la distribuci'on de supergigantes es notable en esta galaxia,
principalmente en lo que respecta a las diferencias entre las regiones NGC~4214--I
y  NGC~4214--II.

En  NGC~4214--IA y --IB se observa claramente que las supergigantes son azules, mientras
que las supergigantes  rojas se distribuyen solamente en la periferia. 
En cambio, en  NGC~4214--II se puede
apreciar una distribuci'on m'as homog'enea. Esta diferencia marcada
estar'ia  relacionada 'intimamente con la diferencia de edad entre ambas regiones ($3.0-4.0$ Ma 
para NGC~4214--I y   $1.0-2.0$ Ma 
para NGC~4214--II). La regiones I--A y I--B son m'as evolucionadas y las estrellas 
supergigantes han desaparecido mientras que son visibles en las regi'on m'as joven NGC~4214--II.

A pesar de la dificultad en el conteo de RSG, hallamos un muy buen 
acuerdo entre nuestros datos y los modelos te'oricos de baja metalicidad de Ginebra.

\begin{table}[htbp]
\centering
\begin{minipage}[u]{115mm}
\caption[Cociente de B/R]{{\sl  Valores del cociente de supergigantes  B/R. En esta tabla compramos
los valores te'oricos con los observados.} \label{tbl010}
\vspace{5mm}}
\end{minipage}

\begin{tabular}{  ccccccc  }  
\multicolumn{7}{c}{ \rule{120mm}{0.8mm}}      \\
           &\multicolumn{2}{c}{SMC  $v_{ini} =0$}      &
           \multicolumn{2}{c}{SMC  $v_{ini} =300$}
            & \multicolumn{2}{c}{LMC  $v_{ini} =0$}        \\   
\vspace{-2mm} 
   &\multicolumn{2}{c}{\rule[2mm]{25mm}{0.2mm}}      &
   \multicolumn{2}{c}{\rule[2mm]{25mm}{0.2mm}}           & 
   \multicolumn{2}{c}{\rule[2mm]{25mm}{0.2mm}}         \\   
\vspace{-2mm} Rango de masas  & teor'ia  &   obs   &    teor'ia   &   obs   &  teor'ia   &   obs     \\   \multicolumn{7}{c}{ \rule{120mm}{0.2mm}}      \\

    $   15-20 $   & 24 & $34 \pm 10$ & 0.4 & $29 \pm 10$& 1.0 & $13 \pm 4$   \\
    $  20-25 $   & 47 & $46 \pm 23$ & 0.4 & $45 \pm 31$& 2.7 & $67  \pm 66$  \\ \multicolumn{7}{c}{ \rule{120mm}{0.8mm}}      \\

\end{tabular}

\end{table}


%
Estos modelos predicen un valor de B/R de 24 en el rango de masas  $15-20 M_{\sun}$
y nosotros medimos $34 \pm 10$. Estamos suponiendo que algunas de 
las estrellas ubicadas
por debajo de la trayectoria de 15 $M_{\sun}$  podrian ser supergigantes rojas,
con lo cual estariamos subestimando el n'umero de las mismas en un factor de dos 
entre aquellas observadas en $ 15-20 M_{\sun}$.
En el rango de masas  $20-25 M_{\sun}$ 
la predicci'on te'orica es 47 y nosotros observamos $46 \pm 23$.
En ambos casos, nuestros resultados  est'an cerca de  los te'oricos
dentro de un error de Poisson.


Existen dos problemas relacionados con este resultado que deben mencionarse:
Primero, la peque\~na cantidad de supergigantes rojas en nuestra muestra
determina la presencia de efectos estoc'asticos debido a estad'isticas de numeros peque\~nos.
Sin embargo, dada la gran diferencia entre los valores de B/R te'oricos que se presentan en la 
Tabla~\ref{tbl010}, este efecto parecer'ia ser insignificante.
El segundo problema est'a relacionado con la conversi'on entre colores observados 
y la temperatura efectiva y la  correcci'on bolom'etrica.

\begin{figure*}[ht!]
\begin{minipage}[u]{6cm}
\centering
\caption[Correcci'on de la posici'on de estrellas supergigantes rojas]{\sl\footnotesize
Diagramas 
[$\log(T_{\mathrm{eff}}) $, $M_{\mathrm{bol}}$ ] 
con las estrellas supergigantes rojas 
de la Galaxia. Se despliegan adem'as las trayectorias evolutivas
de  \cite{MaedMeyn03}. 
Las lineas llenas denotan los modelos sin rotaci'on, mientras que las 
punteadas muestran las trayectorias construidas con
modelos  calculados con una velocidad de  rotaci'on inicial de 
 300 \kms. 
Las trayectorias con rotaci'on aparecen por encima de
las  carentes de rotaci'on.
En (a) se muestran la ubicaci'on de las supergigantes rojas
de acuerdo  a   \cite{Hump78}, usando las temperaturas efectivas y
correcciones  bolom'etricas de  \cite{HumpMcel84}. 
En (b)  se muestra la ubicaci'on de las supergigantes rojas
usando los nuevos modelos de atm'osferas MARCS. Ver  \citet{Leveetal05} 
para los detalles. 
  \label{fig026}}
\end{minipage}%
\hspace{1cm}
\begin{minipage}[c]{10cm}
\centering
 \includegraphics[width=10cm]{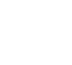}
\end{minipage}
\end{figure*}

\cite{Leveetal05} presentan una nueva escala de  temperatura efectiva para supergigantes
rojas de la  Galaxia obtenida al ajustar modelos de atm'osferas MARCS 
\citep{Gustetal75, Plezetal92} que incluyen un tratamiento mejorado de la opacidad molecular.
Esta nueva escala fue obtenida con 74 supergigantes rojas  gal'acticas 
$(Z=0.020)  $ 
 de distancia conocida. 
Usaron modelos te'oricos de \cite{MaedMeyn03} y hallaron un acuerdo 
mucho mejor entre teor'ia y observaci'on. 
La Figura~\ref{fig026}  muestra las mejoras obtenidas
usando los modelos de atm'osfera  MARCS.
En la  Figura~\ref{fig026} (b) el desacuerdo entre la teor'ia y su ubicaci'on en el \hrd\
ha desaparecido.
Si este efecto fuese el mismo  para las estrellas de NGC~4214, hubi'esemos tenido 
este desacuerdo entre teor'ia y observaci'on.
El resultado m'as importante es que las RSG parecen ser m'as calientes que lo pensado 
anteriormente. Este efecto desplaza a las estrellas hacia la izquierda de los diagramas
haciendo que  su posici'on coincida con la posici'on de las trayectorias
evolutivas.

Ser'ia importante comparar el cambio de temperatura proveniente del ajuste 
de los colores observados usando estos modelos de atm'osferas para
verificar
si explica la discrepancia en $350-400$ K que detectamos en NGC~4214.
Adem'as, tal cambio producir'ia un desplazamiento de las
estrellas hacia abajo del diagrama de Hertzsprung-Russell, pues una mayor 
temperatura, implica una menor correcci'on bolom'etrica.
La versi'on que empleamos de \cho\ no ten'ia la capacidad
de elegir el modelo de atm'osfera deseado, por lo que tuvimos que usar
los modelos de Kurucz. El empleo de los modelos de Lejeune
dio resultados comparables. 

N'otese  que el recurso que empleamos  de extender artificialmente las 
trayectorias evolutivas a  
$M_{\mathrm{bol}}$ constante es equivalente a desplazar 
 a las estrellas hacia la izquierda en el  diagrama que es el efecto 
producido al emplear los modelos  MARCS.


\chapter{Extinci'on en NGC~4214}
\thispagestyle{empty}
 \hfill  {\em  I remember
there was mist   ... }
\label{cha:ext}

\newpage
\section{El medio interestelar}
\subsection{Gas y polvo}

El medio interestelar (ISM, del ingl'es: InsterStellar Medium) es  el material distribu'ido en el espacio
entre las estrellas  y  est'a compuesto por gas (mol'eculas, 'atomos, iones) y polvo (peque\~nos granos
formados por hielos de varios tipos, grafitos, silicatos y posiblemente metales). 
El gas se pone de manifiesto  a trav'es de l'ineas de absorci'on caracter'isticas, tanto en el 'optico como
en radio, de los espectros de estrellas y fuentes de radio. Tambi'en se hace presente con l'ineas en emisi'on
en un amplio rango de longitudes de onda: desde el 'optico emitido por
nebulosas difusas  hasta  en ondas de radio provenientes de l'ineas moleculares. 
El polvo se observa en nebulosas oscuras de absorci'on y en nebulosas de reflexi'on que reflejan
la luz proveniente de estrellas cercanas. 
La presencia del ISM tiene dos efectos sobre la luz de las estrellas: produce una disminuci'on en la intensidad
producida por la absorci'on y dispersi'on por parte del polvo ({\em extinci'on}) y 
un efecto  selectivo en funci'on de la longitud de onda ({\em enrojecimiento}).

El material gaseoso est'a compuesto por varios componentes, que van desde simples part'iculas
como electrones libres, protones, 'atomos neutros de H, mol'eculas de H$_2$ hasta
algunas mol'eculas org'anicas complicadas.
Los granos de polvo has sido estudiados en profundidad \citep{Math90}.
El efecto principal del polvo interestelar
es la extinci'on de la luz de estrellas lejanas. La extinci'on en la regi'on 'optica del espectro
se debe principalmente a la dispersi'on de la luz, pero tambi'en 
puede haber absorci'on. 
La extinci'on se hace m'as pronunciada en longitudes de onda cortas y este 
efecto selectivo del polvo es conocido como {\em enrojecimiento interestelar}
debido a que en el espectro visible la parte menos afectada es el rojo.
El polvo interestelar es un componente muy importante de las galaxias. El polvo
oscurece la luz proveneniente de las estrellas absorbi'endola y reemiti'endola 
en longitudes de onda largas principalmente el infrarrojo (IR). 
El polvo es fundamental en lo que respecta a la qu'imica interestelar, pues reduce 
la radiaci'on UV  que causa la disociaci'on de mol'eculas y provee
de un lugar para la formaci'on de la mol'ecula interestelar m'as abundante (H$_2$).

De acuerdo a los trabajos de   \cite{Math90}  y \cite{LameCass99}, la mayor parte del polvo 
presente en el ISM  proviene de estrellas en la rama asint'otica de las
gigantes (AGB), ricas en C y/o O y variables tipo Mira. Las supernovas podr'ian ser otra fuente 
de enriquecimiento del medio interestelar con elementos pesados. 
Las novas y las estrellas WR (del tipo tard'io WC) son fuentes menos importantes.
Ambas inyectan polvo rico en C. Por 'ultimo, las nebulosas
planetarias son fuentes de gas pero no son una fuente importante de polvo, debido
a que tienen un bajo contenido del mismo.

\subsection{Extinci'on Gal'actica}

La {\it extinci'on (absorci'on + dispersi'on) } es la propiedad mejor estudiada
del polvo, pues puede ser determinada con precisi'on sobre
un amplio rango de longitudes de onda y sobre 
l'ineas de la visual apuntando  hacia regiones con diferentes
caracter'isticas f'isicas.
Cada l'inea de la visual tiene su propia ley de extinci'on, o variaci'on de la 
extinci'on con la longitud de onda, usualmente expresada mediante  $A(\lambda)/A(V)$.
El uso de $A(V)$ como referencia es  completamente arbitrario.
La extinci'on $A(\lambda)$ a una determinada longitud de onda se la define
como la magnitud aparente medida menos la magnitud aparente que se medir'ia en ausencia de polvo.
El exceso de color o enrojecimiento se define como  $ E(B-V) \equiv A_B-A_V$, o sea, la
extinci'on en el filtro $B$ menos la extinci'on en el filtro $V$.

El estudio de la dependencia de la extinci'on con la longitud
de onda es importante por dos razones: Primero, la extinci'on depende 
de las propiedades 'opticas de los granos de polvo
y puede revelar informaci'on sobre la composici'on y la distribuci'on de los mismos.
Segundo, la dependencia con la longitud de onda se necesita para remover
los efectos del debilitamiento de la energ'ia proveniente de los objetos astron'omicos
de inter'es. 
La extinci'on en el 'optico y en el IR se estudia tradicionalmente con 
fotometr'ia de banda ancha e intermedia desde Tierra. Para estudiar la extinci'on en el
UV, se emplean observaciones desde sat'elites.
Varios investigadores han empleado el  International Ultraviolet Explorer (IUE) para
hacer estudios detallados de las leyes de extinci'on en el UV.

 \cite{Cardetal89} han usado  las observaciones en el UV de \cite{FitzMass90}, y
 las combinaron con observaciones en el  'optico de las mismas
 estrellas para determinar la relaci'on entre las  diferentes leyes a lo largo de 
 un amplio rango de longitudes de onda.
  \cite{Cardetal89}  usaron el cociente  $R_V \equiv A(V)/E(B-V) $  como par'ametro. 
  El valor de $R_V$ depende de las caracter'isticas del medio a lo largo de
  la l'inea de la visual y da una medida del tama\~no del grano de polvo. 
 Regiones est'andar del ISM (donde la densidad es baja  $ n \approx 10^{-3} cm^{-3}$)
 presentan un valor bajo de $R_V$ ($\sim$ 3.1).
 Cuando las l'ineas de la visual penetran regiones densas ($n > 10^3 cm^{-3}$) se mide
 $4<R_V<6$. 
 \citeauthor{Cardetal89}  ajustaron varias curvas  a  $A(\lambda)/A(V)-R_V^{-1}$
 mediante una f'ormula anal'itica que representa la extinci'on media como una funci'on de 
 $R_V$.
  Esta f'ormula se puede aplicar a regiones difusas y densas del medio interestelar
  y se emplea para calcular la extinci'on y para desenrojecer observaciones.
  La f'ormula tiene la siguiente forma:

\begin{equation}
\label{eqn-0}  
\frac{A(\lambda)}{A(V)} = a(x)+ \frac{b(x)}{R_V},
\end{equation}

donde $x \equiv 1/ \lambda$ (en $\mu$m$^{-1}$), $a$ y $b$ son dos funciones definidas 
en el rango de inter'es y  $R_V \equiv A(V)/E(B-V) $

 
\begin{figure*}[ht!]
\begin{minipage}[c]{15cm}
\centering
\centerline{\includegraphics*[  width=0.47\linewidth]{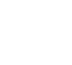}
                 \includegraphics*[  width=0.47\linewidth]{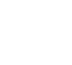}}
\caption[Leyes de extinci'on]{ {\sl\footnotesize
[Izquierda] Las leyes de extinci'on de   \cite{Cardetal89} dan el valor de la absorci'on
$A(\lambda)/A(V)$ como funci'on de la longitud de onda y est'an parametrizadas 
por medio de la cantidad $R_V$. Los gr'aficos representan la ley para
$R_V =1, 2, 3, 4, 5.$   
 [Derecha] Comparaci'on del comportamiento de las leyes de extinci'on Gal'acticas 
[$R_V=3.1$  \citep{Cardetal89}; F99 \citep{Fitz99} ] y
 de las Nubes de Magallanes  [AVGLMC, LMC2 \citep{Missetal99}; SMC \citep{GordClay98}]   \label{fig051}}}
\end{minipage}
\end{figure*}


La Figura~\ref{fig051} (derecha) muestra la forma
de la curva de extinci'on Gal'actica promedio   desde el IR lejano hasta el UV. Las curvas
se presentan dando $A(\lambda)/A(V)$ en funci'on de  $1/ \lambda$. T'ipicamente,
la extinci'on  aumenta en el IR con una dependencia en forma de ley de potencias, decrece en la zona
'optica del espectro y muestra una caracter'istica sobresaliente en 2175  \AA\ 
en el UV cercano.

  \citeauthor{Cardetal89} encontraron que la forma de la extinci'on  
  se correlaciona muy bien con el par'ametro  $R_V$ y adoptaron a este par'ametro 
  para hacer su familia de leyes. 
 La esencia de la ley de   \citeauthor{Cardetal89} est'a ilustrada en Figura~\ref{fig051} (izquierda), donde
 hemos graficado cinco  curvas representativas, cada
 una con un valor de $R_V$. Las leyes planas o ``grises'' en el UV se caracterizan
 por altos valores de $R_V$; mientras que las que presentan una pendiente empinada en el UV 
 se mantienen empinadas en el IR y se caracterizan
 por valores  bajos de $R_V$.
  
 Las leyes de extinci'on de  \cite{SaMa79} y  \cite{Seat79} 
 son usadas corrientemente para corregir por la presencia de polvo
 en la Via L'actea. Ambas leyes
 pueden reproducirse tomando $R_V=3.2$  en las leyes de \cite{Cardetal89}. 
 
 La caracter'istica fundamental en el espectro observado es el  lomo o  ``bump''   situado a 
 2175 \AA\   o 4.6 $\mu$m$^{-1}$.
 Este lomo est'a presente en todos los valores de $R_V$. Su origen no est'a perfectamente
 determinado, pero se cree que est'a causado principalmente por el grafito o alguna
 estructura molecular derivada del carbono y se lo conoce como ``bump del grafito''.

\subsection{Extinci'on extragal'actica}

Existen algunos trabajos sobre la extinci'on en las Nubes de Magallanes.
En la Nube Mayor, se encuentra que $R_V \approx 3.2 \pm 0.2$. En el UV, las estrellas 
cerca de la regi'on \hii gigante 30 Doradus presentan 
lomos muy d'ebiles, mientras que las estrellas  en las afueras de esa regi'on
tiene leyes de extinci'on semejantes a las de la Galaxia. 
En la Nube Menor, casi no se encuentran estrellas enrojecidas. En general, el valor de 
$R_V$ es bajo, no se observa el bump del grafito y en el UV la pendiente es muy grande.

{\bf Nube Menor de Magallanes} 

La ley de extinci'on de la Nube Menor de Magallanes fue estudiada por  \cite{GordClay98}.
Estos autores han mejorado considerablemente nuestro conocimiento de esta ley
respecto al estudio previo de \cite{Prevetal84}. 
Combinando un  estudio cuidadoso de los trabajos previos con
nuevas observaciones  IUE
de estrellas de tipos espectrales entre O9 y B3, encuentran un total de cuatro estrellas
apropiadas para estudios de extinci'on. Tres de estas estrellas poseen una curva casi lineal.
Las l'ineas de la visual de las mismas pasan por regiones de formaci'on estelar activas.
La cuarta estrella  presenta una curva con el bump y una pendiente
en el UV lejano menos empinada. Su l'inea de la visual pasa tambi'en por una 
regi'on de formaci'on estelar, pero no tan activa como en el caso de las otras tres.

{\bf Nube Mayor de Magallanes} 

Los estudios realizados sobre la Nube Mayor de Magallanes llegan todos a conclusiones similares:
la ley de extinci'on promedio est'a caracterizada por un lomo m'as d'ebil que  el  de la Galaxia, y
presenta una pendiente mayor en el UV lejano. 
\cite{Fitz85} analiz'o 19 objetos en total incluyendo 7 fuera de la
regi'on 30~Dor y encontr'o diferencias significativas entre ambas regiones.

\cite{Missetal99}  reanalizaron los datos de esta galaxia empleando datos del IUE. Encontraron que 
existe un grupo de estrellas con un lomo muy d'ebil que se encuentran en o cerca de la regi'on
ocupada por la regi'on  LMC2 al sudeste de 30~Dor. Las leyes de extinci'on promedio  dentro y
fuera de la regi'on  LMC2
muestran  una significativa diferencia en la intensidad del lomo en 2175 \AA, pero
su comportamiento en el UV lejano es similar. 
Esto dio origen a dos leyes muy usadas para estudiar la  Nube Mayor: la ley 
definida para la regi'on LMC2  (que denominamos LMC2) y la ley promedio sobre el resto de 
la galaxia (que denominamos AVGLMC). La Figura~\ref{fig051}  ilustra el comportamiento de
estas leyes como funci'on de la longitud de onda.

{\bf Comparaci'on entre leyes} 


\cite{Gordetal03}  realizaron una comparaci'on de todas las curvas
de extinci'on en las Nubes de Magallanes teniendo en cuenta al
comportamiento general de  las leyes gal'acticas.
Encontraron que s'olo unas pocas de las leyes de la Nube Mayor son consistentes con las 
leyes de  \cite{Cardetal89}. 
Estas leyes Gal'acticas fueron obtenidas a partir de un grupo de objetos 
de la Via L'actea y parece ser una muy buena descripci'on de la extinci'on en la Galaxia.
 
 \cite{Gordetal03}  combinaron espectros  UV de archivo del IUE, fotometr'ia 'optica y 
 fotometr'ia en el IR cercano de 2MASS y DENIS con espectroscop'ia UV de STIS a bordo del
 HST para determinar las curvas de extinci'on de 24 objetos en las Nubes de Magallanes.
 Los valores promedio obtenidos para estas leyes son: para la Nube Menor $R_V = 2.74 \pm 0.13$,
 para la regi'on LMC2 $R_V = 2.76 \pm 0.09$, y para la Nube Mayor en promedio $R_V = 3.41 \pm 0.06$.
 
Su conclusi'on final es que las leyes de extinci'on de las Nubes son muy diferentes
de las de la Galaxia, lo cual implica que los ambientes  y la metalicidad 
en esas tres galaxias deben
ser muy diferentes.
Notan adem'as que las leyes de    \cite{Cardetal89} est'an generadas a partir
de estrellas OB de secuencia principal, lejos de zonas
de formaci'on estelar, mientras que las l'ineas de la visual medidas  para las Nubes
tienen un efecto de selecci'on en cuanto a que se han empleado estrellas supergigantes OB
ubicadas en regiones de alta actividad.

\section{La extinci\'on  en NGC 4214}

Comparando modelos de atm'osferas estelares y colores 
fotom'etricos observados es posible inferir  la extinci'on 
de estrellas individuales.
\cho\  lee una lista de espectros estelares, los extingue  variando tanto 
$R_V $ como $E(B-V)$ y luego
obtiene la fotometr'ia sint'etica que mejor se ajusta a los datos de entrada.
Como resultado de los ajustes que hicimos  y explicamos en 
cap'itulos anteriores, obtuvimos la ley de extinci'on 
(\rv) ~que mejor se ajusta  a nuestros datos, asi como el valor del 
enrojecimiento (\ecc) ~de cada estrella en nuestras
listas.

\subsection{Estudios previos } 
En estudios previos de NGC~4214, la correcci'on por extinci'on 
fue realizada usando diferentes valores de  
  $E(B-V)$:  \citet{Maizetal98}  midieron un  $E(B-V)$ variable 
  con valores entre 0.0 y 0.6 mag a partir del cociente
de   las lineas de emisi'on \ha\ y \hb.

\citet{Maizetal02a} midieron una valor medio de $E(B-V)=0.09$ mag 
a partir de los colores 'opticos de la poblaci'on estelar joven ubicada 
lejos de las regiones de formaci'on estelar.

\citet{Droz02} y 
\citet{Calzetal04}  adoptaron el valor  $E(B-V)=0.02$ mag  provisto por
el mapa IRAS DIRBE  
de  \citet{Schletal98}. Para hacer estas estimaciones, supusieron siempre 
como v'alida a la ley de extinci'on de  \citet{Cardetal89} con $R_{5495}=3.1$.

\subsection{Mapa de extinci'on}

A partir de las estimaciones  de la extinci'on estelar realizados con \cho\
construimos  un mapa de extinci'on de la regi'on estudiada de NGC~4214.
El mapa fue construido seleccionando algunos objetos de la
 {\sc\footnotesize  LISTA336}. Los criterios de selecci'on de objetos son:

Del conjunto de estrellas con $E(4405-5495)  \leqslant 0.4 $ mag elegimos las 
de menor error ($\sigma_{E(4405-5495)} \\ \leqslant~0.1~$mag).
Para los objetos con   $E(4405-5495)  > 0.4 $  mag
consideramos adem'as su magnitud en el filtro F336W
(F336W $< 21$ mag)  y su distribuci'on  en la galaxia.
Es importante hacer notar que nuestra muestra tiene un sesgo
relacionado con la selecci'on de objetos y que podr'iamos estar
perdiendo algunos objetos azules con la selecci'on realizada.
La extinci'on es un efecto tridimensional muy complicado que 
depende de la distribuci'on de estrellas, gas y polvo y que var'ia
en escalas peque\~nas ($  \lesssim 1$ pc). Todos estos factores 
contribuyen a hacer que el modelado de la extinci'on sea 
bastante complicado.

\begin{figure*}[ht!]
\centering
\includegraphics[width=\textwidth]{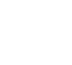}
\caption[Mapa de estrellas usadas en el estudio de la extinci'on]{  {\sl\footnotesize  
Mosaico de NGC~4214 creado a partir 
de im'agenes obtenidas con el filtro F656N.
Se ha marcado la selecci'on de estrellas empleadas para
el presente estudio de extinci'on.  La escala de colores representa valores de 
  \ecc\ expresados en magnitudes.
  \label{fig037}}}
\end{figure*}

Finalmente, agregamos a nuestra lista de objetos 
los resultados del an'alisis de extinci'on (ver Cap'itulo~9)
hecho sobre los \cus   I--As, I--Es, IIIs y IVs.  
 La Figura~\ref{fig037} representa  un mosaico constru'ido con el filtro F656N  
de la regi'on estudiada de  NGC~4214, donde marcamos los objetos 
de nuestra lista de extinci'on con un esquema de colores
para indicar la distribuci'on de la extinci'on a trav'es de
la galaxia.
Para construir el mosaico en \ha\ (filtro F656N), empleamos 
las im'agenes u3n8010fm+gm de la propuesta 6569
con un tiempo total de exposici'on de 1600 segundos.
Este mosaico pone en evidencia la estructura  variable de la extinci'on.

\begin{figure*}[ht!]
\centering
\includegraphics[width=\textwidth]{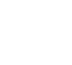}
\caption[Mapa de extinci'on]{ {\sl\footnotesize Mapa de extinci'on de la regi'on central
de NGC~4214.  La orientaci'on es Norte   hacia arriba y Este  
hacia la izquierda. Se han marcado las aperturas de algunos c'umulos 
para guiar al ojo. Ver el texto para la explicaci'on de c'omo se cre'o este mapa.
La escala de colores representa valores de 
  \ecc\ expresados en magnitudes.
 \label{fig038}}}
\end{figure*}

El procedimiento empleado para generar al  mapa de extinci'on comienza con 
la creaci'on de una matriz
$M$  que mapea a toda la regi'on de inter'es. A cada elemento  $M(i,j)$ de la matriz le  asignamos 
un valor pesado de la extinci'on dado por:

\begin{equation}
\label{eqn-0}  
M(i,j) = \frac{\sum_k E_k \cdot w_k}{\sum_k w_k},
\end{equation}

donde la suma se extiende sobre todas las estrellas de la lista y $ E =E(4405-5495). $
El peso $w_k $ asignado a cada estrella est'a dado por:

\begin{equation}
\label{eqn-0}  
w_k = \frac{\exp\left( - \frac{ d_{ijk}^2}{30}    \right)}{\sigma_{E_k}^2}  ,
\end{equation}

siendo  $d_{ijk}$ la distancia (en pixeles) entre
cada elemento  de  la matriz  $M(i,j)$ y todas las estrellas $k$ de nuestra lista y $\sigma_{E_k}$
es el error en $ E =E(4405-5495)$ de la estrella $k$.
El valor 30 en el denominador es un simple factor de escala arbitrario. 
La Figura~\ref{fig038} muestra el mapa de extinci'on final.


\subsection{An'alisis de la extinci'on}

El resultado  m'as importante en este mapa es el hecho de que
NGC~4214 est'a caracterizada por valores bajos de la extinci'on, excepto
en algunas regiones espec'ificas con valores mayores que el promedio,
lo cual est'a en acuerdo con 
\cite{Maizetal98}, \cite{Droz02} y  \cite{Calzetal04}.

\cite{Maizetal98} y \citet{Maiz00} emplearon el cociente de  Balmer  
(H$\alpha$/H$\beta$) para rastrear el enrojecimiento 
que afecta al gas ionizado y podujeron mapas de este cociente.
El cociente de Balmer se estudia mediante la intensidad de la l'ineas 
H$\alpha$  y H$\beta$, definiendo la cantidad

\begin{equation}
\label{eqn-0}  
r_B \equiv \frac{I(\mathrm{H}\alpha)}{I(\mathrm{H}\beta)},
\end{equation}

En una regi'on \hii  con $n_e=100$ cm$^{-3}$ y $T_\mathrm{eff} =10\,000$ K,
se espera $r_B=2.86$.
Cualquier medida de valores de $r_B$ diferentes de 2.86 indicar'ia  que la radiaci'on ha experimentado 
extinci'on. 

Estos autores muestran que el cociente tiene valores relativamente bajos  $r_B \leqslant 3.25$
con la excepci'on de algunas regiones donde la presencia de nubes de polvo lo lleva a valores superiores a 4.25.
Estas nubes se encuentran en el complejo NGC~4214--II y en 
varios puntos del extremo oriental del complejo  NGC~4214--I.

Su an'alisis muestra una diferencia significativa entre los dos complejos 
principales: \\ NGC~4214--I y   --II.
Concluyen  que las  emisiones  nebular y estelar son producidos
en forma co--espacial en NGC~4214--II, mientras que el gas en emisi'on
se encuentra claramante  desplazado de los \cus estelares en las
regiones m'as brillantes del complejo NGC~4214--I. 

Tambi'en encuentran que el enrojecimiento en NGC 4214--II es, en promedio,
mayor que en NGC 4214--I. Nuestros resultados, derivados a partir de colores
estelares, est'an de acuerdo con estos resultados. Las dos cavidades principales
de NGC 4214--I muestran  una baja extinci'on rodeada de una zona de 
alta extinci'on, mientras que en NGC 4214--II  la extinci'on es en general mayor.
 
\begin{figure*}[ht!]
\begin{minipage}[c]{16cm}
\centering
\centerline{\includegraphics*[  width=0.47\linewidth]{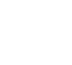}
                 \includegraphics*[  width=0.47\linewidth]{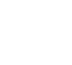}}
\caption[ Im'agenes de NGC~4214 en \hi y \ha\ ]{ {\sl\footnotesize
[Izquierda]  Imagen de NGC~4214 en \hi obtenida de \citet{McIn97}. N'otese la presencia
 de brazos espirales incipientes. El cuadro marcado representa la zona estudiada por  
 \cite{Waltetal01}  en CO que se muestra ampliada a la derecha.
 [Derecha] Imagen de NGC~4214 construida con  
 H$\alpha ~+$ [N~II] sobre
 la que se han dibujado los contornos de CO. Los
 c'irculos grandes marcan las regiones abarcadas por el mosaico
 de CO de  \cite{Waltetal01}. Se observa una regi'on \ha\ muy d'ebil
 cerca del contorno de CO superior. La emisi'on \ha\ es muy intensa en
 las regiones NGC~4214--I y NGC~4214--II.  \label{fig040}}}
\end{minipage}
\end{figure*}

\begin{wrapfigure}[20]{l}[0mm]{100mm}
\includegraphics[width=100mm]{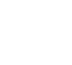}
\caption[Contornos de CO superpuestos sobre un mosaico WFPC2  de NGC~4214]{ {\sl\footnotesize 
Contornos de CO superpuestos sobre un mosaico WFPC2  de NGC~4214
construido con varios filtros (F336W, F502N, F555W, F656N, F702W y
F814W) proveniente del Hubble Heritage Team \citep{Waltetal01}.  \label{fig042}}}
\end{wrapfigure}

\cite{Waltetal01}  presentan un estudio del gas molecular de NGC~4214. 
Detectaron tres regiones
de emisi'on molecular, en el noroeste, en el sudeste y en el centro de
la galaxia que se muestran en la 
Figura~\ref{fig040}. Estos autores compararon  la estructura de CO (un indicador
de estructuras moleculares) con la estructura de \ha\ (un indicador de regiones de 
formaci'on estelar).
Dos de los tres complejos de CO est'an asociados directamente con regiones
de formaci'on estelar. El complejo sudeste es co--espacial con
NGC 4214--II. Ver Figura~\ref{fig042}.
El m'aximo de intensidad de CO est'a casi sobre uno de los c'umulos.
Nuestro mapa de extinci'on muestra  una regi'on de alta extinci'on co--espacial
con la regi'on m'as grande de emisi'on de \ha, la cual abarca la mayor 
parte del complejo NGC 4214--I.
Este complejo de CO se encuentra esparcido, en lugar de concentrado, y 
la emisi'on m'axima de CO est'a desplazada hacia el oeste  respecto al pico de emisi'on
de \ha, observ'andose poco CO en la ubicaci'on de las dos cavidades principales.
A grandes rasgos,  nuestro mapa de extinci'on mapea las nubes moleculares en
esta parte de la galaxia, aunque con menos detalle que en NGC 4214--II.

\cite{MacKetal00} estudiaron la extinci'on diferencial
del gas en la estructura de   NGC~4214 y hallaron que 
las componentes estelares est'an concentradas principalmente
en regiones donde el gas presenta baja extinci'on.
En el presente trabajo complementamos dicho estudio
al estudiar la extinci'on diferencial proveniente del continuo estelar
de la galaxia.
A partir del mapa construido concluimos que la extinci'on 
derivada a partir del gas y la extinci'on derivada a partir 
del estudio de las estrellas individuales 
proveen valores similares punto a punto. Esta coincidencia es 
notable sobre el campo de la galaxia estudiado en esta tesis.


%
%
%

\chapter{La funci'on inicial  de masa}
\label{cha:imf}
\thispagestyle{empty}
\begin{flushright} 
  {\em  Stranger
than you dreamt it, \\
can you even 
dare to look? }
 \end{flushright}

\newpage
 \section{La funci'on inicial  de masa}
La f'isica  que subyace tras   la formaci'on estelar determina la conversi'on
de gas en estrellas. El producto de la formaci'on estelar son estrellas 
con un amplio rango de masas. 
Un resultado elemental de la teor'ia de evoluci'on estelar es que 
la estructura y la evoluci'on de una estrella de una dada composici'on qu'imica
est'a controlada principalmente por su masa entre otros par'ametros (metalicidad, momentum angular). 
La distribuci'on de masas en el momento del nacimiento de las estrellas
es una funci'on muy importante denominada {\em funci'on inicial
de masa}  (IMF, del ingl'es: Initial Mass Function). \index{ funci'on}
Esta  funci'on determina la proporci'on  de estrellas de diferentes tipos y masas.

Una vez que la masa de una estrella queda definida, 
se pueden determinar cantidades como su
luminosidad, radio y espectro  en cualquier momento de su historia.

La determinaci'on observacional de la IMF de una dada poblaci'on estelar
provee de restricciones fundamentales 
sobre las teor'ias de formaci'on estelar. La mayor'ia de las estrellas se forman en
grupos dentro de nubes moleculares \citep{LadaLada03}, y la
IMF
contiene informaci'on sobre de qu'e manera las masa original de la nube
se divide en fragmentos y se distribuye  en nubes protoestelares. 

\section{Definiciones: convirtiendo el n'umero de estrellas 
en una IMF}

 Siguiendo la notaci'on de \cite{Scalo86}, definimos al espectro de
 masas $f(m)$  de manera tal que $f(m)\cdot dm$ es el n'umero de
estrellas formadas en el mismo momento en cierto volumen de espacio con masas
en el rango   $m$ a $m+dm$.
En general, el momento de ``formaci'on'' se refiere  al estado en
el que la estrella quema  hidr'ogeno  en  su estad'ia en la secuencia
principal.

 Las unidades de $f(m)$  se definen mediante una normalizaci'on adecuada.
 Usualmente se trata a $f(m)$  como una densidad de probabilidad, de manera tal que

 \begin{equation}
\label{eqn-00} 
  \int_{M_{up}}^{M_{low}}  f(m)   \cdot dm =1,
\end{equation}

donde $M_{up}$ y $M_{low}$  son los l'imites superior e inferior de las masas,
cuyos valores, se piensa en la actualidad, var'ian alrededor de
$M_{low} \approx 0.05-0.1 \, M_{\odot} $   y    $M_{up} \approx 120-200 \, M_{\odot}$.
En nuestro estudio nos vamos a concentrar en el an'alisis de 
la IMF asociada a estrellas de alta masa $(> 15 \, M_{\odot})$ en NGC~4214.

En el caso de estrellas masivas en regiones de formaci'on estelar,
podemos contar la cantidad de 'estas como funci'on de su masa:
esta es la funci'on de masa presente, y se puede
obtener directamente.
La manera en que esta cantidad est'a relacionada con la IMF depende
de la historia de formaci'on estelar para la regi'on que se est'a estudiando.
Nosotros consideramos dos casos extremos: (1) cuando todas las estrellas de una 
regi'on fueron formadas simult'aneamente, y (2) cuando el proceso de formaci'on estelar fue
continuo y la tasa de formaci'on constante.

El n'umero de estrellas formadas con masas entre  $m$  y  $m+dm$  en el 
intervalo de tiempo  entre $t$ y $t+dt$  est'a dado por  $f(m) \psi(t) \cdot dm \cdot dt$,
donde $\psi(t)$ es la tasa  de formaci'on estelar.
El n'umero de estrellas observadas en el momento $t$ con masas entre 
 $m_1$ y $m_2$ est'a dado por:
 
\begin{equation}
\label{eqn-00} 
N(t) |^{m_2}_{m_1} = \int_{m_1}^{m_2} \int_{t_0}^{t} f(m) \psi(t')  \cdot dm \cdot dt',
\end{equation}

si la formaci'on estelar comenz'o en el instante $t=t_0,$ y si la
edad promedio de una estrella en la \sp  $\tau_{MS}$  con masa
 $m_2$ es menor que  $t.$
 
La forma del espectro de masas m'as com'unmente empleado es la
ley de potencias  \citep{Scalo86,Masse98}:

\begin{equation}
\label{eqnpotencias}  
\frac{dN}{dm} = f(m) = A \cdot m^{\gamma}
\end{equation}

 La cantidad $A$
es un par'ametro de normalizaci'on.

{\bf Formaci'on estelar simult'anea: }
Cuando todas las estrellas de una regi'on se forman en un
mismo evento (en el instante  $t=t_0$),
la velocidad de formaci'on de estrellas es simplemente una funci'on delta
y la integral se simplifica a :

\begin{equation}
\label{eqn-02} 
\int_{m_1}^{m_2} \frac{dN}{dm} dm = A \cdot \int_{m_1}^{m_2} m^{\gamma} dm
\end{equation}

\begin{equation}
\label{eqn-03} 
N |^{m_2}_{m_1} = N(m_2) - N(m_1) = \frac{A}{\gamma + 1} \cdot \left[ {m_2}^{\gamma + 1}- 
{m_1}^{\gamma + 1} \right]  \hspace{1cm}; \gamma \neq -1
\end{equation}

Cuando la formaci'on estelar es simult'anea, la funci'on de masa actual {\it es}
la funci'on inicial  de masa, teniendo en cuenta que las estrellas
m'as masivas habr'an desaparecido.

{\bf Formaci'on estelar continua:  }
Otro caso posible es aquel en el que la formaci'on estelar 
ha sido continua en la regi'on estudiada. La tasa de
formaci'on estelar es constante y para una masa en particular,
el n'umero de estrellas que nacen cada a~no es el mismo que el n'umero 
de estrellas que dejan la \sp cada a~no.
Luego de un tiempo $t$ mayor que la edad sobre la \sp 
de las estrellas de menor masa consideradas, el n'umero de 
estrellas observadas como funci'on de su masa est'a dado por
el producto de la funci'on  inicial  de masa
y la edad en la \sp   $\tau_{MS}$ a esa dada
masa $m$:

\begin{equation}
\label{eqn-19} 
N |^{m_2}_{m_1} = N(m_2) - N(m_1) \propto   \left[ {m}^{\gamma + 1} \cdot  
\tau_{MS}(m)  \right]^{m_2}_{m_1}  
\end{equation}

Este caso de formaci'on estelar
puede ser aplicado cuando se promedia sobre
varias regiones de formaci'on estelar de diferentes edades,
como es el caso de una galaxia entera.

\section{M'etodos para determinar la IMF}

El m'etodo m'as directo y confiable de obtener la IMF de una dada poblaci'on estelar
se basa en  conteos de estrellas como una funci'on de su luminosidad/masa \citep{Lequ79}.
Sin embargo, existen otros m'etodos indirectos que no emplean conteos de estrellas,
pero que pueden dar informaci'on acerca de la IMF  \citep{Scalo86}.
Se han empleado cantidades observables como colores fotom'etricos, relaciones
entre masa y luminosidad, abundancia de elementos qu'imicos, intensidad de l'ineas
espectrales, etc. para inferir la IMF de nuestra Galaxia  y de otras cercanas.
El problema principal  de estos m'etodos indirectos es que los observables
en los que se basa la determinaci'on de la IMF dependen de una o varias funciones 
desconocidas como la velocidad de formaci'on estelar, la abundancia de 
los productos de la nucleos'intesis, las trayectorias evolutivas detalladas de 
las estrellas gigantes, etc. 

\cite{Scalo86} presenta una discusi'on detallada con ejemplos de diferentes 
formas de inferir la IMF de una poblaci'on estelar dada.
Uno de 'estos es el m'etodo de poblaciones sint'eticas, que consiste  en  hallar la mejor mezcla
de estrella de diferentes tipos espectrales y clases de luminosidad que coincidan con
 los colores observados, el espectro o bien  la intensidad
de ciertas l'ineas espectrales de la galaxia. 
El problema principal que surge con este m'etodo  es la unicidad de la soluci'on y 
los tipos de restricciones astrof'isicas  que se necesitan imponer a  los datos.

\section{La IMF de la poblaci'on resuelta}

\subsection{La V'ia L'actea}
La historia de la determinaci'on de la IMF de poblaciones estelares
comenz'o en 1955 en la Australian National University
cuando Edwin E. Salpeter public'o la primera estimaci'on de la IMF de las estrellas en la vecindad
del sol \citep{Salp55}.
En el caso de estrellas  con masas en el rango $0.4$ a 10 $M_{\odot}$ encontr'o que 
la IMF pod'ia ser descripta mediante una ley de potencias como la 
Ecuaci'on~\ref{eqnpotencias} con un 'indice $\gamma = -2.35$.
En \citet{Masse98} el autor demuestra mediante un extenso estudio espectrosc'opico
que el resultado original de Slapeter se extiende hasta los rangos de estrellas m'as masivas
de nuestra Galaxia. 
Hoy sabemos que la IMF en las cercan'ias del sol se aplana significativamente 
para estrellas de masas por debajo de   $0.5 \, M_{\odot}$ \citep{Krou02}.

El estudio de estrellas masivas es complicado ya que la mayor parte de su energ'ia
es emitida en el UV lejano que no puede detectarse desde Tierra, y porque
su vida en la secuencia principal es muy corta ($\lesssim$ 4 Ma). 
Estudios basados en fotometr'ia 
'optica dan un valor de $\gamma = -3.0$ en lugar del valor $\gamma = -2.2$
obtenido usando espectroscop'ia \citep{Garmetal82,HumpMcel84,Krou02}.

Las estrellas de masa intermedia tiene una vida en la secuencia principal 
del mismo orden que la edad del disco de la Galaxia. La determinaci'on de
la IMF en este rango de masas es sensible a la historia de formaci'on estelar
en la cercan'ia del Sol  as'i como de la edad y estructura del disco. Ninguno de estos
par'ametros se conoce muy bien y por este motivo es dif'icil determinar
la IMF en este rango. Lo que se hace es suponer que la IMF es una funci'on continua 
y diferenciable entre la parte correspondiente a las estrellas masivas y las de baja masa
 \citep{Scalo86,Krou02}.
 
Las estrellas de muy baja masa  ($< 1.0 \,  M_{\odot}$) tienen una edad promedio de $ \gtrsim 10$ Ga
y representan una mezcla de muchos eventos de formaci'on estelar.  En este caso, la IMF
se puede aproximar mediante una ley de potencias en dos partes con:
 $\gamma = -1.3$ en  $0.08 < M / M_{\odot} < 0.5 $
y $\gamma = -2.3$ en  $0.5 < M / M_{\odot} < 1.0 $.

\subsection{Las Nubes de Magallanes }

La mayor parte de la informaci'on extragal'actica corresponde
a los dos galaxias irregulares sat'elite de la V'ia L'actea: 
 las Nubes de Magallanes. 
Estas galaxias son relativamente peque\~nas (radios de $\sim$ 12 kpc y 8 kpc
para la Mayor y Menor respectivamente), azules,  ricas en gas y pobres en metales
($Z=0.008$  y $Z=0.004$ respectivamente en comparaci'on con
$Z=0.02$  en las cercanias del Sol).  
La formaci'on estelar de las mismas no se concentra el brazos espirales,
sino que est'a distribuida en forma irregular en un n'umero peque~no de
regiones muy activas. 
En las 'ultimas d'ecadas se ha puesto mucho esfuerzo en obtener 
la funci'on inicial  de masa    de las Nubes de Magallanes.
El conteo de estrellas en c'umulos y asociaciones en esas galaxias muestran que
la IMF tiene una pendiente cercana a Salpeter $(\gamma = -2.35) $ 
por encima de  $\sim 1 \, M_{\odot}$  \citep{Scha03}.

\cite{Masse98}  llega a  la conclusi'on de que no hay diferencias
significativas entre los promedios de las pendientes de la IMF halladas en la Galaxia y  las 
dos Nubes de Magallanes.
El promedio pesado de las pendientes de la IMF de la Via L'actea es $\gamma = -2.1\pm0.1$;
y aquella de las Nubes es   $\gamma = -2.3\pm0.1$. Esto demostrar'ia
que la metalicidad no afecta la pendiente de la IMF de estrellas masivas.

Se sabe que algunas estrellas muy masivas no pertenecen a c'umulos  ni asociaciones,
sino que forman parte del campo  \citep{Masse98}.
Estudios de las Nubes de Magallanes muestran que estrellas muy masivas 
pueden hallarse en regiones muy aisladas, y pueden pensarse como el 
resultado de un evento de formaci'on estelar peque\~no.
Un estudio de  la IMF de estos objetos de campo  produjo resultados 
sorprendentes: la pendiente de la IMF es muy empinada, con  $\gamma \sim -5.0$  \citep{Massetal95b}.
Estos valores se encuentran en las Nubes de Magallanes as'i como en
la Galaxia \citep{Bergh04}.
Actualmente se est'a debatiendo si la diferencia en pendientes es debida a 
diferencias entre estilos de formaci'on estelar in situ o al hecho de que 
la mayor'ia de estrellas O del campo sean errantes o fugitivas (runaways)  \citep{deWietal05}.

\subsection{Otras galaxias}

?`Qu'e es lo que sabemos acerca de la IMF estelar por encima de
$\sim 1 \, M_{\odot}$ de galaxias ubicadas m'as all'a de las Nubes de Magallanes?
Hay varios problemas involucrados en la determinaci'on de la IMF en
sistemas extragal'acticos. S'olo algunas  pocas galaxias est'an lo suficientemente cerca
como para que se puedan realizar conteos de estrellas en detalle, e incluso
'estas son tan distantes que s'olo los objetos de la parte m'as brillante de
la funci'on de luminosidad puede ser estudiada.
Adem'as,  hay problemas pr'acticos con el conteo de estrellas en galaxias externas,
incluyendo contaminaci'on por otras fuentes (crowding), la 
incompletitud de los datos y la correcci'on por estrellas  en la misma l'inea de la visual.
Algunas galaxias sobre las que existen trabajos sobre 
la IMF son:  IC~1613, NGC~6822, M~31, M~33, Sextans A, Sextans B y Leo A. 

Entre otros estudios, \cite{Veltetal04} analiz'o la IMF de M~31 y obtuvo 
una pendiente  de  $\gamma = -2.59\pm0.09$; \cite{Jameetal04} 
encontraron  una pendiente de $\gamma  = -2.37 \pm  0.16$ 
en el c'umulo ionizante de NGC~588 en las afueras de la galaxia cercana M~33;
\cite{Annietal03} analizaron la historia de la formaci'on estelar de 
NGC~1705 e infirieron una pendiente cercana a Salpeter.

\section{La IMF de NGC 4214}

\subsection{Descripci'on del m'etodo y elecci'on de la muestra}
La determinaci'on de la IMF de una poblaci'on estelar
con una mezcla de edades es un problema dif'icil de estudiar.
Las masas estelares no pueden ser medidas directamente en la mayor'ia de los casos,
por lo cual deben ser estimadas en forma indirecta midiendo la luminosidad
de las estrellas y su estado evolutivo.

Nosotros empleamos el m'etodo fotom'etrico/espectrosc'opico 
propuesto por \cite{Lequ79}: 
estimamos las magnitudes bolom'etricas ($M_\mathrm{bol}$), y  las 
temperaturas efectivas  ($T_\mathrm{eff}$) de cada estrella en nuestra muestra.
La funci'on de masa presente fue estimada contando el n'umero de estrellas en el 
\hrd ~ entre las trayectorias evolutivas calculadas para modelos de distintas
masas.
En una regi'on de formaci'on estelar  donde las
estrellas se forman simult'aneamente,  la funci'on obtenida
{\em es} la IMF modificada, por supuesto, por la evoluci'on de las estrellas
de mayor masa.
En una regi'on con formaci'on estelar continua, la IMF se obtiene dividiendo 
el n'umero de objetos en cada rango de masas por la edad promedio de
una estrella en la secuencia principal  $(\tau_{SP})$ de la 
masa correspondiente.
 
Para la determinaci'on de la IMF, empleamos las trayectorias evolutivas
de   \cite{LeSc01} en el rango de masas  $5  - 120 \, M_{\odot}$.
Estos  modelos te'oricos proveen 51 valores de  $T_\mathrm{eff}$, $L$ y edad
para cada trayectoria evolutiva de una dada metalicidad.
Los primeros 11 puntos de cada trayectoria definen la 
secci'on que consideramos como secuencia principal.
Una linea que conecta el primer punto en cada trayectoria define la secuencia principal de 
edad cero (ZAMS, del ingl'es: Zero Age Main Sequence). 
Esta l'inea es el l'imite izquierdo de nuestra secuencia principal.
Una l'inea que conecta el punto onceavo de cada trayectoria 
define  el limite derecho de la secuencia principal.

En este trabajo presentamos un estudio detallado de la IMF de NGC~4214 
usando dos hip'otesis: (i) s'olo seleccionamos estrellas de nuestra muestra
que est'an inclu'idas en la secuencia principal; (ii) suponemos que la IMF ha permanecido constante
como funci'on del tiempo.

La  Figura~\ref{fig027}  muestra tres regiones de emisi'on intensa en H$\alpha$, que pueden asociarse
con regiones de formaci'on estelar intensa  reciente.
Estas regiones son I--A, I--B y II \citep{MacKetal00}.

En   la  {\sc\footnotesize  LISTA336}  hallamos un total de   3061  objetos 
ubicados en la secuencia principal.
Consideramos tres listas de masas para calcular la IMF:

\begin{description}
\item[{\sc\footnotesize  LISTA1}:] una lista con todos los objetos. 
\item[{\sc\footnotesize  LISTA2}:] una lista con los objetos de las regiones de formaci'on
estelar intensa (I--A, I--B y II.)
\item[{\sc\footnotesize  LISTA3}:] una lista con todos los objetos excluidos aquellos
de las regiones de formaci'on estelar intensa.

\end{description}

\begin{figure*}[hb!]
 \begin{minipage}[c]{10cm}
\centering
 \includegraphics*[width=0.99\linewidth]{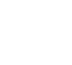}
\end{minipage}
\begin{minipage}[c]{6cm}
 \caption[Distribuci'on de masas estelares
 en un \hrd ]{   {\sl\footnotesize Este \hrd  representa la 
 distribuci'on de masas  estelares a partir de
 la interpolaci'on realizada sobre una famila de trayectorias evolutivas
 de \cite{LeSc01}.       \label{fig029}}}
\end{minipage}
\end{figure*}

Comenzamos determinando las masas de todas las estrellas y sus errores
usando los valores estimados de \mbol ~y \teff ~obtenidos con \cho.
Graficamos un \hrd  ~[$\log$($T_\mathrm{eff}$), \mbol ]  sobre el que superpusimos
las trayectoras evolutivas obtenidas de  \cite{LeSc01}  en el rango de masas
$5-120 \, M_{\odot}$. Para hacer esto, consideramos solamente la parte de las
trayectorias que corresponde a la secuencia principal, con lo cual hemos podido 
descartar aquellas partes donde las trayectorias se superponen y tienen
formas complicadas (loops). El efecto de la rotaci'on 
estelar  sobre las trayectorias  no fue considerado, pero como se explica m'as adelante,
nuestros resultados no dependen de los modelos evolutivos empleados.  
La Figura~\ref{fig029} muestra c'omo 
aislamos las regiones correspondientes a la secuencia principal
y obtuvimos valores interpolados de los valores
de la masa entre las trayectorias usando {\sc\footnotesize  SPLINES}. 
Desarrollamos un c'odigo IDL
que triangula la matriz completa e interpola los datos para obtener 
una matriz fina de masas que cubre todo el  diagrama de Hertzsprung-Russell.

Este procedimiento nos dio una funci'on de dos variables
$M = f(T_\mathrm{eff}$, \mbol) que asigna un valor de la masa $(M)$ 
a cada par de valores de entrada \mbol ~y  $T_\mathrm{eff}$.
Las masas de las estrellas individuales 
$(M_i)$ se obtuvieron directamente a partir de esa funci'on.
Para obtener las incertezas en las masas  $(\sigma_{M_i})$
creamos  10\,000 puntos  distribuidos en forma aleatoria alrededor de los valores promedio
de  \teff~   y \mbol~
de cada estrella e inferimos sus masas individuales.
Estos puntos fueron generados de acuerdo a una  gaussiana bidimensional
y empleando los valores de salida de {\sc\footnotesize  CHORIZOS}.
La desviaci'on est'andar de estas masas nos dio un valor 
confiable del error $(\sigma_{M_i})$ en la masa $(M_i)$  de cada estrella.

?`Qu'e tan dependientes son nuestros 
resultados de los modelos evolutivos empleados?
Encontramos que la posici'on de una trayectoria evolutiva de una
dada masa en el   diagrama de Hertzsprung-Russell, cambia con condiciones iniciales
como la metalicidad, la p'erdida de masa, la rotaci'on estelar, etc. 
La mayor'ia de ellos  tiene una estructura (loops) que les dan una forma muy complicada.
Sin embargo, estos cambios son peque\~nos y pueden ser despreciados
cuando consideramos las incertezas en la determinaci'on de las masas  de los
objetos en nuestras listas, especialmente pues estamos considerando
objetos que se encuentran entre los l'imites de la secuencia principal donde esos
cambios son menos significativos.
La elipse de covarianza de cada objeto
cubre generalmente un amplio rango de trayectorias evolutivas (debido 
a los errores intr'insecos en la determinaci'on de \teff\ y \mbol\ a partir de \cho.)
Esto no cambiar'ia si hubi'esemos usado un grupo diferente de modelos
te'oricos.
Es importante notar que los objetos m'as masivos de nuestras listas tienen errores 
relativos mas 
peque\~nos que los objetos de menor masa; ver por ejemplo la  Figura~\ref{fig001}.

?`C'omo afecta esto la determinaci'on del error individual de cada masa?
Aunque los objetos m'as masivos tiene errores m'as peque\~nos en \teff\ y \mbol\,
sus errores en masa pueden ser grandes debido a que sus elipses de covarinza
se extienden sobre un rango de masas muy amplio.
Por otro lado, los objetos menos masivos tienen errores mas grandes
en \teff\ y \mbol\, pero su ubicaci'on en el \hrd\ es tal que sus elipses cubren un rango de masas
mas peque\~no.

Nosotros tratamos a las regiones I--A, I--B y II  como regiones j'ovenes de diferente edad
pero donde podemos  asegurar que la funci'on de masa presente  $es$ la 
IMF (donde no estamos teniendo en cuenta la posible evoluci'on de
las estrellas m'as masivas).
El resto de la galaxia est'a compuesta de un gran n'umero de  regiones de formaci'on estelar
de  diferentes edades, y una buena aproximaci'on ser'ia  considerarla como una regi'on de
formaci'on estelar continua.
En este caso, corregimos a las cuentas de estrellas teniendo en cuenta
la edad promedio de las estrellas en la secuencia principal para cada masa, empleando
los valores provistos en la  Tabla~\ref{tbl011}.

\begin{table}[htbp]
\centering
\begin{minipage}[u]{15cm}
\caption[Valores de la completitud 
y de la  edad sobre la secuencia principal]{Valores de completitud 
calculados para cada trayectoria evolutiva
de   \cite{LeSc01} y su correspondiente edad sobre la secuencia principal ($\tau_{SP}$)
definida como en el texto.
 \label{tbl011} \vspace{5mm}}
 \end{minipage}

\begin{tabular}{ lcccccccccc  } 
\multicolumn{11}{c}{ \rule{160mm}{0.8mm}}      \\

Trayectoria    ($M_{\odot}$) {\rule [-2mm]{0mm}{7mm} }
  &  7    &  10 & 12 & 15 & 20 & 25 & 40 & 60 & 85 & 120     \\ 
  \multicolumn{11}{c}{ \rule{160mm}{0.2mm}}      \\

  Completitud     
   &  0.04   &  0.26 & 0.50 & 0.71 & 0.86 & 0.90 & 0.93 & 0.95 & 0.95 & 1.00   \\  
   $\tau_{SP}$    (Ma)  
   &  46.50  &  23.98 & 17.84 & 12.95 & 9.12 & 7.17 & 4.82 & 3.71 & 3.48 & 3.01      \\ 
   \multicolumn{11}{c}{ \rule{160mm}{0.8mm}}      \\

\end{tabular}
\end{table}

\subsection{Correcci'on de los datos por efectos sistem'aticos}

Para determinar la IMF de nuestra muestra de estrellas, estudiamos cuatro fuentes
de errores sistem'aticos, como fue presentado por  \cite{Maizetal05}: 
(i) Incompletitud de los datos, (ii) selecci'on apropiada del tama~no del bin,
(iii) an'alisis de objetos m'ultiples y (iv) difusi'on de masas.

\subsubsection{Incompletitud de los datos}
 
Al derivar una IMF, es necesario hacer una evaluaci'on cuantitativa de
la incompletitud de los datos.
Para obtener los valores de completitutd a lo largo de la secci'on de cada
trayectoria evolutiva que se corresponde a la 
secuencia principal, calculamos su valor en cada punto que define la \sp 
usando los resultados de la Secci'on~\ref{sec:compl}; adem'as,  calculamos un peso
como la diferencia entre la edad del punto sobre la trayectoria evolutiva, y la edad
del punto correspondiente en la trayectoria inmediatamente m'as masiva disponible. 
El valor de completitud adoptado es el promedio pesado en cada intervalo.
Los valores se resumen en la  Tabla~\ref{tbl011}.
Tambien determinamos la edad promedio en la \sp para cada trayectoria
disponible interpolando entre las edades que proveen los modelos.
La  Tabla~\ref{tbl011} da estos valores de edad promedio sobre la  secuencia principal.

\subsubsection{Elecci'on apropiada del tama\~no del bin}

Para calcular la pendiente de la IMF empleamos la t'ecnica
sugerida por \cite{MaizUbed05}  
y usamos un ajuste por cuadrados m'inimos pesados 
para ajustar una ley de potencias de la forma  $\frac{dN}{dm}  = A \cdot m^{\gamma} $
para la cual usamos 5 bines de tama\~no variable, de manera tal
que el n'umero de estrellas en cada bin fuera aproximadamente constante.
Este m'etodo garantiza que los errores producidos
por la elecci'on del tama~no de los bines sea m'inimo.
En cada bin usamos un peso $w_i$ derivado a partir de
un  distribuci'on binomial como fue sugerido por   \cite{MaizUbed05}.
Este art'iculo forma parte de este trabajo de Tesis y sus resultados 
han sido incluidos en  el Ap'endice~A donde se discute
la elecci'on apropiada del  tama\~no del bin     en forma completa.

\subsubsection{An'alisis de objetos m'ultiples}

En sistemas lejanos como las Nubes de Magallanes  o NGC~4214, encontramos un serio problema:
la presencia de objetos no resueltos en las  im'agenes  hacen que la determinaci'on
de la IMF sea complicada.
A la distancia de NGC~4214 (2.94 Mpc), 1 segundo de arco corresponde
aproximadamente a 14 pc. Por esta raz'on esperamos que algunas regiones de la galaxia 
contengan objetos no resueltos. Estos objetos m'ultiples pueden 
surgir por varias razones: o son sistemas f'isicos o bien aparecen m'ultiples
por la forma en que estan orientados respecto al observador.

La confiabilidad de las estrellas de mayor masa es cuestionable debido 
a efectos estoc'asticos por estad'istica de numeros peque\~nos  y a efectos
evolutivos. Sin duda los objetos de mayor masa han evolucionado y algunos de ellos
pueden estar fuera de la secuencia principal. 

El efecto producido por objetos m'ultiples en una muestra de estrellas
para determinar la IMF ha sido analizado por varios autores.
\cite{SaRi91}  estudiaron el efecto de binarias
en el rango de masas  $2-14 \, M_{\odot}$  realizando algunos experimentos 
Monte--Carlo usando valores iniciales de  $\gamma = -3.5$,$ -2.5$, y $ -1.5$
y considerando diferentes fracciones de binarias.
Llegan a la conclusi'on de que el efecto principal es un aplanamiento de la funci'on de masa.
Cuando consideraron la pendiente  $\gamma = -3.5$ encontraron muy poca variaci'on
en la pendiente de la IMF; cuando consideraron  $\gamma = -2.5$ demostraron
que  la pendiente
sufre un aplanamiento de $0.34$ si la fracci'on de binarias alcanza el valor 50\%.
En todos los casos, estos autores consideraron la hip'otesis de que las masas de las estrellas
est'an distribuidas en forma aleatoria entre las componentes.
Obviamente el efecto ser'a mas pronunciado en el dominio de estrellas de baja masa,
debido a que una compa\~nera d'ebil de una estrella brillante no altera 
considerablemente la luminosidad total
del objeto compuesto. 

\cite{Krou91}  demostraron que las estrellas binarias no resueltas afectan 
significativamante la determinaci'on de la funci'on de luminosidad.
Su investigaci'on  establece que cuando no se considera el efecto de binarias
existe una subestimaci'on del n'umero de estrellas de baja masa, lo cual se
traduce en un aplanamiento de la pendiente de la IMF.
Este efecto debe ser significante para estrellas de alta masa debido a 
que la mayoria de las estrellas OB se observan en sistemas m'ultiples \citep{Masoetal98,Maizetal05}.

El problema que se nos presenta es que la fracci'on de binarias  en nuestra
muestra es desconocida y por lo tanto no es posible corregir por la
presencia de estas. 
Empleamos el valor $M_{up} = 100  \,M_{\odot}   $ como el l'imite superior
de la integraci'on para obtener la pendiente de la IMF. Esta elecci'on nos
permiti'o descartar algunos de los objetos no resueltos dentro de nuestra muestra.
Es importante hacer notar que este efecto es muy dif'icil de corregir en el caso de una
galaxia  tan distante como NGC~4214 pues, a pesar de usar las im'agenes de m'as alta
resoluci'on disponibles, estamos seguros de la presencia de objetos no resueltos.

Una fuente adicional de objetos  m'ultiples son aquellos
que se originan
por la forma en que dos o m'as objetos
no relacionados f'isicamente  est'an orientados respecto al observador.
Cuanto mayor es  la distancia al objeto de estudio, mayor es
la probabilidad de encontrarnos con este tipo
de objetos m'ultiples.

 Nuestro estudio del efecto que producen los sistemas m'ultiples
 nos lleva a concluir que el valor real de la pendiente de la IMF debe ser
 m'as empinado que el valor que obtenemos en nuestro an'alisis.
 
Para el  l'imite inferior de la integraci'on usamos varios valores de masa en el rango \\
$6, 6.5, ..., 25  \, M_{\odot}   $. 
El mayor problema relacionado con el l'imite inferior de la masa
es la incompletitud de los datos, lo cual tuvimos en cuenta
al usar los  valores de la  Tabla~\ref{tbl011}  
como se explic'o mas arriba. Los valores de la pendiente de la IMF
que presentamos fueron calculados   para  $M_{low} = 20 \, M_{\odot} $ 
donde los valores de completitud son mayores que el 85 \%.

\subsubsection{Difusi'on de masa}

Existe un efecto muy sutil que debemos tener en cuenta al estimar la
pendiente de la IMF de una poblaci'on estelar dada:
Supongamos que observamos varias veces una estrella de magnitud real $m$.
Debido al ruido de Poisson, al ruido del detector y al ruido de fondo, observaremos 
una magnitud diferente $m'$.
Si nuestro detector est'a bien calibrado, esperamos que 
 una 'unica medici'on de su magnitud nos dar'a un error $\sigma_m$
de manera tal que $m' \pm \sigma_m$ responde a una distribuci'on Gaussiana.
Ahora bien, en el rango de masas que estamos estudiando la IMF tiene
una pendiente negativa, lo cual significa que existen m'as estrellas de baja masa
que estrellas de alta masa. Esto implica que si uno mide una estrella de magnitud $m'$
debe haber una mayor probabilidad de que su magnitud real $m$ sea m'as d'ebil
que m'as brillante.
\cite{Maizetal05}  denomina a este efecto sobre los datos
``difusi'on de masa''  pues act'ua de manera an'aloga a un proceso de difusi'on:
suavizando un gradiente de masas en este caso, al desplazar objetos de un lugar donde
son m'as abundantes a otro donde son menos abundantes.

\begin{figure*}[ht!]
 \centering
  \begin{minipage}[c]{78mm}
 \includegraphics*[width=\linewidth]{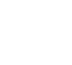}
  \caption[Errores en las masas]{  {\sl\footnotesize 
  Gr'afico de $\sigma_{M}(M)$ obtenido a partir de los valores
 promedio de masa y error en masa calculados en 44 intervalos
 de masa. Los puntos azules representan los valores medidos y la curva roja
 representa el polinomio de tercer grado que mejor se ajusta
 a los datos.    \label{fig030}}}
 \end{minipage}
 \hfill
  \begin{minipage}[c]{78mm}
  \includegraphics[width=\linewidth]{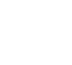}
 \caption[Relaci'on entre $\gamma_{fit} $  y  $\gamma_{real} $]{ {\sl\footnotesize 
   Polinomio que relaciona a los 
 valores   observados  $\gamma_{fit} $  con los valores
 reales de un ajuste  $\gamma_{real} $. Notar que la
 correcci'on es menor cuanto menor es  $\gamma_{fit} $  en
 valor absoluto. El ejemplo presentado
 corresponde a   $M_{low} = 20 \, M_{\sun}$ y
  $M_{up} = 100 \, M_{\sun}$. \label{fig031}}}
 \end{minipage}
\end{figure*}

Para realizar la correcci'on por este efecto, procedimos de la siguiente manera:
Primero, calculamos la funci'on que relaciona a la masa de
una estrella con el error en la masa  $\sigma_{M}(M)$. Para eso, dividimos al rango de
masas en 44 intervalos y en cada uno determinamos
el promedio de las  masas individuales de las estrellas  $(M_i )$
y de sus errores $( \sigma_{M_i})$.
Luego,  
ajustamos un polinomio
de tercer grado a los promedios. El resultado se 
muestra en la Figura~\ref{fig030}.

Luego, generamos 50 listas de  masas aleatorias en el rango $  10 - 200  \, M_{\odot}$ 
usando diferentes valores de la pendiente $\gamma_{real}$ en el rango 
 desde $-1.2 $ a $-4.0$ con paso $0.1$.
A estas masas aleatorias les asignamos errores aleatorios para suavizar
los valores usando la funci'on  $\sigma_{M}(M)$, y simulamos la incompletitud de los datos 
con los valores de la  Tabla~\ref{tbl011}.
Procedimos luego a ajustar cada una de estas listas artificiales con una ley de potencias
de la forma  $\frac{dN}{dm}  = A \cdot m^{\gamma} $
usando 5 bines de tama\~no variable de manera tal que
el n'umero de objetos en cada bin es aproximadamente constante. Para la 
integraci'on, empleamos los mismos l'imites que usamos con los datos reales.
Este proceso nos permiti'o  ajustar un poliniomio a  $\gamma_{fit} - \gamma_{real}$
como funci'on de $\gamma_{fit}$  que es la correcci'on que uno necesita aplicar 
para obtener   $ \gamma_{real}$  dado el valor estimado $\gamma_{fit} $.
 La Figura~\ref{fig031} muestra un ejemplo de un polinomio de ajuste 
que empleamos para corregir  $\gamma_{fit} $  cuando  $M_{low} = 20 \,M_{\sun}$.
Es importante notar que la correcci'on que proveen los polinomios es 
una funci'on decreciente de $M_{low}$. Esta correcci'on es mayor para los menores valores de $M_{low}$
donde  la incompletitud de los datos juega un papel m'as fuerte.

\subsection{An'alisis de los resultados}

Los resultados del  proceso completo de ajuste  de la pendiente de la IMF mediante una ley 
de potencias 
se muestran en las Tablas~\ref{tbl012}  y ~\ref{tbl013}.
 Para cada 
conjunto de estrellas, damos los valores calculados de 
$\gamma_{fit}$ y $ \gamma_{real}$
que obtuvimos para varios  valores del l'imite inferior de integraci'on $M_{low}$. 
Tambi'en presentamos la incerteza 
en $\gamma_{real}$ derivada a partir de ajuste $\chi ^2$
y el n'umero  $N$ de estrellas empleado en cada ajuste.

\subsubsection{Formaci'on estelar continua}

La Tabla~\ref{tbl012} da los resultados obtenidos cuando ajustamos un escenario de
formaci'on estelar continua. Presentamos los valores
calculados para las tres listas de objetos {\sc\footnotesize  LISTA1,2,3}, variando
el l'imite inferior de la integraci'on $M_{low}$ entre 15 y 20 $M_{\odot}$. En todos
los casos  $M_{up} = 100 \, M_{\odot}$.

\begin{table*}[ht!] 
\centering
\begin{minipage}[u]{140mm}
\centering
\caption[Pendientes  de la IMF calculadas 
 en el caso de  formaci'on estelar continua]{Pendiente de la 
IMF obtenida para las tres listas {\sc\footnotesize  LISTA1,2,3}.
Los errores en  $\gamma_{real}$ se calcularon a partir  del ajuste.
Se presentan los valores calculados considerando seis l'imites inferiores
de la integraci'on $M_{low}$ en masas solares. Estos resultados se obtienen suponiendo una
formaci'on estelar continua.  \label{tbl012}}
\end{minipage}

\begin{tabular}{ lccccccccccc}
\multicolumn{12}{c}{ \rule{171mm}{0.8mm}}      \\
& \multicolumn{3}{c}{ I--A +  I--B +  II {\sc\footnotesize  (LISTA2)} }&   
& \multicolumn{3}{c}{Resto  {\sc\footnotesize  (LISTA3)}}&   
& \multicolumn{3}{c}{NGC~4214 {\sc\footnotesize  (LISTA1)}}  \\
  &    \multicolumn{3}{c}{ \rule{45mm}{0.2mm}} &
  &    \multicolumn{3}{c}{ \rule{45mm}{0.2mm}} &
  &    \multicolumn{3}{c}{ \rule{45mm}{0.2mm}}  \\ 

$M_{low}  $    & $\gamma_{fit}$ & $\gamma_{real}$  &$N$  & &
$\gamma_{fit}$ &  $\gamma_{real}$  &  $N$  & &
 $\gamma_{fit}$ &  $\gamma_{real}$  & $N$    \\ 
 \multicolumn{12}{c}{ \rule{171mm}{0.2mm}}      \\

  15 &    -2.17 &   $ -2.67 \pm     0.08 $&  752 &  &  -2.15 &  $  -2.64 \pm     0.06 $& 1680 &  &  -2.15 & $   -2.64 \pm      0.05  $ & 2432   {\rule [0mm]{0mm}{0mm} }\\
   16 &    -2.20 &  $  -2.61 \pm     0.09 $&  655 &   & -2.28 &  $  -2.72 \pm     0.06 $& 1494 &  &  -2.26 & $   -2.70 \pm      0.05 $  & 2149 {\rule [0mm]{0mm}{0mm} } \\
   17 &    -2.24 &    $-2.60 \pm     0.10 $&  573 &   & -2.43 &  $  -2.83 \pm    0.07 $& 1357 &  &  -2.39 & $   -2.78 \pm         0.06 $  & 1930 {\rule [0mm]{0mm}{0mm} } \\
   18 &    -2.21 &   $ -2.51 \pm     0.10 $&  501 &   & -2.59 &$    -2.95 \pm     0.08 $& 1216 &  &  -2.48 & $   -2.82  \pm        0.06  $ & 1717 {\rule [0mm]{0mm}{0mm} } \\
   19 &    -2.30 &   $ -2.57 \pm     0.11 $&  449 &  &  -2.72 & $   -3.03 \pm     0.08$ & 1098 &  &  -2.56 &$    -2.86 \pm     0.06  $ & 1547 {\rule [0mm]{0mm}{0mm} }  \\
   20 &    -2.36 &   $ -2.60 \pm     0.12$ &  397 &  &  -2.71 &$    -2.98 \pm     0.09 $&  961 &  &  -2.58 & $   -2.83 \pm        0.07$   & 1358  {\rule [0mm]{0mm}{0mm} }  \\ 
   \multicolumn{12}{c}{ \rule{171mm}{0.8mm}}      \\

\end{tabular}

\end{table*}
 
\begin{figure*}[hb!]
 \begin{minipage}[c]{10cm}
\centering
 \includegraphics[width=0.9\linewidth]{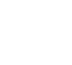}
\end{minipage}
\begin{minipage}[c]{6cm}
 \caption[Ejemplo de ajuste de la IMF ]{ {\sl\footnotesize   Histograma del n'umero de estrellas en funci'on de la masa
 usado para el ajuste  de la IMF en el rango  $20-100 \, M_{\odot}$.  
 Se observa el tama~no variable de los cinco bines. El ajuste 
 da el valor   $\gamma_{fit} =-2.58 \pm 0.07$, 
 donde el error se obtiene a partir del ajuste $\chi ^2$.  \label{fig032}}}
\end{minipage}
\end{figure*}

La Figura~\ref{fig032} muestra un ejemplo del histograma de las cuentas como funci'on de la masa
para la  {\sc\footnotesize  LISTA1}. Este es un  ajuste lineal para  $N=1358$ estrellas
con masas en el rango  $20-100 \, M_{\odot}$. El ajuste da el resultado $\gamma_{fit} =-2.58 $;
que se traduce a $ \gamma_{real} = -2.83$ una vez que se aplica el m'etodo descripto aqu'i.

Adoptamos el valor   $ \gamma_{real} = -2.83 \pm 0.07$  como un valor representativo
de la pendiente de la IMF de  NGC~4214,  el cual fue calculado usando
 $M_{low} = 20 \,  M_{\odot}   $ y $ M_{up} = 100  \, M_{\odot}$.

\subsubsection{Formaci'on estelar en starburst}

El modelo de formaci'on estelar por brote s'olo puede aplicarse
a las estrellas inclu'idas en las regiones   I--A, I--B y II ({\sc\footnotesize  LISTA2}),
debido a que  'estas son las regiones de formaci'on estelar en NGC~4214.

La Tabla~\ref{tbl013} da los resultados obtenidos cuando ajustamos un escenario de
formaci'on estelar en starburst. 

 \begin{wraptable}{l}[1mm]{80mm}
   \begin{minipage}[u]{7cm}
    \topcaption[Pendientes  de la IMF calculadas 
  para las regiones de formaci'on estelar activa]{Pendientes 
  calculadas y reales de la IMF calculadas 
  para las regiones de formaci'on estelar activa, empleando
  diferentes l'imites de integraci'on ($M_{low}$ y $M_{up}$) expresados en masas solares.
   \label{tbl013}}
   \end{minipage}

\begin{tabular}{ lcccc }  
\multicolumn{5}{c}{ \rule{70mm}{0.8mm}}      \\

 &   & \multicolumn{3}{c}{ I--A +  I--B +  II}  \\ 
  &   & \multicolumn{3}{c}{ \rule{45mm}{0.2mm}}  \\ 

 $M_{low} $   & 
  $M_{up}  $  & $\gamma_{fit}$ &  $\gamma_{real}$    & $N$  \vspace{1mm} \\ 
  \multicolumn{5}{c}{ \rule{70mm}{0.2mm}}      \\

   15 & 100 &    -3.30 &$    -4.41 \pm     0.09 $&  752       \\ 
   16 & 100 &    -3.25 &  $  -4.06 \pm     0.10 $&  655       \\
   17 &  100 &   -3.24 & $   -3.87 \pm    0.10 $&  573      \\
   18 & 100 &    -3.14 & $   -3.61 \pm    0.11 $&  501     \\
   19 & 100 &    -3.20 & $   -3.59 \pm     0.12 $&  449      \\
   20 &  100 &   -3.23 & $   -3.56 \pm     0.13 $&  397       \\ 
    40 & 100 &   -2.89 & $   -2.97 \pm     0.37 $&  85      \\
   20 & 40 &   -3.73 & $   -4.41 \pm     0.28 $&  312     \\ 
   \multicolumn{5}{c}{ \rule{70mm}{0.8mm}}      \\
 \end{tabular} 
\end{wraptable}


Cuando uno considera los objetos de la  {\sc\footnotesize  LISTA2}, se enfrentan dos problemas:
(1) la incertidumbre  en la correcci'on por objetos m'ultiples como se describi'o
arriba para la {\sc\footnotesize  LISTA1}, y (2) el hecho de que estas regiones
est'an compuestas por una mezcla de estrellas con edades en el rango 
$0-5$ Ma.
Esto significa que algunas de las estrellas m'as masivas han evolucionado y han desaparecido
en la forma de supernovas vaciando los bines de mayor masa y produciendo
as'i una pendiente mas elevada. 
Analizamos este problema teniendo en cuenta los siguientes rangos de masas 
(en unidades de masas solares):  [ 20,  40],  [ 20,  100] y  [ 40,  100].
Los valores  de la pendiente de la IMF obtenidos son:
 $ \gamma_{real} = -4.41 \pm 0.28$,  $ \gamma_{real} = -3.56 \pm 0.13$
y  $ \gamma_{real} = -2.97 \pm 0.37$  respectivamente.
Estos diferentes valores podr'ian explicarse  sosteniendo que 
la fracci'on de sistemas m'ultiples es mayor en el rango  [ 40,  100]  que en  [ 20,  100].
Esto hace que la pendiente aparezca m'as aplanada que cuando se usa $ M_{up} = 40\,  M_{\odot}$.
En resumen, $ \gamma_{real} = -3.56$ es una cota inferior a la pendiente de la IMF
cuando se estudian las regiones activas de formaci'on estelar.

 
\subsubsection{Comparaci'on de resultados con otros trabajos}
  
De acuerdo a   \cite{Leitetal96}  las regiones starburst j'ovenes presentan espectros
UV similares en un amplio rango de metalicidades y edades. Esto les permiti'o 
modelar el espectro UV de NGC~4214-I empleando la t'ecnica de s'intesis evolutiva de 
\cite{Leitetal95b}. Usaron una biblioteca de espectros de alta resoluci'on de estrellas O y WR
y una biblioteca de espectros de baja resoluci'on de estrellas de tipo  espectral B.
Combinaron esas bibliotecas con un c'odigo de s'intesis evolutiva para computar 
un espectro sint'etico en el rango $1\,200 - 1\,800 $ \AA\  y usaron como
par'ametros de entrada a la tasa de formaci'on estelar, la edad del starburst, la IMF y la 
metalicidad. Entre sus resultados, hallaron que, si la IMF de este starburst sigue una ley
de potencias, entonces el exponente debe estar comprendido entre  $ \gamma = -2.35$ 
y $ \gamma = -3.00$. Esta estimaci'on  es bastante conservadora ya que se presentan
varias incertidumbres asociadas a los ajustes del continuo, la presencia de l'ineas 
interestelares y efectos de la metalicidad. N'otese  que   \cite{Leitetal96}   emplean
una  distancia de 4.1 Mpc a NGC~4214 y una metalicidad de $Z=0.01$.

\cite{MasHKunt99} 
emplean codigos sint'eticos \citep{MasHKunt91a} 
basados en las trayectorias evolutivas provistas por
el grupo de Ginebra \citep{Schaetal93b,Charetal93}.
Sus modelos generan a la IMF usando el m'etodo Monte Carlo con masas
en el rango $2- 120  \, M_{\odot}$ y una ley de potencias de la forma $dN/dm \sim m^{-\alpha}$.
Supusieron tres valores de la pendiente: Salpeter ($\alpha = 2.35$), y dos extremos ($\alpha = 1, 3$).
Emplearon intervalos de tiempo muy cortos (0.05 Ma) para  poder modelar
fases de r'apida evoluci'on estelar. Consideraron dos mecanismos de formaci'on estelar:
un burst instant'aneo y  formaci'on estelar continua.
Los  espectros IUE  empleados  incluyen claramente estrellas en c'umulos as'i
como estrellas de campo. Debe notarse adem'as que para su estudio 
\cite{MasHKunt99}  consideran la secci'on superior de la IMF $(M> 10 \, M_{\odot})$.
La aplicaci'on de este c'odigo a NGC~4214 dio como resultado uno de los  valores 
extremos de la pendiente  ($\alpha =  3$ o su equivalente, $ \gamma = -3$)
La pendiente de la IMF que ellos estiman  coincide con el valor determinado por 
 \cite{Leitetal96}. 

 \cite{Chanetal05}
emplearon un espectro de NGC~4214 obtenido 
con  STIS en el UV lejano con el objeto de  estudiar el contenido estelar
de las regiones localizadas entre c'umulos. Compararon al espectro extra'ido con
los modelos provistos por \s99  y encontraron que las regiones estudiadas  tienen
un espectro dominado por estrellas  de tipo espectral B. Sus observaciones indican que
las regiones entre c'umulos contienen estrellas menos masivas que los c'umulos 
cercanos. 
\cite{Chanetal05}  estimaron el valor  $ \gamma = -3.5$
para las estrellas de campo en NGC~4214   usando  
$M_{low} = 1  \,M_{\odot} $ y 
$M_{up} = 100 \, M_{\odot}$ en un escenario  de formaci'on estelar continua.
Las regiones analizadas incluyen c'umulos estelares,
y por 
lo tanto, deben contener algunas estrellas masivas que no pertenecen al campo.
Esto implica que el valor obtenido de la pendiente representa un l'imite inferior.

Nosotros abordamos el estudio de la funci'on inicial de masa de NGC~4214 en forma diferente,
pues usamos al m'etodo m'as directo y confiable que existe para  obtener la IMF de una dada poblaci'on estelar
y que se basa en  conteos de estrellas como una funci'on de su luminosidad/masa \citep{Lequ79}.
Nuestro estudio detallado y exhaustivo nos permite adoptar 
el valor   $ \gamma_{real} = -2.83 \pm 0.07$  como un valor representativo
de la pendiente de la IMF de  NGC~4214,  el cual fue calculado usando
 $M_{low} = 20  \, M_{\odot}   $ y $ M_{up} = 100  \,  M_{\odot}$ en  un escenario  de formaci'on estelar continua.
Debemos mencionar que el error citado en la pendiente (0.07) es un valor formal que sale de los 
ajustes por cuadrados m'inimos. 
Debido a que la correcci'on por objetos m'ultiples es incierta (dada la gran distancia
a la que se encuentra la galaxia) y sabiendo que los objetos no resueltos producen 
un aplanamiento sistem'atico de la pendiente,
concluimos que el valor real de la pendiente de la IMF 
calculada para la  {\sc\footnotesize  LISTA1} debe ser m'as empinada que $ \gamma= -2.83$.

Resulta interesante comparar los resultados  de la estimaci'on de la pendiente
de la IMF de NGC~4214 mencionados anteriormente
con los de la Nube Mayor de Magallanes que tiene una  metalicidad parecida y su cercan'ia 
permite determinaciones muy confiables de la IMF.
La mayor parte de los estudios de la IMF de estrellas de campo  en le LMC  fueron realizados 
por P. Massey  y sus colaboradores  (ver como ejemplo a  \cite{Massetal95b}).
Se encuentra que la IMF para masas por encima de $\sim 40\,  M_{\odot}$ es claramente m'as
empinada que Salpeter  ($\gamma \sim -5.0$), resultado 
 que se ha encontrado tambi'en  en la Galaxia y en la Nube Menor 
de Magallanes. 
Las determinaciones de la pendiente de la IMF en c'umulos en la LMC muestran
en cambio un valor cercano a Salpeter sobre un amplio rango de densidades estelares, desde
 asociaciones OB  difusas hasta el SSC R136 donde la densidad es similar a la de un c'umulo globular.

Nosotros estimamos que el valor ``real''  de la pendiente de la IMF est'a comprendido entre el valor 
propuesto por   \cite{Chanetal05}  ($\gamma = -3.5$), y el obtenido en este trabajo ($\gamma =  -2.83$).
Esta pendiente est'a de acuerdo con valores estimados para algunas  asociaciones OB 
en la LMC y en la V'ia L'actea \citep{Massetal95,Massetal95b,Masse98} y con los valores
inferidos por los trabajos que ajustan espectros y que fueron mencionados con anterioridad.

\cite{Gregetal98} realizan un estudio completo de la poblaci'on resuelta
de  NGC~1569 que es considerada la galaxia starburst enana m'as cercana
($2.2  \pm 0.6$ Mpc; \cite{IsraBruy88}). Encuentran que sus simulaciones ajustan
la IMF de NGC~1569 con una pendiente m'as empinada que Salpeter   ($\gamma$ entre  $ -2.6$ y $ -3.0$), 
si consideran estrellas de campo menos masivas que $\sim 30 \, M_{\odot}$.
Como ya dijimos, el valor de la pendiente de la IMF de estrellas de campo
es m'as empinada que Salpeter
en la Galaxia y en las Nubes de Magallanes.

Esto parecer'ia estar de acuerdo con nuestras mediciones en NGC~4214. 
En la  Tabla~\ref{tbl012}  damos los resultados obtenidos cuando ajustamos un escenario de
formaci'on estelar continua. Presentamos los valores
calculados para las tres listas de objetos {\sc\footnotesize  LISTA1,2,3} para un rango de $M_{low}$.
La {\sc\footnotesize  LISTA2} contiene objetos en las regiones de formaci'on estelar intensa y es ah'i
donde se obtiene la pendiente m'as aplanada ($\gamma =  -2.60$).
La {\sc\footnotesize  LISTA3} contiene a todos los objetos salvo los 
de las regiones de formaci'on estelar intensa, lo cual implica una mayor concentraci'on de 
objetos de campo (aunque seguramente  existen estrellas de c'umulos). En este caso,
  obtenemos  la pendiente m'as empinada ($\gamma =  -2.98$).
Estos resultados nos est'an diciendo que la mayor'ia de las estrellas masivas se  han formado
en los c'umulos en NGC~4214.
Ser'ia interesante poder extender a nuestro trabajo para incluir a la poblaci'on estelar menos masiva
($ < 20 \, M_{\odot}$)
de NGC~4214.

\chapter{Objetos extendidos  en NGC~4214}
\thispagestyle{empty}
\begin{flushright} 
   {\em   Past the point 
of no return \\
the final threshold, \\
the bridge 
is crossed, so stand \\
and watch it burn  ...  \\
We've passed the point \\
of no return. }

 \end{flushright}
 
\newpage


El estudio de algunos  objetos extendidos en
NGC~4214 se llev'o a cabo mediante el empleo del c'odigo \cho\
aplicado a la fotometr'ia de apertura previamente calculada y que se presenta
en las Tablas~\ref{tbl006}  y ~\ref{tbl014}.
Estudiamos con detalle  13 \cus estelares: 
 NGC~4214--I--As, I--Es, IIIs, IVs, I--A, I--B, I--Ds, II, II--A, II--B, II--C, II--D y II--E
usando los modelos  te'oricos  de s'intesis evolutiva  de \s99 como fue descripto en la Secci'on~\ref{sec:models}.
En este Cap'itulo  analizamos los resultados.

\section{C'umulos no resueltos   NGC~4214--I--As, I--Es, IIIs y IVs}
\subsection{C'umulo  NGC~4214--I--As}

Comenzamos nuestro an'alisis con el \cu  I--As. 
Ejecutamos  \cho\  dejando todos los
par'ametros [$\log({\rm edad/a}) $, $E(4405-5495)$,  $R_{5495}$] libres,
lo cual nos dio valores muy bien definidos de la edad del c'umulo:
 $\log(\rm {edad/a})=6.60     \pm 0.06 $, y un valor muy bajo de la extinci'on 
$E(4405-5495) =  0.07 \pm 0.02$ mag.
\begin{figure*}[b!]
\includegraphics*[width=0.99\linewidth]{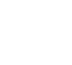}
 \caption[Diagramas de contornos  de c'umulo I--As]{  {\sl\footnotesize [Izquierda] 
 Diagrama de contornos del plano  [$\log({\rm edad})$, \ecc] para el
 c'umulo I--As,  que se obtiene
 de {\sc\footnotesize  STATPLOTS} 
 sin imponer ninguna restricci'on en los par'ametros 
 durante la  ejecuci'on de {\sc\footnotesize  CHORIZOS}. 
[Derecha] Diagrama de contornos del plano  [$\log({\rm edad})$, \rv] en el cual se
observa claramente la degeneraci'on en \rv\ debida a los bajos valores
de \ecc\  determinados por el ajuste. 
Como se explic'o en el Cap'itulo~5, \cho\ incluye 
las leyes de extinci'on Gal'acticas de  \cite{Cardetal89}, 
las leyes promedio   de la Nube Mayor de Magallanes y la ley LMC2
provistas por  \citet{Missetal99}; y por 'ultimo la ley asociada a la
Nube Menor de Magallanes de  \cite{GordClay98}. El valor de \rv\
en \cho\ se define como el valor verdadero  de las leyes Gal'acticas
mientras que para las leyes de las Nubes de Magallanes se asignan los valores
$1.0$, $0.0$ y $-1.0$ para la ley promedio, LMC2 y SMC respectivamente.
\label{fig058}}}
\end{figure*}

El mapa de contornos de [$\log({\rm edad})$, \ecc]  (ver Figura~\ref{fig058}) muestra
claramente que hay una 'unica soluci'on (se observan dos picos
pero uno de ellos es el m'as intenso) compatible con nuestra fotometr'ia.
Sin embargo, el bajo valor del enrojecimiento implica una degeneraci'on en los
valores de \rv\, ya que el efecto del enrojecimiento se hace independiente del tipo
de ley de extinci'on  cuando  \ecc $\ll 1.0$. 
Esto implica que aunque seamos capaces de restringir la edad
y el enrojecimiento del c'umulo, el valor apropiado de \rv\ queda indefinido
usando el conjunto de nuestras magnitudes observadas.

Volvimos a ejecutar el c'odigo con las mismas magnitudes, pero
restringiendo la ley de extinci'on a aquellas de las  nubes de Magallanes
(LMC2 y SMC como se explica en el Cap'itulo~\ref{cha:ext}) 
y obtuvimos valores comparables en ambos
casos, como se muestra en 
la  Tabla~\ref{tbl019}.   
En este problema estamos ajustando dos par'ametros
y usando cuatro colores (derivados a partir de cinco magnitudes)
lo cual implica que el problema tiene dos grados de libertad.
El mejor ajuste es el de la Nube Menor  con la ley SMC, de acuerdo al
valor de $\chi^2$.


 \begin{table*}[ht!] 
\centering
\begin{minipage}[u]{15cm}
\caption[Resultados de los c'umulos  I--As y  I--Es]{Resultados del ajuste de   \cho\  para los c'umulos  I--As y  I--Es. El 
n'umero de estrellas es un resultado provisto por \s99.
\label{tbl019} }
\end{minipage}

 \begin{tabular}{ llllll }
 \multicolumn{6}{l}{ \rule{150mm}{0.8mm}}      \\
 \multicolumn{1}{l}{Par\'ametro}     
&  \multicolumn{2}{c}{C\'umulo I--As} &     
&  \multicolumn{2}{c}{C\'umulo I--Es}  \\ 

  &    \multicolumn{2}{c}{ \rule{40mm}{0.2mm}}  
  & & \multicolumn{2}{c}{ \rule{40mm}{0.2mm}}  \\ 

& \multicolumn{1}{c}{LMC2}    
&  \multicolumn{1}{c}{SMC}    
& 
&  \multicolumn{1}{c}{Joven}    
& \multicolumn{1}{c}{Viejo}      \\ 
   \multicolumn{6}{l}{ \rule{150mm}{0.2mm}}      \\

Edad (Ma)  &      $4.0 \pm 0.6$ &  $4.2 \pm 0.6$ &  &   $ 7.1\pm0.1$    & $189\pm53$   
               \\ 
\ecc ~(mag)  &   $0.05 \pm 0.01$    &    $0.04 \pm 0.01$  &   &  $0.46\pm0.01$   & $0.32\pm0.03$   
              \\ 
Ley de extinci'on   / \rv  &   LMC2    &  SMC    & &     $4.21\pm0.25$  & $5.72\pm0.26$   
              \\ 
Masa ($10^3 M_{\odot} $)   & $ 27\pm   4  $ & $ 27 \pm   4  $ & &    $6\pm1$     &  $129\pm26$   
                              \\ 
$\chi^2  $por grado de libertad   &    1.33   &   0.91  &  &  2.58 &  2.19    \\ 
Estrellas O+B, de tipos  I y II  &     11  &  11 &   &2 &  0   \\ 
Estrellas K+M, de tipos  I y II &   0    & 0  &  &  1 & 9    \\ 
Estrellas WR    & 1     & 1  & &   0 &  0      \\  
 \multicolumn{6}{l}{ \rule{150mm}{0.8mm}}      \\

 \end{tabular}
\end{table*}

\begin{figure*}[h]
\centering
\includegraphics*[width=0.99\linewidth]{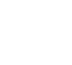}
 \caption[Espectros te'oricos 
  del c\'umulo I--As]{ {\sl\footnotesize Espectro te'orico 
 que mejor se ajusta a las
 magnitudes medidas del c\'umulo I--As obtenido empleando
  {\sc\footnotesize  STATPLOTS} 
 con las leyes de extinci'on de la Nube Mayor de Magallanes   (izquierda)
 y la de la Nube Menor (derecha).  
\label{fig060}}}
\end{figure*}

\begin{figure*}[h]
\centering
\includegraphics*[width=0.99\linewidth]{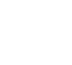}
 \caption[Diagrama de contornos del   c'umulo I--As]{  {\sl\footnotesize Diagrama 
 de contornos del plano  [$\log({\rm edad})$, \ecc] para el
 c'umulo I--As,  que se obtiene
 de {\sc\footnotesize  STATPLOTS} 
 usando las leyes de extinci'on de la Nube Mayor de Magallanes   (izquierda)
 y la de la Nube Menor (derecha).  
\label{fig059}}}
\end{figure*}

La Figura~\ref{fig060} muestra los modelos te'oricos  
que mejor se ajustan a los datos usando las leyes de la SMC y LMC2.
En la Figura~\ref{fig059}   mostramos los diagramas de contornos  [$\log({\rm edad})$, \ecc]
usando las leyes de la SMC y LMC2. Ambos diagramas muestran que la edad de este
\cu est'a en el rango   $6.50 \le   \log(\rm {edad/a}) \le  6.70 $.

En el Ap'endice~B se explica en forma detallada el m'etodo empleado para
la determinaci'on de la masa de los c'umulos. En particular, 
  la masa a edad cero del c'umulo I--As es   $\approx 27\,000 M_{\odot}.$ 
N'otese que en todo este trabajo suponemos que la masa m'as aproximada 
es la que se obtiene con una IMF Kroupa calculada con el rango de masas
 $M_{low} =0.1 \, M_{\odot} $   y    $M_{up} =100 \,  M_{\odot}$.

Una vez que la masa aproximada del \cu se conoce, usamos
\s99 para inferir el contenido estelar de un \cu con esta masa, edad y metalicidad.
Nos interesa principalmente el contenido de supergigantes tempranas (estrellas
  de tipo espectral O$+$B y clase de luminosidad I y II) y tard'ias 
   (estrellas
  de tipo espectral K$+$M y clase de luminosidad I y II).
  Tambi'en consideramos las estrellas de tipo WR.
Los resultados de los modelos evolutivos te'oricos
est'an en la Tabla~\ref{tbl019}. Se sabe que el  \cu  I--As
es un SSC con un n'ucleo compacto y un halo 
masivo muy intenso, con una estructura muy similar a la de 30 Doradus en la LMC,
como fue mostrado por \cite{Maiz01b} y por  \cite{Leitetal96}. Esto implica que los efectos de apertura son 
importantes en el momento de realizar mediciones fotom'etricas
de este \cu y seguramente las cantidades derivadas como la masa y la poblaci'on estelar
van a depender de la apertura elegida.

Usando una ley de extinci'on de la SMC, nuestros modelos dan una edad de $4.0 \pm 0.6$ Ma
para el \cu  I--As  a partir de nuestra fotometr'ia 'optica--UV.
Este valor est'a de acuerdo con los obtenidos por 
 \citet{Leitetal96} usando espectroscop'ia UV  (4--5 Ma) y
por \citet{Maizetal98} usando el ancho equivalente $W$(H$\beta$)  
y la intensidad de la banda  WR $\lambda$4686 \AA\ (3--4 Ma).
Los modelos que mejor ajustan nuestros datos dan el valor\footnote{Notar que el valor
de  \ecc ~es levemente diferente de aquel obtenido sin restringir la 
ley de extinci'on, pero ambos valores est'an separados por solo
 $\approx 1 \sigma.$} $A_V = 0.19  \pm 0.06$ mag, 
y  
$E(4405-5495) =  0.04  \pm 0.01 $ mag.   Esta baja extinci'on era de esperar,
ya que I--As est'a localizado dentro de una cavidad con forma de coraz'on creada
por la introducci'on de  energ'ia cin'etica y momentum provistos
por el \cu dentro del medio interestelar \citep{Maizetal99a,MacKetal00} y por tal raz'on 
uno hubiese esperado poco gas y polvo en la  linea de la visual.
El alto n'umero de estrellas masivas formadas en esta regi'on, han 
hecho desaparecer las part'iculas de polvo de su entorno, 
dejando un paso libre a  trav'es del cual podemos ver el continuo estelar.
Los efectos combinados de vientos estelares y supernovas han creado
una burbuja, cuyo interior est'a casi limpio de gas, una situaci'on
semejante a la que se encuentra en la asociaci'on OB~78 en M~31  
\citep{Leitetal96}.
La baja extinci'on observada est'a de acuerdo con el valor medido en
esta parte de la galaxia por  \citet{Maizetal98}  a partir del cociente
nebular de \ha\  sobre \hb\ como ya vimos en el Cap'itulo~\ref{cha:ext}.

\begin{figure*}[hb!]
 \begin{minipage}[c]{10cm}
\centering
\includegraphics*[width=0.99\linewidth]{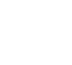}
\end{minipage}
\begin{minipage}[c]{6cm}
 \caption[Espectro del c'umulo I--As 
 obtenido por  \cite{SargFili91}]{  {\sl\footnotesize Espectro del c'umulo I--As  (Knot 1 en su notaci'on)
 obtenido por  \cite{SargFili91}
 con el telescopio de 5 m de Monte Palomar.  Se nota claramente la intensa l'inea
 de emisi'on atribuida a estrellas WR.
\label{fig061}}}
\end{minipage}
\end{figure*}

Varios autores 
\citep{Leitetal96,SargFili91,MasHKunt91a,Maizetal98} indican  la presencia 
de estrellas WR dentro del \cu   I--As a partir de la 
detecci'on de lineas de emisi'on caracter'isticas de estos objetos en los espectros.
\cite{Leitetal96}  reexamin'o los resultados de  \cite{SargFili91} que presentamos en 
la Figura~\ref{fig061}.


Sus resultados fueron calculados a partir de espectros UV de  I--As  
obtenidos con el instrumento  
Faint Object Spectrograph a bordo del HST en 1993. Usaron una apertura
circular de  $1  \arcs 0  $  y supusieron una distancia de 
4.1 Mpc a NGC~4214.
Estimaron que en esta regi'on hay aproximadamente 15  estrellas WN y 15 estrellas WC  
presentes, teniendo en cuenta incertidumbres  en un factor de dos.
Estos resultados  corresponden a 5 estrellas WR empleando la distancia 
mejorada de  2.94 Mpc.
\cite{Maizetal98}  estudiaron cuatro regiones con estrellas WR en  NGC~4214, 
incluyendo   I--As. 
Nosotros empleamos los valores observados de anchos equivalentes 
de las lineas de emision caracter'istica de las WR medidos por \cite{Maizetal98}
y estimamos la presencia de 60 estrellas WR dentro de un radio de 
$1  \arcs 8 $ a una distancia de 4.1 Mpc, lo cual se corresponde  a 3 estrellas 
WR usando nuestra apertura y distancia.
Los resultados de \cho\ combinados con los de \s99\ predicen 
la presencia de 4  WR usando la metalicidad en las cercan'ias 
del Sol $(Z=0.02)$, y 1 WR usando  $(Z=0.004)$,
lo cual es consistente con las observaciones anteriores.
Es importante notar que las estrellas WR pueden estar presentes tanto 
en el n'ucleo del c'umulo como en el halo masivo. 
Esto implica que los efectos de apertura son muy importantes al estudiar este c'umulo
y por tal raz'on se hace dif'icil comparar  trabajos realizados con
diferentes m'etodos. A pesar de esto, nuestros resultados 
dan valores similares a los publicados previamente dentro de errores de Poisson y estamos 
de alguna manera confirmando la presencia de estrellas WR en el c'umulo I--As
empleando un m'etodo  que se basa en mediciones estrictamente fotom'etricas.

\subsection{C'umulo   NGC~4214--I--Es}

\begin{figure*}[hc!]
\begin{minipage}[c]{8cm}
\centering
\includegraphics*[width=\linewidth]{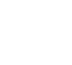}
\end{minipage}
\begin{minipage}[c]{8cm}
 \caption[ Diagrama de contornos   para el
 c'umulo I--Es]{  {\sl\footnotesize   Diagrama de
  contornos del plano  [$\log({\rm edad})$, \ecc] para el
 c'umulo I--Es,  que se obtiene
 de {\sc\footnotesize  STATPLOTS} 
 sin imponer ninguna restricci'on en los par'ametros 
 durante la  ejecuci'on de {\sc\footnotesize  CHORIZOS}. Se observan dos
 soluciones compatibles con la fotometr'ia. 
\label{fig062}}}
\end{minipage}
\end{figure*}

Usando las siete magnitudes medidas (seis colores) del \cu 
I--Es  listadas en  la Tabla~\ref{tbl006} y dejando todos los
par'ametros [$\log({\rm edad)},  R_{5495}$ y  \ecc]  libres,
ejecutamos \cho\ y {\sc\footnotesize  STATPLOTS}.
El mapa de  [$\log({\rm edad})$, \ecc] (ver Figura~\ref{fig062}) muestra dos soluciones: una joven
en $\log({\rm edad/a}) \approx 6.85$, y una vieja en  $\log({\rm edad/a}) \approx 8.25$.

\begin{figure*}[hc!]
\centering
\includegraphics*[width=0.99\linewidth]{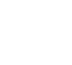}
 \caption[Salida gr'afica de {\sc\footnotesize  STATPLOTS}:
 soluci'on vieja  del  c'umulo I--Es]{ {\sl\footnotesize Salida 
 gr'afica de {\sc\footnotesize  STATPLOTS}  correspondiente
 a la soluci'on vieja (edad  189  Ma)  del  c'umulo I--Es. Se presentan el espectro te'orico que mejor
 ajusta a las magnitudes observadas (arriba izquierda) y los contornos de probabilidad
  [$\log({\rm edad})$, \ecc]  (arriba derecha),  [  \ecc, \rv] (abajo izquierda) y
   [$\log({\rm edad})$, \rv] (abajo derecha).  
\label{fig063}}}
\end{figure*}

\begin{figure*}[hc!]
\centering
\includegraphics*[width=0.99\linewidth]{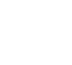}
 \caption[Salida gr'afica de {\sc\footnotesize  STATPLOTS}: 
 soluci'on joven  del  c'umulo I--Es]{ {\sl\footnotesize  
  Salida gr'afica de {\sc\footnotesize  STATPLOTS}  correspondiente
 a la soluci'on joven (edad  7  Ma)  del  c'umulo I--Es. Se presentan el espectro te'orico que mejor
 ajusta a las magnitudes observadas (arriba izquierda) y los contornos de probabilidad
  [$\log({\rm edad})$, \ecc]  (arriba derecha),  [  \ecc, \rv] (abajo izquierda) y
   [$\log({\rm edad})$, \rv] (abajo derecha). 
\label{fig064}}}
\end{figure*}

Volvimos a ejecutar el c'odigo para aislar estas dos soluciones, y mostramos las 
propiedades de ambas     en la Tabla~\ref{tbl019}.
La soluci'on vieja  (edad  $189\pm53$  Ma) es la de mayor probabilidad, pero
eso no implica que debamos 
 descartar a la soluci'on joven inmediatamente.
Las Figuras~\ref{fig063} y \ref{fig064}  muestran los mapas de contornos  
producidos por  {\sc\footnotesize  STATPLOTS}   para
este c'umulo en ambas soluciones. 
La soluci'on m'as vieja produce un valor   alto de \rv\ como se ve en la  Figura~\ref{fig063}.
Tambi'en presentamos los espectros que mejor
se ajustan  a nuestras observaciones.
Ambas soluciones muestran que el \cu  I--Es  es m'as viejo que el  I--As 
lo cual est'a de acuerdo con \cite{MacKetal00}  quienes indican que los colores
de I--Es   son significativamente m'as rojos que los de I--As, lo cual
indica una edad m'as avanzada.  

Luego de calcular la masa de este \cu en ambas soluciones,
ejecutamos \s99\ para estimar el contenido estelar 
de un \cu\ con masa, edad y metalicidad compatibles con ambas soluciones. 
La soluci'on joven  ($\approx 6\,000 \, M_{\odot}$ )
predice la presencia de una estrella supergigante roja y dos azules, mientras
que la soluci'on vieja  ($\approx 129\,000  \,M_{\odot}$ )   predice
la presencia de nueve    supergigantes rojas y ninguna azul. Ver Tabla~\ref{tbl019}.

Al tratar de discernir  entre las dos soluciones, debemos dar una explicaci'on sobre
las predicciones de los modelos de s'intesis evolutiva: en general, 
los modelos te'oricos tienen incertezas intr'insecas, especialmente cuando contienen
fases de r'apida evoluci'on como la fase de las supergigantes rojas.
Adem'as, existe un efecto m'as sutil producido por la incompletitud de los datos.
La mayor'ia de los codigos de s'intesis actuales predicen los 
valores promedio de los observables para una poblaci'on infinita de estrellas, la cual 
abarca todas las masas disponibles dentro de una IMF. 

\begin{figure*}[hc!]
\centering
\includegraphics*[width=0.99\linewidth]{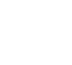}
 \caption[Im'agenes del c'umulo I--Es  en los filtros
 F336W, F555W  y  F814W]{  {\sl\footnotesize 
  Im'agenes del c'umulo I--Es obtenidas con HST/WFPC2 en los filtros
 F336W, F555W  y  F814W. Las tres im'agenes tienen la misma escala
 lineal con un tama~no de   $15.9  \times 15.9$ pc$^2$, y la misma orientaci'on,
 con el Norte   hacia arriba y el Este hacia la izquierda.
 N'otese  los objetos ubicados en las afueras del centro del c'umulo. La discusi'on
 se presenta en el texto. 
\label{fig065}}}
\end{figure*}

\cite{CervVall03} muestran que para un observable determinado a partir
de un n'umero $N$ de estrellas, existe un valor cr'itico de  $N\approx 10$ 
por debajo del cual los resultados del c'odigo deben aceptarse con cuidado,
ya que pueden estar mal determinados.
El \cu  I--Es es un ejemplo de este problema, ya que su magnitud F814W 
est'a dominada por la presencia de unas pocas supergigantes rojas
(ambas soluciones predicen esto.) Por tal motivo, los colores integrados
no nos permiten diferenciar entre las dos soluciones, ya que los efectos estoc'asticos 
pueden ser muy grandes.
?`De qu'e manera se puede  resolver este problema ? Una soluci'on posible ser'ia 
observar la poblaci'on estelar resuelta del \cu que es lo que hicimos
en la Figura~\ref{fig065} al representar un detalle del  \cu en los filtros   F336W,
F555W y  F814W. Estas figuras muestran que el c'umulo posee una subestructura
no radial. Se pueden apreciar dos objetos rodeando el centro del c'umulo,
localizados dentro del radio de apertura que empleamos para la 
fotometr'ia marcada en verde en la Figura~\ref{fig065}.
El objeto ubicado al norte es visible en
los tres filtros y es posiblemente una supergigante azul, mientras que el
otro objeto no se detecta en la imagen F336W y podemos 
argumentar que se trata de una supergigante roja.

La presencia de un objeto rojo favorece a la soluci'on j'oven sobre la vieja,
ya que la primera predice una supergigante roja y la vieja nueve. Por otro lado,
este objeto podria no estar asociado con el c'umulo y encontrarse 
en la misma linea de la visual ( la Figura~\ref{fig066} muestra que este caso no 
es imposible.) Por estos motivos, debemos concluir que los datos disponibles
no nos permiten discernir entre las dos soluciones que proveen los modelos.

Con respecto a la extinci'on que afecta a I--Es, ambas soluciones dan valores 
mucho mayores que en el \cu  I--As.
Esto es consistente con los resultados de  \ha/\hb\ provistos por 
 \citet{Maizetal98} y \citet{Maiz00}  y con la existencia de un pico
  en la distribuci'on de CO cerca de la posici'on del \cu\  \citep{Waltetal01}.
Tambi'en es interesante  notar que el valor obtenido de $R_{5495}$  es mayor que 
el est'andar 3.1. Sin embargo, dadas las incertezas en las propiedades deducidas de este \cu\
a partir de sus colores integrados debido a la poca cantidad de estrellas, debemos 
considerar al valor de $R_{5495}$  como bastante incierto.

\subsection{C'umulo  NGC~4214--IIIs}
\begin{figure*}[hc!]
\begin{minipage}[c]{8cm}
\centering
\includegraphics*[width=\linewidth]{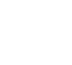}
\end{minipage}
\begin{minipage}[c]{8cm}
 \caption[Diagrama de contornos para el
 c'umulo IIIs]{  {\sl\footnotesize   Diagrama de 
 contornos del plano  [$\log({\rm edad})$, \ecc] para el
 c'umulo IIIs,  que se obtiene
 de {\sc\footnotesize  STATPLOTS} 
 sin imponer ninguna restricci'on en los par'ametros 
 durante la  ejecuci'on de {\sc\footnotesize  CHORIZOS} Se observan dos
 soluciones compatibles con la fotometr'ia. 
\label{fig067}}}
\end{minipage}
\end{figure*}

Para analizar al \cu\ IIIs, usamos las siete magnitudes de la 
Tabla~\ref{tbl006}.  Ejecutamos \cho\ dejando todos los par'ametros libres
y tomando al filtro F336W como referencia.
En este caso, tenemos seis colores y tres par'ametros libres.
Hallamos un excelente ajuste del espectro, con dos posibles soluciones como
puede verse en la Figura~\ref{fig067}:
una soluci'on joven en 
$\log({\rm edad/a}) \approx 7.20$   y otra m'as vieja  
en $\log({\rm edad/a}) \approx 8.20$.

\begin{figure*}[hc!]
\centering
\includegraphics*[width=0.99\linewidth]{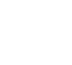}
 \caption[Salida gr'afica de {\sc\footnotesize  STATPLOTS}:
  soluci'on joven    del  c'umulo IIIs]{ {\sl\footnotesize   
 Salida gr'afica de {\sc\footnotesize  STATPLOTS}  correspondiente
 a la soluci'on joven (edad  16  Ma)  del  c'umulo IIIs. Se presentan el espectro te'orico que mejor
 ajusta a las magnitudes observadas (arriba izquierda) y los contornos de probabilidad
  [$\log({\rm edad})$, \ecc]  (arriba derecha),  [  \ecc, \rv] (abajo izquierda) y
   [$\log({\rm edad})$, \rv] (abajo derecha). 
\label{fig068}}}
\end{figure*}

\begin{figure*}[hc!]
\centering
\includegraphics*[width=0.99\linewidth]{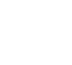}
 \caption[Salida gr'afica de {\sc\footnotesize  STATPLOTS}:
 soluci'on vieja    del  c'umulo IIIs]{ {\sl\footnotesize  
  Salida gr'afica de {\sc\footnotesize  STATPLOTS}  correspondiente
 a la soluci'on vieja (edad  168  Ma)  del  c'umulo IIIs. Se presentan el espectro te'orico que mejor
 ajusta a las magnitudes observadas (arriba izquierda) y los contornos de probabilidad
  [$\log({\rm edad})$, \ecc]  (arriba derecha),  [  \ecc, \rv] (abajo izquierda) y
   [$\log({\rm edad})$, \rv] (abajo derecha). 
\label{fig069}}}
\end{figure*}

Ejecutamos el c'odigo dos veces para aislar estas dos soluciones y presentamos los 
resultados de ambos modelos en la Tabla~\ref{tbl020}.
Las  Figura~\ref{fig068}  y \ref{fig069}  presentan
 los mapas de contornos de ambas soluciones, as'i como
 los espectros ajustados.
La diferencia sutil entre los ajustes  de los espectros
  est'a en la profundidad de la l'inea asociada al grafito
 y localizada alrededor de  2175 \AA. Nuestro estudio no puede  
discernir, a priori,  entre ambas soluciones, ya que no poseemos ninguna magnitud medida
en esta parte del espectro.  Sin embargo, la posici'on 
relativa del \cu\ en la galaxia sugiere que debemos obtener un valor bajo 
para el enrojecimiento.


 \begin{table*}[ht!] 
\centering
\begin{minipage}[u]{15cm}

\caption[Resultados de los c'umulos  IIIs y IVs]{Resultados 
del ajuste de   \cho\  para los c'umulos    IIIs y IVs. El 
n'umero de estrellas es un resultado provisto por \s99.
\label{tbl020} }
\end{minipage}

\begin{tabular}{ lllllll}
 \multicolumn{6}{l}{ \rule{150mm}{0.8mm}}      \\

 \multicolumn{1}{l}{Par\'ametro}     
&  \multicolumn{2}{c}{C\'umulo IIIs}  &    
&   \multicolumn{2}{c}{ C\'umulo IVs}   {\rule [-2mm]{0mm}{7mm} } \\ 

  &    \multicolumn{2}{c}{ \rule{40mm}{0.2mm}}  
  & & \multicolumn{2}{c}{ \rule{40mm}{0.2mm}}  \\ 

& \multicolumn{1}{c}{Joven}    
&  \multicolumn{1}{c}{Viejo}    
&  
&  \multicolumn{1}{c}{Joven}       
& \multicolumn{1}{c}{Viejo}    \\ 
   \multicolumn{6}{l}{ \rule{150mm}{0.2mm}}      \\

Edad (Ma)  &        
               $15.8\pm 0.2$ &  $168 \pm 61$ & &   $15.8 \pm 0.0$    & $150 \pm 34$       \\ 
\ecc ~(mag)  &   
           $0.44 \pm 0.01$    &    $0.26 \pm 0.03$  &   &  $0.22\pm0.01$   & $0.06\pm0.03$   \\ 
Ley de extinci\'on   / \rv  
             &           $3.95\pm0.20$     &  $4.98\pm0.60$   &   &     $3.39\pm0.52$  & $4.35\pm1.12$   \\ 
Masa ($10^3 M_{\odot} $)   & $ 134\pm   18  $ & $ 626 \pm   175  $ & &    $27\pm4$     &  $114\pm20$   \\ 
$\chi^2  $por grado de libertad   &      0.39   &   0.66   & &  1.77  &  2.18   \\ 
Estrellas O+B, de tipos  I y II  &       31  &  0 &   &7 &  0  \\ 
Estrellas K+M, de tipos  I y II &       6   &  45  & &  1 & 26 \\ 
Estrellas WR        & 0     & 0&  &  0 &  0  \\   
 \multicolumn{6}{l}{ \rule{150mm}{0.8mm}}      \\

\end{tabular}
\end{table*}


Adem'as, la imagen del \cu\ tiene una estructura suave  (ver  Figura~\ref{fig106}), 
sin presentar  caracter'istica sobresaliente
alguna. Esto sugiere que la soluci'on mas vieja
(edad $168 \pm 61$  Ma, masa   $(6 \pm 2) \times 10^5$  $M_{\odot}$ )
deber'ia ser la correcta.

\subsection{C'umulo  NGC~4214--IVs} 
Para analizar al \cu IVs usamos las siete magnitudes
de la Tabla~\ref{tbl006}. Ejecutamos \cho\ dejando libres a todos los par'ametros
y usando a F555W como filtro de referencia.
Nuevamente, encontramos dos soluciones diferentes compatibles con nuestras
observaciones: una soluci'on d'ebil y joven en
 $\log({\rm edad/a}) \approx 7.20$ 
y otra mas intensa  y vieja en
 $\log({\rm edad/a}) \approx 8.20$. Esto se pone de manifiesto en 
 la Figura~\ref{fig070}.

\begin{figure*}[t!]
\centering
\includegraphics*[width=0.99\linewidth]{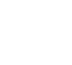}
 \caption[Detalle de las im'agenes de IIIs y IVs]{ {\sl\footnotesize  
Im'agenes de los c'umulos IIIs y IVs obtenidas con HST/WFPC2
en los filtros F336W y F814W.  Las im'agenes tienen la misma escala lineal y un tama\~no 
de $4 \arcs 1 \times 4 \arcs 1 $  y la misma orientaci'on con el Norte hacia arriba
y el Este hacia la Izquierda. Notar la estructura suave de estos objetos 
en contraposici'on a la estructura del c'umulo I--Es que se muestra
en la Figura~\ref{fig065}.
\label{fig106}}}
\end{figure*}

\begin{figure*}[hc!]
\begin{minipage}[c]{8cm}
\centering
\includegraphics*[width=\linewidth]{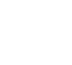}
\end{minipage}
\begin{minipage}[c]{8cm}
 \caption[ Diagrama de contornos   para el
 c'umulo IVs]{  {\sl\footnotesize   Diagrama de contornos del plano  [$\log({\rm edad})$, \ecc] para el
 c'umulo IVs,  que se obtiene
 de {\sc\footnotesize  STATPLOTS} 
 sin imponer ninguna restricci'on en los par'ametros 
 durante la  ejecuci'on de {\sc\footnotesize  CHORIZOS}. Se observan dos
 soluciones compatibles con la fotometr'ia. 
\label{fig070}}}
\end{minipage}
\end{figure*}

\begin{figure*}[hc!]
\centering
\includegraphics*[width=0.99\linewidth]{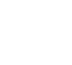}
 \caption[ Salida gr'afica de {\sc\footnotesize  STATPLOTS}:
  soluci'on vieja   del  c'umulo IVs]{ {\sl\footnotesize  
  Salida gr'afica de {\sc\footnotesize  STATPLOTS}  correspondiente
 a la soluci'on vieja (edad  150  Ma)  del  c'umulo IVs. Se presentan el espectro te'orico que mejor
 ajusta a las magnitudes observadas (arriba izquierda) y los contornos de probabilidad
  [$\log({\rm edad})$, \ecc]  (arriba derecha),  [  \ecc, \rv] (abajo izquierda) y
   [$\log({\rm edad})$, \rv] (abajo derecha). 
\label{fig071}}}
\end{figure*}

\begin{figure*}[hc!]
\centering
\includegraphics*[width=0.99\linewidth]{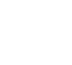}
 \caption[Salida gr'afica de {\sc\footnotesize  STATPLOTS}:
 soluci'on joven   del  c'umulo IVs]{ {\sl\footnotesize  
  Salida gr'afica de {\sc\footnotesize  STATPLOTS}  correspondiente
 a la soluci'on joven (edad  16  Ma)  del  c'umulo IVs. Se presentan el espectro te'orico que mejor
 ajusta a las magnitudes observadas (arriba izquierda) y los contornos de probabilidad
  [$\log({\rm edad})$, \ecc]  (arriba derecha),  [  \ecc, \rv] (abajo izquierda) y
   [$\log({\rm edad})$, \rv] (abajo derecha). 
\label{fig072}}}
\end{figure*}

Las    Figuras~\ref{fig071}  y \ref{fig072}  contienen a los espectros que mejor ajustan a 
los datos y   los mapas de probabilidad de las
soluciones  vieja y  joven  respectivamente.

Con las magnitudes disponibles, no nos fue posible
discenir entre ambos modelos. Sin embargo, dada la ubicaci'on de
este \cu\ dentro de la galaxia (una regi'on de
poco polvo y gas), esperamos un bajo enrojecimiento.
Adem'as, como sucedi'o con el \cu\ IIIs, su apariencia sin estructura  (ver  Figura~\ref{fig106})
favorece a la soluci'on m'as vieja (edad $150 \pm 34$  Ma) por sobre la 
m'as joven.

Al igual que con el \cu  I--As, encontramos que $ R_{5495}$
abarca valores en el rango $2.0  \leqslant R_{5495}  \leqslant 6.0$,
lo cual indica que para un enrojecimiento bajo, \rv\ est'a degenerado.
Para resolver este problema, ser'ia muy 'util realizar la medici'on de una magnitud
cerca del lado derecho de la discontinuidad de Balmer como por ejemplo
con el filtro F439W (WFPC2 $B$).

En la Tabla~\ref{tbl020}  presentamos el n'umero de 
estrellas  K+M de clases de luminosidad I y II que se obtienen a partir
de la salida de \s99\  para  los c'umulos IIIs y IVs.
Es notable que en los \cus IIIs y IVs, la soluci'on m'as joven provee
una menor cantidad de supergigantes rojas que en sus
respectivas soluciones  viejas.
La mayor cantidad de supergigantes rojas explicar'ia la apariencia
carente de estructura de los \cus\ IIIs y IVs que observamos en sus im'agenes.
Este es otro hecho que favorece la selecci'on de la soluci'on vieja por sobre la j'oven.

\cite{Billeetal02}   estudiaron a los c'umulos IIIs y IVs como parte
de un relevamiento  de c'umulos  estelares compactos 
en galaxias cercanas que incluy'o a NGC~4214. Desafortunadamente,
su fotometr'ia de apertura incluye errores como no tener en cuenta el efecto de la 
contaminaci'on en el filtro F336W, y la conversi'on de magnitudes del HST al sistema de
Johnson--Cousins, lo cual no es recomendado debido a la introducci'on de errores
durante las transformaciones \citep{Gonzetal03,deGretal05}. 
Usando una traza evolutiva calculada con los modelos de 
s'intesis evolutiva \s99 que desplegaron sobre un diagrama color--color, 
\citeauthor{Billeetal02} tratan de inferir la edad de estos c'umulos. 


\begin{figure*}[hc!]
\begin{minipage}[c]{8cm}
\centering
\includegraphics*[width=\linewidth]{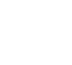}
\end{minipage}
\begin{minipage}[c]{8cm}
 \caption[ Diagrama color--color  de \cite{Billeetal02}]{   {\sl\footnotesize  
  Diagrama color--color [$(U-V)_0$ , $(V-I)_0$] de \cite{Billeetal02} donde 
 se muestra la trayectoria evolutiva de un c'umulo de acuerdo a 
 los modelos te'oricos de \s99. Claramente la trayectoria es degenerada. 
 \citeauthor{Billeetal02} marcaron los c'umulos estelares de NGC~4214
 que estudiaron  con el objeto de determinar sus edades. Hemos marcado en
 rojo a los c'umulos NGC~4214--IIIs y IVs.
 La edad estimada para IIIs es $700-1000$ Ma. La edad estimada para 
 IVs es 220 Ma.
\label{fig083}}}
\end{minipage}
\end{figure*}

 
La trayectoria evolutiva   es claramente degenerada
como puede observarse en la Figura~\ref{fig083} y muestra varias soluciones
posibles para la edad de estos c'umulos.
Estos autores emplearon solamente dos colores fotom'etricos, lo cual se 
traduce en una muy pobre determinaci'on de la edad de los \cus  \citep{deGretal03a}. 
Usando la fotometr'ia generada por  \cite{Billeetal02} y los modelos
de s'intesis evolutiva de \cite{BruzChar03},  
\cite{Larsetal04} estiman que la edad de estos
dos  \cus es  $\sim (200 \pm 52 ) $ Ma, lo cual es consistente con nuestros resultados.

\section{C'umulos resueltos  NGC~4214--I--Ds y  II--ABCDE}
Analizamos las propiedades de los c'umulos I--Ds, II-ABCDE en NGC~4214.
Estos c'umulos presentan cierta estructura en las im'agenes del HST
que empleamos y aparecen mayormente resueltos en estrellas individuales.

\begin{figure*}[hc!]
\centering
\includegraphics*[width=0.99\linewidth]{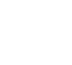}
 \caption[Salida gr'afica de {\sc\footnotesize  STATPLOTS}: c'umulo II--A]{ {\sl\footnotesize Salida gr'afica de {\sc\footnotesize  STATPLOTS}  correspondiente
 al  c'umulo II--A. Se presentan el espectro te'orico que mejor
 ajusta a las magnitudes observadas (arriba izquierda) y los contornos de probabilidad
  [$\log({\rm edad})$, \ecc]  (arriba derecha),  [  \ecc, \rv] (abajo izquierda) y
   [$\log({\rm edad})$, \rv] (abajo derecha).  
\label{fig073}}}
\end{figure*}

\begin{figure*}[hc!]
\centering
\includegraphics*[width=0.99\linewidth]{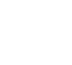}
 \caption[Salida gr'afica de {\sc\footnotesize  STATPLOTS}: c'umulo II--B]{ {\sl\footnotesize Salida gr'afica de {\sc\footnotesize  STATPLOTS}  correspondiente
 al  c'umulo II--B. Se presentan el espectro te'orico que mejor
 ajusta a las magnitudes observadas (arriba izquierda) y los contornos de probabilidad
[$\log({\rm edad})$, \ecc]  (arriba derecha),  [  \ecc, \rv] (abajo izquierda) y
[$\log({\rm edad})$, \rv] (abajo derecha).  En el texto se explica la raz'on del 
desplazamiento del punto blanco respecto a la m'axima probabilidad.
\label{fig074}}}
\end{figure*}

\begin{figure*}[hc!]
\centering
\includegraphics*[width=0.99\linewidth]{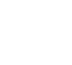}
 \caption[Salida gr'afica de {\sc\footnotesize  STATPLOTS}:   
  c'umulo II--C]{ {\sl\footnotesize Salida gr'afica de {\sc\footnotesize  STATPLOTS}  correspondiente
 al  c'umulo II--C. Se presentan el espectro te'orico que mejor
 ajusta a las magnitudes observadas (arriba izquierda) y los contornos de probabilidad
  [$\log({\rm edad})$, \ecc]  (arriba derecha),  [  \ecc, \rv] (abajo izquierda) y
   [$\log({\rm edad})$, \rv] (abajo derecha).  En el texto se explica la raz'on del 
desplazamiento del punto blanco respecto a la m'axima probabilidad.
\label{fig075}}}
\end{figure*}

\begin{figure*}[hc!]
\centering
\includegraphics*[width=0.99\linewidth]{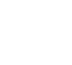}
 \caption[Salida gr'afica de {\sc\footnotesize  STATPLOTS}:   
  c'umulo II--D]{ {\sl\footnotesize Salida gr'afica de {\sc\footnotesize  STATPLOTS}  correspondiente
 al  c'umulo II--D. Se presentan el espectro te'orico que mejor
 ajusta a las magnitudes observadas (arriba izquierda) y los contornos de probabilidad
  [$\log({\rm edad})$, \ecc]  (arriba derecha),  [  \ecc, \rv] (abajo izquierda) y
   [$\log({\rm edad})$, \rv] (abajo derecha).  En el texto se explica la raz'on del 
desplazamiento del punto blanco respecto a la m'axima probabilidad.
\label{fig076}}}
\end{figure*}

\begin{figure*}[hc!]
\centering
\includegraphics*[width=0.99\linewidth]{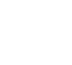}
 \caption[ Salida gr'afica de {\sc\footnotesize  STATPLOTS}:   
 c'umulo II--E]{ {\sl\footnotesize Salida gr'afica de {\sc\footnotesize  STATPLOTS}  correspondiente
 al  c'umulo II--E. Se presentan el espectro te'orico que mejor
 ajusta a las magnitudes observadas (arriba izquierda) y los contornos de probabilidad
  [$\log({\rm edad})$, \ecc]  (arriba derecha),  [  \ecc, \rv] (abajo izquierda) y
   [$\log({\rm edad})$, \rv] (abajo derecha).  En el texto se explica la raz'on del 
desplazamiento del punto blanco respecto a la m'axima probabilidad.
\label{fig077}}}
\end{figure*}

Los c'umulos que forman parte de la estructura del complejo NGC~4214--II
(Figura~\ref{fig027})
fueron analizados usando las magnitudes listadas en 
la Tabla~\ref{tbl014} con excepci'on de las magnitudes en los filtros 
F555W y F702W. Estas magnitudes no fueron considerados 
durante la ejecuci'on de \cho\
debido a que  las mismas  est'an contaminadas por emisi'on nebular proveniente
del rico medio interestelar que  caracteriza a esta regi'on de la galaxia.
La ejecuci'on de \cho\ nos dio 
 ajustes aceptables  (m'aximo $\chi^2 = 2.53 $ ) de los espectros 
 que presentamos en las   Figuras~\ref{fig073}, ~\ref{fig074},~\ref{fig075},~\ref{fig076} y ~\ref{fig077},
 junto con sus respectivos diagramas de contornos. 
N'otese  que los puntos que corresponden a las magnitudes 
en los filtros 
F555W y F702W 
est'an marcados a pesar de no haber sido considerados para los ajustes.
 
\begin{figure*}[h]
 \begin{minipage}[c]{10cm}
\centering
\includegraphics*[width=0.99\linewidth]{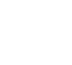}
\end{minipage}
\begin{minipage}[c]{6cm}
 \caption[Predicciones de  $W$(H$\alpha$)
en funci'on del tiempo]{  {\sl\footnotesize Predicciones de  $W$(H$\alpha$)
en funci'on del tiempo de acuerdo a varios modelos de s'intesis evolutiva. 
Se comparan los modelos de \cite{Cervetal02} (CMHK) y \s99 [\citep{Leitetal99},
SB99]. El primer modelo CMHK usa trayectorias evolutivas
calculadas con una tasa de  p'erdida de masa normal \citep{Schaetal93b}
mientras que los otros dos modelos usan las  trazas de \cite{Meynetal94}.
todos los modelos fueron calculados para $Z=0.004$ y empleando una
IMF de Salpeter. Figura obtenida de \cite{MacKetal00}.  
\label{fig078}}}
\end{minipage}
\end{figure*}

Una manera de estimar la edad de una regi'on de formaci'on estelar
es mediante el uso del ancho equivalente de las lineas de Balmer
\citep{CervMasH94}. 
Los valores esperados de $W$(H$\alpha$)
en \cus muy jovenes (edad $ \lesssim 3$ Ma) 
son  $1000-2500$ \AA\
y  $500-1000$ \AA\ \,  para \cus con edades en el rango $3-4$ Ma.
\cite{MacKetal00} estimaron este par'ametro usando tres leyes
de extinci'on para varios c'umulos de 
NGC~4214. La Figura~\ref{fig078} muestra la evoluci'on de
 $W$(H$\alpha$) con el tiempo.


\subsection{C\'umulo  NGC~4214--II--A}

La primera ejecuci'on de \cho\ sobre 
las magnitudes del \cu  II--A  muestra dos posibles soluciones
compatibles con nuestros datos: una en el rango 
$0.1  \leqslant   E(4405-5495)     \leqslant 0.26$ mag
y la  otra en el rango     $  0.26  \leqslant   E(4405-5495)   \leqslant 0.6$ mag. 
Volvimos a ejecutar el c'odigo para aislar las soluciones en forma individual,
y presentamos s'olo los diagramas correspondientes al segundo caso en
la Figura~\ref{fig073}.  
\cite{MacKetal00} proveen valores de  $W$(H$\alpha$)  para todos 
los \cus dentro del complejo  NGC~4214--II.  Estos valores 
est'an cerca de o son mayores que  $1000$ \AA\  lo cual indica 
que el complejo es muy joven.
Este factor nos llev'o a favorecer a la soluci'on con el mayor $E(4405-5495)$.

Con este m'etodo, pudimos restringir la edad del \cu II--A  
a $3.1 \pm 1.4  $ Ma, lo cual implica que dentro de su estructura podr'ia haber 
estrellas WR. Obtuvimos un valor alto de \rv, compatible con el calculado 
para el complejo II.  Los resultados correspondientes a este c'umulo se
presentan en la Tabla~\ref{tbl021}.

El  gr'afico de abajo izquierda en la Figura~\ref{fig073} 
muestra el  mapa de contornos de  [  \ecc, \rv].
Un s'imbolo blanco 
 muestra  los valores de los par'ametros usados para la gr'afica del espectro.
 Se puede apreciar que este punto no coincide con la zona marcada
 de m'axima probabilidad. Esto es as'i debido a que el punto blanco
se obtiene en un espacio de tres dimensiones [$\log({\rm edad})$, \rv,  \ecc]  
y lo que estamos graficando es la proyecci'on sobre cada plano  en dos dimensiones.

  \begin{table*}[ht!] 
\centering
\begin{minipage}[u]{115mm}
\caption{Resultados del ajuste de   \cho\  de los c'umulos   II--A, II--B y II--C
\label{tbl021} }
\end{minipage}

\begin{tabular}{ llll }
 \multicolumn{4}{l}{ \rule{120mm}{0.8mm}}      \\
\multicolumn{1}{l}{Par'ametro}   
   
&  \multicolumn{1}{c}{C\'umulo II--A}    
&  \multicolumn{1}{c}{C\'umulo II--B}    
&  \multicolumn{1}{c}{C\'umulo II--C}    
        \\ 
        \multicolumn{4}{l}{ \rule{120mm}{0.2mm}}      \\

Edad (Ma)                                             &  $3.1 \pm 1.4$           &   $ 2.0\pm0.8$    & $1.7\pm0.6$    \\ 
\ecc ~(mag)                                       &    $0.35 \pm 0.08$     &  $0.35\pm0.01$   & $0.28\pm0.01$     \\  
Ley de extinci\'on   / \rv                                   & $3.79\pm0.33$          &   $4.11\pm0.11$     &  $4.82\pm0.13$         \\ 
Masa ($10^3 M_{\odot} $)                      & $ 34 \pm   23  $          &    $43\pm10$     &  $63\pm14$      \\ 
$\chi^2  $por grado de libertad                        &   0.41                           &     2.02 &  2.53       \\ 
 \multicolumn{4}{l}{ \rule{120mm}{0.8mm}}      \\

\end{tabular}
\end{table*}

\subsection{C\'umulo  NGC~4214--II--B}

Al ejecutar \cho\ sobre el conjunto de magnitudes
del \cu II--B, sucedi'o algo parecido al \cu II--A, pues hallamos dos 
potenciales soluciones con el mismo valor de \rv.
Ac'a presentamos los gr'aficos que corresponden a la
soluci'on m'as joven en la Figura~\ref{fig074}.
Siguiendo el mismo razonamiento que para II--A,
debemos favorecer a la soluci'on mas joven debido a los
altos valores de  $W$(H$\alpha$)  medidos por \cite{MacKetal00} 
en esta regi'on.
La edad medida para este \cu es: $2.0 \pm 0.8  $ Ma.
Los resultados correspondientes a este c'umulo se
presentan en la Tabla~\ref{tbl021}.

\subsection{C\'umulos  NGC~4214--II--CDE}

Los resultados de los mejores ajustes de los otros tres \cus 
est'an listados en las 
Tablas~\ref{tbl021} y \ref{tbl022}, 
y sus diagramas de contornos y espectros  se presentan en las
Figuras~\ref{fig075},~\ref{fig076} y ~\ref{fig077}.
El \cu II--C  da una 'unica soluci'on, mientras que  II--D
y II--E  tienen salidas gr'aficas bastante mas complicadas;
sin embargo, sus edades y enrojecimientos son compatibles entre s'i.

\subsection{C\'umulo  NGC~4214--I--Ds}
\begin{figure*}[hc!]
\centering
\includegraphics*[width=0.99\linewidth]{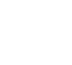}
 \caption[ Salida gr'afica de {\sc\footnotesize  STATPLOTS}:   
  c'umulo I--Ds]{ {\sl\footnotesize Salida gr'afica de {\sc\footnotesize  STATPLOTS}  correspondiente
 al  c'umulo I--Ds. Se presentan el espectro te'orico que mejor
 ajusta a las magnitudes observadas (arriba izquierda) y los contornos de probabilidad
  [$\log({\rm edad})$, \ecc]  (arriba derecha),  [  \ecc, \rv] (abajo izquierda) y
   [$\log({\rm edad})$, \rv] (abajo derecha).   En el texto se explica la raz'on del 
desplazamiento del punto blanco respecto a la m'axima probabilidad.
\label{fig079}}}
\end{figure*}

Usando  las magnitudes de la   Tabla~\ref{tbl006}, ejecutamos 
 \cho\  para analizar las propiedades del \cu   I--Ds. 
 Dejamos a todos los par'ametros 
[  $\log({\rm edad}) $, $E(4405-5495)$  y  $R_{5495}$ ] sin restringir, 
lo cual significa que 
s'olo dispusimos de  un grado de libertad en este caso.
La Figura~\ref{fig079}  muestra el espectro que mejor se ajusta a los datos fotom'etricos
y los diagramas de contornos del ajuste para todas las combinaciones
de los par'ametros arriba mencionados.

Parece haber dos posibles soluciones, ambas muy j'ovenes, con 
un m'aximo de probabilidad en $\log({\rm edad/a}) \approx 6.5 $.  
La salida del c'odigo  da un valor medio para la edad de  $2.6 \pm 1.5  $ Ma,
y un enrojecimiento promedio de $ E(4405-5495) =  0.28 \pm 0.05$ mag.
Los resultados del ajuste est'an listados en la Tabla~\ref{tbl022}.

  \begin{table*}[ht!] 
\centering
\begin{minipage}[u]{115mm}
\caption{Resultados del ajuste de   \cho\  de los c'umulos    II--D, II--E, I--Ds
\label{tbl022} }
\end{minipage}

\begin{tabular}{ llll }
 \multicolumn{4}{l}{ \rule{120mm}{0.8mm}}      \\

\multicolumn{1}{l}{Par'ametro}   
& \multicolumn{1}{c}{C\'umulo II--D}    
 & \multicolumn{1}{c}{C\'umulo II--E}    
& \multicolumn{1}{c}{C\'umulo I--Ds}      \\ 
        \multicolumn{4}{l}{ \rule{120mm}{0.2mm}}      \\

Edad (Ma)                                         &       $4.0\pm 4.0$ &  $3.1 \pm 1.4$    &$2.6 \pm 1.5$    \\ 
\ecc ~(mag)                                       &   $0.22 \pm 0.16$    &    $0.20 \pm 0.05$   &$0.28 \pm 0.05$   \\  
Ley de extinci\'on   / \rv                                &       $4.05\pm0.91$  & $2.92\pm0.43$   &$4.05\pm0.19$  \\ 
Masa ($10^3 M_{\odot} $)                   &  $ 7\pm  5  $ & $ 7 \pm 3  $     & $ 20 \pm   9  $  \\ 
$\chi^2  $por grado de libertad                      &       0.15   &   0.22   &    1.31    \\ 
 \multicolumn{4}{l}{ \rule{120mm}{0.8mm}}      \\

\end{tabular}
\end{table*}

\section{Complejos grandes NGC~4214--I--A, I--B y II}

Cuando estudiamos a los complejos   I--A, I--B y II
ejecutamos  \cho\  dejando a 
 $\log({\rm edad}) $, $E(4405-5495)$  y   $R_{5495}$ como par'ametros sin restringir.
Usamos  las magnitudes de la 
Tabla~\ref{tbl006}, pero en el caso del complejo II, las magnitudes 
en los filtros  F555W y F702W fueron listadas, pero no fueron consideradas
para el ajuste de cuadrados m'inimos por encontrarse seguramente
contaminadas por el alto contenido de emisi'on nebular del medio 
interestelar.
El agregar las magnitudes 2MASS no cambi'o significativamente
la salida del c'odigo en ning'un caso.

\begin{figure*}[hc!]
\centering
\includegraphics*[width=0.99\linewidth]{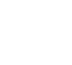}
 \caption[Salida gr'afica de {\sc\footnotesize  STATPLOTS}:   
  c'umulo I--A]{ {\sl\footnotesize Salida gr'afica de {\sc\footnotesize  STATPLOTS}  correspondiente
 al  c'umulo I--A. Se presentan el espectro te'orico que mejor
 ajusta a las magnitudes observadas (arriba izquierda) y los contornos de probabilidad
  [$\log({\rm edad})$, \ecc]  (arriba derecha),  [  \ecc, \rv] (abajo izquierda) y
   [$\log({\rm edad})$, \rv] (abajo derecha).   En el texto se explica la raz'on del 
desplazamiento del punto blanco respecto a la m'axima probabilidad.
\label{fig080}}}
\end{figure*}

\begin{figure*}[hc!]
\centering
\includegraphics*[width=0.99\linewidth]{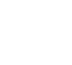}
 \caption[Salida gr'afica de {\sc\footnotesize  STATPLOTS}:   
  c'umulo I--B]{ {\sl\footnotesize Salida gr'afica de {\sc\footnotesize  STATPLOTS}  correspondiente
 al  c'umulo I--B. Se presentan el espectro te'orico que mejor
 ajusta a las magnitudes observadas (arriba izquierda) y los contornos de probabilidad
  [$\log({\rm edad})$, \ecc]  (arriba derecha),  [  \ecc, \rv] (abajo izquierda) y
   [$\log({\rm edad})$, \rv] (abajo derecha).   En el texto se explica la raz'on del 
desplazamiento del punto blanco respecto a la m'axima probabilidad.
\label{fig081}}}
\end{figure*}

\begin{figure*}[hc!]
\centering
\includegraphics*[width=0.99\linewidth]{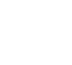}
 \caption[Salida gr'afica de {\sc\footnotesize  STATPLOTS}:   
 c'umulo II]{ {\sl\footnotesize Salida gr'afica de {\sc\footnotesize  STATPLOTS}  correspondiente
 al  c'umulo II. Se presentan el espectro te'orico que mejor
 ajusta a las magnitudes observadas (arriba izquierda) y los contornos de probabilidad
  [$\log({\rm edad})$, \ecc]  (arriba derecha),  [  \ecc, \rv] (abajo izquierda) y
   [$\log({\rm edad})$, \rv] (abajo derecha).   En el texto se explica la raz'on del 
desplazamiento del punto blanco respecto a la m'axima probabilidad.
\label{fig082}}}
\end{figure*}

Las salidas gr'aficas de  {\sc\footnotesize  STATPLOTS}  de estos \cus 
se presentan en las  Figuras~\ref{fig080}, ~\ref{fig081} y ~\ref{fig082}.
Los ajustes de I--A y  I--B
son excelentes. La magnitud calculada en el filtro F170W es la que
contribuye m'as al valor $\chi^2 = 4.25$ en el complejo 
 I--B.
Los gr'aficos muestran que tanto para I--A como para  I--B
existe una 'unica soluci'on compatible con nuestra fotometr'ia. Hallamos
la misma edad  (5.0 Ma) para ambos y, como era de esperarse comparando
con nuestro estudio del \cu I--As,
un bajo valor de la extinci'on: $ E(4405-5495) =  0.07$ mag.
Este valor es muy cercano al obtenido para el c'umulo  I--As, que se
encuentra dentro de la estructura de I--A.

Las aperturas empleadas para analizar los complejos  I--A y  I--B
son  lo suficientemente grandes como para incluir una parte de
la galaxia que contiene un mezcla de diferentes tipos de estrellas.
La Figura~\ref{fig027}  muestra claramente la presencia de estrellas tard'ias
y tempranas dentro de las aperturas elegidas.  Esto nos lleva a concluir
que  los espectros de los complejos  I--A y  I--B son espectros 
compuestos por una poblaci'on estelar compleja.


Una estimaci'on de la masa total de estos c'umulos 
provee los valores: $(156 \pm 19) \times 10^3$  
$M_{\odot}$ para  I--A 
y  $(34 \pm 4) \times 10^3$  $M_{\odot}$  para I--B.

Para el \cu II usamos la apertura m'as grande: 200 p'ixeles de la PC.
Esto significa que estamos abarcando una pobalci'on mezclada
de estrellas dentro de un medio rico en gas. Como resultado, obtuvimos 
un espectro compuesto caracterizado por una edad de $1.9 \pm 0.9$ Ma
y una masa estimada de $(9 \pm 3) \times 10^5$  $M_{\odot}$. 
 El gr'afico  [$\log({\rm edad})$, \ecc]  indica que   este \cu es m'as joven 
que I--A, y que esta extinguido. Los diagramas  [$\log({\rm edad})$, \rv] implican que la
ley de extinci'on que mejor se ajusta a nuestra fotometr'ia tiene
un alto valor de \rv\ en la familia de leyes Gal'acticas de \citet{Cardetal89}.

Se espera que el complejo II contenga objetos de alta extinci'on
dentro de su estructura, ya que 'estos est'an ubicados dentro o muy cerca de 
regiones \ha\ que muestran poca evidencia de burbujas creadas por
supernovas o vientos estelares \citep{Maizetal98,MacKetal00}.
El mapa de extinci'on (Figura~\ref{fig038}) 
representa muy bien esta parte de NGC~4214,  y  muestra los 
mayores valores de extinci'on en la zona sur del \cu II--A y
entre los \cus  II--B y  II--C.
Los valores medidos  de \ecc\ correspondientes a las estrellas est'an de acuerdo
con aquellos obtenidos a partir de  \ha/\hb\ \citep{Maizetal98}.
Tambi'en calculamos el valor medio de la extinci'on estelar en cada \cu individual
(II--ABCDE)  dentro del complejo  NGC~4214--II  y encontramos
que son muy parecidos a los valores que se obtienen al ajustar
modelos de \s99  a las magnitudes integradas.

\begin{table*}[ht!] 
\centering
\begin{minipage}[u]{115mm}

\caption{Resultados del ajuste de   \cho\  de los c'umulos   I--A, I--B y II
\label{tbl023} }
\end{minipage}

\begin{tabular}{ llll }
 \multicolumn{4}{c}{ \rule{115mm}{0.8mm}}      \\
\multicolumn{1}{l}{Par'ametro}   
   
&  \multicolumn{1}{c}{C\'umulo I--A}    
&  \multicolumn{1}{c}{C\'umulo I--B}    
&  \multicolumn{1}{c}{C\'umulo II}    \\
        \multicolumn{4}{c}{ \rule{115mm}{0.2mm}}      \\
  
Edad (Ma)                                             &  $5.0 \pm 0.0$           &   $ 5.0\pm0.0$    & $1.9\pm0.9$    \\ 
\ecc ~(mag)                                       &    $0.07 \pm 0.01$     &  $0.07\pm0.01$   & $0.43\pm0.03$     \\  
Ley de extinci\'on   / \rv                                   & $5.88\pm0.16$          &   $5.81\pm0.19$     &  $4.99\pm0.11$         \\ 
Masa ($10^3 M_{\odot} $)                      & $ 156 \pm   19  $          &    $34\pm4$     &  $923\pm331$      \\ 
$\chi^2  $por grado de libertad                        &   0.95                           &     4.25 &  0.21      \\ 
 \multicolumn{4}{c}{ \rule{115mm}{0.8mm}}      \\

\end{tabular}
\end{table*}

Los resultados de los ajustes de estos tres c'umulos 
se presentan en la    Tabla~\ref{tbl023}. Un resultado importante 
es que nuestros modelos dan valores altos de \rv\ cuando analizamos a los complejos grandes.
Analizamos los valores de \rv\ de las estrellas mas brillantes en estas regiones y 
hallamos que las estrellas individuales presentan altos valores de extinci'on, 
especialmente aquellos del \cu II--B. 
Se sabe que el valor de \rv\ depende del medio a lo largo de la l'inea de la visual.
Una l'inea que pasa a trav'es de una zona de baja densidad presenta bajos
valores de la extinci'on  ($\sim$ 3.1).
En cambio, si se observa a trav'es de una nube molecular densa como Ori'on,
Ofiuco o Tauro, los valores pueden ser significativamente m'as grandes
obteni'endose $4 < R_V< 6$  \citep{Math90}. Por ejemplo,
la estrella de tipo O de la ZAMS Her~36 en M~8 tiene $R_{5495} = 5.39 \pm 0.09$ \citep{Ariaetal05}.

\chapter{Resumen y conclusiones}
\label{cha:conclusions}

\thispagestyle{empty}
 \begin{flushright}
{\em It's over now, the music of the night! }

 \end{flushright}
\newpage

Las galaxias cercanas nos proveen de laboratorios ideales donde es
posible analizar, entre otras cosas, c'omo se forman las estrellas, qu'e procesos 
disparan la formaci'on estelar, c'omo interact'uan las estrellas
con el medio interestelar y de qu'e manera se crean las galaxias.
El estudio de galaxias starburst como NGC~4214 es de sumo inter'es en Astrof'isica
ya que la mayor'ia de los resultados que inferimos sobre las galaxias
lejanas (con alto corrimiento al rojo) 
 se basa en lo que observamos en las galaxias del Universo local.
Por tal motivo, es importante comprender con detalle lo que sucede en las galaxias 
cercanas para saber qu'e sucede en las galaxias m'as alejadas.

NGC~4214 es una galaxia de baja metalicidad $(Z=0.006)$, y su estudio  nos dio 
la posibilidad de analizar las condiciones f'isicas en un ambiente de
alto inter'es en la Astrof'isica actual.

En esta Tesis de Doctorado en Astronom'ia estudiamos en detalle
diversos aspectos de  la galaxia 
enana  NGC~4214 ubicada a una distancia de $2.94$~Mpc, 
empleando im'agenes de alta resoluci'on obtenidas en observaciones 
realizadas con el Telescopio Espacial Hubble. Los resultados presentados 
son originales y complementan una gran variedad de estudios previos realizados 
sobre NGC~4214 usando   diferentes t'ecnicas y    todo 
tipo de im'agenes y espectros. 

Se presenta por primera vez un estudio de las supergigantes rojas y
azules de la galaxia; se describe la distribuci'on de la extinci'on debida al 
continuo estelar y se la relaciona con la emision nebular; se estima la pendiente de
la funci'on inicial de masa empleando un m'etodo fotom'etrico y se
determinan par'ametros f'isicos de trece complejos estelares distribuidos 
a lo largo de su estructura principal. A continuaci'on,  resumimos los 
principales resultados de este trabajo. 

\section{Extinci'on}
Empleando los valores de extinci'on estelar   de los ajustes realizados con \cho\  pudimos 
construir 
un mapa que muestra la  extinci'on 
variable a trav'es de la estructura de NGC~4214.
Lo que se observa como rasgo sobresaliente,
es que NGC~4214 presenta valores muy bajos ($E(4405-5495) \lesssim 0.1$) de la
extinci'on en la mayor parte de su estructura,
con la excepci'on de algunas pocas regiones bien definidas que muestran altos valores
de extinci'on ($E(4405-5495) \gtrsim 0.3$). Estas son las regiones de formaci'on estelar 
reciente. 
Comparamos nuestro mapa de extinci'on con los resultados hallados por 
\cite{Maizetal98} y \citet{Maiz00}  quienes usaron el cociente de Balmer 
(H$\alpha$/H$\beta$) como un 
trazador  del enrojecimiento.   
Estos autores encontraron que el enrojecimiento en   NGC 4214--II 
es, en promedio, mayor que en NGC 4214--I. Nuestros resultados, derivados 
a partir de los colores fotom'etricos de las estrellas, est'an de acuerdo con este hecho.
Las dos cavidades principales en NGC 4214--I muestran una regi'on de baja extinci'on
rodeada de valores m'as altos mientras que NGC 4214--II  tiene una extinci'on alta en
general. 

Nuestro mapa de extinci'on representa bastante bien las nubes moleculares 
de CO estudiadas por \cite{Waltetal01} los cuales est'an directamente asociadas
a las regiones de formaci'on estelar activa en NGC~4214.

\cite{Faneetal88} 
hallaron una inconsistencia entre la extinci'on calculada a partir del continuo UV
de siete galaxias starburst y la extinci'on calculada mediant las l'ineas de Balmer.
Observaron que la extinci'on derivada a partir del continuo UV era sistem'aticamente menor. Esto llev'o a \cite{Calzetal94}  a analizar espectros UV del  IUE  y espectros 'opticos
de 39 galaxias starburst con el objeto de estudiar  las propiedades promedio 
de la extinci'on. Derivaron una ley de extinci'on en el rango de longitudes de onda
que abarca desde el UV hasta el 'optico bajo la hip'otesis de que el polvo 
forma una pantalla en frente de la fuente.
Las caracter'isticas de esta  ley de extinci'on son diferentes a las que presentamos en el 
Cap'itulo~7 para las leyes de la Galaxia y las Nubes de Magallanes: la pendiente es m'as
gris en general y, lo m'as importante, el lomo o ``bump'' del grafito est'a ausente dentro de los
errores observacionales. 
\cite{Calzetal94}  usaron a las diferencias en las pendientes para 
explicar las diferencias observadas entre la extinci'on nebular y la estelar.

Sin embargo,  \cite{MasHKunt99}  mostraron que los espectros observados 
entre el UV y el 'optico de 17 galaxias starburst pueden ser reproducidos 
perfectamente enrrojeciendo  a los espectros sint'eticos usando una de las tres leyes
de extinci'on (V'ia L'actea, Nubes de Magallanes) sin necesidad de invocar 
otra ley de extinci'on adicional.
Como fue propuesto por  \cite{MasHKunt99} y \cite{MacKetal00}, 
el efecto es meramente geom'etrico: mientras el continuo proviene de las
estrellas, la emisi'on nebular se origina en regiones extendidas adyacentes a la
nube molecular original. 
Los vientos estelares y las explosiones de supernova pueden hacer desaparecer el polvo
de las cercan'ias de las estrellas masivas y concentrarlo en filamentos y concentraciones de polvo 
localizadas dentro de la regi'on de material ionizado.

Dependiendo de la distribuci'on geom'etrica de las estrellas y el polvo,
la extinci'on puede afectar principalmente la emisi'on del gas pero s'olo levemente
al continuo estelar. Esto implica que 
la ley de atenuaci'on de \cite{Calzetal94} se aplica s'olo a las
regiones  como  I--A en  NGC~4214, donde las estrellas han evolucionado 
lo suficiente como para haber modificado al ISM; por otro lado, esta ley no
puede aplicarse a NGC~4214--II, ya que la distribuci'on de las estrellas y el polvo son 
co--espaciales en esta parte de la galaxia.

 \cite{Maizetal98} realizaron en escaneo mediante espectroscop'ia de
 alta resoluci'on de NGC~4214 con el  cual pudieron crear mapas
 de varios par'ametros f'isicos incluyendo a la extinci'on. Encuentran que la regi'on de 
 formaci'on estelar m'as importante (NGC~4214--I) presenta
 un medio interestelar afectado por vientos estelares y/o 
 explosiones de supernovas. Esto muestra claramente diferencias en 
 la distribuci'on espacial entre estrellas, gas y polvo. Esto se pone claramente
 en evidencia en nuestro mapa de extinci'on, donde las dos cavidades 
 principales en NGC 4214--I muestran una regi'on de baja extinci'on
rodeada de valores m'as altos. El polvo estar'ia concentrado en el borde de estas
regiones afectando principalmente a la emisi'on nebular y d'ebilmente al
continuo estelar.

Tanto  \cite{Maizetal98} como \citet{MacKetal00} 
encontraron que el enrojecimiento en   NGC 4214--II 
es, en promedio, mayor que en NGC 4214--I. Nuestros resultados, derivados 
a partir de los colores fotom'etricos de las estrellas, est'an de acuerdo con este hecho.
mientras que NGC 4214--II  tiene una extinci'on alta en
general. 
Ellos muestran diferencias importantes en la estructura de estas dos regiones
que nosotros confirmamos con nuestro estudio de extinci'on estelar. El continuo estelar y la
emisi'on nebular se producen  de manera co--espacial en NGC 4214--II, mientras
que, seg'un estos autores, el gas en emisi'on se encuentra claramente desplazado
con respecto a los c'umulos estelares en NGC 4214--I.

En resumen, en la presente Tesis   estudiamos la extinci'on del continuo estelar en 
un campo de observaci'on que abarca tanto a NGC 4214--I como a NGC 4214--II.
Nuestra conclusi'on es que la extinci'on derivada a partir del 
estudio del continuo estelar es similar a la extinci'on derivada del an'alisis de las l'ineas 
nebulares por   \cite{Maizetal98} y por  \citet{Maiz00}.
  Encontramos adem'as que esto se verifica para cada una de las estrellas en el 
  campo de observaci'on.
La coincidencia es bastante buena en toda la estructura de la galaxia.

\section{El cociente B/R de supergigantes}
Analizamos la presencia de estrellas supergigantes rojas y azules
con el objeto de determinar por vez primera
el cociente B/R para NGC~4214.
Este cociente es un observable sumamente importante en Astrof'isica
debido principalmente a las restricciones fundamentales que puede imponer
sobre los modelos te'oricos de evoluci'on estelar.  
Se sabe que el cociente B/R es una cantidad sensible a la p'erdida de masa, la rotaci'on, la convecci'on
y los procesos de mezclado en el interior estelar; por tal motivo,
la determinaci'on de B/R constituye una prueba  muy sensible e importante sobre los modelos de evoluci'on estelar
\citep{MaedCont94,LaMa94,Eggeetal02}.

El cociente B/R es muy dependiente de c'omo se cuentan las estrellas, y por lo tanto,
las comparaciones  con las predicciones de los modelos evolutivos tiene que ser realizados
con mucho cuidado. Nosotros
seguimos el m'etodo sugerido por \cite{MaedMeyn01}.

Comparamos nuestros resultados con tres grillas de modelos te'oricos
calculados con la metalicidad de las Nubes de Magallanes ($Z=0.004$ y $Z=0.008$)
y arribamos a la conclusi'on de que  el mejor ajuste es el del modelo de Ginebra
con $Z=0.004  $  \citep{MaedMeyn01}.

Estos modelos predicen un  valor B/R de 24 en el intervalo de masas  $ 15-20\, M_{\sun}$,
y nosotros medimos $34 \pm 10$.
En el rango de masas $ 20-25 \, M_{\sun}$ la predicci'on te'orica
da el valor 47 y nosotros obtuvimos   $46 \pm 23$.
En ambos casos, nuestros resultados est'an de acuerdo con los te'oricos 
dentro de posibles errores de Poisson.  Discutimos dos posibles errores en nuestra
determinaci'on del valor de B/R: los efectos estoc'asticos debido a la peque~na cantidad
de estrellas supergigantes rojas en nuestra muestra (mayor fuente de error), 
y la conversi'on entre 
colores observados y temperatura efectiva y magnitud bolom'etrica.
Hallamos una discrepancia de $350-400$ K entre  nuestros datos 
y los modelos te'oricos que podr'ian ser explicados ajustando a los 
colores fotom'etrico observados con los modelos de atm'osferas MARCS
en lugar de las atm'osferas de Kurucz.

Encontramos adem'as que la morfolog'ia de la distribuci'on de supergigantes es notable en esta galaxia,
principalmente en lo que respecta a las diferencias entre las regiones NGC~4214--I
y  NGC~4214--II:
en  NGC~4214--IA y --IB se observa claramente que las supergigantes son azules, mientras
que las supergigantes  rojas se distribuyen solamente en la periferia. 
En cambio, en  NGC~4214--II se puede
apreciar una distribuci'on m'as homog'enea. Esta diferencia marcada
estar'ia  relacionada 'intimamente con la diferencia de edad entre ambas regiones ($3.0-4.0$ Ma 
para NGC~4214--I y   $1.0-2.0$ Ma 
para NGC~4214--II). La regiones I--A y I--B son m'as evolucionadas y las estrellas 
supergigantes rojas han desaparecido mientras que son visibles en la regi'on m'as joven NGC~4214--II.

Nuestras observaciones est'an de acuerdo con un mapa que muestra
la distribuci'on espacial de las estrellas rojas y azules m'as brillantes en
\cite{Droz02}. Ellos muestran que las estrellas m'as j'ovenes se agrupan
en los dos complejos principales de formaci'on estelar, con la mayor densidad 
en la cavidad dentro de NGC~4214--I--A.

\section{La funci'on inicial  de masa} 
El estudio de la IMF de NGC~4214 fue encarado en varias investigaciones previas
empleando el ajuste de espectros en el UV  con modelos de s'intesis evolutiva.

 \cite{Leitetal96}    hallaron que, si la IMF de este starburst sigue una ley
de potencias de la forma $\frac{dN}{dm} = f(m) = A \cdot m^{\gamma}$, 
entonces el exponente debe estar comprendido entre  $ \gamma = -2.35$ 
y $ \gamma = -3.00$. Esta estimaci'on  es bastante conservadora ya que se presentan
varias incertidumbres asociadas a los ajustes del continuo, la presencia de l'ineas 
interestelares y efectos de la metalicidad. N'otese  que   \cite{Leitetal96}   emplean
una  distancia de 4.1 Mpc a NGC~4214 y una metalicidad de $Z=0.01$, diferentes
a las adoptadas en este trabajo.

\cite{MasHKunt99} 
emplean codigos sint'eticos \citep{MasHKunt91a} 
para ajustar espectros  UV de NGC~4214 obtenidos con el  IUE  que 
 incluyen claramente estrellas en c'umulos as'i
como estrellas de campo. Encontraron que el mejor ajuste
se da para la pendiente   $ \gamma = -3.0$.
La pendiente de la IMF que ellos estiman es muy aproximada al valor determinado por 
 \cite{Leitetal96}. 

 \cite{Chanetal05}
emplearon un espectro de NGC~4214 obtenido 
con  STIS en el UV lejano con el objeto de  estudiar el contenido estelar
de las regiones localizadas entre c'umulos. Compararon al espectro extra'ido con
los modelos provistos por \s99  y encontraron que las regiones estudiadas  tienen
un espectro dominado por estrellas  de tipo espectral B. Sus observaciones indican que
las regiones entre c'umulos contienen estrellas menos masivas que los c'umulos 
cercanos. 
\cite{Chanetal05}  estimaron el valor  $ \gamma = -3.5$
para las estrellas de campo en NGC~4214   usando  
$M_{low} = 1 \, M_{\odot} $ y 
$M_{up} = 100\,  M_{\odot}$ en un escenario  de formaci'on estelar continua.
Las regiones analizadas incluyen c'umulos estelares,
y por 
lo tanto, deben contener algunas estrellas masivas que no pertenecen al campo.
Esto implica que el valor obtenido de la pendiente representa un l'imite inferior.

Nosotros abordamos el estudio de la funci'on inicial de masa de NGC~4214 en forma diferente,
pues usamos al m'etodo m'as directo y confiable que existe para  obtener la IMF de una dada poblaci'on estelar
y que se basa en  conteos de estrellas como una funci'on de su luminosidad/masa \citep{Lequ79},
A este enfoque le agregamos  varias mejoras al tener  en cuenta
cuatro fuentes de efectos sistem'aticos (incompletitud de los datos,
tama~no 'optimo del bin en el ajuste, difusi'on de masa y sistemas no resueltos)
presentados en  \cite{Maizetal05}.

Nuestro estudio detallado y exhaustivo nos permite adoptar 
el valor   $ \gamma_{real} = -2.83 \pm 0.07$  como un valor representativo
de la pendiente de la IMF de  NGC~4214,  el cual fue calculado usando
 $M_{low} = 20 \, M_{\odot}   $ y $ M_{up} = 100\,  M_{\odot}$ en  un escenario  de formaci'on estelar continua.
Debemos mencionar que el error citado en la pendiente (0.07) es un valor formal que sale de los 
ajustes por cuadrados m'inimos. 
Debido a que la correcci'on por objetos m'ultiples es incierta (dada la gran distancia
a la que se encuentra la galaxia) y sabiendo que los objetos no resueltos producen 
un aplanamiento sistem'atico de la pendiente,
concluimos que el valor real de la pendiente de la IMF 
  debe ser m'as empinada que $ \gamma= -2.83$.

Nosotros estimamos que el valor ``real''  de la pendiente de la IMF 
de NGC~4214 est'a comprendido entre el valor 
propuesto por   \cite{Chanetal05}  ($\gamma = -3.5$), y el obtenido en este trabajo ($\gamma =  -2.83$).

El valor estimado para la pendiente de la IMF en
NGC~4214 es comparable al calculado por  \cite{Gregetal98} en un  estudio
 de la poblaci'on resuelta
de  NGC~1569 que es considerada la galaxia starburst enana m'as cercana
($2.2  \pm 0.6$ Mpc). Sus simulaciones ajustan
la IMF de NGC~1569 con una pendiente m'as empinada que Salpeter   ($\gamma$ entre  $ -2.6$ y $ -3.0$), 
si consideran estrellas de campo menos masivas que $\sim 30 \, M_{\odot}$.
La pendiente de la IMF que presentamos
en este trabajo  est'a de acuerdo con valores estimados para algunas  asociaciones OB 
en la LMC y en la V'ia L'actea \citep{Massetal95,Massetal95b,Masse98} y con los valores
inferidos por los trabajos que ajustan espectros y que fueron mencionados con anterioridad.

\section{Objetos extendidos}
 
\subsection{Determinaci'on de edad, extinci'on,  masa y enrojecimiento}

Buscamos los mejores ajustes de 
modelos de s'intesis evolutiva provistos por \s99 \citep{Leitetal99} a los colores fotom'etricos
de 13 c'umulos en NGC~4214.
Analizamos cuatro c'umulos compactos no resueltos, tres complejos grandes que incluyen
c'umulos m'as peque~nos y seis c'umulos resueltos.

Encontramos que los modelos que mejor se ajustan 
con nuestras observaciones  del c'umulo I--As son los que tienen
la extinci'on de la Nube Menor de Magallanes  \citep{GordClay98}.
Estos modelos predicen que  I--As es un  c'umulo
estelar joven  ($4.0 \pm 0.6$ Ma), masivo  ($\approx 27\,000 M_{\odot}$)
y con baja extinci'on ($E(4405-5495) = 0.04 $ mag ).
Con nuestro estudio estamos confirmando la presencia de estrellas
WR dentro de su estructura, con una cantidad cercana a la 
determinada  previamente 
 por  \cite{Leitetal96}
y \cite{Maizetal98}.

\cite{MacKetal00} estimaron el ancho  equivalente $W$(H$\alpha$)  de la linea
de Balmer H$\alpha$ usando tres correcciones por extinci'on de todos
los c'umulos en NGC~4214, para determinar sus edades.
Para los c'umulos  NGC~4214--I--A y  I--B estimaron una edad
en el rango $4-5$ Ma, la cual concuerda con nuestra determinaci'on
(5 Ma).
La edad que estos autores estiman para el \cu I--Ds es 7 Ma (sin dar una estimaci'on de
la incertidumbre), mientras nosotros
obtenemos un valor m'as chico: $2.6 \pm 1.5$ Ma.
La fotometr'ia disponible para el  c'umulo I--Es  
nos permite inferir dos posibles soluciones muy diferentes en edad y masa
y no pudimos hallar informaci'on adicional que nos permitiera 
elegir a una soluci'on por sobre la otra.

\cite{MacKetal00}  encuentran que todos los c'umulos 
dentro de la estructura del complejo NGC~4214--II
tienen anchos equivalentes cercanos o mayores que 1000 \AA,
excepto II--C, lo cual indica que este complejo es muy joven. 

 Nuestras observaciones dan edades consistentes en todas
 las componentes, incluyendo a II--C.
El valor  de  $W$(H$\alpha$)  del complejo NGC~4214--II como un todo  es bastante menor que 
 1000 \AA\ seg'un \cite{MacKetal00}. Estos autores explican que se podr'ia deber a
 la existencia de una poblaci'on estelar subyacente.
Teniendo en cuenta el efecto producido por esta poblaci'on estiman el valor $2.5-3.0$ Ma
para la edad del c'umulo II.
Nuestra estimaci'on de la edad es  $1.9 \pm 0.9$ Ma lo cual
est'a de acuerdo con la determinaci'on previa de \cite{MacKetal00}.
El valor de $W$(H$\alpha$) calculado para NGC~4214--I es en general menor
que el de las componentes de NGC~4214--II
y \cite{MacKetal00} estiman una edad promedio en el rango $3.0-4.0 $ Ma.

Nuestro estudio de los c'umulos IIIs y IVs considera seis colores fotom'etricos
que cubren longitudes de onda desde el UV hasta el IR con im'agenes de 2MASS. 
Esto nos permiti'o hallar restricciones a la edad y masas de estos c'umulos.
Empleamos adem'as varios argumentos (extinci'on, cantidad de 
estrellas supergigantes rojas, apariencia de la imagen) para decidir que la soluci'on 
m'as vieja debe ser elegida como la correcta por sobre la soluci'on joven.

\cite{Billeetal02}   estudiaron a los c'umulos IIIs y IVs como parte
de un relevamiento  de c'umulos  estelares compactos 
en galaxias cercanas que incluy'o a NGC~4214.
Usando una traza evolutiva calculada con los modelos de 
s'intesis evolutiva \s99 que desplegaron sobre un diagrama color--color, 
\cite{Billeetal02} tratan de inferir la edad de estos c'umulos. 
La trayectoria evolutiva   es claramente degenerada
como puede observarse en la Figura~\ref{fig083} y muestra varias soluciones
posibles para la edad de estos c'umulos.
Estos autores emplearon solamente dos colores fotom'etricos, lo cual se 
traduce en una muy pobre determinaci'on de la edad de los \cus  \citep{deGretal03a}.

\subsection{M'etodo fotom'etrico}

La funci'on de luminosidad de los \cus y la distribuci'on de sus colores
son las herramientas m'as importantes en el estudio de las poblaciones de \cus en galaxias
cercanas.  Obtener espectroscop'ia individual de los \cus
es un proceso que requiere de mucho tiempo de observaci'on, ya que 
se requiere de un gran n'umero de \cus para disponer de resultados estad'isticos significativos.
 \cite{deGretal03a} analizaron las incertezas sistem'aticas en edad, extinci'on  y metalicidad
 de \cus estelares j'ovenes, que provienen del uso de colores
 fotom'etricos y modelos de s'intesis evolutiva.
Sus trabajos llevan a la conclusi'on de que se necesitan al menos  cuatro 
filtros que incluyan tanto una banda azul como una roja para poder
hacer determinaciones confiables de esos par'ametros.
Esta elecci'on llevar'a a la distribuci'on de edad m'as representativa. 
Otros trabajos en la misma linea de an'alisis pueden verse
en  \cite{Andeetal04} y  \cite{deGretal05}.

En nuestro estudio de los objetos extendidos de NGC~4214, 
nosotros empleamos siempre al menos cinco filtros en cada caso y determinamos 
todos los par'ametros libres [$\log({\rm edad/a}) $, $E(4405-5495)$,  $R_{5495}$ y masa]
 individualmente para cada \cu como
fue sugerido por \cite{deGretal03b}.
La precisi'on con que se pueden determinar las edades depende
del n'umero de filtros en diferentes bandas empleados y, principalmente,
en cu'al es el rango de  longitudes de onda
cubierto por las observaciones.

En nuestro an'alisis de los \cus nos encontramos con algunas degeneraciones
que podr'ian haber sido resueltas si hubi'esemos tenido una medici'on
de la magnitud cerca  del lado derecho de la discontinuidad de Balmer
como por ejemplo el filtro F439W (WFPC2 $B$).
Sin embargo, estamos seguros de que nuestras estimaciones de 
edad de los \cus de NGC~4214 que analizamos son correctas.

\subsection{Mortalidad infantil}

Encontramos que  NGC~4214--IIIs y --IVs son \cus masivos
compactos muy viejos, con edades   mayores que $100$ Ma.
Al estudiar otros  diez \cus (un c'umulo compacto no resuelto, tres complejos grandes
y seis \cus resueltos)  obtuvimos edades menores que $10$ Ma.

\cite{Falletal05} estudiaron la distribuci'on de edades de la poblaci'on de
c'umulos estelares  en las galaxias Antenae, usando im'agenes obtenidas
en varios filtros.
Estimaron las edades de los \cus ajustando espectros sint'eticos
de los modelos de  \cite{BruzChar03} 
y encontraron que la distribuci'on de edades decrece r'apidamente, comenzando
a edades muy tempranas.
La edad media de los \cus que encontraron  es $\sim10$ Ma. De acuerdo a su 
estudio, luego de $\sim10$ Ma, el brillo superficial de los c'umulos
se hace 5 mag m'as d'ebil que el original (a $\sim1$ Ma), y por tal motivo,
el c'umulo desaparece dentro de las fluctuaciones estad'isticas
entre  las estrellas de campo y 
el fondo.

 \citeauthor*{Falletal05}  denominaron a este efecto  la ``moratilidad infantil'' del c'umulo.
Nuestra muestra de \cus en NGC~4214 incluye 10 objetos m'as j'ovenes que
10 Ma, y dos de ellos  (NGC~4214--IA y --IB)
ya muestran efectos de ruptura.
?`Qu'e pudo haber causado la ruptura de estos c'umulos?
\citeauthor*{Falletal05} explican que 
el momentum proveniente de las estrellas masivas viene en la forma de radiaci'on ionizante,
vientos estelares, jets y supernovas, y todos estos procesos podr'ian f'acilmente
desplazar e incluso destruir al ISM de la nube original dejando a las estrellas 
dentro del mismo gravitacionalmente libres. 

 \cite{Claretal05} proveen otra explicaci'on: estos autores
realizaron simulaciones y mostraron que nubes moleculares gigantes 
no ligadas  logran formar varios \cus estelares y luego se dispersan 
en $\sim10$ Ma, haciendo de los  mismos, objetos transitorios. 
Luego de $\sim10$ Ma estos objetos no podr'ian ser reconocidos como
c'umulos. 
Este resultado es confirmado por varios  trabajos tanto te'oricos como
emp'iricos. Ver por ejemplo \cite{LadaLada03} donde se hace
un estudio  completo y las referencias ah'i citadas.
\cite{Lameetal05a}  presentan una descripci'on anal'itica
muy simple
de la ruptura de los \cus estelares y encuentran que aproximadamente 
la mitad de los \cus en las cercan'ias del Sol se convierten en  
gravitacionalmente libres  en 
10 Ma. 

?`C'omo se explica la existencia de c'umulos con edades mayores que
10 Ma como IIIs y IVs?
Los argumentos que \citeauthor*{Falletal05} presentan
sugieren una manera de explicar la ruptura y desintegraci'on
de los c'umulos, no su supervivencia. 
Esto parece ser m'as dif'icil de explicar.
La supervivencia de un \cu parece depender de otros factores 
como cu'ando y d'onde se han formado las estrellas masivas
dentro de la nube molecular original.

Se supone que los n'ucleos internos (y m'as densos) de 
las nubes que originan a los c'umulos tienen m'as posibilidades de
sobrevivir que sus envolturas.
Podemos suponer entonces que lo que estamos observando al ver
a los c'umulos NGC~4214--IIIs y --IVs son los n'ucleos densos
originales que han perdido las estrellas que formaban su envoltura
y que se las han arreglado para mantenerse ligados gravitacionalmente 
por bastante m'as que 10 Ma. Esto los har'ia candidatos a 
progenitores de c'umulos globulares  \citep{deGretal05b}, 
como en el caso de R136 en la LMC.

\begin{figure*}[h!]
\centering
\includegraphics[width=\linewidth]{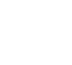}
\vspace{-4mm}

\begin{minipage}[c]{8.0cm}
\includegraphics[width=79mm]{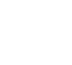}
\end{minipage}
\hfill
\begin{minipage}[u]{75mm}
\caption[Destrucci'on del ISM
en los  c'umulos NGC~4214--I--A,  LMC--N11 y LMC--30~Dor]{{\sl\footnotesize   La destrucci'on del medio interestelar por 
la interacci'on del mismo con las estrellas masivas es evidente en 
estas tres im'agenes: [Arriba izquierda] detalle del \cu I--A. Se observa c'omo 
el ISM fue moldeado en forma de coraz'on y barrido completamente
en la linea de la visual. El \cu I--B tambi'en muestra se~nales 
de ruptura. El c'umulo central es el I--As que podr'ia haber 
producido una segunda etapa de formaci'on estelar en 
I--A1n y I--A3n.  
Cr'edito de la imagen:  John MacKenty (STScI), the Hubble Heritage Team (AURA/ STScI/ NASA).
[Arriba derecha] Im'agen de N11 en la Nube Mayor de Magallanes. 
La morfolog'ia es comparable con la de I--A en NGC~4214. El c'umulo
central es la asociaci'on LH9. 
Cr'edito de la imagen:  Aguilera, C. Smith,  S. Points/NOAO/AURA/NSF.
[Izquierda] Mosaico de 30~Dor en la Nube Mayor de Magallanes. El c'umulo
central que dispar'o la formaci'on estelar en la periferia es    Radcliffe  136.
Cr'edito de la imagen: Jes\'us Ma'iz-Apell\'aniz  (STScI), Nolan R. Walborn (STScI), Rodolfo H. Barb\'a
(Observatorio Astron'omico de La Plata). Las im'agenes no est'an a escala.
   \label{fig084}}}
\end{minipage}
\end{figure*}

\subsection{Formaci'on estelar de segunda generaci'on}
 Nuestras im'agenes de alta resoluci'on de NGC~4214
 muestran la evoluci'on en la interacci'on entre 
 estrellas masivas y el ISM, a trav'es de la morfolog'ia de
 varias regiones:  Las regiones \hii alrededor de 
 los \cus II--A y II--B son compactas y estan llenas,
 mientras que las regiones ubicadas alrededor de 
 los \cus m'as viejos como I--As y I--Bs son de mayor
 tama~no y su parte central ha sido evacuada de materia.
Un caso intermedio podr'ia ser el observado en II--C, y en
I--Ds.  
Estos diferentes estados pueden explicarse mediante la transferencia de 
energ'ia entre las estrellas masivas y el ISM.

La ruptura del medio interestelar da origen a otro efecto sumamente interesante
que fue descripto por \cite{ElmeLada77} y por \cite{WalbPark92} y que se denomina
``two--stage starburst'' o formaci'on estelar en dos etapas.
\citeauthor*{WalbPark92} estudiaron las regiones
N11 y 30~Dor en la Nube Mayor de Magallanes.
30~Dor es un objeto que ha sido estudiado en profundidad 
\citep{Walb86,Campetal92,Hunt95b,WalbBlad97,Rubietal98a,Walbetal99c}
(por nombrar s'olo algunos pocos trabajos) y que sorprendentemente
presenta evidencia sustancial de formaci'on estelar en las cercan'ias de los 
filamentos de la Nebulosa Tar'antula que rodean al \cu central R136,
especialmente hacia el norte y oeste.
Se han observado varias estrellas de tipo espectral O muy tempranas
que estar'ian todav'ia encerradas dentro de la nube molecular que
les dio origen \citep{WalbBlad97,WalbPark92}. Estudios en el IR 
\citep{Hylaetal92,Rubietal92} secundan estos descubrimientos.

Todos estos estudios implican la presencia de procesos secuenciales de formaci'on estelar.  
La morfolog'ia y  el contenido estelar de tanto 30~Dor como N11   
  sugieren la hip'otesis de formaci'on estelar en dos etapas. 
 La asociaci'on en el centro de  N11 (LH9) tiene una edad  5 Ma. 
 El \cu central de 30~Dor tiene $2-3$ Ma y la nueva poblaci'on que la rodea 
debe ser menor que $\sim 1$ Ma.  
Evidentemente, en ambas regiones   un starburst central ha producido  luego de
2 Ma,
un starburst secundario
en su periferia.
El proceso en N11 se encuentra un poco m'as avanzado por lo que 
la apariencia actual de N11 es la que 30~Dor tendr'a en unos 2 Ma \citep{WalbPark92}.
Este trabajo original es secundado por varios trabajos posteriores
 \citep{Rubietal98a,Walbetal99c,Branetal01,Walbetal02,Barbetal03a}

Podemos argumentar que NGC~4214 presenta este mismo 
tipo de comportamiento en su estructura, principalmente en el \cu I--A.
Una segunda generaci'on de estrellas puede ser producida $2-3$ Ma luego de 
la formaci'on del primer starburst de $3-5$ Ma. 
\cite{MacKetal00} sostienen que las regiones I--A1n y I--A3n (ver Figura~\ref{fig018})
est'an siendo ionizadas in situ y no por al \cu I--As, lo cual 
se asemeja enormemente a la morfolog'ia de N11.
Estos autores proponen que I--As + (I--A1n + I--A3n)
podr'ia ser un ejemplo de formaci'on estelar en dos fases presente 
en NGC~4214  como se puede apreciar en la Figura~\ref{fig084}.
La confirmaci'on de esta hip'otesis requiere de espectroscop'ia UV de estas 
regiones.


\renewcommand{\theequation}{A-\arabic{equation}}
\setcounter{equation}{0}  

\appendix
\addtocontents{toc}{\protect\vspace{-0.125in}}
\label{appendixa}

\chapter{Elecci\'on del tama\~no del bin}
\thispagestyle{empty}
\newpage

\section{La funci'on inicial de masa (IMF)}

Varios estudios en Astronom'ia necesitan de  la determinaci'on de
la funci'on inicial  de masa (IMF) de     una dada  poblaci'on estelar.
Lo que se suele hacer es
 medir la luminosidad de  cada estrella, convertir los resultados en masas,
y ajustar una ley de potencias o funci'on similar a las masas ordenadas
en bines usando el m'etodo de cuadrados m'inimos. 
Nosotros estudiamos diferentes posibles maneras
de realizar este ajuste mediante  varios experimentos num'ericos
con  datos artificiales. Los resultados fueron publicados en  \cite{MaizUbed05}.

 Siguiendo la notaci'on de \cite{Scalo86}, definimos al espectro de
 masas $f(m)$  de manera tal que $f(m)\cdot dm$ es el n'umero de
estrellas formadas en el mismo momento en cierto volumen de espacio con masas
en el rango   $m$ a $m+dm$.
 
 Las unidades de $f(m)$  se definen mediante una normalizaci'on adecuada.
 Usualmente se trata a $f(m)$  como una densidad de probabilidad, de manera tal que

 \begin{equation}
\label{eqn-00} 
  \int_{M_{up}}^{M_{low}}  f(m)   \cdot dm =1,
\end{equation}

donde $M_{up}$ y $M_{low}$  son los l'imites superior e inferior de las masas,
cuyos valores, se piensa en la actualidad, var'ian alrededor de
$M_{low} \approx 0.05-0.1 M_{\odot} $   y    $M_{up} \approx 120-200 M_{\odot} $.


En el caso de estrellas masivas en regiones de formaci'on estelar,
podemos contar la cantidad de 'estas como funci'on de su masa:
esta es la funci'on de masa presente, y se la puede
obtener directamente.
La manera en que esta cantidad est'a relacionada con la IMF depende
de la historia de formaci'on estelar para la regi'on que se est'a estudiando.
Nosotros consideramos dos casos extremos: (1) cuando todas las estrellas de una 
regi'on fueron formadas simult'aneamente, y (2) cuando el proceso de formaci'on estelar fue
continuo.

El n'umero de estrellas formadas con masas entre  $m$  y  $m+dm$  en el 
intervalo de tiempo  entre $t$ y $t+dt$  est'a dado por  $f(m) \psi(t) \cdot dm \cdot dt$,
donde $\psi(t)$ es la tasa de formaci'on estelar.
El n'umero de estrellas observadas en el momento $t$ con masas entre 
 $m_1$ y $m_2$ est'a dado por:
 
\begin{equation}
\label{eqn-00} 
N(t) |^{m_2}_{m_1} = \int_{m_1}^{m_2} \int_{t_0}^{t} f(m) \psi(t')  \cdot dm \cdot dt',
\end{equation}

si la formaci'on estelar comenz'o en el instante $t=t_0,$ y si la
edad promedio de una estrella en la \sp  $\tau_{MS}$  con masa
 $m_2$ es menor que  $t.$
 
La forma del espectro de masas m'as com'unmente empleado es la
ley de potencias  \citep{Scalo86,Masse98}:

\begin{equation}
\label{aa}  
\frac{dN}{dm} = f(m) = A \cdot m^{\gamma},
\end{equation}

donde la cantidad $A$
es un par'ametro de normalizaci'on.

Cuando todas las estrellas de una regi'on se forman en un
mismo evento (en el instante  $t=t_0$),
la tasa de formaci'on de estrellas es simplemente una func'on delta
y la integral se simplifica a :

\begin{equation}
\label{eqn-02} 
\int_{m_1}^{m_2} \frac{dN}{dm} dm = A \cdot \int_{m_1}^{m_2} m^{\gamma} dm
\end{equation}

\begin{equation}
\label{eqn-03} 
N |^{m_2}_{m_1} = N(m_2) - N(m_1) = \frac{A}{\gamma + 1} \cdot \left[ {m_2}^{\gamma + 1}- 
{m_1}^{\gamma + 1} \right]  \hspace{1cm}; \gamma \neq -1
\end{equation}

Definiendo  $\Delta m = m_2 - m_1$  y $x= \frac{m_1+m_2}{2}$ es posible reescribir
a la Ecuaci'on~\ref{eqn-03} como

\begin{equation}
\label{eqn-04} 
N_i = N(m_2) - N(m_1) = \frac{A}{\gamma + 1} \cdot \left[   \left( x + \frac{\Delta m}{2} 
\right)^{\gamma + 1}-  \left( x-\frac{\Delta m}{2} \right)^{\gamma + 1} \right]
\end{equation}

Aplicando logaritmos en ambos miembros obtenemas una expresi'on
de la forma $y_i$ ( logaritmo del n'umero de estrellas por bin $i$)
como funci'on de  $x_i$  (masa en el centro del bin $i$
definida en la escala lineal):

\begin{equation}
\label{eqn-05} 
y_i = \log N_i = \log \left(  \frac{A}{\gamma + 1} \cdot \left[   \left( x + \frac{\Delta m}{2} 
\right)^{\gamma + 1}-  \left( x-\frac{\Delta m}{2} \right)^{\gamma + 1} \right]  \right).
\end{equation}

\section{Experimentos num'ericos}

Para nuestros experimentos num'ericos usamos un generador de n'umeros aleatorios
para producir 1000 listas con 1000 estrellas cada una, distribuidas de 
acuerdo a la ecuaci'on~(\ref{aa}) con una pendiente de Salpeter  ($\gamma = -2.35$)
entre   $m = 6.31 M_{\odot} \,  (\log_{10} m/M_{\odot}  = 0.8) $  y 
$m = 158.49 M_{\odot}  \,  (\log_{10} m/M_{\odot}  = 2.2) $.
Los extremos fueron seleccionados como los t'ipicos 
en estudios de estrellas masivas, donde la IMF
medida  dio valores cercanos a Salpeter 
bajo diferentes circunstancias,
y supusimos que nuestros resultados no depender'ian de estos valores.

Las 1000 masas de estrellas fueron generadas usando el 
siguiente procedimiento de   \cite{Bevi92}:

Consideremos la distribuci'on uniforme

\begin{equation}
\label{eqn-06} 
p(r) =  \left\{  \begin{array}{rcc}
1& :&0 \le r < 1 \\ 
0 & :& \mbox{\rm  si no}
\end{array} \right.
\end{equation}

normalizada de manera tal que

\begin{equation}
\label{eqn-07} 
\int_{-\infty}^{\infty} p(r)   dr = 1
\end{equation}

La relaci'on general para obtener una variable aleatoria
$x$ a partir de una distribuci'on de probabilidad  $f(x)$ 
en t'erminos de una variable $r$ sacada de una distribuci'on uniforme  $p(r)$ 
est'a dada por la siguiente ecuaci'on:

\begin{equation}
\label{eqn-08} 
r=\int_{x_{min}}^{m} f(x) dx
\end{equation}

Luego, para hallar $m$, seleccionada aleatoriamente de la distribuci'on de probabilidad 
 $f(x)$, generamos un n'umero aleatorio $r$ a partir de la distribuci'on uniforme
 y hallamos el l'imite $m$ que satisface la ecuaci'on integral ~(\ref{eqn-08}).

Como ejemplo, consideremos la ley de potencias con pendiente Salpeter

\begin{equation}
\label{eqn-09}  
f(x) =  \left\{  \begin{array}{rcc}
A \cdot x^{-2.35} &:& x_{min} \le x <  x_{max} \\ 
0 &:& \mbox{\rm  si no}
\end{array} \right.
\end{equation}

Obtenemos $A$ suponiendo que:

\begin{equation}
\label{eqn-10} 
\int_{-\infty}^{\infty} f(x) dx = 1
\end{equation}

Lo cual implica que 

\begin{equation}
\label{eqn-10} 
A = \frac{1.35}{x_{min}^{-1.35}   - x_{max}^{-1.35}    }
\end{equation}

Reemplazando en la Ecuaci'on~\ref{eqn-08}:

\begin{equation}
\label{eqn-11} 
r=\int_{x_{min}}^{m} f(x) dx = \int_{x_{min}}^{m} A \cdot x^{-2.35} dx = 
 A \cdot  \frac{m^{-1.35}   - x_{min}^{-1.35}    }{-1.35}  
\end{equation}

\begin{equation}
\label{eqn-12} 
r= \frac{m^{-1.35}   - x_{min}^{-1.35}    }{x_{max}^{-1.35} - x_{min}^{-1.35}    }
\end{equation}

Con la Ecuaci'on~\ref{eqn-12}  podemos generar una lista de masas
( $m$) que obedecen la una ley de potencias
con pendiente Salpeter asignando a  $r$ valores aleatorios obtenidos de
una distribuci'on uniforme.

Consideramos varios n'umeros de objetos seleccionando los primeros   30, 100, 300 y 1000 estrellas. 
Tambi'en experimentamos con agrupar a 
las estrellas en  3, 5, 10, 30 o 50 bines  para estudiar si
existe un tama'no 'optimo.
 
Desarrollamos experimentos usando tres tipos de bines:

\begin{description}
\item[experimento 1] Bin de tama~no uniforme en escala logar'itmica con el extremo izquierdo
del  primer bin, $m_{\mbox{\tiny down}}$, igual a
${10}^{0.8} M_{\odot}  $ y el extremo derecho del 'ultimo bin, $m_{\mbox{\tiny up}}$, igual a
 ${10}^{2.2} M_{\odot}  $ (i.e. los mismos valores usados
 para la generaci'on de las listas).

\item[experimento 2]  N'umero de estrellas aproximadamente constante en cada bin con 
$m_{\mbox{\tiny down}} =  {10}^{0.8} M_{\odot}   $ 
 y $m_{\mbox{\tiny up}} =  {10}^{2.2} M_{\odot}$. Esta regla no puede 
 hacerse exacta siempre, ya que no es posible
 dividir, por ejemplo,  100 estrellas en  30 bines con el mismo n'umero de estrellas
 cada uno.
 En tal caso, dise~namos a nuestros bines para que 
 contengan e.g. 3, 4, 3, 3, 4, 3. . . estrellas.

\item[experimento 3]  Lo mismo que en el caso anterior pero con
 $m_{\mbox{\tiny  down}}$ y  $m_{\mbox{\tiny up}}$  determinados
 a partir de los datos en cada lista.

\end{description}

El n'umero de objetos por bin  ($N_i$) sigue una distribuci'on binomial caracterizada
por $N$ y $p_i$, donde: 

\begin{equation}
\label{eqn-13} 
p_i= \frac{  \int_{x_i-\Delta m_i /2}^{x_i+\Delta m_i /2}   f(m) dm }{ \int_{m_{\mbox{\tiny down}} }^{m_{\mbox{\tiny up}}}  
    f(m) dm  }
\end{equation}

El problema ac'a es que el verdadero $p_i$ debe determinarse
a partir de 
la verdadera  $f(m)$, que se desconoce. Una alternativa consiste en sustituir
$p_i$  por su aproximaci'on $N_i/N$, que se deriva de los datos, de manera tal que
la incerteza asociada con 
$ y_i = N_i$ es $\sqrt{N_i ( N - N_i)/N}  $, y la de $y_i =  \log N_i$  es 
 $\sqrt{\frac{N - N_i}{N_i \cdot N \cdot (\log_{10} e )^2}}  $

Al usar un algoritmo de cuadrados m'inimos para ajustar datos 
de la forma $y_i  \pm s_i $, el peso $w_i$ 
asociado es $w_i = 1/s_{i}^{2}$. En nuestro caso, tenemos:

\begin{equation}
\label{eqn-14} 
w_i= \frac{ N_i N }{ (N - N_i) \cdot  (\log_{10} e )^2   }
\end{equation}

Es importante notar que esta expresi'on da peso cero en los bines sin estrellas,
y aquellos casos con $N_i=N$ deben descartarse, pues tratar de 
ajustar una funci'on a un histograma que tiene todos los datos
en un mismo bin es un problema mal definido.

Para cada una de las  $k  \hspace{3mm} (k = 1,1000)$ listas  en cada combinaci'on de  
$N$ + tama~no-bin, ajustamos una ley de potencias con un programa
de cuadrados m'inimos desarrollado en IDL para obtener $\gamma_k$  a partir de
 $(x_i, Æm_i, y_i, w_i) $, as'i como su incertidumbre  $\sigma_k$.
 El programa tambi'en nos da $A_k$  y su incertidumbre, que no 
 necesitamos.
 
En cada experimento, calculamos las siguientes cantidades:

(1) el valor medio del exponente en la ley de potencias:

\begin{equation}
\label{eqn-15} 
\overline{\gamma} = \frac{ 1 }{ 1000  } \sum_{k=1}^{1000} \gamma_k
\end{equation}

(2) la incerteza promedio en el exponente de la ley de potencias:

\begin{equation}
\label{eqn-16} 
\overline{\sigma} = \frac{ 1 }{ 1000  } \sum_{k=1}^{1000} \sigma_k
\end{equation}

(3)  y el valor del sesgo  normalizado respecto a la incerteza:

\begin{equation}
\label{eqn-17} 
b = \frac{ 1 }{ 1000  } \sum_{k=1}^{1000} \frac{\gamma_k +2.35 }{ \sigma_k  } 
\end{equation}

El valor de $b$ es una forma de medir la existencia de sesgos. Si  $|b| \ll 1 $,  luego, el ajuste
no tiene sesgo. Si, por otro lado,  $|b| \sim 1$ o mayor, entonces el m'etodo
tiene un gran sesgo.

Tambi'en analizamos la distribuci'on de la cantidad   $(\gamma_k+2.35) / \sigma_k$, cuyo promedio
est'a dado por $b$ y cuya desviaci'on  est'andar es:

\begin{equation}
\label{eqn-18} 
\beta = \sqrt{ \frac{ 1 }{ 1000  } \sum_{k=1}^{1000}  \left( \frac{\gamma_k +2.35 }{ \sigma_k  } -b \right) ^2 }
\end{equation}

Se espera que una  t'ecnica sin sesgos  y que d'e una estimaci'on correcta de las incertezas
debe dar una distribuci'on de $(\gamma_k+2.35) / \sigma_k$ que se parezca a una Gaussiana
con  $b = 0$ y $\beta  = 1$.

\section{Resultados}
Las Tablas~\ref{tbl027}, ~\ref{tbl028} y ~\ref{tbl029} muestran 
estos valores para los tres experimentos realizados.

De estos resultados, podemos concluir que cuando uno usa 
bines de tama~no uniforme y un n'umero variable de objetos por bin, 
se obtienen los mayores errores, incluso cuando  cada bin tiene
al menos una estrella; esto nos hizo  desalentar el uso de 
bines de tama~no uniforme en el c'alculo de  una IMF.

Del segundo experimento pudimos concluir que elegir bines de tama~no variable 
ayuda a reducir los errores.

El tercer experimento da resultados similares al anterior.
La determinaci'on de los l'imites de integraci'on a partir de los datos
es una manera pr'actica de minimizar los errores. En este experimento,
los sesgos  son mayores debido a que desconocemos los extremos de la distribuci'on.


  \begin{table*}[ht!] 
 \begin{minipage}{8cm}
\hspace{2mm}\begin{minipage}{63mm}
\caption{ Valores de $\overline{\gamma}$ 
obtenidos en los tres experimentos.   \label{tbl027}} 
\end{minipage} 

 {\footnotesize
 \begin{tabular}{lcccccc}    
\multicolumn{7}{c}{ \rule{75mm}{0.8mm}}      \\

&&    \multicolumn{5}{c}{ $\overline{\gamma}$}    \\

&& 3&5& 10 &30&50     \\ 
 \multicolumn{7}{c}{ \rule{75mm}{0.2mm}}      \\
  \multicolumn{7}{c}{ Experimento 1}  \\  
  \multicolumn{7}{c}{ \rule{75mm}{0.2mm}}     \\

  30  &    &  
         -2.31        &         -2.21     &           -2.05    &            -1.69       &        -1.51 \\             
  100 &  &          
    -2.34       &         -2.31           &   -2.24          &    -2.04   &            -1.90   \\    
   300 &       &       
 -2.35  &              -2.34  &              -2.32  &              -2.23  &              -2.16\\
 1000& &  
           -2.35  &              -2.34  &              -2.34  &              -2.31  &              -2.29\\
 \multicolumn{7}{c}{ \rule{75mm}{0.2mm}}      \\
  \multicolumn{7}{c}{ Experimento 2}  \\  
  \multicolumn{7}{c}{ \rule{75mm}{0.2mm}}     \\

 30  &   &   
  -2.40         &         -2.40  &                -2.40     &             -2.39     &             \dotfill   \\     
 100 &  & 
   -2.37  &              -2.37  &              -2.37  &              -2.36  &              -2.37 \\
   300 &    & 
   -2.35 &              -2.35  &              -2.35  &              -2.36  &              -2.35\\
 1000&    & 
 -2.35  &              -2.35  &              -2.35  &              -2.35  &              -2.35  \\  
 
  \multicolumn{7}{c}{ \rule{75mm}{0.2mm}}      \\
  \multicolumn{7}{c}{ Experimento 3}  \\  
  \multicolumn{7}{c}{ \rule{75mm}{0.2mm}}     \\

30  &    &  
   -2.37  &              -2.35  &              -2.33  &              -2.30  &                \dotfill   \\
  100 &   &
    -2.36  &              -2.36  &              -2.35  &              -2.34  &              -2.34  \\
 300 &  &
   -2.35  &              -2.35  &              -2.35  &              -2.35  &              -2.35 \\
    1000&   &
    -2.35  &              -2.35  &              -2.35  &              -2.35  &              -2.35 \\
   
       \multicolumn{7}{c}{ \rule{75mm}{0.8mm}}      \\

\end{tabular}
}

\end{minipage} 
\hfill
 \begin{minipage}{8cm}
\hspace{2mm} \begin{minipage}{63mm}
\caption{  Valores de $\overline{\sigma}$
obtenidos en los tres experimentos.  \label{tbl028}} 
\end{minipage} 

 {\footnotesize
 \begin{tabular}{lcccccc}    
\multicolumn{7}{c}{ \rule{65mm}{0.8mm}}      \\

&&    \multicolumn{5}{c}{ $\overline{\sigma}$}    \\

&& 3&5& 10 &30&50     \\ 
 \multicolumn{7}{c}{ \rule{65mm}{0.2mm}}      \\
  \multicolumn{7}{c}{ Experimento 1}  \\  
  \multicolumn{7}{c}{ \rule{65mm}{0.2mm}}     \\

  30  &    &  
          0.30  &               0.28  &               0.28  &               0.29  &               0.30  \\             
  100 &  &          
   0.16  &               0.14 &               0.15 &               0.15  &               0.16  \\    
   300 &       &       
0.09  &               0.08  &               0.09  &               0.09  &               0.09 \\
 1000& &  
        0.05  &               0.05  &               0.05  &               0.05  &               0.05  \\
 \multicolumn{7}{c}{ \rule{65mm}{0.2mm}}      \\
  \multicolumn{7}{c}{ Experimento 2}  \\  
  \multicolumn{7}{c}{ \rule{65mm}{0.2mm}}     \\

 30  &   &   
   0.29  &               0.29  &               0.29  &               0.29  &               \dotfill   \\     
 100 &  & 
   0.15  &               0.15  &               0.15  &               0.16  &               0.16 \\
   300 &    & 
   0.09  &               0.09  &               0.09  &               0.09  &               0.09\\
 1000&    & 
 0.05  &               0.05  &               0.05  &               0.05  &               0.05  \\  
 
  \multicolumn{7}{c}{ \rule{65mm}{0.2mm}}      \\
  \multicolumn{7}{c}{ Experimento 3}  \\  
  \multicolumn{7}{c}{ \rule{65mm}{0.2mm}}     \\

30  &    &  
   0.31  &               0.31  &               0.31 &               0.30  &                \dotfill   \\
  100 &   &
     0.16  &               0.16  &               0.16  &               0.16  &               0.16  \\
 300 &  &
   0.09  &               0.09  &               0.09  &               0.09  &               0.09 \\
    1000&   &
   0.05  &               0.05  &               0.05  &               0.05  &               0.05\\
   
       \multicolumn{7}{c}{ \rule{65mm}{0.8mm}}      \\

\end{tabular}
}
 
\end{minipage} 
\end{table*}


 
 \begin{table*}[ht!] 
\hspace{2mm}
\begin{minipage}{67mm}
\caption{  Valores de $b$
obtenidos en los tres experimentos.  \label{tbl029}} 
\end{minipage} 

 {\footnotesize
 \begin{tabular}{lcccccc}    
\multicolumn{7}{c}{ \rule{70mm}{0.8mm}}      \\

&&    \multicolumn{5}{c}{ $b$}    \\

&& 3&5& 10 &30&50     \\ 
 \multicolumn{7}{c}{ \rule{70mm}{0.2mm}}      \\
  \multicolumn{7}{c}{ Experimento 1}  \\  
  \multicolumn{7}{c}{ \rule{70mm}{0.2mm}}     \\

  30  &    &  
    0.38  &               0.66  &               1.18  &               2.39  &               2.99  \\ 
      100 &  &          
0.18  &               0.38  &               0.77  &               2.06  &               2.99  \\  
   300 &       &       
 0.12  &               0.22  &               0.43  &               1.38  &               2.20  \\ 
 1000& &  
0.15  &               0.16  &               0.26  &               0.77  &               1.28  \\ 
 \multicolumn{7}{c}{ \rule{70mm}{0.2mm}}      \\
  \multicolumn{7}{c}{ Experimento 2}  \\  
  \multicolumn{7}{c}{ \rule{70mm}{0.2mm}}     \\

 30  &   &   
 0.01  &              -0.02  &              -0.05  &              -0.03  &               \dotfill  \\   
 100 &  & 
  0.02  &              -0.02  &              -0.02  &              -0.02  &              -0.03  \\ 
   300 &    & 
 0.03  &               0.03  &               0.01  &              -0.01  &              -0.01  \\   
 1000&    & 
 0.04  &               0.06  &               0.05  &               0.04  &               0.05  \\ 
 
  \multicolumn{7}{c}{ \rule{70mm}{0.2mm}}      \\
  \multicolumn{7}{c}{ Experimento 3}  \\  
  \multicolumn{7}{c}{ \rule{70mm}{0.2mm}}     \\

30  &    &  
     0.12  &               0.13  &               0.18  &               0.26  &                \dotfill   \\ 
  100 &   &
   0.05  &               0.05  &               0.07  &               0.14  &               0.16  \\
 300 &  &
     0.05  &               0.07  &               0.07  &               0.07  &               0.08  \\     
    1000&   &
    0.04  &               0.08  &               0.07  &               0.07  &               0.09  \\  
   
       \multicolumn{7}{c}{ \rule{70mm}{0.8mm}}      \\

\end{tabular}
}
\end{table*}
 

\begin{figure*}[h!]
\centering
\includegraphics[width=\linewidth]{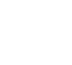}
\vspace{-5mm}

\begin{minipage}[c]{81mm}
\includegraphics[width=81mm]{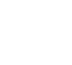}
\end{minipage}
\hfill
\begin{minipage}[u]{75mm}
\caption[Ajuste de cuadrados m'inimos para el c'alculo de la IMF]{\sl\footnotesize 
Histogramas
que muestran el ajuste de cuadrados m'inimos correspondiente a una de
las realizaciones donde se ajusta 
un total de  30 estrellas tomando 10 bines.
[Arriba izquierda] Resultado del primer experimento. 
[Arriba derecha] Resultado del segundo experimento
[Izquierda] Resultado del tercer experimento.
 \label{fig098}}
\end{minipage}
\end{figure*}


\renewcommand{\theequation}{B-\arabic{equation}}
\setcounter{equation}{0}  

\addtocontents{toc}{\protect\vspace{-0.125in}}
\label{appendixb}

\chapter{Determinaci\'on de las masas de los c\'umulos}
\thispagestyle{empty}
\newpage

\section{C'alculo de la masa a partir de la fotometr'ia}
 
Para la estimaci'on de las masas de los c'umulos  a partir de la fotometr'ia medida,
empleamos 
la relaci'on

\begin{equation}
{\rm masa} = 10^{0.4 \,  (M_{F336W,0}-M_{F336W,t}} \, M_{\odot}
\end{equation}

donde   $M_{F336W,0}$ es la magnitud absoluta de un c'umulo de $1\, M_{\odot}$
a tiempo $\log({\rm edad/a}) =6.0 $ y cuyo valor es  $M_{F336W,0}=-1.944$ para 
una metalicidad $Z=0.008$. La elecci'on del filtro F336W fue arbitraria y se la utiliz'o
para el c'alculo de masas de todos los c'umulos, excepto para el c'umulo NGC~4214--IVs
en cuyo caso se uso al filtro de referencia F555W con $M_{F555W,0} = -0.113 $.

$M_{F336W,t}$ es la magnitud absoluta en el filtro F336W del c'umulo
corregida por  extinci'on y referida a edad cero de la siguiente manera:

 \begin{equation}
M_{F336W,t} = m_{F336W} - C(t) - A_{F336W} +5-5\log(d)
\end{equation}

donde $m_{F336W} - C(t) - A_{F336W} $  es la magnitud aparente del c'umulo 
corregida por extinci'on y referida a edad cero. El factor de correcci'on por edad $C(t)$
tiene en cuenta el debilitamiento en el brillo del c'umulo con el transcurso del tiempo.

$m_{F336W} - C(t) - A_{F336W} $   forma parte de los archivos de  salida 
de {\sc\footnotesize  STATPLOTS}.  
Para realizar este c'alculo empleamos la distancia $d=2.94 \pm 0.18 $ Mpc dada por
 \cite{Maizetal02a} para NGC~4214.

Los c'alculos en \cho\  est'an realizados en base a los resultados de 
\s99\  donde se define a la edad cero como 1~Ma y donde se emplea
una funci'on incial de masa de Salpeter   \citep{Salp55}  
$\frac{dN}{dm} = f(m) = A \cdot m^{-2.35}$
 con $M_{low} =1 \, M_{\odot} $   y    $M_{up} =100 \, M_{\odot}$.
Este rango de masas es apropiado para la  determinaci'on 
de edades de los c'umulos, de sus  magnitudes  y de su extinci'on entre otros
par'ametros, pero no es el adecuado para la estimaci'on de la masa ya que
la mayor parte de la masa de un c'umulo estar'ia contenida en su poblaci'on
de estrellas de menor masa.
Esto significa que usando los resultados directos 
de \s99\ obtendr'iamos una subestimaci'on de la masa real de un c'umulo.
Por otro lado, una extrapolaci'on hacia  $M_{low} =0.1 \, M_{\odot} $
producir'ia una sobreestimaci'on de la masa  \citep{Leit98,Leitetal99}. 
Una situaci'on de compromiso se alcanza   utilizando la funci'on 
inicial de masa de Kroupa \citep{Krou02}.

A los efectos de estimar la masa m'as apropiada para cada c'umulo 
estudiado, usamos    dos expresiones de la  IMF:

\begin{description}
\item[IMF Salpeter] $ \frac{dN}{dm} =  f(m)=   A \cdot m^{-2.35}    $  
\item[IMF Kroupa ]  $ \frac{dN}{dm} = f(m)=A \cdot \left\{ \begin{array}{lr}
m^{-1.3}  &   0.1  \le m/ M_{\odot} \le 0.5 \\
m^{-2.35}  &   0.5  \le m/M_{\odot} \le 100 
\end{array} \right.    $
  
\end{description}

Estas leyes nos permitieron transformar a las masas de los c'umulos calculadas
con una IMF  con $M_{low} =1\,  M_{\odot} $   y    $M_{up} =100 \,  M_{\odot}$ ( $ M_{1 \rightarrow 100} $)
en masas calculadas con una IMF  con $M_{low} =0.1 \, M_{\odot} $   y    $M_{up} =100 \,  M_{\odot}$ 
( $ M_{0.1 \rightarrow 100} $).
Los factores de transformaci'on  \citep{MilleScal79,Krou02} usados en cada caso son:

\begin{description}
\item[IMF Salpeter]  $ M_{0.1 \rightarrow 100} = M_{1 \rightarrow 100} \times  \frac{\int^{100}_{0.1} m \, m^{-2.35}dm}{\int^{100}_{1} m \,  m^{-2.35} dm} = M_{1 \rightarrow 100} \times  2.547$
\item[IMF Kroupa ] $  M_{0.1 \rightarrow 100} = M_{1 \rightarrow 100} \times    \frac{\int^{0.5}_{0.1} m \, m^{-1.3}dm + \int^{100}_{0.5} m \, m^{-2.35}dm }{\int^{100}_{1} m \,  m^{-2.35} dm}=M_{1 \rightarrow 100} \times 1.603$
\end{description}

La Tabla~\ref{tbl032} presenta los valores calculados de las masas de los  trece c'umulos
estudiados, empleando tres  funciones  de masa:  Salpeter en los rangos   $1-100\, M_{\odot}$ y
$0.1-100\, M_{\odot}$ y Kroupa en el rango $0.1-100\, M_{\odot}$. Las masas y sus errores 
est'an expresados en  unidades de  $10^3 \, M_{\odot}$.

\begin{table}[htbp]
\begin{tabular}[h]{lcccc} 
\multicolumn{5}{l}{ \rule{95mm}{0.8mm}}  \\
 C'umulo & Salpeter    & Salpeter   & Kroupa    \\
                &  $1-100\, M_{\odot}$  &  $0.1-100\, M_{\odot}$ &   $0.1-100\, M_{\odot}$  \\
 \multicolumn{5}{l}{ \rule[2mm]{95mm}{0.2mm}}  \\
 I--As LMC2 \dotfill& $17\pm3 $ & $43 \pm 7$ &$ 27 \pm 4$ \\
I--As SMC \dotfill & $17\pm3 $ & $43 \pm 7$ &$ 27 \pm 4$ \\
 I--Es joven \dotfill & $4\pm1 $ & $10 \pm 1$ &$ 6 \pm 1$ \\
I--Es  viejo \dotfill& $80\pm16$ & $205 \pm 41$ &$ 129 \pm 26$ \\
 IIIs joven  \dotfill& $83\pm11 $ & $213 \pm 29$ &$ 134 \pm 18$ \\
IIIs  viejo \dotfill& $391\pm109$ & $996 \pm 278$ &$ 626 \pm 175$ \\
 IVs joven \dotfill & $17\pm2 $ & $44 \pm 6$ &$ 27 \pm 4$ \\
IVs viejo \dotfill& $71\pm12$ & $181 \pm 31$ &$ 114\pm 20$ \\
II--A \dotfill& $22\pm15$ & $55 \pm 37$ &$ 34 \pm 23$ \\
II--B \dotfill& $27\pm7$ & $69 \pm 17$ &$ 43 \pm 10$ \\
II--C \dotfill& $40\pm9$ & $101 \pm 22$ &$ 63 \pm 14$ \\
II--D \dotfill & $5\pm3$ & $11 \pm 8$ &$ 7 \pm 5$ \\
II--E \dotfill & $5\pm2$ & $12 \pm 5$ &$ 7 \pm 3$ \\
I--Ds \dotfill& $13\pm6$ & $31 \pm 15$ &$ 20 \pm 9$ \\
I--A \dotfill& $98\pm12$ & $248 \pm 30$ &$ 156 \pm 19$ \\
I--B \dotfill& $22\pm3$ & $55 \pm 7$ &$ 34 \pm 4$ \\
II \dotfill& $576\pm202$ & $1468 \pm 526$ &$ 923 \pm 331$ \\
\multicolumn{5}{l}{ \rule{95mm}{0.8mm}}  \\
 \end{tabular}
 
 \begin{minipage}[u]{92mm}
\caption[Masas de los c'umulos]{Estimaci'on de las masas de los c'umulos
estudiados  expresadas en $10^3 \, M_{\odot}$. Se usaron tres IMF para incluir a 
las estrellas de m'as baja masa que son las que contribuyen principalmente
a la masa total de cada c'umulo.}
\label{tbl032}
 \end{minipage}
\end{table}
 

\renewcommand{\theequation}{C-\arabic{equation}}
\setcounter{equation}{0}  

\addtocontents{toc}{\protect\vspace{-0.125in}}
\label{appendixc}

\chapter{Acr'onimos y t'erminos en ingl'es}
\thispagestyle{empty}
\newpage

 

\begin{table}[htbp]
\begin{tabular}[h]{ll} 
\multicolumn{2}{l}{ \rule{150mm}{0.8mm}}  \\
 Acr'onimo & Descripci'on    \\
 \multicolumn{2}{l}{ \rule[2mm]{150mm}{0.2mm}}  \\
 
  2MASS & 2 Micron All-Sky Survey\\ 
ACS & Advanced Camera for Surveys \\
 ADU & Analog-to-Digital Units  \\
 AGB &  Asymptotic Giant Branch (Rama Asint'otica de las Gigantes)\\
AURA &  Association of Universities for Research in Astronomy\\
CCD & Charge-Coupled Device \\
  CHORIZOS &  CHi--square cOde for parameterRized modelling  \\
   & and characterIZation of   phOtometry and Spectrophotmetry \\

COS & Cosmic Origins Spectrograph \\
COSTAR &  Corrective Optics Space Telescope Axial Replacement   \\
CTE &  Charge Transfer Efficiency \\
DENIS & Deep Near Infrared Survey  of the Southern Sky \\
DIRBE & Diffuse Infrared Background Experiment  \\
DN&  Data Number (salida de un convertidor A/D) \\
DSS & Digitized Sky Survey \\

ESA & European Space Agency \\
ESO & European    Southern Observatory\\
FGS & Fine Guidance  Sensors\\
 FITS & Flexible Image Transport System \\
FOC & Faint Object Camera \\
FOS & Faint Object Spectrograph\\
FOV& Field of View \\
FUV & Far UltraViolet  (Ultra Violeta Lejano) \\ 

FWHM & Full Width at Half Maximum  \\
GEIS&Generic Edited Information Set \\
 GHRS & Goddard High Resolution Spectrograph\\
GSFC &Goddard Space Flight Center \\

 GMC& Giant Molecular Cloud  (Nube Molecular Gigante)\\
GSC2 & The Guide Star Catalog Version 2   \\

HD & Henry Draper \\
 HST &Hubble Space Telescope (Telescopio Espacial Hubble) \\
 IC & Index Catalogue \\
 IDL & Interactive Data Language \\
 IMF & Initial Mass Function (Funci'on Inicial de Masa) \\
 
 IPAC & Infrared Processing and Analysis Center \\ 
 
IR & Infrared (Infrarrojo)\\
IRAF & Image Reduction and Analysis Facility  \\

IRAS &Infrared Astronomical Satellite  \\
 ISM &  InsterStellar Medium  (Medio InterEstelar)\\

ISR &Instrument Science Report \\
IUE &  International Ultraviolet Explorer \\
JPL & Jet Propulsion Laboratory \\
 JWST & James Webb Space Telescope \\

KPNO & Kitt Peak National Observatory \\

\multicolumn{2}{l}{ \rule{150mm}{0.8mm}}  \\
 \end{tabular}
 
\end{table}

\begin{table}[htbp]
\begin{tabular}[h]{ll} 
\multicolumn{2}{l}{ \rule{150mm}{0.8mm}}  \\
 Acr'onimo & Descripci'on    \\
 \multicolumn{2}{l}{ \rule[2mm]{150mm}{0.2mm}}  \\

 LBV & Luminous Blue Variable \\
 
  LMC  &   Large Magellanic Cloud (Nube Mayor de Magallanes) \\
LTE & Local Thermodynamic Equilibrium (Equilibrio Termodin'amico Local) \\
MAMA & Multianode Microchannel Array   \\

MYC & Massive Young  Cluster  (C'umulo Masivo Joven)\\
NASA & National Aeronautics and Space Administration \\
NGC & New General Catalog \\
 NICMOS &Near-IR Camera and Multi-Object Spectrograph \\
 
 NOAO & National Optical Astronomy Observatory  \\
 NRAO & National Radio Astronomy Observatory \\
 NSF & National Science Foundation \\
NTT & New Technology Telescope\\
NUV & Near UltraViolet  (Ultra Violeta Cercano) \\ 
OTA &Optical Telescope Assembly \\

 OTFR & On--the--Fly Reprocessing \\
 OVRO  & Owens Valley Radio Observatory  \\

PC &Planetary Camera \\
PI & Principal Investigator (Investigador Principal) \\ 
PODPS &Post Observation Data Processing System  \\
PSF & Point Spread Function \\

QE & Quantum Efficiency (Eficiencia Cu'antica) \\

RSG & Red Super Giant  (Supergigante Roja) \\
SED & Spectral Energy Distribution \\
SFH  & Star Formation History (Historia de Formaci'on Estelar) \\

  SFR &  Star Formation Rate (Tasa de Formaci'on de Estrellas)\\

 SNR & Signal-to-Noise Ratio (Se'nal Ruido)\\
   SMC  &  Small Magellanic Cloud (Nube Menor de Magallanes) \\

 SOBA & Scaled OB Association  (Asociaci'on OB a Gran Escala)\\

 SOFA & Selectable Optical  Filter Assembly \\
SM & Service Mission \\
SSC & Super Star Cluster (Superc'umulo Estelar)\\

STIS& Space Telescope Imaging Spectrograph \\

STOCC  & Space Telescope Operations Control Center   \\
STS &Space Transportation System \\
STScI& Space Telescope Science Institute \\
STSDAS  & Space Telescope Science Data Analysis System \\
TRGB& Tip of the Red Giant Branch \\

UCLA & University of California Los Angeles \\
 UIT & Ultraviolet Imaging Telescope  \\
UMASS & University of Massachusetts Amherst \\

UV &Ultraviolet (Ultravioleta)\\
 VLA & Very Large Array \\

WF/PC--1& Wide Field and Planetary Camera \\
WFC &Wide Field Camera \\

\multicolumn{2}{l}{ \rule{150mm}{0.8mm}}  \\
 \end{tabular}
 
\end{table}

\begin{table}[t]
\begin{tabular}[t]{ll} 
\multicolumn{2}{l}{ \rule{150mm}{0.8mm}}  \\
 Acr'onimo & Descripci'on    \\
 \multicolumn{2}{l}{ \rule[2mm]{150mm}{0.2mm}}  \\
WFPC2& Wide Field and Planetary Camera 2 \\
WFC3& Wide Field Camera 3 \\

WIRO   & Wyoming Infrared Observatory   \\
WR & Wolf--Rayet Star \\

ZAMS & Zero Age Main Sequence (Secuencia Principal de Edad Cero)\\ 
\multicolumn{2}{l}{ \rule{150mm}{0.8mm}}  \\
 & \\
 & \\
  & \\
 & \\
 & \\
 & \\
 & \\
 & \\
 & \\
 & \\
 & \\
 & \\
 & \\
 & \\
 & \\
 & \\
 & \\
 & \\
 & \\
 & \\
 & \\
 & \\
 & \\
 & \\
 & \\
 & \\
 & \\
 & \\
 & \\
 & \\
 & \\
 & \\
 & \\
 & \\
 & \\
 & \\
 & \\
 \end{tabular}
\end{table}


\newpage
\thispagestyle{empty}

\begin{figure*}[h!]
 \centering
\includegraphics[width=\linewidth]{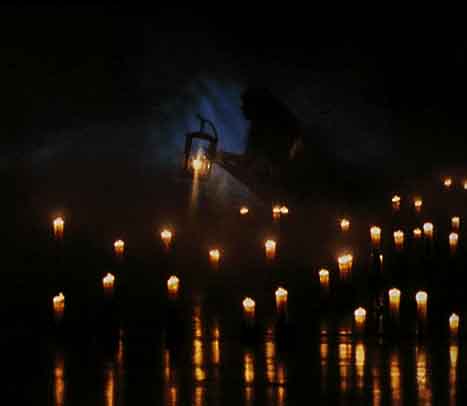}
\end{figure*}
\begin{center}
 \begin{minipage}[u]{92mm}
Sarah Brightman as  Christine  entering the Phantom's lair
in The Phantom of the Opera. All quotes are taken from the libretto 
of  The Phantom of the 
Opera.
\end{minipage}
\end{center}

\end{document}